\documentclass[12pt]{article}

\usepackage{epsfig}

\def\comp{{\rm C}\llap{\vrule height7.1pt width1pt depth-.4pt\phantom t}}
\def\Fint{\rlap{$\Biggl\rfloor$}\Biggl\lceil}
\def\p{\prime}
\def\ltwid{\mathrel{\raise.3ex\hbox{$<$\kern-.75em\lower1ex\hbox{$\sim$}}}}
\def \be{\begin{equation}}
\def \ee{\end{equation}}
\def \bea{\begin{eqnarray}}
\def \eea{\end{eqnarray}}

\def \del{\partial}
\def \a{\alpha}
\def \b{\beta}
\def \f{\frac}

\def \g{\gamma}
\def \nn{\nonumber}
\def \trid{\triangle}
\def \Lam{\Lambda}

\def \pa{\parallel}
\def \sta{\stackrel}
\def \Om{\Omega}

\def \si{\sigma}

\def \ka{\kappa}
\def \d{\delta}
\def \D{\Delta}
\def \e{\eta}

\begin{document}

\begin{titlepage}

\begin{flushright}
UFIFT-QG-07-02
\end{flushright}

\vspace{2cm}

\begin{center}
{\bf The Quantum-Corrected Fermion Mode Function during Inflation}
\end{center}

\vspace{.5cm}

\begin{center}
S. P. Miao$^{\dagger}$
\end{center}

\vspace{.5cm}

\begin{center}
\it{Department of Physics \\
University of Florida \\
Gainesville, FL 32611, USA}
\end{center}

\vspace{1cm}

\begin{center}
ABSTRACT
\end{center}
My project computed the one loop fermion self-energy for massless
Dirac $+$ Einstein in the presence of a locally de Sitter
background. I employed dimensional regularization and obtain a fully
renormalized result by absorbing all divergences with Bogliubov,
Parasiuk, Hepp and Zimmermann (BPHZ) counterterms. An interesting
technical aspect of my computation was the need for a noninvariant
counterterm, owing to the breaking of de Sitter invariance by our
gauge condition. I also solved the effective Dirac equation for
massless fermions during inflation in the simplest gauge, including
all one loop corrections from quantum gravity. At late times the
result for a spatial plane wave behaves as if the classical solution
were subjected to a time-dependent field strength renormalization of
$Z_2(t) = 1 - \frac{17}{4 \pi} G H^2 \ln(a) + O(G^2)$. I showed that
this also follows from making the Hartree approximation, although
the numerical coefficients differ.

\begin{flushleft}
$^{\dagger}$ e-mail: miao@phys.ufl.edu
\end{flushleft}

\end{titlepage}

\hbox{}

\vspace{4.25in}

\begin{center}
To my dearest aunt, Hsiu-Lian Chuang
\end{center}
\newpage

\centerline{ACKNOWLEDGEMENTS}

I am indebted to a great number of people. Without them I never
could have completed this achievement. First of all, I would like to
thank my advisor, Professor Richard Woodard. He is a very intense,
hard-working but rather patient person. Without his direction, I
could not have overcome all of the obstacles. He has a mysterious
ability to extract the best in people due to his optimism and
generous character. It is very enjoyable to work with him. I also
want to thank him for spending an enormous amount of time to correct
my horrible ``Chin-English.'' Secondly, I would like to thank
Professor Pei-Ming Ho. He was my advisor at National Taiwan
University. He motivated my interest in the fundamental physics
which I never knew I could do before. After I got my master's
degree, I was trapped in the position of administrant assistant at
National Taiwan Normal University. At that time I was too busy to
think of applying for Ph.D program. Without his encouragement and
guidance, I would never have studied abroad. In my academic career I
am an extremely lucky person to have two great physicists as my
mentors.

I would like to thank my parents, Lin-Sheng Miao and Hsiu-Chu
Chuang. They always respected my decision, especially my mother,
even though they really didn't understand what I was doing because
theoretical physics was never part of their lives. I want to thank
my two old roommates, Mei-Wen Huang and Chin-Hsin Liu, for their
selfless support throughout my Ph.D. career. I also want to thank
Dr. Robert Deserio and Charles Parks for giving me a hand through
the tough time of being a TA.

I am grateful to Professor Charles Thorn for improving me during
independent study with him, and for serving on my dissertation
committee. I gratefully acknowledge Professor Pierre Sikivie and
Professor James Fry for writing letters of recommendation on my
behalf. Finally, I would like to express my gratitude to Professor
Stanley Deser, who doesn't really know me at all, for intervening to
help me take a French course. Without this course, I would have a
hard time when I attended the general relativity advanced school in
Paris.

\newpage

\section{INTRODUCTION}
My research focussed on infer how quantum gravity affects massless
fermions at one loop order in the inflationary background geometry
which corresponds to a locally de Sitter space. In the following
sections, we will discuss what inflation is, why it enhances the
effect of quantum gravity, how one can study this enhancement and
why reliable conclusions can be reached in spite of the fact that a
completely consistent theory of quantum gravity is not yet known.

\subsection{Inflation} On the largest scales our universe is
amazingly homogeneous and isotropic. It also seems to have nearly
zero spatial curvature \cite{DN}. Based on these three features our
universe can be described by the following geometry,
\begin{equation}
ds^2=-dt^2+a^2(t)d\vec{x}\cdot d\vec{x}\;.\label{metr}
\end{equation}
The coordinate $t$ is physical time.  The function $a(t)$ is called
the scale factor. This is because it converts Euclidean coordinate
distance $\|\vec{x}-\vec{y}\|$ into physical distance
$a(t)\|\vec{x}-\vec{y}\|$.

From the scale factor we form the redshift $z(t)$, the Hubble
parameter $H(t)$ as well as the deceleration parameter $q(t)$. Their
definitions are:
\begin{equation}
z(t) \equiv \f{a_0}{a(t)}-1 \;\;\;,\;\;\; H(t)\equiv
\f{\dot{a}}{a}\;\;\;,\;\;\;
q(t)\equiv-\f{a\ddot{a}}{\dot{a}^2}=-1-\f{\dot{H}}{H^2}\;.\label{def}
\end{equation}
The Hubble parameter $H(t)$ tells us the rate at which the universe
is expanding. The deceleration parameter measures the fractional
acceleration rate ($\ddot{a}/a$) in units of Hubble parameter. The
current value of Hubble parameter is , $H_0=(71^{+4}_{-3}){\rm
\f{Km/s}{Mpc}}\simeq 2.3\times 10^{-18} {\rm Hz}$ \cite{DN}. From
the observation of Type Ia supernovae one can infer $q_{0}
\simeq-0.6$ \cite{cos}, which is consistent with a universe which is
currently about $30\%$ matter and $70\%$ vacuum energy.

Inflation is defined as accelerated expansion, that is, $q(t)<0$ as
well as $H(t)>0$. During the epoch of primordial inflation the
Hubble parameter may have been as large as $H_{I}\sim 10^{37}{\rm
Hz}$ and the deceleration parameter is thought to  have been
infinitesimally greater than $-1$. The current values of the
cosmological parameters are consistent with inflation, however, the
phenomenological interest in my calculation concerns primordial
inflation.

\subsection{Uncertainty Principle during Inflation}
To understand quantum effects during inflation it is instructive to
review the energy-time uncertainty principle,
\begin{equation}
\trid E \trid t \sta {>}{\sim} 1\; .\label{Et}
\end{equation}
Consider the process of a pair of virtual particles emerging from
the vacuum. This process can conserve 3-momentum if the particles
have $\pm\vec{k}$ but it must violate energy conservation. If the
particles have mass $m$ then each of them has energy,
\begin{equation}
E(\vec{k})=\sqrt{m^2+\pa\vec{k}\pa^2}\;.\label{Ef}
\end{equation}
The energy-time uncertainty principle restricts how long a virtual
pair of such particles with $\pm\vec{k}$ can exist. If the pair was
created at time $t$, it can last for a time $\Delta t$ given by the
inequality,
\begin{equation}
2E(\vec{k})\trid t \sta{<}\sim1\; .
\end{equation}
The lifetime of the pair is therefore
\begin{equation}
\trid t =\f{1}{2E(\vec{k})}\;.\label{Etf}
\end{equation}
One can see that in flat spacetime all particles with $\vec{k}\neq
0$ have a finite lifetime, and that massless particles live longer
than massive particles with the same $\vec{k}$.

How does this change during inflation? Because the homogeneous and
isotropic geometry shown by Equation \ref{metr} possesses spatial
translation invariance it follows that particles are still labeled
by constant wave numbers $\vec{k}$, just as in flat space. However,
because $\vec{k}$ involves an inverse length, which must be
multiplied by the scale factor $a(t)$ to give the physical length,
the physical wave number is $\vec{k}/a(t)$. Therefore the physical
energy is not Equation \ref{Ef} but rather,
\begin{equation}
E(t,\vec{k})=\sqrt{m^2+\pa\vec{k}\pa^2/a^2(t)}\;.\label{Ec}
\end{equation} The
left-hand side of the previous inequality becomes an integral:
\begin{equation}
\int^{t+\trid
t}_{t}dt^{\p}2E(t^{\p},\vec{k})\sta{<}{\sim}1\label{Etc}
\end{equation}

Obviously anything that reduces $E(t^{\p},\vec{k})$ increases $\trid
 t$. Therefore let us consider zero mass. Zero mass will simplify
the integrand in Equation \ref{Etc} to $2\|\vec{k}\|/a(t^{\p})$. If
the scale factor $a(t)$ grows fast enough, the quantity
$2\|\vec{k}\|/a(t^{\p})$ becomes so small that the integral will be
dominated by the lower limit and the inequality of Equation
\ref{Etc} can remain satisfied even though $\trid t$ goes to
infinity. Under these conditions with $m=0$ and $ a(t)=a_I e^{Ht}$,
Equation \ref{Etc} gives,
\begin{equation}
\f{2\pa\vec{k}\pa}{Ha(t)}(1-e^{-H\trid t})\sta{<}{\sim}
1\;.\label{forever}
\end{equation}
 From this discussion we conclude that massless
virtual particles can live forever during inflation
if they emerge with
$\pa\vec{k}\pa\sta{<}{\sim} Ha(t)$.

\subsection{Crucial Role of Conformal Invariance}%
One might think that the big obstacle to inflationary particle
production is nonzero mass. However, the scale of primordial
inflation is so high that a lot of particles are effectively
massless and they nevertheless experience little inflationary
production. The reason is that they possess a symmetry called
``conformal invariance.''

A simple conformally invariant theory is electromagnetism in $D=4$
spacetime dimensions. Consider $D$ dimensional electromagnetism,
\begin{equation}
{\cal{L}}_{EM}=-\f{1}{4}F_{\a\b}F_{\rho\sigma}
g^{\a\rho}g^{\b\sigma}\sqrt{-g}\;,\label{em}
\end{equation}
where $F_{\mu\nu}\equiv\del_{\mu}A_{\nu}-\del_{\nu}A_{\mu}$. Under a
conformal transformation $g^{\p}_{\mu\nu}=\Om^{2}(x)g_{\mu\nu}$ and
$A^{\p}_{\mu}= A_{\mu}$ the Lagrangian becomes,
\begin{equation}
{\cal{L^{\p}}}= F_{\a\b}F_{\rho\sigma}\Om^{-2}g^{\a\rho}
\Om^{-2}g^{\b\sigma}\Om^D\sqrt{-g}={\cal{L}}\Om^{D-4}\label{emm}
\end{equation}
Hence electromagnetism is conformally invariant in $D=4$. Other
conformally invariant theories are the massless conformally coupled
scalar,
\begin{eqnarray}
{\cal{L}}=-\f{1}{2}\del_{\mu}\phi\del_{\nu}\phi
g^{\mu\nu}\sqrt{-g}-\f{1}{8}\Bigl(\f{D-2}{D-1}\Bigr)\phi^2
R\sqrt{-g}\;.
\end{eqnarray}
and massless fermions,
\begin{eqnarray}
{\cal{L}}&=&\overline{\psi}e^{\mu}\,_{b}\g^{b}
(i\del_{\mu}-\f{1}{2}A_{\mu cd}J^{cd}) \psi\sqrt{-g}\,.\label{La}
\end{eqnarray}
Here $\phi^{\p}=\Om^{1-\f{D}{2}}\phi$ and
$\psi^{\p}=\Om^{\f{1-D}{2}}\psi$ under a conformal transformation.

If the theory possesses conformal invariance, it is much more
convenient to express the homogeneous and isotropic geometry of
Equation \ref{metr} in conformal coordinates,
\begin{eqnarray}
dt=a(t)d{\eta}\Longrightarrow ds^2&=&-dt^2+a^2(t)d\vec{x}\cdot
d\vec{x}\nn\\&=& a^2(t)(-d{\eta}^2+d\vec{x}\cdot
d\vec{x}\:)\;.\label{cfm}
\end{eqnarray}
Here $t$ is physical time and $\eta$ is conformal time. In the
$(\eta,\vec{x})$ coordinates, conformally invariant theories are
locally identical to their flat space cousins. The rate at which
virtual particles emerge from the vacuum per unit conformal time
must be the same constant --- call it $\Gamma$ --- as in flat space.
Hence the rate of emergence per unit physical time is,
\begin{equation}
\frac{dN}{dt} = \frac{dN}{d\eta} \frac{d\eta}{dt} =
\frac{\Gamma}{a(t)} \; .\label{rate}
\end{equation}
One can see that the emergence rate in a locally de Sitter
background is suppressed by a factor of $1/a$ ($a\sim
e^{Ht}\;,H>0$). Therefore any conformally invariant, massless virtual
particles with $\pa\vec{k}\pa\sta{<}{\sim}Ha(t)$ can live forever
but the problem is that they don't have much chance to emerge from
the vacuum.

\subsection{Gravitons and Massless Minimally Coupled Scalars}
Not every massless particle is
conformally invariant. Two exceptions are gravity and the massless
minimally coupled (MMC) scalar,
\begin{equation}
{\cal{L}}=\f{1}{16\pi G}(R-2\Lam)\sqrt{-g}\;,\label{graviton}
\end{equation}
\begin{equation}
{\cal{L}}=-\f{1}{2}\del_{\mu}\phi\del_{\nu}\phi
g^{\mu\nu}\sqrt{-g}\;.\label{scalar}
\end{equation}
Here $R$ is the Ricci scalar and $\Lam$ is the cosmological
constant. From previous sections one can conclude that big quantum
effects come from combining
\begin{itemize}
\item{Inflation;}
\item{Massless particles; and}
\item{The absence of invariance.}
\end{itemize}
\noindent Therefore one can conclude that gravitons and MMC scalars
have the potential to mediate vastly enhanced quantum effects during
inflation because they are simultaneously massless and not
conformally invariant.

To see that the production of gravitons and MMC scalars is not
suppressed during inflation note that each polarization and wave
number behaves like a harmonic oscillator \cite{LP,LPG},
\begin{equation}
L = \frac12 m \dot{q}^2 - \frac12 m \omega^2 q^2 \; ,
\end{equation}
with time dependent mass $m(t) = a^3(t)$ and frequency $\omega(t) =
\frac{k}{a(t)}$. The Heisenberg equation of motion can be solved in
terms of mode functions $u(t,k)$ and canonically normalized raising
and lowering operators $\alpha^{\dagger}$ and $\alpha$,
\begin{equation}
\ddot{q} + 3 H \dot{q} + \frac{k^2}{a^2} q = 0 \quad \Longrightarrow
\quad q(t) = u(t,k) \alpha + u^*(t,k) \alpha^{\dagger} \quad {\rm
with} \quad [\alpha,\alpha^{\dagger}] = 1 \; ,
\end{equation}
The mode functions $u(t,k)$ are quite complicated for a general
scale factor $a(t)$ \cite{TW12} but they take a simple form for de
Sitter,
\begin{equation}
u(t,k) = \frac{H}{\sqrt{2 k^3}} \Bigl[1 - \frac{i k}{H a(t)}\Bigr]
\exp\Bigl[\frac{ik}{Ha(t)}\Bigr] \; . \label{uform}
\end{equation}

The (co-moving) energy operator for this system is,
\begin{equation}
E(t) = \frac12 m(t) \dot{q}^2(t) + \frac12 m(t) \omega^2(t) q^2(t)
\; .
\end{equation}
Owing to the time dependent mass and frequency, there are no
stationary states for this system. At any given time the minimum
eigenstate of $E(t)$ has energy $\frac12 \omega(t)$, but which the
state changes for each value of time. The state $\vert \Omega
\rangle$ which is annihilated by $\alpha$ has minimum energy in the
distant past. The expectation value of the energy operator in this
state is,
\begin{equation}
\Bigl\langle \Omega \Bigl\vert E(t) \Bigr\vert \Omega \Bigr\rangle =
\frac12 a^3(t) \vert \dot{u}(t,k) \vert^2 + \frac12 a(t) k^2 \vert
u(t,k) \vert^2 \Bigl\vert_{\rm de\ Sitter} = \frac{k}{2a} +
\frac{H^2 a}{4k} \; .
\end{equation}
If one thinks of each particle having energy $k/a(t)$, it follows
that the number of particles with any polarization and wave number
$k$ grows as the square of the inflationary scale factor,
\begin{equation}
N(t,k) = \Bigl(\frac{H a(t)}{2 k}\Bigr)^2 \; . \label{create}
\end{equation}

Quantum field theoretic effects are driven by essentially classical
physics operating in response to the source of virtual particles
implied by quantization. On the basis of Equation \ref{create} one
might expect inflation to dramatically enhance quantum effects from
MMC scalars and gravitons, and explicit studies over a quarter
century have confirmed this. The oldest results are of course the
cosmological perturbations induced by scalar inflatons \cite{MC} and
by gravitons \cite{AAS2}. More recently it was shown that the
one-loop vacuum polarization induced by a charged MMC scalar in de
Sitter background causes super-horizon photons to behave like
massive particles in some ways \cite{PTW1,PTW2,PW3}. Another recent
result is that the one-loop fermion self-energy induced by a MMC
Yukawa scalar in de Sitter background reflects the generation of a
nonzero fermion mass \cite{PW,GP}.

\subsection{Overview}
One naturally wonders how interactions with these quanta affect
themselves and other particles. The first step in answering this
question on the linearized level is to compute the one particle
irreducible (1PI) 2-point function for the field whose behavior is
in question. This has been done at one loop order for gravitons in
pure quantum gravity \cite{TW6}, for photons \cite{PTW1,PTW2} and
charged scalars \cite{KW} in scalar quantum electrodynamics (SQED),
for fermions \cite{PW,GP} and Yukawa scalars \cite{DW} in Yukawa
theory, for fermions in Dirac + Einstein \cite{MW1} and, at two loop
order, for scalars in $\phi^4$ theory \cite{BOW}.

In the first part of my dissertation we compute and renormalize the
one loop quantum gravitational corrections to the self-energy of
massless fermions in a locally de Sitter background. The physical
motivation for this exercise is to check for graviton analogues of
the enhanced quantum effects seen in this background for
interactions which involve one or more undifferentiated, massless,
minimally coupled (MMC) scalars. Those effects are driven by the
fact that inflation tends to rip virtual, long wavelength scalars
out of the vacuum and thereby lengthens the time during which they
can interact with themselves or other particles. Gravitons possess
the same crucial property of masslessness without classical
conformal invariance that is responsible for the inflationary
production of MMC scalars. One might therefore expect a
corresponding strengthening of quantum gravitational effects during
inflation.

Of particular interest to us is what happens when a MMC scalar is
Yukawa coupled to a massless Dirac fermion for non-dynamical
gravity. The one loop fermion self-energy has been computed for this
model and used to solve the quantum-corrected Dirac equation
\cite{PW},
\begin{equation}
\sqrt{-g} \, i\hspace{-.1cm}\not{\hspace{-.15cm} \mathcal{D}}_{ij}
\psi_{j}(x) - \int d^4x^{\p} \, \Bigl[\mbox{}_i \Sigma_j
\Bigr](x;x^{\p}) \, \psi_{j}(x^{\p}) = 0 \; . \label{Diraceq}
\end{equation}
Powers of the inflationary scale factor $a = e^{Ht}$ play a crucial
role in understanding this equation for the Yukawa model and also
for what we expect from quantum gravity. The Yukawa result for the
self-energy \cite{PW} consists of terms which were originally
ultraviolet divergent and which end up, after renormalization,
carrying the same number of scale factors as the classical term. Had
the scalar been conformally coupled these would be the only
contributions to the one loop self-energy. However, minimally
coupled scalars also give contributions due to inflationary particle
production. These are ultraviolet finite from the beginning and
possesses an extra factor of $a \ln(a)$ relative to the classical
term. Higher loops can bring more factors of $\ln(a)$, but no more
powers of $a$, so it is consistent to solve the equation with only
the one loop corrections. The result is a drop in wave function
which is consistent with the fermion developing a mass that grows as
$\ln(a)$. A recent one loop computation of the Yukawa scalar
self-mass-squared indicates that the scalar which catalyzes this
process cannot develop a large enough mass quickly enough to inhibit
the process \cite{DW}.

Analogous graviton effects should be suppressed by the fact that the
$h_{\mu\nu} \overline{\psi} \psi$ interaction of Dirac + Einstein
carries a derivative, as opposed to the undifferentiated
$\phi\bar{\psi}\psi$ interaction of Yukawa theory. What we expect is
that the corresponding quantum gravitational self-energy will
consist of two terms. The most ultraviolet singular one will require
higher derivative counterterms and will end up, after
renormalization, possessing {\it one less} factor of $a$ than the
classical term. The less singular term due to inflationary particle
production should require only lower derivative counterterms and
will be enhanced from the classical term by a factor of $\ln(a)$.
This would give a much weaker effect than the analogous term in the
Yukawa model, but it would still be interesting. And note that any
such effect from gravitons would be universal, independent of
assumptions about the existence or couplings of unnaturally light
scalars.

The second part of my dissertation consists of  using the 1PI
2-point function to correct the linearized equation of motion from
Equation \ref{Diraceq} for the field in question. We employ the
Schwinger-Keldysh formalism to solve for the loop corrected fermion
mode function.  In the late time limit we find that the one loop
corrected, spatial plane mode functions behave as if the tree order
mode functions were simply subject to a time-dependent field
strength renormalization. The same result pertains for the Hartree
approximation in which the expectation value of the quantum Dirac
equation is taken in free graviton vacuum.

\subsection{The Issue of Nonrenormalizability}
Dirac $+$ Einstein is not perturbatively renormalizable \cite{DVN},
however, ultraviolet divergences can always be absorbed in the BPHZ
sense \cite{BP,H,Z1,Z2}. A widespread misconception exists that no
valid quantum predictions can be extracted from such an exercise.
This is false: while nonrenormalizability does preclude being able
to compute {\it everything}, that not the same thing as being able
to compute {\it nothing}. The problem with a nonrenormalizable
theory is that no physical principle fixes the finite parts of the
escalating series of BPHZ counterterms needed to absorb ultraviolet
divergences, order-by-order in perturbation theory. Hence any
prediction of the theory that can be changed by adjusting the finite
parts of these counterterms is essentially arbitrary. However, loops
of massless particles make nonlocal contributions to the effective
action that can never be affected by local counterterms. These
nonlocal contributions typically dominate the infrared. Further,
they cannot be affected by whatever modification of ultraviolet
physics ultimately results in a completely consistent formalism. As
long as the eventual fix introduces no new massless particles, and
does not disturb the low energy couplings of the existing ones, the
far infrared predictions of a BPHZ-renormalized quantum theory will
agree with those of its fully consistent descendant.

It is worthwhile to review the vast body of distinguished work that
has exploited this fact. The oldest example is the solution of the
infrared problem in quantum electrodynamics by Bloch and Nordsieck
\cite{BN}, long before that theory's renormalizability was
suspected. Weinberg \cite{SW} was able to achieve a similar
resolution for quantum gravity with zero cosmological constant. The
same principle was at work in the Fermi theory computation of the
long range force due to loops of massless neutrinos by Feinberg and
Sucher \cite{FS,HS}. Matter which is not supersymmetric generates
nonrenormalizable corrections to the graviton propagator at one
loop, but this did not prevent the computation of photon, massless
neutrino and massless, conformally coupled scalar loop corrections
to the long range gravitational force \cite{CDH,CD,DMC1,DL}. More
recently, Donoghue \cite{JFD1,JFD2} has touched off a minor industry
\cite{MV,HL,ABS,KK1,KK2} by applying the principles of low energy
effective field theory to compute graviton corrections to the long
range gravitational force. Our analysis exploits the power of low
energy effective field theory in the same way, differing from the
previous examples only in the detail that our background geometry is
locally de Sitter rather than flat.\footnote{For another recent
example in a nontrivial cosmology see D. Espriu, T. Multam\"aki and
E. C. Vagenas, Phys. Lett. B628 (2005) 197, gr-qc/0503033.}

\section{FEYNMAN RULES}
When the geometry is Minkowski, we work in momentum space because of
spacetime translation invariance. This symmetry is broken in de
Sitter background so propagators and vertices are no longer simple
in momentum space. Therefore we require Feynman rules in position
space. We start from the general Dirac Lagrangian which is
conformally invariant. We exploit this by conformally rescaling the
fields to obtain simple expressions for the fermion propagator and
the vertex operators. However, there are several subtleties for the
graviton propagator. First of all, the Einstein theory is not
conformally invariant. Secondly, there is a poorly understood
obstacle to adding a de Sitter invariant gauge-fixing term to the
action. We avoid this by adding a gauge-fixing term which breaks de
Sitter invariance. That gives correct physics but it leads to the
third problem, which is the possibility of noninvariant
counterterms. Fortunately, only one of these occurs.

\subsection{Fermions in Quantum Gravity}

The coupling of gravity to particles with half integer spin is
usually accomplished by shifting the fundamental gravitational field
variable from the metric $g_{\mu\nu}(x)$ to the vierbein $e_{\mu
m}(x)$.\footnote{ For another approach see H. A. Weldon, Phys. Rev.
D63 (2001) 104010,      gr-qc/0009086.} Greek letters stand for
coordinate indices and Latin letters denote Lorentz indices, and
both kinds of indices take values in the set
$\{0,1,2,\dots,(D\!-\!1)\}$. One recovers the metric by contracting
two vierbeins into the Lorentz metric $\eta^{bc}$,
\begin{equation}
g_{\mu\nu}(x) = e_{\mu b}(x) e_{\nu c}(x) \eta^{bc} \; .
\end{equation}
The coordinate index is raised and lowered with the metric
($e^{\mu}\,_{b} = g^{\mu\nu} e_{\nu b}$), while the Lorentz index is
raised and lowered with the Lorentz metric ($e_{\mu}\,^{b} =
\eta^{bc} e_{\mu c}$). We employ the usual metric-compatible and
vierbein-compatible connections,
\begin{eqnarray}
g_{\rho\sigma ; \mu} = 0 & \Longrightarrow & \Gamma^{\rho}_{~\mu\nu}
= \frac12 g^{\rho\sigma} \Bigl(g_{\sigma \mu , \nu} + g_{\nu \sigma
, \mu}
- g_{\mu\nu , \sigma}\Bigr) \; , \\
e_{\beta b ; \mu} = 0 & \Longrightarrow & A_{\mu c d} = e^{\nu}_{~c}
\Bigl( e_{\nu d, \mu} - \Gamma^{\rho}_{~\mu\nu} e_{\rho d}\Bigr) \;
. \label{spin}
\end{eqnarray}

Fermions also require gamma matrices, $\gamma^b_{ij}$. The
anti-commutation relations,
\begin{equation}
\Bigl\{\gamma^b , \gamma^c\Bigr\} \equiv \Bigl(\gamma^b \gamma^c +
\gamma^c \gamma^b\Bigr) = -2 \eta^{bc} I \; ,
\end{equation}
imply that only fully anti-symmetric products of gamma matrices are
actually independent. The Dirac Lorentz representation matrices are
such an anti-symmetric product,
\begin{equation}
J^{bc} \equiv \frac{i}4 \Bigl(\gamma^b \gamma^c - \gamma^c
\gamma^b\Bigr) \equiv \frac{i}2 \gamma^{[b} \gamma^{c]} \; .
\end{equation}
They can be combined with the spin connection of Equation \ref{spin}
to form the Dirac covariant derivative operator,
\begin{equation}
\mathcal{D}_{\mu} \equiv \partial_{\mu} + \frac{i}2 A_{\mu cd}
J^{cd} \; .
\end{equation}
Other identities we shall often employ involve anti-symmetric
products,
\begin{eqnarray}
\gamma^b \gamma^c \gamma^d & = & \gamma^{[b} \gamma^c \gamma^{d]} -
\eta^{bc}
\gamma^d + \eta^{db} \gamma^c - \eta^{cd} \gamma^b \; , \\
\gamma^b J^{cd} & = & \frac{i}2 \gamma^{[b} \gamma^c \gamma^{d]} +
\frac{i}2 \eta^{bd} \gamma^c - \frac{i}2 \eta^{bc} \gamma^d \; .
\label{Jred}
\end{eqnarray}
We shall also encounter cases in which one gamma matrix is
contracted into another through some other combination of gamma
matrices,
\begin{eqnarray}
\gamma^b \gamma_b & = & -D I \; , \\
\gamma^b \gamma^c \gamma_b & = & (D\!-\!2) \gamma^c \; , \\
\gamma^b \gamma^c \gamma^d \gamma_b & = & 4 \eta^{cd} I -(D\!-\!4)
\gamma^c
\gamma^d \; , \\
\gamma^b \gamma^c \gamma^d \gamma^e \gamma_b & = & 2 \gamma^e
\gamma^d \gamma^c + (D\!-\!4) \gamma^c \gamma^d \gamma^e \; .
\end{eqnarray}

The Lagrangian of massless fermions is,
\begin{equation}
\mathcal{L}_{\rm Dirac} \equiv \overline{\psi} e^{\mu}_{~b}
\gamma^{b} i \mathcal{D}_{\mu} \psi \sqrt{-g} \; . \label{Dirac}
\end{equation}
Because our locally de Sitter background is conformally flat it is
useful to rescale the vierbein by an arbitrary function of spacetime
$a(x)$,
\begin{equation}
e_{\beta b} \equiv a \, \widetilde{e}_{\beta b} \qquad
\Longrightarrow \qquad e^{\beta b} = a^{-1} \, \widetilde{e}^{\beta
b} \; .
\end{equation}
Of course this implies a rescaled metric $\widetilde{g}_{\mu\nu}$,
\begin{equation}
g_{\mu\nu} = a^2 \, \widetilde{g}_{\mu\nu} \qquad \Longrightarrow
\qquad g^{\mu\nu} = a^{-2} \, \widetilde{g}^{\mu\nu} \;.
\label{confg}
\end{equation}
The old connections can be expressed as follows in terms of the ones
formed from the rescaled fields,
\begin{eqnarray}
\Gamma^{\rho}_{~\mu\nu} & = & a^{-1} \Bigl(\delta^{\rho}_{~\mu} \,
a_{,\nu} \!+\!  \delta^{\rho}_{~\nu} \, a_{,\mu} \!-\!
\widetilde{g}^{\rho\sigma} \, a_{,\sigma} \,
\widetilde{g}_{\mu\nu}\Bigr) + \widetilde{\Gamma}^{\rho}_{
~\mu\nu} \label{confG} \; \\
A_{\mu cd} & = &-a^{-1} \Bigl(\widetilde{e}^{\nu}_{~c} \,
\widetilde{e}_{\mu d} \!-\!  \widetilde{e}^{\nu}_{~d} \,
\widetilde{e}_{\mu c} \Bigr) a_{,\nu} + \widetilde{A}_{\mu cd} \; .
\end{eqnarray}
We define rescaled fermion fields as follows,
\begin{equation}
\Psi \equiv a^{\frac{D-1}2} \psi \qquad {\rm and} \qquad
\overline{\Psi} \equiv a^{\frac{D-1}2} \overline{\psi} \; .
\end{equation}
The utility of these definitions stems from the conformal invariance
of the Dirac Lagrangian,
\begin{equation}
\mathcal{L}_{\rm Dirac} = \overline{\Psi} \,
\widetilde{e}^{\mu}_{~b} \, \gamma^b \, i
\widetilde{\mathcal{D}}_{\mu} \Psi \sqrt{-\widetilde{g}} \; ,
\label{Diract}
\end{equation}
where $\widetilde{\mathcal{D}}_{\mu} \equiv \partial_{\mu} \!+\!
\frac{i}2 \widetilde{A}_{\mu cd} J^{cd}$.

One could follow early computations about flat space background
\cite{BG1,BG2} in defining the graviton field as a first order
perturbation of the (conformally rescaled) vierbein. However, so
much of gravity involves the vierbein only through the metric that
it is simpler to instead take the graviton field to be a first order
perturbation of the conformally rescaled metric,
\begin{equation}
\widetilde{g}_{\mu\nu} \equiv \eta_{\mu\nu} + \kappa h_{\mu\nu}
\qquad {\rm with} \qquad \kappa^2 = 16 \pi G \; .
\end{equation}
We then impose symmetric gauge ($e_{\beta b} = e_{b \beta}$) to fix
the local Lorentz gauge freedom, and solve for the vierbein in terms
of the graviton,
\begin{equation}
\widetilde{e}[\widetilde{g}]_{\beta b} \equiv
\Bigl(\sqrt{\widetilde{g} \eta^{-1}} \, \Bigr)_{\!\beta}^{~\gamma}
\, \eta_{\gamma b} = \eta_{\beta b} + \frac12 \kappa h_{\beta b} -
\frac18 \kappa^2 h_{\beta}^{~\gamma} h_{\gamma b} + \dots
\end{equation}
It can be shown that the local Lorentz ghosts decouple in this gauge
and one can treat the model, at least perturbatively, as if the
fundamental variable were the metric and the only symmetry were
diffeomorphism invariance \cite{RPW1}. At this stage there is no
more point in distinguishing between Latin letters for local Lorentz
indices and Greek letters for vector indices. Other conventions are
that graviton indices are raised and lowered with the Lorentz metric
($h^{\mu}_{~\nu} \equiv \eta^{\mu\rho} h_{\rho\nu}$, $h^{\mu\nu}
\equiv \eta^{\mu\rho} \eta^{\nu\sigma} h_{\rho\sigma}$) and that the
trace of the graviton field is $h \equiv \eta^{\mu\nu} h_{\mu\nu}$.
We also employ the usual Dirac ``slash'' notation,
\begin{equation}
\hspace{-.1cm}\not{\hspace{-.05cm} V}_{ij} \equiv V_{\mu}
\gamma^{\mu}_{ij} \; .
\end{equation}

It is straightforward to expand all familiar operators in powers of
the graviton field,
\begin{eqnarray}
\widetilde{e}^{\mu}_{~b} & = & \delta^{\mu}_{~b} - \frac12 \kappa
h^{\mu}_{~b}
+ \frac38 \kappa^2 h^{\mu\rho} h_{\rho b} + \dots \; , \\
\widetilde{g}^{\mu\nu} & = & \eta^{\mu\nu} - \kappa h^{\mu\nu} +
\kappa^2
h^{\mu}_{~\rho} h^{\rho\nu} - \dots \; , \\
\sqrt{-\widetilde{g}} & = & 1 + \frac12 \kappa h + \frac18 \kappa^2
h^2 - \frac14 \kappa^2 h^{\rho\sigma} h_{\rho\sigma} + \dots
\end{eqnarray}
Applying these identities to the conformally rescaled Dirac
Lagrangian gives,
\begin{eqnarray}
\lefteqn{\mathcal{L}_{\rm Dirac} = \overline{\Psi} i
\hspace{-.1cm}\not{ \hspace{-.1cm} \partial} \Psi + \frac{\kappa}2
\Biggl\{h \overline{\Psi} i \hspace{-.1cm}\not{\hspace{-.1cm}
\partial} \Psi \!-\! h^{\mu\nu} \overline{\Psi} \gamma_{\mu} i
\partial_{\nu} \Psi \!-\! h_{\mu\rho , \sigma} \overline{\Psi}
\gamma^{\mu} J^{\rho\sigma} \Psi
\Biggr\} } \nonumber \\
& & + \kappa^2 \Biggl\{ \Bigl[\frac18 h^2 \!-\! \frac14
h^{\rho\sigma} h_{\rho\sigma}\Bigr] \overline{\Psi} i
\hspace{-.1cm}\not{\hspace{-.1cm}
\partial} \Psi \!+\! \Bigl[-\frac14 h h^{\mu\nu} \!+\! \frac38 h^{\mu\rho}
h_{\rho}^{~\nu}\Bigr] \overline{\Psi} \gamma_{\mu} i \partial_{\nu}
\Psi
+ \Biggl[-\frac14 h h_{\mu \rho , \sigma} \nonumber \\
& & \hspace{1.5cm} + \frac18 h^{\nu}_{~\rho} h_{\nu \sigma , \mu} +
\frac14 (h^{\nu}_{~\mu} h_{\nu\rho})_{,\sigma} \!+\! \frac14
h^{\nu}_{~ \sigma} h_{\mu\rho ,\nu}\Biggr] \overline{\Psi}
\gamma^{\mu} J^{\rho\sigma} \Psi \Biggr\} + O(\kappa^3) \; . \qquad
\label{Dexp}
\end{eqnarray}
From the first term we see that the rescaled fermion propagator is
the same as for flat space,
\begin{equation}
i\Bigl[\mbox{}_i S_j \Bigr](x;x^{\p}) =
\frac{\Gamma(\frac{D}2\!-\!1)}{4\pi^{ \frac{D}2}} \, i
\hspace{-.1cm}\not{\hspace{-.1cm} \partial}_{ij} \Big(\frac1{\Delta
x^2}\Big)^{\frac{D}2-1} , \label{fprop}
\end{equation}
where the coordinate interval is $\Delta x^2(x;x^{\p}) \equiv \Vert
\vec{x} \!-\! \vec{x}^{\p} \Vert^2 - (\vert \eta\!-\!\e^{\p}\vert -i
\delta)^2$.

We now represent the various interaction terms in Equation
\ref{Dexp} as vertex operators acting on the fields. At order
$\kappa$ the interactions involve fields, $\overline{\Psi}_i$,
$\Psi_j$ and $h_{\alpha\beta}$, which we number ``1'', ``2'' and
``3'', respectively. Each of the three interactions can be written
as some combination $V_{I ij}^{ \alpha\beta}$ of tensors, spinors
and a derivative operator acting on these fields. For example, the
first interaction is,
\begin{equation}
\frac{\kappa}2 h \overline{\Psi} i \hspace{-.1cm}\not{\hspace{-.1cm}
\partial} \Psi = \frac{\kappa}2 \eta^{\alpha \beta} i
\hspace{-.1cm}\not{\hspace{-.1cm}
\partial}_{2 ij} \times \overline{\Psi}_i \Psi_j h_{\alpha\beta} \equiv
V_{1ij}^{\alpha\beta} \times \overline{\Psi}_i \Psi_j
h_{\alpha\beta} \; .
\end{equation}
Hence the 3-point vertex operators are,
\begin{equation}
V_{1ij}^{\alpha\beta} = \frac{\kappa}2 \eta^{\alpha \beta} i
\hspace{-.1cm} \not{\hspace{-.1cm} \partial}_{2 ij} \quad , \quad
V_{2ij}^{\alpha\beta} = -\frac{\kappa}2 \gamma^{(\alpha}_{ij}
i\partial_2^{\beta)} \quad , \quad V_{3ij}^{\alpha\beta} =
-\frac{\kappa}2 \Bigl(\gamma^{(\alpha} J^{\beta)\mu} \Bigr)_{ij}
\partial_{3 \mu} \; . \label{3VO}
\end{equation}
The order $\kappa^2$ interactions define 4-point vertex operators
$U_{I ij}^{ \alpha\beta\rho\sigma}$ similarly, for example,
\begin{equation}
\frac18 \kappa^2 h^2 \overline{\Psi}
i\hspace{-.1cm}\not{\hspace{-.1cm}
\partial} \Psi = \frac18 \kappa^2 \eta^{\alpha \beta} \eta^{\rho\sigma} i
\hspace{-.1cm} \not{\hspace{-.1cm} \partial}_{2 ij} \times
\overline{\Psi}_i \Psi_j h_{\alpha\beta} h_{\rho\sigma} \equiv
U_{1ij}^{\alpha\beta\rho\sigma} \times \overline{\Psi}_i \Psi_j
h_{\alpha\beta} h_{\rho\sigma} \; .
\end{equation}
The eight 4-point vertex operators are given in Table \ref{v4ops}.
Note that we do not bother to symmetrize upon the identical graviton
fields.

\begin{table}

\caption{Vertex operators $U_{I ij}^{\alpha\beta\rho\sigma}$
contracted into $\overline{\Psi}_i \Psi_j h_{\alpha\beta}
h_{\rho\sigma}$.}

\vbox{\tabskip=0pt \offinterlineskip
\def\tablerule{\noalign{\hrule}}
\halign to390pt {\strut#& \vrule#\tabskip=1em plus2em& \hfil#&
\vrule#& \hfil#\hfil& \vrule#& \hfil#& \vrule#& \hfil#\hfil&
\vrule#\tabskip=0pt\cr \tablerule
\omit&height4pt&\omit&&\omit&&\omit&&\omit&\cr &&\omit\hidewidth \#
&&\omit\hidewidth {\rm Vertex Operator}\hidewidth&& \omit\hidewidth
\#\hidewidth&& \omit\hidewidth {\rm Vertex Operator} \hidewidth&\cr
\omit&height4pt&\omit&&\omit&&\omit&&\omit&\cr \tablerule
\omit&height2pt&\omit&&\omit&&\omit&&\omit&\cr && 1 && $\frac18
\kappa^2 \eta^{\alpha\beta} \eta^{\rho\sigma} i \hspace{-.1cm}
\not{\hspace{-.1cm} \partial}_{2 ij}$ && 5 && $-\frac14 \kappa^2
\eta^{\alpha\beta} (\gamma^{\rho} J^{\sigma\mu})_{ij}
\partial_{4\mu}$ &\cr \omit&height2pt&\omit&&\omit&&\omit&&\omit&\cr
\tablerule \omit&height2pt&\omit&&\omit&&\omit&&\omit&\cr && 2 &&
$-\frac14 \kappa^2 \eta^{\alpha\rho} \eta^{\sigma\beta} i
\hspace{-.1cm} \not{\hspace{-.1cm} \partial}_{2 ij}$ && 6 &&
$\frac18 \kappa^2 \eta^{\alpha\rho} (\gamma^{\mu}
J^{\beta\sigma})_{ij}
\partial_{4\mu}$ &\cr
\omit&height2pt&\omit&&\omit&&\omit&&\omit&\cr \tablerule
\omit&height2pt&\omit&&\omit&&\omit&&\omit&\cr && 3 && $-\frac14
\kappa^2 \eta^{\alpha\beta}\gamma^{\rho}_{ij} i\partial^{\sigma}_2$
&& 7 && $\frac14 \kappa^2 \eta^{\alpha\rho} (\gamma^{\beta}
J^{\sigma\mu})_{ij} (\partial_3 + \partial_4)_{\mu}$ &\cr
\omit&height2pt&\omit&&\omit&&\omit&&\omit&\cr \tablerule
\omit&height2pt&\omit&&\omit&&\omit&&\omit&\cr && 4 && $\frac38
\kappa^2 \eta^{\alpha\rho} \gamma^{\beta}_{ij} i\partial^{\sigma}_2$
&& 8 && $\frac14 \kappa^2 (\gamma^{\rho} J^{\sigma\alpha})_{ij}
\partial_4^{\beta}$ &\cr
\omit&height2pt&\omit&&\omit&&\omit&&\omit&\cr \tablerule}}

\label{v4ops}

\end{table}

%\begin{figure}
 %\centering
  %\includegraphics[width=3.0in, height=2.5in in]{test.ps}
 %\caption{Polarization plots for corrosion and hydrogen
  %evolution. The dashed lines indicate the corrosion potential \Vcorrs and the corrosion current
  %$i_{\rm{corr}}$.}
   %\label{fig:test}
%\end{figure}

%\begin{uflistb}

%\bitem \url{http://sea.am.ub.es/Latex/ltx-2.html}\\

%\bitem \url{http://www.staff.uni-mainz.de/pseelig/latex/}\\

%\bitem \url{http://makingtexwork.sourceforge.net/mtw/}\\

%\end{uflistb}

\subsection {The Graviton Propagator}
The gravitational Lagrangian of low energy effective field theory
is,
\begin{equation}
\mathcal{L}_{\rm Einstein} \equiv \frac1{16\pi G} \Bigl( R -
(D\!-\!2) \Lambda\Bigr) \sqrt{-g} \; . \label{Einstein}
\end{equation}
The symbols $G$ and $\Lambda$ stand for Newton's constant and the
cosmological constant, respectively. The unfamiliar factor of
$D\!-\!2$ multiplying $\Lambda$ makes the pure gravity field
equations imply $R_{\mu\nu} = \Lambda g_{\mu\nu}$ in any dimension.
The symbol $R$ stands for the Ricci scalar where our metric is
spacelike and our curvature convention is,
\begin{equation}
R \equiv g^{\mu\nu} R_{\mu\nu} \equiv g^{\mu\nu}
\Bigl(\Gamma^{\rho}_{~\nu\mu , \rho} - \Gamma^{\rho}_{~\rho \mu ,
\nu} + \Gamma^{\rho}_{~\rho \sigma} \Gamma^{\sigma}_{~\nu\mu} -
\Gamma^{\rho}_{~\nu\sigma} \Gamma^{\sigma}_{~\rho \mu} \Bigr)  .
\end{equation}
Unlike massless fermions, gravity is not conformally invariant.
However, it is still useful to express it in terms of the rescaled
metric of Equation \ref{confg} and connection of Equation
\ref{confG},
\begin{eqnarray}
\lefteqn{\mathcal{L}_{\rm Einstein} = \frac1{16 \pi G} \Biggl\{
a^{D-2} \widetilde{R} \!-\! 2 (D\!-\!1) a^{D-3}
\widetilde{g}^{\mu\nu} \Bigl(a_{,\mu\nu} \!-\!
\widetilde{\Gamma}^{\rho}_{~\mu\nu} a_{,\rho}\Bigr) }
\nonumber \\
& & \hspace{3cm} - (D\!-\!4) (D\!-\! 1) a^{D-4}
\widetilde{g}^{\mu\nu} a_{,\mu} a_{,\nu} \!-\! (D\!-\!2) \Lambda a^D
\Biggr\} \sqrt{-\widetilde{g}} \; . \qquad \label{Econf}
\end{eqnarray}
The factors of $a$ which complicate this expression are the ultimate
reason there is interesting physics in this model!

None of the fermionic Feynman rules depended upon the functional
form of the scale factor $a$ because the Dirac Lagrangian is
conformally invariant. However, we shall need to fix $a$ in order to
work out the graviton propagator from the Einstein Lagrangian in
Equation \ref{Econf}. The unique, maximally symmetric solution for
positive $\Lambda$ is known as de Sitter space. In order to regard
this as a paradigm for inflation we work on a portion of the full de
Sitter manifold known as the open conformal coordinate patch. The
invariant element for this is,
\begin{equation}
ds^2 = a^2 \Bigl( -d\eta^2 + d\vec{x} \!\cdot\! d\vec{x}\Bigr)
\qquad {\rm where} \qquad a(\eta) = -\frac1{H\eta} \; ,
\end{equation}
and the $D$-dimensional Hubble constant is $H \equiv
\sqrt{\Lambda/(D\!-\!1)}$. Note that the conformal time $\eta$ runs
from $-\infty$ to zero. For this choice of scale factor we can
extract a surface term from the invariant Lagrangian and write it in
the form \cite{TW1},
\begin{eqnarray}
\lefteqn{\mathcal{L}_{\rm Einstein} \!-\! {\rm Surface} =
{\scriptstyle (\frac{D}2 - 1)} H a^{D-1} \sqrt{-\widetilde{g}}
\widetilde{g}^{\rho\sigma} \widetilde{g}^{\mu \nu} h_{\rho\sigma
,\mu} h_{\nu 0} + a^{D-2} \sqrt{-\widetilde{g}}
\widetilde{g}^{\alpha\beta} \widetilde{g}^{\rho\sigma}
\widetilde{g}^{\mu\nu} } \nonumber \\
& & \hspace{2cm} \times \Bigl\{{\scriptstyle \frac12} h_{\alpha\rho
,\mu} h_{\beta\sigma ,\nu} \!-\! {\scriptstyle \frac12}
h_{\alpha\beta ,\rho} h_{\sigma\mu ,\nu} \!+\! {\scriptstyle
\frac14} h_{\alpha\beta ,\rho} h_{\mu\nu ,\sigma} \!-\!
{\scriptstyle \frac14} h_{\alpha\rho ,\mu} h_{\beta\sigma ,\nu}
\Bigr\} . \quad \label{Linv}
\end{eqnarray}

Gauge fixing is accomplished as usual by adding a gauge fixing term.
However, it turns out not to be possible to employ a de Sitter
invariant gauge for reasons that are not yet completely understood.
One can add such a gauge fixing term and then use the well-known
formalism of Allen and Jacobson \cite{AJ} to solve for a fully de
Sitter invariant propagator \cite{AT,AM,HHT,HK,HW}. However, a
curious thing happens when one uses the imaginary part of any such
propagator to infer what ought to be the retarded Green's function
of classical general relativity on a de Sitter background. The
resulting Green's function gives a divergent response for a point
mass which also fails to obey the linearized invariant Einstein
equation \cite{AM}! We stress that the various propagators really do
solve the gauge-fixed, linearized equations with a point source. It
is the physics which is wrong, not the math. There must be some
obstacle to adding a de Sitter invariant gauge fixing term in
gravity.

The problem seems to be related to combining constraint equations
with the causal structure of the de Sitter geometry. Before gauge
fixing the constraint equations are elliptic, and they typically
generate a nonzero response throughout the de Sitter manifold, even
in regions which are not future-related to the source. Imposing a de
Sitter invariant gauge results in hyperbolic equations for which the
response is zero in any region that is not future-related to the
source. This feature of gauge theories on de Sitter space was first
noted by Penrose in 1963 \cite{RP} and has since been studied for
gravity \cite{TW1} and electromagnetism \cite{BK}.

One consequence of the causality obstacle is that no completely de
Sitter invariant gauge field propagator can correctly describe even
classical physics over the entire de Sitter manifold. The confusing
point is the extent of the region over which the original, gauge
invariant field equations are violated. For electromagnetism it
turns out that a de Sitter invariant gauge can respect the gauge
invariant equations on the submanifold which is future-directed from
the source \cite{RPW2}. For gravity there seem to be violations of
the Einstein equations everywhere \cite{AM}. The reason for this
difference is not understood.

Quantum corrections bring new problems when using de Sitter
invariant gauges. The one loop scalar self-mass-squared has recently
been computed in two different gauges for scalar quantum
electrodynamics \cite{KW}. With each gauge the computation was made
for charged scalars which are massless, minimally coupled and for
charged scalars which are massless, conformally coupled. What goes
wrong is clearest for the conformally coupled scalar, which should
experience no large de Sitter enhancement over the flat space result
on account of the conformal flatness of the de Sitter geometry. This
is indeed the case when one employs the de Sitter breaking gauge
that takes maximum account of the conformal invariance of
electromagnetism in $D\!=\!3 \!+\!1$ spacetime dimensions. However,
when the computation was done in the de Sitter invariant analogue of
Feynman gauge the result was on-shell singularities! Off shell
one-particle-irreducible functions need not agree in different
gauges \cite{RJ} but they should agree on shell \cite{Lam}. In view
of its on-shell singularities the result in the de Sitter invariant
gauge is clearly wrong.

The nature of the problem may be the apparent inconsistency between
de Sitter invariance and the manifold's linearization instability.
Any propagator gives the response (with a certain boundary
condition) to a single point source. If the propagator is also de
Sitter invariant then this response must be valid throughout the
full de Sitter manifold. But the linearization instability precludes
solving the invariant field equations for a single point source on
the full manifold! This feature of the invariant theory is lost when
a de Sitter invariant gauge fixing term is simply added to the
action so it must be that the process of adding it was not
legitimate. In striving to attain a propagator which is valid
everywhere, one invariably obtains a propagator that is not valid
anywhere!

Although the pathology has not be identified as well as we should
like, the procedure for dealing with it does seem to be clear. One
can avoid the problem either by working on the full manifold with a
noncovariant gauge condition that preserves the elliptic character
of the constraint equations, or else by employing a covariant, but
not de Sitter invariant gauge on an open submanifold \cite{TW1}. We
choose the latter course and employ the following analogue of the de
Donder gauge fixing term of flat space,
\begin{equation}
\mathcal{L}_{GF} = -\frac12 a^{D-2} \eta^{\mu\nu} F_{\mu} F_{\nu} \;
, \; F_{\mu} \equiv \eta^{\rho\sigma} \Bigl(h_{\mu\rho , \sigma} -
\frac12 h_{\rho \sigma , \mu} + (D \!-\! 2) H a h_{\mu \rho}
\delta^0_{\sigma} \Bigr) . \label{GR}
\end{equation}

Because our gauge condition breaks de Sitter invariance it will be
necessary to contemplate noninvariant counterterms. It is therefore
appropriate to digress at this point with a description of the
various de Sitter symmetries and their effect upon Equation
\ref{GR}. In our $D$-dimensional conformal coordinate system the
$\frac12 D(D\!+\!1)$ de Sitter transformations take the following
form:
\begin{enumerate}
\item{Spatial translations --- comprising $(D\!-\!1)$ transformations.}
\begin{eqnarray}
\eta^{\p} & = & \eta \label{homot} \; , \\
x^{\prime i} & = & x^i + \epsilon^i \label{homox} \; .
\end{eqnarray}
\item{Rotations --- comprising $\frac12 (D\!-\!1) (D\!-\!2)$ transformations.}
\begin{eqnarray}
\eta^{\p} & = & \eta \; , \label{isot} \\
x^{\prime i} & = & R^{ij} x^j \label{isox} \; .
\end{eqnarray}
\item{Dilatation --- comprising $1$ transformation.}
\begin{eqnarray}
\eta^{\p} & = & k \, \eta \; , \label{dilt} \\
x^{\prime i} & = & k \, x^i \label{dilx} \; .
\end{eqnarray}
\item{Spatial special conformal transformations --- comprising $(D\!-\!1)$
transformations.}
\begin{eqnarray}
\eta^{\p} & = & \frac{\eta}{1 \!-\! 2 \vec{\theta} \!\cdot\! \vec{x}
\!+\! \Vert \vec{\theta} \Vert^2 x\!\cdot\! x} \; , \label{ssct} \\
x^{\prime i} & = & \frac{x^i - \theta^i x\!\cdot\! x}{1 \!-\! 2
\vec{\theta} \!\cdot\! \vec{x} \!+\! \Vert \vec{\theta} \Vert^2
x\!\cdot\! x} \; . \label{sscx}
\end{eqnarray}
\end{enumerate}
It is easy to check that our gauge condition respects all of these
but the spatial special conformal transformations. We will see that
the other symmetries impose important restrictions upon the BPHZ
counterterms which are allowed.

It is now time to solve for the graviton propagator. Because its
space and time components are treated differently in our coordinate
system and gauge it is useful to have an expression for the purely
spatial parts of the Lorentz metric and the Kronecker delta,
\begin{equation}
\overline{\eta}_{\mu\nu} \equiv \eta_{\mu\nu} + \delta^0_{\mu}
\delta^0_{\nu} \qquad {\rm and} \qquad \overline{\delta}^{\mu}_{\nu}
\equiv \delta^{\mu}_{\nu} - \delta_0^{\mu} \delta^0_{\nu} \; .
\end{equation}
The quadratic part of $\mathcal{L}_{\rm Einstein} +
\mathcal{L}_{GF}$ can be partially integrated to take the form
$\frac12 h^{\mu\nu} D_{\mu\nu}^{~~\rho \sigma} h_{\rho\sigma}$,
where the kinetic operator is,
\begin{eqnarray}
\lefteqn{D_{\mu\nu}^{~~\rho\sigma} \equiv \left\{ \frac12
\overline{\delta}_{ \mu}^{~(\rho} \overline{\delta}_{\nu}^{~\sigma)}
- \frac14 \eta_{\mu\nu} \eta^{\rho\sigma} - \frac1{2(D\!-\!3)}
\delta_{\mu}^0 \delta_{\nu}^0
\delta_0^{\rho} \delta_0^{\sigma} \right\} D_A } \nonumber \\
& & \hspace{3cm} + \delta^0_{(\mu} \overline{\delta}_{\nu)}^{(\rho}
\delta_0^{\sigma)} \, D_B + \frac12
\Bigl(\frac{D\!-\!2}{D\!-\!3}\Bigr) \delta_{\mu}^0 \delta_{\nu}^0
\delta_0^{\rho} \delta_0^{\sigma} \, D_C \; , \qquad
\end{eqnarray}
and the three scalar differential operators are,
\begin{eqnarray}
D_A & \equiv & \partial_{\mu} \Bigl(\sqrt{-g} g^{\mu\nu}
\partial_{\nu}\Bigr)
\; , \\
D_B & \equiv & \partial_{\mu} \Bigl(\sqrt{-g} g^{\mu\nu}
\partial_{\nu}\Bigr)
- \frac1{D} \Bigl(\frac{D\!-\!2}{D\!-\!1}\Bigr) R \sqrt{-g} \; , \\
D_C & \equiv & \partial_{\mu} \Bigl(\sqrt{-g} g^{\mu\nu}
\partial_{\nu}\Bigr) - \frac2{D} \Bigl(\frac{D\!-\!3}{D\!-\!1}\Bigr)
R \sqrt{-g} \; .
\end{eqnarray}

The graviton propagator in this gauge takes the form of a sum of
constant index factors times scalar propagators,
\begin{equation}
i\Bigl[{}_{\mu\nu} \Delta_{\rho\sigma}\Bigr](x;x^{\p}) =
\sum_{I=A,B,C} \Bigl[{}_{\mu\nu} T^I_{\rho\sigma}\Bigr]
i\Delta_I(x;x^{\p}) \; . \label{gprop0}
\end{equation}
The three scalar propagators invert the various scalar kinetic
operators,
\begin{equation}
D_I \times i\Delta_I(x;x^{\p}) = i \delta^D(x - x^{\p}) \qquad {\rm
for} \qquad I = A,B,C \; , \label{sprops}
\end{equation}
and we will presently give explicit expressions for them. The index
factors are,
\begin{eqnarray}
\Bigl[{}_{\mu\nu} T^A_{\rho\sigma}\Bigr] & = & 2 \,
\overline{\eta}_{\mu (\rho} \overline{\eta}_{\sigma) \nu} -
\frac2{D\!-\! 3} \overline{\eta}_{\mu\nu}
\overline{\eta}_{\rho \sigma} \; , \label{A}\\
\Bigl[{}_{\mu\nu} T^B_{\rho\sigma}\Bigr] & = & -4 \delta^0_{(\mu}
\overline{\eta}_{\nu) (\rho} \delta^0_{\sigma)} \; , \label{B} \\
\Bigl[{}_{\mu\nu} T^C_{\rho\sigma}\Bigr] & = & \frac2{(D \!-\!2) (D
\!-\!3)} \Bigl[(D \!-\!3) \delta^0_{\mu} \delta^0_{\nu} +
\overline{\eta}_{\mu\nu}\Bigr] \Bigl[(D \!-\!3) \delta^0_{\rho}
\delta^0_{\sigma} + \overline{\eta}_{\rho \sigma}\Bigr] \;
\label{C}.
\end{eqnarray}
With these definitions and Equation \ref{sprops} for the scalar
propagators it is straightforward to verify that the graviton
propagator of Equation \ref{gprop0} indeed inverts the gauge-fixed
kinetic operator,
\begin{equation}
D_{\mu\nu}^{~~\rho\sigma} \times i\Bigl[{}_{\rho\sigma}
\Delta^{\alpha\beta} \Bigr](x;x^{\p}) = \delta_{\mu}^{(\alpha}
\delta_{\nu}^{\beta)} i \delta^D(x-x^{\p}) \; .
\end{equation}

The scalar propagators can be expressed in terms of the following
function of the invariant length $\ell(x;x^{\p})$ between $x^{\mu}$
and $x^{\prime \mu}$,
\begin{eqnarray}
y(x;x^{\p}) & \equiv & 4 \sin^2\Bigl(\frac12 H \ell(x;x^{\p})\Bigr)
=  a a^{\p} H^2 {\Delta x }^2(x;x^{\p}) \; , \\
& = & a a^{\p} H^2 \Bigl( \Vert \vec{x} - \vec{x}^{\p}\Vert^2 -
(\vert \eta \!-\! \eta^{\p}\vert \!-\! i\delta)^2 \Bigr) \; .
\label{fully}
\end{eqnarray}
The most singular term for each case is the propagator for a
massless, conformally coupled scalar \cite{BD},
\begin{equation}
{i\Delta}_{\rm cf}(x;x^{\p}) = \frac{H^{D-2}}{(4\pi)^{\frac{D}2}}
\Gamma\Bigl( \frac{D}2 \!-\! 1\Bigr)
\Bigl(\frac4{y}\Bigr)^{\frac{D}2-1} \; .
\end{equation}
The $A$-type propagator obeys the same equation as that of a
massless, minimally coupled scalar. It has long been known that no
de Sitter invariant solution exists \cite{AF}. If one elects to
break de Sitter invariance while preserving homogeneity of Equations
\ref{homot}-\ref{homox} and isotropy of Equations
\ref{isot}-\ref{isox}
--- this is known as the ``E(3)'' vacuum \cite{BA}
--- the minimal solution is \cite{OW1,OW2},
\begin{eqnarray}
\lefteqn{i \Delta_A(x;x^{\p}) =  i \Delta_{\rm cf}(x;x^{\p}) } \nonumber \\
& & + \frac{H^{D-2}}{(4\pi)^{\frac{D}2}} \frac{\Gamma(D \!-\!
1)}{\Gamma( \frac{D}2)} \left\{\! \frac{D}{D\!-\! 4}
\frac{\Gamma^2(\frac{D}2)}{\Gamma(D \!-\! 1)}
\Bigl(\frac4{y}\Bigr)^{\frac{D}2 -2} \!\!\!\!\!\! - \pi
\cot\Bigl(\frac{\pi}2 D\Bigr) + \ln(a a^{\p}) \!\right\} \nonumber \\
& & + \frac{H^{D-2}}{(4\pi)^{\frac{D}2}} \! \sum_{n=1}^{\infty}\!
\left\{\! \frac1{n} \frac{\Gamma(n \!+\! D \!-\! 1)}{\Gamma(n \!+\!
\frac{D}2)} \Bigl(\frac{y}4 \Bigr)^n \!\!\!\! - \frac1{n \!-\!
\frac{D}2 \!+\! 2} \frac{\Gamma(n \!+\!  \frac{D}2 \!+\!
1)}{\Gamma(n \!+\! 2)} \Bigl(\frac{y}4 \Bigr)^{n - \frac{D}2 +2}
\!\right\} \! . \quad \label{DeltaA}
\end{eqnarray}
Note that this solution breaks dilatation invariance of Equations
\ref{dilt}-\ref{dilx} in addition to the spatial special conformal
invariance of Equations \ref{ssct}-\ref{sscx} broken by the gauge
condition. By convoluting naive de Sitter transformations with the
compensating diffeomorphisms necessary to restore our gauge
condition of Equation \ref{GR} one can show that the breaking of
dilatation invariance is physical whereas the apparent breaking of
spatial special conformal invariance is a gauge artifact \cite{GK}.

The B-type and $C$-type propagators possess de Sitter invariant (and
also unique) solutions,
\begin{eqnarray}
\lefteqn{i \Delta_B(x;x^{\p}) =  i \Delta_{\rm cf}(x;x^{\p}) -
\frac{H^{D-2}}{(4 \pi)^{\frac{D}2}} \! \sum_{n=0}^{\infty}\!
\left\{\!  \frac{\Gamma(n \!+\! D \!-\! 2)}{\Gamma(n \!+\!
\frac{D}2)} \Bigl(\frac{y}4 \Bigr)^n \right. }
\nonumber \\
& & \hspace{6.5cm} \left. - \frac{\Gamma(n \!+\!
\frac{D}2)}{\Gamma(n \!+\! 2)} \Bigl( \frac{y}4 \Bigr)^{n -
\frac{D}2 +2} \!\right\} \! , \qquad
\label{DeltaB} \\
\lefteqn{i \Delta_C(x;x^{\p}) =  i \Delta_{\rm cf}(x;x^{\p}) +
\frac{H^{D-2}}{(4\pi)^{\frac{D}2}} \! \sum_{n=0}^{\infty} \left\{\!
(n\!+\!1) \frac{\Gamma(n \!+\! D \!-\! 3)}{\Gamma(n \!+\!
\frac{D}2)}
\Bigl(\frac{y}4 \Bigr)^n \right. } \nonumber \\
& & \hspace{4.5cm} \left. - \Bigl(n \!-\! \frac{D}2 \!+\!  3\Bigr)
\frac{ \Gamma(n \!+\! \frac{D}2 \!-\! 1)}{\Gamma(n \!+\! 2)}
\Bigl(\frac{y}4 \Bigr)^{n - \frac{D}2 +2} \!\right\} \! . \qquad
\label{DeltaC}
\end{eqnarray}
They can be more compactly, but less usefully, expressed as
hypergeometric functions \cite{CR,DC},
\begin{eqnarray}
i\Delta_B(x;x^{\p}) & = & \frac{H^{D-2}}{(4\pi)^{\frac{D}2}}
\frac{\Gamma(D\!-\!2) \Gamma(1)}{\Gamma(\frac{D}2)} \,
\mbox{}_2F_1\Bigl(D\!-\!2,1;\frac{D}2;1 \!-\!
\frac{y}4\Bigr) \; , \label{FDB} \\
i\Delta_C(x;x^{\p}) & = & \frac{H^{D-2}}{(4\pi)^{\frac{D}2}}
\frac{\Gamma(D\!-\!3) \Gamma(2)}{\Gamma(\frac{D}2)} \,
\mbox{}_2F_1\Bigl(D\!-\!3,2;\frac{D}2;1 \!-\! \frac{y}4\Bigr) \; .
\label{FDC}
\end{eqnarray}
These expressions might seem daunting but they are actually simple
to use because the infinite sums vanish in $D=4$, and each term in
these sums goes like a positive power of $y(x;x^{\p})$. This means
the infinite sums can only contribute when multiplied by a divergent
term, and even then only a small number of terms can contribute.
Note also that the $B$-type and $C$-type propagators agree with the
conformal propagator in $D=4$.

In view of the subtle problems associated with the graviton
propagator in what seemed to be perfectly valid, de Sitter invariant
gauges \cite{AM,TW1}, it is well to review the extensive checks that
have been made on the consistency of this noninvariant propagator.
On the classical level it has been checked that the response to a
point mass is in perfect agreement with the linearized, de
Sitter-Schwarzchild geometry \cite{TW1}. The linearized
diffeomorphisms which enforce the gauge condition have also been
explicitly constructed \cite{TW2}. Although a tractable,
$D$-dimensional form for the various scalar propagators
$i\Delta_I(x;x^{\p})$ was not originally known, some simple
identities obeyed by the mode functions in their Fourier expansions
sufficed to verify the tree order Ward identity \cite{TW2}. The
full, $D$-dimensional formalism has been used recently to compute
the graviton 1-point function at one loop order \cite{TW3}. The
result seems to be in qualitative agreement with canonical
computations in other gauges \cite{LHF,FMVV}. A $D\!=\!3\!+\!1$
version of the formalism --- with regularization accomplished by
keeping the parameter $\delta \neq 0$ in the de Sitter length
function $y(x;x^{\p})$ Equation \ref{fully} --- was used to evaluate
the leading late time correction to the 2-loop 1-point function
\cite{TW4,TW5}. The same technique was used to compute the
unrenormalized graviton self-energy at one loop order \cite{TW6}. An
explicit check was made that the flat space limit of this quantity
agrees with Capper's result \cite{DMC2} for the graviton self-energy
in the same gauge. The one loop Ward identity was also checked in de
Sitter background \cite{TW6}. Finally, the $D\!=\!4$ formalism was
used to compute the two loop contribution from a massless, minimally
coupled scalar to the 1-graviton function \cite{TW7}. The result was
shown to obey an important bound imposed by global conformal
invariance on the maximum possible late time effect.

\subsection {Renormalization and Counterterms}

It remains to deal with the local counterterms we must add,
order-by-order in perturbation theory, to absorb divergences in the
sense of BPHZ renormalization. The particular counterterms which
renormalize the ferm\-i\-on self-energy must obviously involve a
single $\overline{\psi}$ and a single $\psi$.\footnote{Although the
Dirac Lagrangian is conformally invariant, the counterterms required
to renormalize the fermion self-energy will not possess this
symmetry because quantum gravity does not. We must therefore work
with the original fields rather than the conformally rescaled ones.}
At one loop order the superficial degree of divergence of quantum
gravitational contributions to the fermion self-energy is three, so
the necessary counterterms can involve zero, one, two or three
derivatives. These derivatives can either act upon the fermi fields
or upon the metric, in which case they must be organized into
curvatures or derivatives of curvatures. We will first exhaust the
possible invariant counterterms for a general renormalized fermion
mass and a general background geometry, and then specialize to the
case of zero mass in de Sitter background. We close with a
discussion of possible noninvariant counterterms.

All one loop corrections from quantum gravity must carry a factor of
$\kappa^2 \sim {\rm mass}^{-2}$. There will be additional dimensions
associated with derivatives and with the various fields, and the
balance must be struck using the renormalized fermion mass, $m$.
Hence the only invariant counterterm with no derivatives has the
form,
\begin{equation}
\kappa^2 m^3 \overline{\psi} \psi \sqrt{-g} \; . \label{zero}
\end{equation}
With one derivative we can always partially integrate to act upon
the $\psi$ field, so the only invariant counterterm is,
\begin{equation}
\kappa^2 m^2 \overline{\psi} i \hspace{-.1cm} \not{\hspace{-.15cm}
\mathcal{D}} \psi \sqrt{-g} \; . \label{one}
\end{equation}
Two derivatives can either act upon the fermions or else on the
metric to produce curvatures. We can organize the various
possibilities as follows,
\begin{equation}
\kappa^2 m \overline{\psi} (i \hspace{-.1cm} \not{\hspace{-.15cm}
\mathcal{D}})^2 \psi \sqrt{-g} \quad , \quad \kappa^2 m R
\overline{\psi} \psi \sqrt{-g} \; . \label{two}
\end{equation}
Three derivatives can be all acted on the fermions, or one on the
fermions and two in the form of curvatures, or there can be a
differentiated curvature,
\begin{eqnarray}
\kappa^2 \overline{\psi} \Bigl( (i \hspace{-.1cm}
\not{\hspace{-.15cm} \mathcal{D}})^2 \!+\! \frac{R}{D(D\!-\!1)}
\Bigr) i \hspace{-.1cm} \not{\hspace{-.15cm} \mathcal{D}} \psi
\sqrt{-g} & , & \kappa^2 R \, \overline{\psi} \, i \hspace{-.1cm}
\not{\hspace{-.15cm}
\mathcal{D}} \psi \sqrt{-g} \; , \nonumber \\
\kappa^2 e_{\mu m} \Bigl(R^{\mu\nu} - \frac1{D} g^{\mu\nu} R\Bigr)
\overline{\psi} \gamma^m i \mathcal{D}_{\nu} \psi \sqrt{-g} & , &
\kappa^2 e^{\mu}_{~m} R_{,\mu} \overline{\psi} \gamma^m \psi
\sqrt{-g} \; .\label{three}
\end{eqnarray}

Because mass is multiplicatively renormalized in dimensional
regularization, and because we are dealing with zero mass fermions,
counterterms in Equations \ref{zero}, \ref{one} and \ref{two} are
all unnecessary for our calculation. Although all four counterterms
of Equation \ref{three} are nonzero and distinct for a general
metric background, they only affect our fermion self-energy for the
special case of de Sitter background. For that case $R_{\mu\nu} =
(D\!-\!1) H^2 g_{\mu\nu}$, so the last two counterterms vanish. The
specialization of the invariant counter-Lagrangian we require to de
Sitter background is therefore,
\begin{eqnarray}
\lefteqn{\Delta \mathcal{L}_{\rm inv} = \alpha_1 \kappa^2
\overline{\psi} \Bigl( (i \hspace{-.1cm} \not{\hspace{-.15cm}
\mathcal{D}})^2 \!+\! \frac{R}{D (D\!-\!1)}\Bigr) i \hspace{-.1cm}
\not{\hspace{-.15cm} \mathcal{D}}
 \psi \sqrt{-g} + \alpha_2 \kappa^2 R \, \overline{\psi} \, i \hspace{-.1cm}
\not{\hspace{-.15cm} \mathcal{D}} \psi \sqrt{-g} \; , \label{invctms} } \\
& & \longrightarrow \alpha_1 \kappa^2 \overline{\Psi} \Bigl(i
\hspace{-.1cm} \not{\hspace{-.1cm} \partial} a^{-1} i \hspace{-.1cm}
\not{\hspace{-.1cm}
\partial} a^{-1} \!+\! \frac{R}{D(D\!-\!1)} \Bigr) i \hspace{-.1cm}
\not{\hspace{-.1cm} \partial} \Psi + \alpha_2 (D\!-\!1) D \kappa^2
H^2 \overline{\Psi} i \hspace{-.1cm} \not{\hspace{-.1cm} \partial}
\Psi \; . \qquad
\end{eqnarray}
Here $\alpha_1$ and $\alpha_2$ are $D$-dependent constants which are
dimensionless for $D\!=\!4$. The associated vertex operators are,
\begin{eqnarray}
C_{1 ij} & \equiv & \alpha_1 \kappa^2 \Bigl(i \hspace{-.1cm}
\not{\hspace{-.1cm} \partial} a^{-1} i \hspace{-.1cm}
\not{\hspace{-.1cm}
\partial} a^{-1} i \hspace{-.1cm} \not{\hspace{-.1cm} \partial} \!+\! H^2
i \hspace{-.1cm} \not{\hspace{-.1cm} \partial}\Bigr)_{ij} = \alpha_1
\kappa^2 \Bigl(a^{-1} i \hspace{-.1cm} \not{\hspace{-.1cm} \partial}
\partial^2 a^{-1}
\Bigr)_{ij} \; , \label{C1} \\
C_{2 ij} & \equiv & \alpha_2 (D\!-\!1) D \kappa^2 H^2 i
\hspace{-.1cm} \not{\hspace{-.1cm} \partial}_{ij} \; . \label{C2}
\end{eqnarray}
Of course $C_1$ is the higher derivative counterterm mentioned in
section 1. It will renormalize the most singular terms --- coming
from the $i\Delta_{\rm cf}$ part of the graviton propagator ---
which are unimportant because they are suppressed by powers of the
scale factor. The other vertex operator, $C_2$, is a sort of
dimensionful field strength renormalization in de Sitter background.
It will renormalize the less singular contributions which derive
physically from inflationary particle production.

The one loop fermion self-energy would require no additional
counterterms had it been possible to use the background field
technique in background field gauge \cite{BSD1,BSD2,BSD3,LFA}.
However, the obstacle to using a de Sitter invariant gauge obviously
precludes this. We must therefore come to terms with the possibility
that divergences may arise which require noninvariant counterterms.
What form can these counterterms take? Applying the BPHZ theorem
\cite{BP,H,Z1,Z2} to the gauge-fixed theory in de Sitter background
implies that the relevant counterterms must still consist of
$\kappa^2$ times a spinor differential operator with the dimension
of mass-cubed, involving no more than three derivatives and acting
between $\overline{\Psi}$ and $\Psi$. As the only dimensionful
constant in our problem, powers of $H$ must be used to make up
whatever dimensions are not supplied by derivatives.

Because dimensional regularization respects diffeomorphism
invariance, it is only the gauge fixing term in Equation \ref{GR}
that permits noninvariant counterterms.\footnote{One might think
that the they could come as well from the fact that the vacuum
breaks de Sitter invariance, but symmetries broken by the vacuum do
not introduce new counterterms \cite{SRC}. Highly relevant, explicit
examples are provided by recent computations for a massless,
minimally coupled scalar with a quartic self-interaction in the same
locally de Sitter background used here. The vacuum in this theory
also breaks de Sitter invariance but noninvariant counterterms fail
to arise even at {\it two loop} order in either the expectation
value of the stress tensor \cite{OW1,OW2} or the self-mass-squared
\cite{BOW}. It is also relevant that the one loop vacuum
polarization from (massless, minimally coupled) scalar quantum
electrodynamics is free of noninvariant counterterms in the same
background \cite{PTW2}.} Conversely, noninvariant counterterms must
respect the residual symmetries of the gauge condition. Homogeneity
of Equations \ref{homot}-\ref{homox} implies that the spinor
differential operator cannot depend upon the spatial coordinate
$x^i$. Similarly, isotropy of Equations \ref{isot}-\ref{isox}
requires that any spatial derivative operators $\partial_i$ must
either be contracted into $\gamma^i$ or another spatial derivative.
Owing to the identity,
\begin{equation}
(\gamma^i \partial_i)^2 = - \nabla^2 \; ,
\end{equation}
we can think of all spatial derivatives as contracted into
$\gamma^i$. Although the temporal derivative is not required to be
multiplied by $\gamma^0$ we lose nothing by doing so provided
additional dependence upon $\gamma^0$ is allowed.

The final residual symmetry is dilatation invariance shown by
Equations \ref{dilt}-\ref{dilx}. It has the crucial consequence that
derivative operators can only appear in the form $a^{-1}
\partial_{\mu}$. In addition the entire counterterm must have an
overall factor of $a$, and there can be no other dependence upon
$\eta$. So the most general counterterm consistent with our gauge
condition takes the form,
\begin{equation}
\Delta \mathcal{L}_{\rm non} = \kappa^2 H^3 a \overline{\Psi}
\mathcal{S}\Bigl((H a)^{-1} \gamma^0 \partial_0, (H a)^{-1} \gamma^i
\partial_i\Bigr) \Psi \; , \label{noninv}
\end{equation}
where the spinor function $\mathcal{S}(b,c)$ is at most a third
order polynomial function of its arguments, and it may involve
$\gamma^0$ in an arbitrary way.

Three more principles constrain noninvariant counterterms. The first
of these principles is that the fermion self-energy involves only
odd powers of gamma matrices. This follows from the masslessness of
our fermion and the consequent fact that the fermion propagator and
each interaction vertex involves only odd numbers of gamma matrices.
This principle fixes the dependence upon $\gamma^0$ and allows us to
express the spinor differential operator in terms of just ten
constants $\beta_i$,
\begin{eqnarray}
\lefteqn{\kappa^2 H^3 a \mathcal{S}\Bigl((H a)^{-1} \gamma^0
\partial_0, (H a)^{-1} \gamma^i \partial_i\Bigr) = \kappa^2 a
\Biggl\{ \beta_1 (a^{-1}
\gamma^0 \partial_0)^3 } \nonumber \\
& & \hspace{.5cm} + \beta_2 \Bigl[(a^{-1} \gamma^0 \partial_0)^2
(a^{-1} \gamma^i \partial_i)\Bigr] + \beta_3 \Bigl[(a^{-1} \gamma^0
\partial_0) (a^{-1} \gamma^i \partial_i)^2\Bigr] + \beta_4 (a^{-1}
\gamma^i \partial_i)^3
\nonumber \\
& & \hspace{1.5cm} + H \gamma^0 \Biggl( \beta_5 (a^{-1} \gamma^0
\partial_0)^2 + \beta_6 \Bigl[(a^{-1} \gamma^0 \partial_0) (a^{-1}
\gamma^i \partial_i)
\Bigr] + \beta_7 (a^{-1} \gamma^i \partial_i)^2 \Biggr) \nonumber \\
& & \hspace{3cm} + H^2 \Biggl( \beta_8 (a^{-1} \gamma^0 \partial_0)
+ \beta_9 (a^{-1} \gamma^i \partial_i) \Biggr)+ H^3 \gamma^0
\beta_{10} \Biggr\} . \qquad \label{expS}
\end{eqnarray}
In this expansion, but for the rest of this section only, we define
noncommuting factors within square brackets to be symmetrically
ordered, for example,
\begin{eqnarray}
\lefteqn{\Bigl[(a^{-1} \gamma^0 \partial_0)^2 (a^{-1} \gamma^i
\partial_i) \Bigr] \equiv \frac13 (a^{-1} \gamma^0 \partial_0)^2
(a^{-1} \gamma^i
\partial_i)} \nonumber \\
& & \hspace{1.5cm} + \frac13 (a^{-1} \gamma^0 \partial_0) (a^{-1}
\gamma^i
\partial_i) (a^{-1} \gamma^0 \partial_0) + \frac13 (a^{-1} \gamma^i \partial_i)
(a^{-1} \gamma^0 \partial_0)^2 \; . \qquad
\end{eqnarray}

The second principle is that our gauge condition of Equation
\ref{GR} becomes Poincar\'e invariant in the flat space limit of $H
\rightarrow 0$, where the conformal time is $\eta = -e^{-Ht}/H$ with
$t$ held fixed. In that limit only the four cubic terms of Equation
\ref{expS} survive,
\begin{eqnarray}
\lefteqn{\lim_{H \rightarrow 0} \kappa^2 H^3 a \mathcal{S}\Bigl((H
a)^{-1} \gamma^0 \partial_0, (H a)^{-1} \gamma^i \partial_i\Bigr) =
\kappa^2 \Biggl\{
\beta_1 (\gamma^0 \partial_0)^3 } \nonumber \\
& & \hspace{1.5cm} + \beta_2 \Bigl[(\gamma^0 \partial_0)^2 (\gamma^i
\partial_i) \Bigr] + \beta_3 \Bigl[(\gamma^0 \partial_0) (\gamma^i
\partial_i)^2\Bigr] + \beta_4 (\gamma^i \partial_i)^3 \Biggr\} .
\qquad
\end{eqnarray}
Because the entire theory is Poincar\'e invariant in that limit,
these four terms must sum to a term proportional to $(\gamma^{\mu}
\partial_{\mu})^3$, which implies,
\begin{equation}
\beta_1 = \frac13 \beta_2 = \frac13 \beta_3 = \beta_4 \; .
\end{equation}
But in that case the four cubic terms sum to give a linear
combination of the invariant counterterms of Equation \ref{C1} and
Equation \ref{C2},
\begin{eqnarray}
\lefteqn{\kappa^2 a \Biggl\{ (a^{-1} \gamma^0 \partial_0)^3 + 3
\Bigl[(a^{-1}
\gamma^0 \partial_0)^2 (a^{-1} \gamma^i \partial_i)\Bigr] } \nonumber \\
& & \hspace{1.5cm} + 3 \Bigl[(a^{-1} \gamma^0 \partial_0) (a^{-1}
\gamma^i
\partial_i)^2\Bigr] + (a^{-1} \gamma^i \partial_i)^3 \Biggr\} = \kappa^2
\hspace{-.1cm} \not{\hspace{-.1cm} \partial} \, a^{-1}
\hspace{-.1cm} \not{\hspace{-.1cm} \partial} a^{-1} \hspace{-.1cm}
\not{\hspace{-.1cm}
\partial} \; . \qquad
\end{eqnarray}
Because we have already counted this combination among the invariant
counterterms it need not be included in $\mathcal{S}$.

The final simplifying principle is that the fermion self-energy is
odd under interchange of $x^{\mu}$ and $x^{\prime \mu}$,
\begin{equation}
-i \Bigl[\mbox{}_i \Sigma_j\Bigr](x;x^{\p}) = + i \Bigl[\mbox{}_i
\Sigma_j \Bigr](x^{\p};x) \; . \label{refl}
\end{equation}
This symmetry is trivial at tree order, but not easy to show
generally. Moreover, it isn't a property of individual terms, many
of which violate Equation \ref{refl}. However, when everything is
summed up the result must obey Equation \ref{refl}, hence so too
must the counterterms. This has the immediate consequence of
eliminating the counterterms with an even number of derivatives:
those proportional to $\beta_{5-7}$ and to $\beta_{10}$. We have
already dispensed with $\beta_{1-4}$, which leaves only the linear
terms, $\beta_{8-9}$. Because one linear combination of these
already appears in the invariant of Equation \ref{C2} the sole
noninvariant counterterm we require is,
\begin{equation}
\Delta \mathcal{L}_{\rm non} = \overline{\Psi} C_3 \Psi \qquad {\rm
where} \qquad C_{3ij} \equiv \alpha_3 \kappa^2 H^2 i \,
\hspace{-.1cm} \overline{\not{\hspace{-.1cm} \partial}}_{ij} \; .
\label{nictm}
\end{equation}

\section{COMPUTATIONAL RESULTS}
For one-loop order the big simplification of working in position
space is that it doesn't involve any integrations after all the
delta functions are used. However, even though calculating the one
loop fermion self energy is only a multiplication of propagators,
vertices and derivatives, the computation is still a tedious work
owing to the great number of vertices and the complicated graviton
propagator. Generally speaking, we first contract 4-point and pairs
of 3-point vertices into the full graviton propagator. Then we break
up the graviton propagator into its conformal part plus the
residuals proportional to each of three index factors. The next step
is to act the derivatives and sum up the results. At each step we
also tabulate the results in order to clearly see the potential
tendencies such as cancelations among these terms. Finally, we must
remember that the fermion self energy will be used inside an
integral in the quantum-corrected Dirac equation. For this purpose,
we extract the derivatives with respect to the coordinates
``$x^{\mu}$'' by partially integrating them out. This procedure also
can be implemented so as to segregate the divergence to a delta
function that can be absorbed by the counterterms which we found in
chapter $2$.

\subsection{Contributions from the 4-Point Vertices}

\begin{figure}
%\centerline{\epsfig{file=ffig1.eps}}
\includegraphics[width=4 in]{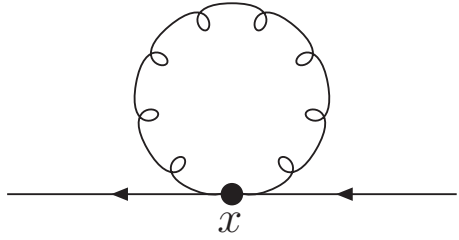}
\caption{Contribution from 4-point vertices.} \label{ffig1}
\end{figure}

In this section we evaluate the contributions from 4-point vertex
operators of Table \ref{v4ops}. The generic diagram topology is
depicted in Figure \ref{ffig1}. The analytic form is,
\begin{equation}
-i\Bigl[\mbox{}_i \Sigma_j^{\rm 4pt}\Bigr](x;x^{\p}) = \sum_{I=1}^8
i U^{\alpha\beta \rho\sigma}_{Iij} \, i\Bigl[\mbox{}_{\alpha\beta}
\Delta_{\rho\sigma} \Bigr](x;x^{\p}) \, \delta^D(x\!-\!x^{\p}) \; .
\label{4ptloop}
\end{equation}

And the generic contraction for each of the vertex operators in
Table~\ref{v4ops} is given in Table~\ref{4con}.
\begin{table}

\caption{Generic 4-point contractions}

\vbox{\tabskip=0pt \offinterlineskip
\def\tablerule{\noalign{\hrule}}
\halign to390pt {\strut#& \vrule#\tabskip=1em plus2em& \hfil#\hfil&
\vrule#& \hfil#\hfil& \vrule#\tabskip=0pt\cr \tablerule
\omit&height4pt&\omit&&\omit&\cr &&$\!\!\!\!{\rm I}\!\!\!\!$ &&
$\!\!\!\! i [\mbox{}_{\alpha\beta} \Delta_{\rho\sigma}](x;x^{\p}) \,
i U_I^{\alpha\beta\rho\sigma} \, \delta^D(x\!-\!x^{\p}) \!\!\!\!$ &
\cr \omit&height4pt&\omit&&\omit&\cr \tablerule
\omit&height2pt&\omit&&\omit&\cr && 1 && $-\frac18 \kappa^2 \, i
[\mbox{}^{\alpha}_{~\alpha} \Delta^{\rho}_{~\rho}](x;x) \,
\hspace{-.1cm} \not{\hspace{-.1cm} \partial} \,
\delta^D(x\!-\!x^{\p})$ & \cr \omit&height2pt&\omit&&\omit&\cr
\tablerule \omit&height2pt&\omit&&\omit&\cr && 2 && $\frac14
\kappa^2 \, i [\mbox{}^{\alpha\beta} \Delta_{\alpha\beta}](x;x) \,
\hspace{-.1cm} \not{\hspace{-.1cm} \partial} \,
\delta^D(x\!-\!x^{\p})$ & \cr \omit&height2pt&\omit&&\omit&\cr
\tablerule \omit&height2pt&\omit&&\omit&\cr && 3 && $\frac14
\kappa^2 \, i [\mbox{}^{\alpha}_{~\alpha} \Delta_{\rho\sigma}](x;x)
\, \gamma^{\rho} \partial^{\sigma} \, \delta^D(x\!-\!x^{\p})$ & \cr
\omit&height2pt&\omit&&\omit&\cr \tablerule
\omit&height2pt&\omit&&\omit&\cr && 4 && $-\frac38 \kappa^2 \, i
[\mbox{}^{\alpha}_{~\beta} \Delta_{\alpha\sigma}](x;x) \,
\gamma^{\beta} \partial^{\sigma} \, \delta^D(x\!-\!x^{\p})$ & \cr
\omit&height2pt&\omit&&\omit&\cr \tablerule
\omit&height2pt&\omit&&\omit&\cr && 5 && $-\frac{i}4 \kappa^2 \,
\partial_{\mu}^{\p} i [\mbox{}^{\alpha}_{~\alpha}
\Delta_{\rho\sigma}](x;x^{\p}) \, \gamma^{\rho} J^{\sigma \mu} \,
\delta^D(x\!-\!x^{\p})$ & \cr \omit&height2pt&\omit&&\omit&\cr
\tablerule \omit&height2pt&\omit&&\omit&\cr && 6 && $\frac{i}8
\kappa^2 \,
\partial_{\mu}^{\p} i [\mbox{}^{\alpha}_{~\beta}
\Delta_{\alpha\sigma}](x;x^{\p}) \, \gamma^{\mu} J^{\beta\sigma} \,
\delta^D(x\!-\!x^{\p})$ & \cr \omit&height2pt&\omit&&\omit&\cr
\tablerule \omit&height2pt&\omit&&\omit&\cr && 7 && $\frac{i}4
\kappa^2 \,
\partial_{\mu} i [\mbox{}^{\alpha}_{~\beta}
\Delta_{\alpha\sigma}](x;x) \, \gamma^{\beta} J^{\sigma\mu} \,
\delta^D(x\!-\!x^{\p})$ & \cr \omit&height2pt&\omit&&\omit&\cr
\tablerule \omit&height2pt&\omit&&\omit&\cr && 8 && $\frac{i}4
\kappa^2 \,
\partial^{\prime \beta} i [\mbox{}_{\alpha \beta}
\Delta_{\rho\sigma}](x;x^{\p}) \, \gamma^{\rho} J^{\sigma \alpha} \,
\delta^D(x\!-\!x^{\p})$ & \cr \omit&height2pt&\omit&&\omit&\cr
\tablerule}}

\label{4con}

\end{table}

From an examination of the generic contractions in Table~\ref{4con}
it is apparent that we must work out how the three index factors
$[\mbox{}_{\alpha \beta} T^I_{\rho \sigma}]$ which make up the
graviton propagator contract into $\eta^{\alpha\beta}$ and
$\eta^{\alpha\rho}$. For the $A$-type and $B$-type index factors the
various contractions give,
\begin{eqnarray}
\eta^{\alpha\beta} \, \Bigl[{}_{\alpha\beta} T^A_{\rho\sigma}\Bigr]
= - \Bigl(\frac4{D\!-\!3}\Bigr) \, \overline{\eta}_{\rho\sigma} & ,
& \eta^{\alpha\rho} \, \Bigl[{}_{\alpha\beta} T^A_{\rho\sigma}\Bigr]
= \Bigl(D\!-\!\frac2{D\!-\!3}\Bigr) \,\overline{\eta}_{\beta\sigma} \; , \\
\hspace{-.8cm}\eta^{\alpha\beta} \, \Bigl[{}_{\alpha\beta}
T^B_{\rho\sigma}\Bigr] = 0 & , & \eta^{\alpha\rho} \,
\Bigl[{}_{\alpha\beta} T^B_{\rho\sigma} \Bigr] = -(D \!-\! 1) \,
\delta^0_{\beta} \delta^0_{\sigma} + \overline{\eta}_{ \beta\sigma}
,\;\;
\end{eqnarray}
For the $C$-type index factor they are,
\begin{eqnarray}
\eta^{\alpha\beta} \, \Bigl[{}_{\alpha\beta} T^C_{\rho\sigma}\Bigr]
& = & \Bigl(\frac4{D-2}\Bigr) \, \delta^0_{\rho} \delta^0_{\sigma} +
\frac4{(D\!-\!2)(D\!-\!3)} \, \overline{\eta}_{\rho\sigma} \; , \nonumber \\
\eta^{\alpha\rho} \, \Bigl[{}_{\alpha\beta} T^C_{\rho\sigma}\Bigr] &
= & -2 \Bigl(\frac{D\!-\!3}{D\!-\!2}\Bigr) \, \delta^0_{\beta}
\delta^0_{\sigma} \!+\! \frac2{(D\!-\!2) (D\!-\!3)} \,
\overline{\eta}_{\beta\sigma} \; . \qquad
\end{eqnarray}
On occasion we also require double contractions. For the $A$-type
index factor these are,
\begin{eqnarray}
\eta^{\alpha\beta} \eta^{\rho\sigma} \, \Bigl[{}_{\alpha\beta}
T^A_{\rho\sigma}\Bigr] & = & -4 \Bigl(\frac{D\!-\!1}{D\!-\!3}\Bigr)
\; ,
\nonumber \\
\eta^{\alpha\rho} \eta^{\beta\sigma} \, \Bigl[{}_{\alpha\beta}
T^A_{\rho\sigma}\Bigr] & = & D (D\!-\!1) - 2
\Bigl(\frac{D\!-\!1}{D\!-\!3}\Bigr) \; .
\end{eqnarray}
The double contractions of the $B$-type and $C$-type index factors
are,
\begin{eqnarray}
\eta^{\alpha\beta} \eta^{\rho\sigma} \, \Bigl[{}_{\alpha\beta}
T^B_{\rho\sigma}\Bigr] = 0 & , & \eta^{\alpha\rho}
\eta^{\beta\sigma} \,
\Bigl[{}_{\alpha\beta} T^B_{\rho\sigma} \Bigr] = 2 (D \!-\! 1) \; , \\
\eta^{\alpha\beta} \eta^{\rho\sigma} \, \Bigl[{}_{\alpha\beta}
T^C_{\rho\sigma}\Bigr] = \frac8{(D\!-\!2)(D\!-\!3)} & , &
\eta^{\alpha\rho} \eta^{\beta\sigma} \, \Bigl[{}_{\alpha\beta}
T^C_{\rho\sigma}\Bigr] = 2 \frac{(D^2 \!-\! 5D \!+\!
8)}{(D\!-\!2)(D\!-\!3)}  .\; \qquad
\end{eqnarray}

Table~\ref{4props} was generated from Table~\ref{4con} by expanding
the graviton propagator in terms of index factors,
\begin{equation}
i\Bigl[{}_{\alpha\beta} \Delta_{\rho\sigma}\Bigr](x;x^{\p}) =
\Bigl[{}_{\alpha\beta} T^A_{\rho\sigma}\Bigr] i\Delta_A(x;x^{\p}) +
\Bigl[{}_{\alpha\beta} T^B_{\rho\sigma}\Bigr] i\Delta_B(x;x^{\p}) +
\Bigl[{}_{\alpha\beta} T^C_{\rho\sigma}\Bigr] i\Delta_C(x;x^{\p}) \;
.
\end{equation}
We then perform the relevant contractions using the previous
identities. Relation \ref{Jred} was also exploited to simplify the
gamma matrix structure.

\begin{table}

\caption{Four-point contribution from each part of the graviton
propagator.}

\vbox{\tabskip=0pt \offinterlineskip
\def\tablerule{\noalign{\hrule}}
\halign to390pt {\strut#& \vrule#\tabskip=1em plus2em& \hfil#\hfil&
\vrule#& \hfil#\hfil& \vrule#& \hfil#\hfil& \vrule#\tabskip=0pt\cr
\tablerule \omit&height4pt&\omit&&\omit&&\omit&\cr &&$\!\!\!\!{\rm
I}\!\!\!\!$ && $\!\!\!\!{\rm J}\!\!\!\!$ && $\!\!\!\! i
[\mbox{}_{\alpha\beta} T^J_{\rho\sigma}]\, i\Delta_J(x;x^{\p}) \, i
U_I^{\alpha\beta\rho\sigma} \, \delta^D(x\!-\!x^{\p}) \!\!\!\!$ &
\cr \omit&height4pt&\omit&&\omit&&\omit&\cr \tablerule
\omit&height2pt&\omit&&\omit&&\omit&\cr && 1 && A && $\frac12
(\frac{D-1}{D-3}) \kappa^2 \, i \Delta_A(x;x) \, \hspace{-.1cm}
\not{\hspace{-.1cm} \partial} \, \delta^D(x\!-\!x^{\p})$ & \cr
\omit&height2pt&\omit&&\omit&&\omit&\cr \tablerule
\omit&height2pt&\omit&&\omit&&\omit&\cr && 1 && B && $0$ & \cr
\omit&height2pt&\omit&&\omit&&\omit&\cr \tablerule
\omit&height2pt&\omit&&\omit&&\omit&\cr && 1 && C &&
$\!\!\!\!-\frac1{(D-2)(D-3)} \kappa^2 \, i \Delta_C(x;x) \,
\hspace{-.1cm} \not{\hspace{-.1cm} \partial} \,
\delta^D(x\!-\!x^{\p})\!\!\!\!$ & \cr
\omit&height2pt&\omit&&\omit&&\omit&\cr \tablerule
\omit&height2pt&\omit&&\omit&&\omit&\cr && 2 && A &&
$\!\!\!\!(\frac{D-1}4) (\frac{D^2 -3D -2}{D-3}) \kappa^2 \, i
\Delta_A(x;x) \, \hspace{-.1cm} \not{\hspace{-.1cm} \partial} \,
\delta^D(x\!-\!x^{\p})\!\!\!\!$ & \cr
\omit&height2pt&\omit&&\omit&&\omit&\cr \tablerule
\omit&height2pt&\omit&&\omit&&\omit&\cr && 2 && B && $(\frac{D-1}2)
\kappa^2 \, i \Delta_B(x;x) \, \hspace{-.1cm} \not{\hspace{-.1cm}
\partial} \, \delta^D(x\!-\!x^{\p})$ & \cr
\omit&height2pt&\omit&&\omit&&\omit&\cr \tablerule
\omit&height2pt&\omit&&\omit&&\omit&\cr && 2 && C && $\frac12
\frac{(D^2 -5D + 8)}{(D-2)(D-3)} \kappa^2 \, i \Delta_C(x;x) \,
\hspace{-.1cm} \not{\hspace{-.1cm} \partial} \,
\delta^D(x\!-\!x^{\p})$ & \cr
\omit&height2pt&\omit&&\omit&&\omit&\cr \tablerule
\omit&height2pt&\omit&&\omit&&\omit&\cr && 3 && A && $-\frac1{D-3}
\kappa^2 \, i \Delta_A(x;x) \, \hspace{-.1cm}
\overline{\not{\hspace{-.1cm} \partial}} \, \delta^D(x\!-\!x^{\p})$
& \cr \omit&height2pt&\omit&&\omit&&\omit&\cr \tablerule
\omit&height2pt&\omit&&\omit&&\omit&\cr && 3 && B && $0$ & \cr
\omit&height2pt&\omit&&\omit&&\omit&\cr \tablerule
\omit&height2pt&\omit&&\omit&&\omit&\cr && 3 && C &&
$\!\!\!\!\frac1{(D-2)(D-3)} \kappa^2 \, i \Delta_C(x;x) [\;
\hspace{-.1cm} \overline{\not{\hspace{-.1cm} \partial}} \!-\!
{\scriptstyle (D-3)} \gamma^0 \partial_0] \delta^D(x\!-\!x^{\p})
\!\!\!\!$ & \cr \omit&height2pt&\omit&&\omit&&\omit&\cr \tablerule
\omit&height2pt&\omit&&\omit&&\omit&\cr && 4 && A && $-\frac38
(\frac{D^2 -3D -2}{D-3}) \kappa^2 \, i \Delta_A(x;x) \,
\hspace{-.1cm} \overline{\not{\hspace{-.1cm} \partial}} \,
\delta^D(x\!-\!x^{\p})$ & \cr
\omit&height2pt&\omit&&\omit&&\omit&\cr \tablerule
\omit&height2pt&\omit&&\omit&&\omit&\cr && 4 && B &&
$\!\!\!\!-\frac38 \kappa^2 \, i \Delta_B(x;x) [\; \hspace{-.1cm}
\overline{\not{\hspace{-.1cm} \partial}} \!+\! {\scriptstyle (D-1)}
\gamma^0 \partial_0] \delta^D(x\!-\!x^{\p})\!\!\!\!$ & \cr
\omit&height2pt&\omit&&\omit&&\omit&\cr \tablerule
\omit&height2pt&\omit&&\omit&&\omit&\cr && 4 && C &&
$\!\!\!\!-\frac34 \frac1{(D-2)(D-3)} \kappa^2 \, i \Delta_C(x;x) [\;
\hspace{-.1cm} \overline{\not{\hspace{-.1cm} \partial}} \!+\!
{\scriptstyle (D-3)}^2 \gamma^0 \partial_0] \delta^D(x\!-\!x^{\p})
\!\!\!\!$ & \cr \omit&height2pt&\omit&&\omit&&\omit&\cr \tablerule
\omit&height2pt&\omit&&\omit&&\omit&\cr && 5 && A &&
$\!\!\!\!\kappa^2 [-\frac1{2(D-3)} \; \hspace{-.1cm} \overline{
\not{\hspace{-.1cm} \partial}}^{\p} \!+\! \frac12 (\frac{D-1}{D-3})
\hspace{-.1cm} \not{\hspace{-.1cm} \partial}^{\p} ] \,
i\Delta_A(x;x^{\p}) \, \delta^D(x\!-\!x^{\p}) \!\!\!\!$ & \cr
\omit&height2pt&\omit&&\omit&&\omit&\cr \tablerule
\omit&height2pt&\omit&&\omit&&\omit&\cr && 5 && B && $0$ & \cr
\omit&height2pt&\omit&&\omit&&\omit&\cr \tablerule
\omit&height2pt&\omit&&\omit&&\omit&\cr && 5 && C && $\!\!\!\!-
\frac1{(D-2)(D-3)} \kappa^2 \, [\frac12\; \hspace{-.1cm}
\overline{\not{\hspace{-.1cm} \partial}}^{\p} \!+\! (\frac{D-1}2)
\gamma^0 \partial_0^{\p}] \, i\Delta_C(x;x^{\p}) \,
\delta^D(x\!-\!x^{\p})\!\!\!\!$ & \cr
\omit&height2pt&\omit&&\omit&&\omit&\cr \tablerule
\omit&height2pt&\omit&&\omit&&\omit&\cr && 6 && A && $0$ & \cr
\omit&height2pt&\omit&&\omit&&\omit&\cr \tablerule
\omit&height2pt&\omit&&\omit&&\omit&\cr && 6 && B && $0$ & \cr
\omit&height2pt&\omit&&\omit&&\omit&\cr \tablerule
\omit&height2pt&\omit&&\omit&&\omit&\cr && 6 && C && $0$ & \cr
\omit&height2pt&\omit&&\omit&&\omit&\cr \tablerule
\omit&height2pt&\omit&&\omit&&\omit&\cr && 7 && A && $\!\!\!\!
(\frac{D^2-3D-2}{D-3}) \kappa^2 [-\frac18\; \hspace{-.1cm}
\overline{\not{\hspace{-.1cm} \partial}} \!+\! (\frac{D-1}8)
\hspace{-.1cm} \not{\hspace{-.1cm} \partial} ] \, i\Delta_A(x;x) \,
\delta^D(x\!-\!x^{\p}) \!\!\!\!$ & \cr
\omit&height2pt&\omit&&\omit&&\omit&\cr \tablerule
\omit&height2pt&\omit&&\omit&&\omit&\cr && 7 && B && $\!\!\!\!
\kappa^2 [(\frac{D-2}8) \; \hspace{-.1cm} \overline{
\not{\hspace{-.1cm} \partial}} \!+\! (\frac{D-1}8) \hspace{-.1cm}
\not{\hspace{-.1cm} \partial} ] \, i\Delta_B(x;x) \,
\delta^D(x\!-\!x^{\p}) \!\!\!\!$ & \cr
\omit&height2pt&\omit&&\omit&&\omit&\cr \tablerule
\omit&height2pt&\omit&&\omit&&\omit&\cr && 7 && C && $\!\!\!\!
\frac14 \kappa^2 [\frac{(D^2-6D+8)}{(D-2)(D-3)} \; \hspace{-.1cm}
\overline{\not{\hspace{-.1cm} \partial}} \!+\! \frac{(D-1)}{
(D-2)(D-3)} \hspace{-.1cm} \not{\hspace{-.1cm} \partial} ] \,
i\Delta_C(x;x) \, \delta^D(x\!-\!x^{\p}) \!\!\!\!$ & \cr
\omit&height2pt&\omit&&\omit&&\omit&\cr \tablerule
\omit&height2pt&\omit&&\omit&&\omit&\cr && 8 && A && $\!\!\!\! -
\kappa^2 \frac{(D-2)(D-1)}{8(D-3)} \; \hspace{-.1cm}
\overline{\not{\hspace{-.1cm} \partial}}^{\p} \, i\Delta_A(x;x^{\p})
\, \delta^D(x\!-\!x^{\p}) \!\!\!\!$ & \cr
\omit&height2pt&\omit&&\omit&&\omit&\cr \tablerule
\omit&height2pt&\omit&&\omit&&\omit&\cr && 8 && B && $\!\!\!\!
-\kappa^2 [\frac18 \; \hspace{-.1cm} \overline{ \not{\hspace{-.1cm}
\partial}}^{\p} \!+\! (\frac{D-1}8) \gamma^0 \partial_0^{\p} ] \,
i\Delta_B(x;x^{\p}) \, \delta^D(x\!-\!x^{\p}) \!\!\!\!$ & \cr
\omit&height2pt&\omit&&\omit&&\omit&\cr \tablerule
\omit&height2pt&\omit&&\omit&&\omit&\cr && 8 && C && $\!\!\!\!
\frac14 \kappa^2 [\frac1{(D-2)(D-3)} \; \hspace{-.1cm}
\overline{\not{\hspace{-.1cm} \partial}}^{\p} \!-\!
(\frac{D-1}{D-2}) \gamma^0
\partial_0^{\p} ] \, i\Delta_C(x;x^{\p}) \, \delta^D(x\!-\!x^{\p}) \!\!\!\!$ & \cr
\omit&height2pt&\omit&&\omit&&\omit&\cr \tablerule}}

\label{4props}

\end{table}

From Table~\ref{4props} it is apparent that we require the
coincidence limits of zero or one derivatives acting on each of the
scalar propagators. For the $A$-type propagator these are,
\begin{eqnarray}
\lim_{x^{\p} \rightarrow x} \, {i\Delta}_A(x;x^{\p}) & = &
\frac{H^{D-2}}{(4\pi)^{ \frac{D}{2}}}
\frac{\Gamma(D-1)}{\Gamma(\frac{D}{2})} \left\{-\pi \cot\Bigl(
\frac{\pi}{2} D \Bigr) + 2 \ln(a) \right\} , \label{Acoin} \\
\lim_{x^{\p} \rightarrow x} \, \partial_{\mu} {i\Delta}_A(x;x^{\p})
& = & \frac{H^{D-2}}{(4\pi)^{\frac{D}{2}}}
\frac{\Gamma(D-1)}{\Gamma(\frac{D}{2})} \times H a \delta^0_{\mu} \;
. \label{Acoin'}
\end{eqnarray}
The analogous coincidence limits for the $B$-type propagator are
actually finite in $D=4$ dimensions,
\begin{eqnarray}
\lim_{x^{\p} \rightarrow x} \, {i\Delta}_B(x;x^{\p}) & = &
\frac{H^{D-2}}{(4\pi)^{
\frac{D}{2}}} \frac{\Gamma(D-1)}{\Gamma(\frac{D}{2})}\times -\frac{1}{D\!-\!2} \; ,\\
\lim_{x^{\p} \rightarrow x} \, \partial_{\mu} {i\Delta}_B(x;x^{\p})
& = & 0 \; .
\end{eqnarray}
The same is true for the coincidence limits of the $C$-type
propagator,
\begin{eqnarray}
\lim_{x^{\p} \rightarrow x} \, {i\Delta}_C(x;x^{\p}) & = &
\frac{H^{D-2}}{(4\pi)^{ \frac{D}{2}}}
\frac{\Gamma(D-1)}{\Gamma(\frac{D}2)}\times \frac{1}{(D\!-\!2)
(D\!-\!3)} \; ,\\
\lim_{x^{\p} \rightarrow x} \, \partial_{\mu} {i\Delta}_C(x;x^{\p})
& = & 0 \; . \label{Ccoin'}
\end{eqnarray}

Our final result for the 4-point contributions is given in
Table~\ref{4fin}. It was obtained from Table~\ref{4props} by using
the previous coincidence limits. We have also always chosen to
re-express conformal time derivatives thusly,
\begin{equation}
\gamma^0 \partial_0 = \;\; \hspace{-.1cm} \not{\hspace{-.1cm}
\partial} - \; \hspace{-.1cm} \overline{\not{\hspace{-.1cm}
\partial}} \; .
\end{equation}
A final point concerns the fact that the terms in the final column
of Table~\ref{4fin} do not obey the reflection symmetry. In the next
section we will find the terms which exactly cancel these.

\begin{table}

\caption{Final 4-point contributions. All contributions are
multiplied by $\frac{\kappa^2 H^{D-2}}{(4 \pi)^{\frac{D}{2}}}
\frac{\Gamma(D-1)}{ \Gamma(\frac{D}{2})}$. We define $A \equiv
\frac{\pi}2 \cot(\frac{\pi D}2) \!-\! \ln(a)$.}

\vbox{\tabskip=0pt \offinterlineskip
\def\tablerule{\noalign{\hrule}}
\halign to390pt {\strut#& \vrule#\tabskip=1em plus2em& \hfil#\hfil&
\vrule#& \hfil#\hfil& \vrule#& \hfil#\hfil& \vrule#& \hfil#\hfil&
\vrule#& \hfil#\hfil& \vrule#\tabskip=0pt\cr \tablerule
\omit&height4pt&\omit&&\omit&&\omit&&\omit&&\omit&\cr
&&$\!\!\!\!{\rm I}\!\!\!\!$ && $\!\!\!\!{\rm J}\!\!\!\!$ &&
$\!\!\!\! \hspace{-.1cm} \not{\hspace{-.1cm} \partial} \,
\delta^D(x\!-\!x^{\p})\!\!\!\!$ && $\!\!\!\! \hspace{-.1cm}
\overline{\not{ \hspace{-.1cm} \partial}} \,
\delta^D(x\!-\!x^{\p})\!\!\!\!$ && $\!\!\!\!a H \gamma^0 \,
\delta^D(x\!-\!x^{\p})\!\!\!\!$ & \cr
\omit&height4pt&\omit&&\omit&&\omit&&\omit&&\omit&\cr \tablerule
\omit&height2pt&\omit&&\omit&&\omit&&\omit&&\omit&\cr && 1 && A &&
$-(\frac{D-1}{D-3}) A$ && $0$ && $0$ & \cr
\omit&height2pt&\omit&&\omit&&\omit&&\omit&&\omit&\cr \tablerule
\omit&height2pt&\omit&&\omit&&\omit&&\omit&&\omit&\cr && 1 && B &&
$0$ && $0$ && $0$ & \cr
\omit&height2pt&\omit&&\omit&&\omit&&\omit&&\omit&\cr \tablerule
\omit&height2pt&\omit&&\omit&&\omit&&\omit&&\omit&\cr && 1 && C &&
$\!\!\!\!-\frac{1}{(D-2)^2 (D-3)^2}\!\!\!\!$ && $0$ && $0$ & \cr
\omit&height2pt&\omit&&\omit&&\omit&&\omit&&\omit&\cr \tablerule
\omit&height2pt&\omit&&\omit&&\omit&&\omit&&\omit&\cr && 2 && A &&
$\!\!\!\! [-\frac{D(D-1)}{2} \!+\! (\frac{D-1}{D-3})] A \!\!\!\!$ &&
$0$ && $0$ & \cr
\omit&height2pt&\omit&&\omit&&\omit&&\omit&&\omit&\cr \tablerule
\omit&height2pt&\omit&&\omit&&\omit&&\omit&&\omit&\cr && 2 && B &&
$-\frac12 (\frac{D-1}{D-2})$ && $0$ && $0$ & \cr
\omit&height2pt&\omit&&\omit&&\omit&&\omit&&\omit&\cr \tablerule
\omit&height2pt&\omit&&\omit&&\omit&&\omit&&\omit&\cr && 2 && C &&
$\frac{1}{2} \frac{(D^2 -5D + 8)}{(D-2)^2(D-3)^2}$ && $0$ && $0$ &
\cr \omit&height2pt&\omit&&\omit&&\omit&&\omit&&\omit&\cr \tablerule
\omit&height2pt&\omit&&\omit&&\omit&&\omit&&\omit&\cr && 3 && A &&
$0$ && $\frac{2}{D-3} A$ && $0$ & \cr
\omit&height2pt&\omit&&\omit&&\omit&&\omit&&\omit&\cr \tablerule
\omit&height2pt&\omit&&\omit&&\omit&&\omit&&\omit&\cr && 3 && B &&
$0$ && $0$ && $0$ & \cr
\omit&height2pt&\omit&&\omit&&\omit&&\omit&&\omit&\cr \tablerule
\omit&height2pt&\omit&&\omit&&\omit&&\omit&&\omit&\cr && 3 && C &&
$\!\!\!\!-\frac{1}{(D-2)^2(D-3)}\!\!\!\!$ &&
$\!\!\!\!\frac{1}{(D-2)(D-3)^2}\!\!\!\!$ && $0$ & \cr
\omit&height2pt&\omit&&\omit&&\omit&&\omit&&\omit&\cr \tablerule
\omit&height2pt&\omit&&\omit&&\omit&&\omit&&\omit&\cr && 4 && A &&
$0$ && $\!\!\!\![\frac{3D}{4} \!-\! \frac{3}{2 (D-3)}] A\!\!\!\!$ &&
$0$ & \cr \omit&height2pt&\omit&&\omit&&\omit&&\omit&&\omit&\cr
\tablerule \omit&height2pt&\omit&&\omit&&\omit&&\omit&&\omit&\cr &&
4 && B && $\frac{3}{8} (\frac{D-1}{D-2})$ && $-\frac{3}{8}$ && $0$ &
\cr \omit&height2pt&\omit&&\omit&&\omit&&\omit&&\omit&\cr \tablerule
\omit&height2pt&\omit&&\omit&&\omit&&\omit&&\omit&\cr && 4 && C &&
$-\frac3{4(D-2)^2}$ && $\!\!\!\!\frac34 \frac{(D^2-6D+8)}{(D-2)^2
(D-3)^2} \!\!\!\!$ && $0$ & \cr
\omit&height2pt&\omit&&\omit&&\omit&&\omit&&\omit&\cr \tablerule
\omit&height2pt&\omit&&\omit&&\omit&&\omit&&\omit&\cr && 5 && A &&
$0$ && $0$ && $\!\!\!\!\frac12 (\frac{D-1}{D-3})\!\!\!\!$ & \cr
\omit&height2pt&\omit&&\omit&&\omit&&\omit&&\omit&\cr \tablerule
\omit&height2pt&\omit&&\omit&&\omit&&\omit&&\omit&\cr && 5 && B &&
$0$ && $0$ && $0$ & \cr
\omit&height2pt&\omit&&\omit&&\omit&&\omit&&\omit&\cr \tablerule
\omit&height2pt&\omit&&\omit&&\omit&&\omit&&\omit&\cr && 5 && C &&
$0$ && $0$ && $0$ & \cr
\omit&height2pt&\omit&&\omit&&\omit&&\omit&&\omit&\cr \tablerule
\omit&height2pt&\omit&&\omit&&\omit&&\omit&&\omit&\cr && 6 && A &&
$0$ && $0$ && $0$ & \cr
\omit&height2pt&\omit&&\omit&&\omit&&\omit&&\omit&\cr \tablerule
\omit&height2pt&\omit&&\omit&&\omit&&\omit&&\omit&\cr && 6 && B &&
$0$ && $0$ && $0$ & \cr
\omit&height2pt&\omit&&\omit&&\omit&&\omit&&\omit&\cr \tablerule
\omit&height2pt&\omit&&\omit&&\omit&&\omit&&\omit&\cr && 6 && C &&
$0$ && $0$ && $0$ & \cr
\omit&height2pt&\omit&&\omit&&\omit&&\omit&&\omit&\cr \tablerule
\omit&height2pt&\omit&&\omit&&\omit&&\omit&&\omit&\cr && 7 && A &&
$0$ && $0$ && $\!\!\!\! \frac{D (D-1)}4 \!-\! \frac12
(\frac{D-1}{D-3})\!\!\!\!$ & \cr
\omit&height2pt&\omit&&\omit&&\omit&&\omit&&\omit&\cr \tablerule
\omit&height2pt&\omit&&\omit&&\omit&&\omit&&\omit&\cr && 7 && B &&
$0$ && $0$ && $0$ & \cr
\omit&height2pt&\omit&&\omit&&\omit&&\omit&&\omit&\cr \tablerule
\omit&height2pt&\omit&&\omit&&\omit&&\omit&&\omit&\cr && 7 && C &&
$0$ && $0$ && $0$ & \cr
\omit&height2pt&\omit&&\omit&&\omit&&\omit&&\omit&\cr \tablerule
\omit&height2pt&\omit&&\omit&&\omit&&\omit&&\omit&\cr && 8 && A &&
$0$ && $0$ && $0$ & \cr
\omit&height2pt&\omit&&\omit&&\omit&&\omit&&\omit&\cr \tablerule
\omit&height2pt&\omit&&\omit&&\omit&&\omit&&\omit&\cr && 8 && B &&
$0$ && $0$ && $0$ & \cr
\omit&height2pt&\omit&&\omit&&\omit&&\omit&&\omit&\cr \tablerule
\omit&height2pt&\omit&&\omit&&\omit&&\omit&&\omit&\cr && 8 && C &&
$0$ && $0$ && $0$ & \cr
\omit&height2pt&\omit&&\omit&&\omit&&\omit&&\omit&\cr \tablerule}}

\label{4fin}

\end{table}

\subsection{Contributions from the 3-Point Vertices}

\begin{figure}
%\centerline{\epsfig{file=ffig2.eps}}
\includegraphics[width=4 in]{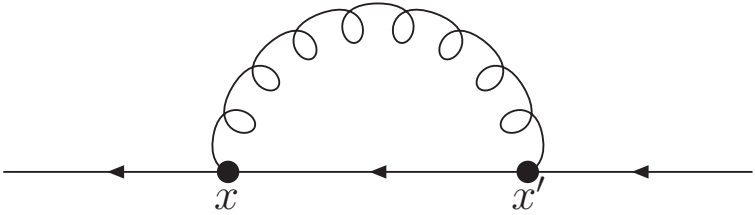}
\caption{Contribution from two 3-point vertices.} \label{ffig2}
\end{figure}

In this section we evaluate the contributions from two 3-point
vertex operators. The generic diagram topology is depicted in Figure
\ref{ffig2}. The analytic form is,
\begin{equation}
-i\Bigl[\mbox{}_i \Sigma_j^{\rm 3pt}\Bigr](x;x^{\p}) = \sum_{I=1}^3
iV^{\alpha\beta }_{Iik}(x) \, i\Bigl[\mbox{}_k
S_{\ell}\Bigr](x;x^{\p}) \sum_{J=1}^3 iV^{\rho\sigma }_{J\ell
j}(x^{\p}) \, i\Bigl[\mbox{}_{\alpha\beta}
\Delta_{\rho\sigma}\Bigr](x;x^{\p}) \; . \label{3ptloop}
\end{equation}

\begin{table}

\caption{Generic contributions from the 3-point vertices.}

\vbox{\tabskip=0pt \offinterlineskip
\def\tablerule{\noalign{\hrule}}
\halign to390pt {\strut#& \vrule#\tabskip=1em plus2em& \hfil#&
\vrule#& \hfil#& \vrule#& \hfil#\hfil& \vrule#\tabskip=0pt\cr
\tablerule \omit&height4pt&\omit&&\omit&&\omit&\cr &&\omit\hidewidth
{\rm I} &&\omit\hidewidth {\rm J} \hidewidth&& \omit\hidewidth
$iV_I^{\alpha\beta}(x) \, i[S](x;x^{\p}) \, i
V_J^{\rho\sigma}(x^{\p}) \, i[\mbox{}_{\alpha\beta}
\Delta_{\rho\sigma}](x;x^{\p})$ \hidewidth&\cr
\omit&height4pt&\omit&&\omit&&\omit&\cr \tablerule
\omit&height2pt&\omit&&\omit&&\omit&\cr && 1 && 1 && $\frac14
\kappa^2 \hspace{-.1cm} \not{\hspace{-.1cm} \partial}
\delta^D(x\!-\!x^{\p}) \,
i[^{\alpha}_{~\alpha}\Delta^{\rho}_{~\rho}](x;x)$ & \cr
\omit&height2pt&\omit&&\omit&&\omit&\cr \tablerule
\omit&height2pt&\omit&&\omit&&\omit&\cr && 1 && 2 && $-\frac14
\kappa^2 \gamma^{\rho} \partial^{\sigma} \delta^D(x\!-\!x^{\p}) \,
i[^{\alpha}_{~\alpha}\Delta_{\rho\sigma}](x;x)$ & \cr
\omit&height2pt&\omit&&\omit&&\omit&\cr \tablerule
\omit&height2pt&\omit&&\omit&&\omit&\cr && 1 && 3 && $\frac14 i
\kappa^2 \gamma^{\rho} J^{\sigma \mu} \delta^D(x\!-\!x^{\p}) \,
\partial^{\prime}_{\mu} i[^{\alpha}_{~\alpha}
\Delta_{\rho\sigma}](x;x^{\p})$ & \cr
\omit&height2pt&\omit&&\omit&&\omit&\cr \tablerule
\omit&height2pt&\omit&&\omit&&\omit&\cr && 2 && 1 && $\frac14
\kappa^2 \partial^{\prime}_{\mu} \{ \gamma^{\alpha}
\partial^{\beta} \, i[S](x;x^{\p}) \, \gamma^{\mu} \, i[\mbox{}_{\alpha\beta}
\Delta^{\rho}_{~\rho}](x;x^{\p}) \}$ & \cr
\omit&height2pt&\omit&&\omit&&\omit&\cr \tablerule
\omit&height2pt&\omit&&\omit&&\omit&\cr && 2 && 2 && $-\frac14
\kappa^2 \partial^{\prime \rho} \{ \gamma^{\alpha}
\partial^{\beta} \, i[S](x;x^{\p}) \, \gamma^{\sigma} \, i[\mbox{}_{\alpha\beta}
\Delta_{\rho\sigma}](x;x^{\p}) \}$ & \cr
\omit&height2pt&\omit&&\omit&&\omit&\cr \tablerule
\omit&height2pt&\omit&&\omit&&\omit&\cr && 2 && 3 && $-\frac14
i\kappa^2 \, \gamma^{\alpha} \partial^{\beta} \, i[S](x;x^{\p}) \,
\gamma^{\rho} J^{\sigma\mu} \partial^{\prime}_{\mu} \,
i[\mbox{}_{\alpha\beta} \Delta_{\rho\sigma}](x;x^{\p})$ & \cr
\omit&height2pt&\omit&&\omit&&\omit&\cr \tablerule
\omit&height2pt&\omit&&\omit&&\omit&\cr && 3 && 1 && $-\frac14 i
\kappa^2 \partial^{\prime}_{\nu} \{ \gamma^{\alpha} J^{\beta\mu} \,
i[S](x;x^{\p}) \, \gamma^{\nu} \partial_{\mu} \, i[\mbox{}_{\alpha
\beta} \Delta^{\rho}_{~\rho}](x;x^{\p}) \}$ & \cr
\omit&height2pt&\omit&&\omit&&\omit&\cr \tablerule
\omit&height2pt&\omit&&\omit&&\omit&\cr && 3 && 2 && $\frac14 i
\kappa^2 \partial^{\prime \rho} \{ \gamma^{\alpha} J^{\beta\mu} \,
i[S](x;x^{\p}) \, \gamma^{\sigma} \partial_{\mu} \,
i[\mbox{}_{\alpha\beta} \Delta_{\rho\sigma}](x;x^{\p}) \}$ & \cr
\omit&height2pt&\omit&&\omit&&\omit&\cr \tablerule
\omit&height2pt&\omit&&\omit&&\omit&\cr && 3 && 3 && $-\frac14
\kappa^2 \, \gamma^{\alpha} J^{\beta\mu} \, i[S](x;x^{\p}) \,
\gamma^{\rho} J^{\sigma\nu} \partial_{\mu} \partial^{\prime}_{\nu}
\, i[\mbox{}_{\alpha\beta} \Delta_{\rho\sigma}](x;x^{\p})$ & \cr
\omit&height2pt&\omit&&\omit&&\omit&\cr \tablerule}}

\label{gen3}

\end{table}

Because there are three 3-point vertex operators of Equation
\ref{3VO}, there are nine vertex products in Equation \ref{3ptloop}.
We label each contribution by the numbers on its vertex pair, for
example,
\begin{equation}
\Bigl[I\!\!-\!\!J\Bigr] \equiv iV_I^{\alpha\beta}(x) \times
i\Bigl[S\Bigr](x;x^{\p}) \times i V_J^{\rho\sigma}(x^{\p}) \times
i\Bigl[\mbox{}_{ \alpha\beta} \Delta_{\rho\sigma}\Bigr](x;x^{\p}) \;
.
\end{equation}
Table \ref{gen3}  gives the generic reductions, before decomposing
the graviton propagator. Most of these reductions are
straightforward but two subtleties deserve mention. First, the Dirac
slash of the fermion propagator gives a delta function,
\begin{equation}
i \hspace{-.1cm} \not{\hspace{-.1cm} \partial} i
\Bigl[S\Bigr](x;x^{\p}) = i \delta^D(x-x^{\p}) \; . \label{fpeqn}
\end{equation}
This occurs whenever the first vertex is $I\!=\!1$, for example,
\begin{eqnarray}
\Bigl[1\!\!-\!\!3\Bigr] & \equiv & \frac{i\kappa}2
\eta^{\alpha\beta} i \hspace{-.1cm} \not{\hspace{-.1cm} \partial}
\times i \Bigl[S\Bigr](x;x^{\p}) \times -\frac{i\kappa}2
\gamma^{\rho} J^{\sigma\mu} \partial^{\prime}_{\mu} \times
i\Bigl[\mbox{}_{\alpha\beta} \Delta_{\rho\sigma}\Bigr](x;x^{\p}) \;
,
\qquad \\
& = & \frac{i \kappa^2}4 \gamma^{\rho} J^{\sigma \mu}
\delta^D(x\!-\!x^{\p}) \,
\partial^{\prime}_{\mu} i \Bigl[ \mbox{}^{\alpha}_{~\alpha} \Delta_{\rho\sigma}
\Bigr](x;x^{\p}) \; .
\end{eqnarray}
The second subtlety is that derivatives on external lines must be
partially integrated back on the entire diagram. This happens
whenever the second vertex is $J\!=\!1$ or $J\!=\!2$, for example,
\begin{eqnarray}
\Bigl[2\!\!-\!\!2\Bigr]  \equiv  -\frac{i\kappa}2 \gamma^{\alpha} i
\partial^{\beta} \times i \Bigl[S\Bigr](x;x^{\p}) \times -\frac{i\kappa}2
\gamma^{\rho} i \partial^{\prime \sigma}_{\rm ext} \times
i\Bigl[\mbox{}_{ \alpha\beta} \Delta_{\rho\sigma}\Bigr](x;x^{\p}) \;
, \qquad\\
\longrightarrow  -\frac{\kappa^2}4
\partial^{\prime \sigma} \Biggl\{ \gamma^{\alpha}
\partial^{\beta} \, i\Bigl[S\Bigr](x;x^{\p}) \, \gamma^{\rho} \, i \Bigl[
\mbox{}_{\alpha\beta} \Delta_{\rho\sigma} \Bigr](x;x^{\p}) \Biggr\}
.
\end{eqnarray}

In comparing Table~\ref{gen3} and Table~\ref{4con} it will be seen
that the 3-point contributions with $I\!=\!1$ are closely related to
three of the 4-point contributions. In fact the $[1\!\!-\!\!1]$
contribution is $-2$ times the 4-point contribution with $I \!=\!
1$; while $[1\!\!-\!\!2]$ and $[1\!\!-\!\!3]$ cancel the 4-point
contributions with $I\!=\!3$ and $I\!=\!5$, respectively. Because of
this it is convenient to add the 3-point contributions with
$I\!=\!1$ to the 4-point contributions from Table~\ref{4fin},
\begin{eqnarray}
\lefteqn{-i \Bigl[\Sigma^{\rm 4pt} + \Sigma^{\rm
3pt}_{I=1}\Bigr](x;x^{\p}) = \frac{\kappa^2 H^{D-2}}{(4
\pi)^{\frac{D}2}} \frac{\Gamma(D\!-\! 1)}{\Gamma(\frac{D}2)}
\Biggl\{ \Bigl[- \frac{(D\!+\!1) (D\!-\!1) (D\!-\!4)}{2 (D\!-\!3)} A
}
\nonumber \\
& & - \frac{(D\!-\!1)(D^3 \!-\! 8 D^2 \!+\! 23 D \!-\! 32)}{8
(D\!-\!2)^2 (D\!-\!3)^2} \Bigr] \hspace{-.1cm} \not{\hspace{-.1cm}
\partial} +
\Bigl[ \frac34 \Bigl(D \!-\! \frac2{D\!-\!3}\Bigr) A \nonumber \\
& & \hspace{.18cm} + \frac{3 (D^2 \!-\! 6 D \!+\! 8)}{4 (D\!-\!2)^2
(D\!-\!3)^2} \!-\! \frac38 \Bigr] \; \hspace{-.1cm}
\overline{\not{\hspace{-.1cm} \partial}} +
\Bigl(\frac{D\!-\!1}4\Bigr) \Bigl(D \!-\! \frac2{D\!-\!3}\Bigr) a H
\gamma^0 \Biggr\} \delta^D(x\!-\!x^{\p}) . \;\qquad \label{1stcon}
\end{eqnarray}
In what follows we will focus on the 3-point contributions with
$I\!=\!2$ and $I\!=\!3$.

\subsection{Conformal Contributions}

The key to achieving a tractable reduction of the diagrams of Fig.~2
is that the first term of each of the scalar propagators
$i\Delta_I(x;x^{\p})$ is the conformal propagator $i\Delta_{\rm
cf}(x;x)$. The sum of the three index factors also gives a simple
tensor, so it is very efficient to write the graviton propagator in
the form,
\begin{eqnarray}
\lefteqn{i\Bigl[{}_{\mu\nu} \Delta_{\rho\sigma}\Bigr](x;x^{\p}) =
\Bigl[2 \eta_{\mu (\rho} \eta_{\sigma) \nu} - \frac2{D\!-\!2}
\eta_{\mu\nu}
\eta_{\rho\sigma}\Bigr] i\Delta_{\rm cf}(x;x^{\p}) } \nonumber \\
& & \hspace{6cm} + \sum_{I=A,B,C} \Bigl[\mbox{}_{\mu\nu}
T^I_{\rho\sigma} \Bigr] \, i{\delta \! \Delta}_I(x;x^{\p}) \; ,
\qquad
\end{eqnarray}
where $i{\delta \! \Delta}_I(x;x^{\p}) \equiv i\Delta_I(x;x^{\p}) -
i\Delta_{\rm cf}(x;x^{\p})$. In this subsection we evaluate the
contribution to Equation \ref{3ptloop} using the 3-point vertex
operators of Equation \ref{3VO} and the fermion propagator of
Equation \ref{fprop} but only the conformal part of the graviton
propagator,
\begin{equation}
i\Bigl[{}_{\mu\nu} \Delta_{\rho\sigma}\Bigr](x;x^{\p})
\longrightarrow \Bigl[2 \eta_{\mu (\rho} \eta_{\sigma) \nu} -
\frac2{D\!-\!2} \eta_{\mu\nu} \eta_{\rho\sigma}\Bigr] i\Delta_{\rm
cf}(x;x^{\p}) \equiv \Bigl[\mbox{}_{\alpha \beta} T^{\rm
cf}_{\rho\sigma}\Bigr] i\Delta_{\rm cf}(x;x) \; . \label{cfpart}
\end{equation}

We carry out the reduction in three stages. In the first stage the
conformal part \ref{cfpart} of the graviton propagator is
substituted into the generic results from Table \ref{gen3} and the
contractions are performed. We also make use of gamma matrix
identities such as Equation \ref{Jred} and,
\begin{equation}
\gamma^{\mu} i\Bigl[S\Bigr](x;x^{\p}) \gamma_{\mu} = (D\!-\!2)
i\Bigl[S\Bigr](x;x^{\p}) \qquad {\rm and} \qquad \gamma_{\alpha}
J^{\alpha \mu} = -\frac{i}2 (D\!-\!1) \gamma^{\mu} \; .
\end{equation}
Finally, we employ relation \ref{fpeqn} whenever $
\;\hspace{-.08cm}\not{ \hspace{-.15cm} \partial}$ acts upon the
fermion propagator. However, we do not at this stage act any other
derivatives. The results of these reductions are summarized in Table
\ref{Dcfcon}. Because the conformal tensor factor $[{}_{\alpha\beta}
T^{\rm cf}_{\rho\sigma}]$ contains three distinct terms, and because
the factors of $\gamma^{\alpha} J^{\beta \mu}$ in Table~\ref{gen3}
can contribute different terms with a distinct structure, we have
sometimes broken up the result for a given vertex pair into parts.
These parts are distinguished in Table~\ref{Dcfcon} and subsequently
by subscripts taken from the lower case Latin letters.

\begin{table}

\caption{Contractions from the $i\Delta_{\rm cf}$ part of the
graviton propagator.}

\vbox{\tabskip=0pt \offinterlineskip
\def\tablerule{\noalign{\hrule}}
\halign to390pt {\strut#& \vrule#\tabskip=1em plus2em& \hfil#\hfil&
\vrule#& \hfil#\hfil& \vrule#& \hfil#\hfil& \vrule#& \hfil#\hfil&
\vrule#\tabskip=0pt\cr \tablerule
\omit&height4pt&\omit&&\omit&&\omit&&\omit&\cr &&$\!\!\!\!{\rm
I}\!\!\!\!$ && $\!\!\!\!{\rm J}\!\!\!\!$ && $\!\!\!\!{\rm sub}
\!\!\!\!$ && $\!\!\!\!iV_I^{\alpha\beta}(x) \, i[S](x;x^{\p}) \, i
V_J^{\rho \sigma}(x^{\p}) \, [\mbox{}_{\alpha\beta} T^{\rm
cf}_{\rho\sigma}] \, i\Delta_{\rm cf}(x;x^{\p}) \!\!\!\!$ & \cr
\omit&height4pt&\omit&&\omit&&\omit&&\omit&\cr \tablerule
\omit&height2pt&\omit&&\omit&&\omit&&\omit&\cr && 2 && 1 && \omit &&
$-\frac1{D-2} \kappa^2 \hspace{-.1cm} \not{\hspace{-.1cm}
\partial}^{\p} \{ \delta^D(x\!-\!x^{\p}) \, i\Delta_{\rm cf}(x;x) \}$ & \cr
\omit&height2pt&\omit&&\omit&&\omit&&\omit&\cr \tablerule
\omit&height2pt&\omit&&\omit&&\omit&&\omit&\cr && 2 && 2 && a &&
$-\frac14 (\frac{D-4}{D-2}) \kappa^2 \hspace{-.1cm}
\not{\hspace{-.1cm} \partial}^{\p} \{ \delta^D(x\!-\!x^{\p}) \,
i\Delta_{\rm cf}(x;x) \}$ & \cr
\omit&height2pt&\omit&&\omit&&\omit&&\omit&\cr \tablerule
\omit&height2pt&\omit&&\omit&&\omit&&\omit&\cr && 2 && 2 && b && $-
(\frac{D-2}4) \kappa^2 \partial_{\mu}^{\prime} \{
\partial^{\mu} i[S](x;x^{\p}) \, i\Delta_{\rm cf}(x;x^{\p}) \}$ & \cr
\omit&height2pt&\omit&&\omit&&\omit&&\omit&\cr \tablerule
\omit&height2pt&\omit&&\omit&&\omit&&\omit&\cr && 2 && 3 && a &&
$\frac18 (\frac{D}{D-2}) \kappa^2 \delta^D(x\!-\!x^{\p})
\hspace{-.1cm} \not{\hspace{-.1cm} \partial}^{\p} \, i\Delta_{\rm
cf}(x;x)$ & \cr \omit&height2pt&\omit&&\omit&&\omit&&\omit&\cr
\tablerule \omit&height2pt&\omit&&\omit&&\omit&&\omit&\cr && 2 && 3
&& b && $+ (\frac{D-2}8) \kappa^2 \partial_{\mu} \, i[S](x;x^{\p})
\partial^{\prime \mu} \, i\Delta_{\rm cf}(x;x^{\p})$ & \cr
\omit&height2pt&\omit&&\omit&&\omit&&\omit&\cr \tablerule
\omit&height2pt&\omit&&\omit&&\omit&&\omit&\cr && 3 && 1 && \omit &&
$\frac12 (\frac{D-1}{D-2}) \kappa^2 \partial^{\prime}_{ \mu} \{ \,
\hspace{-.1cm} \not{\hspace{-.1cm}\partial} \, i\Delta_{\rm cf}(x;x)
\, i[S](x;x^{\p}) \gamma^{\mu} \}$ & \cr
\omit&height2pt&\omit&&\omit&&\omit&&\omit&\cr \tablerule
\omit&height2pt&\omit&&\omit&&\omit&&\omit&\cr && 3 && 2 && a && $-
\frac1{4(D-2)} \kappa^2 \partial^{\prime}_{\mu} \{ \, \hspace{-.1cm}
\not{\hspace{-.1cm} \partial} \, i\Delta_{\rm cf}(x;x) \,
i[S](x;x^{\p}) \gamma^{\mu} \}$ & \cr
\omit&height2pt&\omit&&\omit&&\omit&&\omit&\cr \tablerule
\omit&height2pt&\omit&&\omit&&\omit&&\omit&\cr && 3 && 2 && b &&
$-(\frac{D-2}8) \kappa^2 \partial^{\prime}_{\mu} \{ i[S](x;x^{\p})
\,
\partial^{\mu} i\Delta_{\rm cf}(x;x) \}$ & \cr
\omit&height2pt&\omit&&\omit&&\omit&&\omit&\cr \tablerule
\omit&height2pt&\omit&&\omit&&\omit&&\omit&\cr && 3 && 2 && c && $-
\frac18 \kappa^2 \hspace{-.1cm} \not{\hspace{-.1cm}
\partial}^{\p} \{ i[S](x;x^{\p}) \, \hspace{-.1cm} \not{\hspace{-.1cm} \partial} \,
i\Delta_{\rm cf}(x;x) \}$ & \cr
\omit&height2pt&\omit&&\omit&&\omit&&\omit&\cr \tablerule
\omit&height2pt&\omit&&\omit&&\omit&&\omit&\cr && 3 && 3 && a &&
$(\frac{D-2}{16}) \kappa^2 i[S](x;x^{\p}) \partial \!\cdot\!
\partial^{\p} i\Delta_{\rm cf}(x;x^{\p})$ & \cr
\omit&height2pt&\omit&&\omit&&\omit&&\omit&\cr \tablerule
\omit&height2pt&\omit&&\omit&&\omit&&\omit&\cr && 3 && 3 && b &&
$-\frac18 (\frac{2D-3}{D-2}) \kappa^2 \gamma^{\mu} i[S](x;x^{\p})
\partial_{\mu} \hspace{-.1cm} \not{\hspace{-.1cm} \partial}^{\p} \,
i\Delta_{\rm cf}(x;x)$ & \cr
\omit&height2pt&\omit&&\omit&&\omit&&\omit&\cr \tablerule
\omit&height2pt&\omit&&\omit&&\omit&&\omit&\cr && 3 && 3 && c && $+
\frac1{16} \kappa^2 \gamma^{\mu} i[S](x;x^{\p})
\partial^{\prime}_{\mu} \hspace{-.1cm} \not{\hspace{-.1cm} \partial} \,
i\Delta_{\rm cf}(x;x)$ & \cr
\omit&height2pt&\omit&&\omit&&\omit&&\omit&\cr \tablerule}}

\label{Dcfcon}

\end{table}

In the second stage we substitute the fermion and conformal
propagators,
\begin{equation}
i\Bigl[S\Bigr](x;x^{\p}) =  -\frac{i \Gamma(\frac{D}2)}{2
\pi^{\frac{D}2}} \frac{\gamma^{\mu} \Delta x_{\mu}}{\Delta x^D} \; ,
\end{equation}
\begin{equation}
 i\Delta_{\rm cf}(x;x^{\p}) = \frac{\Gamma(\frac{D}2
\!-\!1)}{4 \pi^{\frac{D}2}} \frac{(a a^{\p})^{1-\frac{D}2}}{\Delta
x^{D-2}} \; .
\end{equation}
At this stage we take advantage of the curious consequence of the
automatic subtraction of dimension regularization that any
dimension-dependent power of zero is discarded,
\begin{equation}
\lim_{x^{\p} \rightarrow x} i\Delta_{\rm cf}(x;x^{\p}) = 0 \qquad
{\rm and} \qquad \lim_{x^{\p} \rightarrow x} \partial^{\prime}_{\mu}
i\Delta_{\rm cf}(x;x^{\p}) = 0 \; .
\end{equation}

In the final stage we act the derivatives. These can act upon the
conformal coordinate separation $\Delta x^{\mu} \equiv x^{\mu} \!-\!
x^{\prime \mu}$, or upon the factor of $(a a^{\p})^{1-\frac{D}2}$
from the conformal propagator. We quote separate results for the
cases where all derivatives act upon the conformal coordinate
separation (Table~\ref{Dcfmost}) and the case where one or more of
the derivatives acts upon the scale factors (Table~\ref{Dcfless}).
In the former case the final result must in each case take the form
of a pure number times the universal factor,
\begin{equation}
\frac{(a a^{\p})^{1-\frac{D}2} \gamma^{\mu} \Delta x_{\mu}}{\Delta
x^{2D}} \; .
\end{equation}

\begin{table}
\caption{Conformal $i\Delta_{\rm cf}$ terms in which all derivatives
act upon $\Delta x^2(x;x^{\p})$. All contributions are multiplied by
$\frac{i \kappa^2}{8 \pi^D} \, \Gamma(\frac{D}2) \Gamma(\frac{D}2
\!-\! 1) (a a^{\p})^{1-\frac{D}2}$.}
\vbox{\tabskip=0pt
\offinterlineskip
\def\tablerule{\noalign{\hrule}}
\halign to390pt {\strut#& \vrule#\tabskip=1em plus2em& \hfil#\hfil&
\vrule#& \hfil#\hfil& \vrule#& \hfil#\hfil& \vrule#& \hfil#\hfil&
\vrule#\tabskip=0pt\cr \tablerule
\omit&height4pt&\omit&&\omit&&\omit&&\omit&\cr &&$\!\!\!\!{\rm
I}\!\!\!\!$ && $\!\!\!\!{\rm J}\!\!\!\!$ && $\!\!\!\!{\rm sub}
\!\!\!\!$ && $\!\!\!\!{\rm Coefficient\ of} \; \frac{\gamma^{\mu}
\Delta x_{\mu}}{\Delta x^{2D}} \!\!\!\!$ & \cr
\omit&height4pt&\omit&&\omit&&\omit&&\omit&\cr \tablerule
\omit&height2pt&\omit&&\omit&&\omit&&\omit&\cr && 2 && 1 && \omit &&
$0$ & \cr \omit&height2pt&\omit&&\omit&&\omit&&\omit&\cr \tablerule
\omit&height2pt&\omit&&\omit&&\omit&&\omit&\cr && 2 && 2 && a && $0$
& \cr \omit&height2pt&\omit&&\omit&&\omit&&\omit&\cr \tablerule
\omit&height2pt&\omit&&\omit&&\omit&&\omit&\cr && 2 && 2 && b &&
$-\frac14 (D\!-\!2)^2 (D\!-\!1)$ & \cr
\omit&height2pt&\omit&&\omit&&\omit&&\omit&\cr \tablerule
\omit&height2pt&\omit&&\omit&&\omit&&\omit&\cr && 2 && 3 && a && $0$
& \cr \omit&height2pt&\omit&&\omit&&\omit&&\omit&\cr \tablerule
\omit&height2pt&\omit&&\omit&&\omit&&\omit&\cr && 2 && 3 && b &&
$\frac18 (D\!-\!2)^2 (D\!-\!1)$ & \cr
\omit&height2pt&\omit&&\omit&&\omit&&\omit&\cr \tablerule
\omit&height2pt&\omit&&\omit&&\omit&&\omit&\cr && 3 && 1 && \omit &&
$-(D\!-\!1)^2$ & \cr \omit&height2pt&\omit&&\omit&&\omit&&\omit&\cr
\tablerule \omit&height2pt&\omit&&\omit&&\omit&&\omit&\cr && 3 && 2
&& a && $\frac12 (D\!-\!1)$ & \cr
\omit&height2pt&\omit&&\omit&&\omit&&\omit&\cr \tablerule
\omit&height2pt&\omit&&\omit&&\omit&&\omit&\cr && 3 && 2 && b &&
$-\frac18 (D\!-\!2)^2 (D\!-\!1)$ & \cr
\omit&height2pt&\omit&&\omit&&\omit&&\omit&\cr \tablerule
\omit&height2pt&\omit&&\omit&&\omit&&\omit&\cr && 3 && 2 && c &&
$\frac14 (D\!-\!2) (D\!-\!1)$ & \cr
\omit&height2pt&\omit&&\omit&&\omit&&\omit&\cr \tablerule
\omit&height2pt&\omit&&\omit&&\omit&&\omit&\cr && 3 && 3 && a && $0$
& \cr \omit&height2pt&\omit&&\omit&&\omit&&\omit&\cr \tablerule
\omit&height2pt&\omit&&\omit&&\omit&&\omit&\cr && 3 && 3 && b &&
$\frac14 (2D\!-\!3) (D\!-\!1)$ & \cr
\omit&height2pt&\omit&&\omit&&\omit&&\omit&\cr \tablerule
\omit&height2pt&\omit&&\omit&&\omit&&\omit&\cr && 3 && 3 && c &&
$-\frac18 (D\!-\!2) (D\!-\!1)$ & \cr
\omit&height2pt&\omit&&\omit&&\omit&&\omit&\cr \tablerule}}

\label{Dcfmost}

\end{table}

The sum of all terms in Table~\ref{Dcfmost} is,
\begin{equation}
-i \Bigl[ \Sigma^{T\ref{Dcfmost}}\Bigr](x;x^{\p}) = \frac{i
\kappa^2}{2^6 \pi^D} \Gamma\Bigl(\frac{D}2\Bigr)
\Gamma\Bigl(\frac{D}2 \!-\! 1\Bigr) (-2 D^2 \!+\! 5D \!-\! 4)
(D\!-\!1) (a a^{\p})^{1-\frac{D}2} \frac{\gamma^{\mu} \Delta
x_{\mu}}{\Delta x^{2D}} \; . \label{most}
\end{equation}
If one simply omits the factor of $(a a^{\p})^{1-\frac{D}2}$ the
result is the same as in flat space. Although Equation \ref{most} is
well defined for $x^{ \prime \mu} \!\neq\! x^{\mu}$ we must remember
that $[\Sigma](x;x^{\p})$ will be used inside an integral in the
quantum-corrected Dirac equation shown by Equation \ref{Diraceq}.
For that purpose the singularity at $x^{\prime \mu} \!=\! x^{\mu}$
is cubicly divergent in $D\!=\!4$ dimensions. To renormalize this
divergence we extract derivatives with respect to the coordinate
$x^{\mu}$, which can of course be taken outside the integral in
Equation \ref{Diraceq} to give a less singular integrand,
\begin{eqnarray}
\lefteqn{\frac{\gamma^{\mu} \Delta x_{\mu}}{\Delta x^{2D}} = \frac{
- \hspace{-.1cm} \not{\hspace{-.1cm}\partial}}{2(D\!-\!1)} \,
\Biggl\{\frac1{\Delta x^{2D-2}} \Biggr\} \; , } \\
& & = \frac{- \hspace{-.1cm} \not{\hspace{-.1cm}\partial} \,
\partial^2}{
4(D\!-\!1) (D\!-\!2)^2} \, \Bigl(\frac1{\Delta x^{2D-4}} \Bigr) \; , \\
& & = \frac{- \hspace{-.1cm} \not{\hspace{-.1cm}\partial} \,
\partial^4}{ 8(D\!-\!1) (D\!-\!2)^2 (D\!-\!3) (D\!-\!4)} \,
\Bigl(\frac1{\Delta x^{2D-6}} \Bigr) \; . \label{ds}
\end{eqnarray}

Expression \ref{ds} is integrable in four dimensions and we could
take $D\!=\!4$ except for the explicit factor of $1/(D\!-\!4)$. Of
course that is how ultraviolet divergences manifest in dimensional
regularization. We can segregate the divergence on a local term by
employing a simple representation for a delta function,
\begin{eqnarray}
\lefteqn{\frac{\partial^2}{D\!-\!4} \, \Bigl(\frac1{\Delta
x^{2D-6}}\Bigr) = \frac{\partial^2}{D\!-\!4} \,
\Biggl\{\frac1{\Delta x^{2D-6}} \!-\! \frac{\mu^{D-4}}{\Delta
x^{D-2}} \Biggr\} \!+\! \frac{i 4 \pi^{\frac{D}2}
\mu^{D-4}}{\Gamma(\frac{D}2 \!-\! 1)} \,
\frac{\delta^D(x\!-\!x^{\p})}{D\!-\!4}  ,\;}
\\
& & \hspace{1.8cm} = -\frac{\partial^2}2 \Biggl\{ \frac{\ln(\mu^2
\Delta x^2)}{ \Delta x^2} \!+\! O(D\!-\!4) \Biggr\} + \frac{i 4
\pi^{\frac{D}2} \mu^{D-4}}{ \Gamma(\frac{D}2 \!-\! 1)} \,
\frac{\delta^D(x\!-\!x^{\p})}{D\!-\!4}  .  \;\qquad
\end{eqnarray}
The final result for Table~\ref{Dcfmost} is,
\begin{eqnarray}
\lefteqn{-i \Bigl[ \Sigma^{T\ref{Dcfmost}}\Bigr](x;x^{\p}) =
-\frac{i \kappa^2}{2^8 \pi^4} \frac1{a a^{\p}} \, \hspace{-.1cm}
\not{\hspace{-.1cm}\partial} \,\partial^4 \Biggl\{ \frac{\ln(\mu^2
\Delta x^2)}{\Delta x^2} \Biggl\} + \, O(D\!-\!4) }
\nonumber \\
& & \hspace{1.6cm} - \frac{\kappa^2 \mu^{D-4}}{2^8 \pi^{\frac{D}2}}
\Gamma\Bigl(\frac{D}2\!-\!1 \Bigr) \frac{(2 D^2 \!-\! 5D \!+\! 4) (a
a^{\p})^{1-\frac{D}2}}{(D\!-\!2) (D\!-\!3) (D\!-\!4)} \hspace{-.1cm}
\not{\hspace{-.1cm} \partial} \, \partial^2 \delta^D(x\!-\!x^{\p})
.\; \qquad \label{2ndcon}
\end{eqnarray}

When one or more derivative acts upon the scale factors a
bewildering variety of spacetime and gamma matrix structures result.
For example, the $[3\!\!-\!\!2]_b$ term gives,
\begin{eqnarray}
\lefteqn{-\Bigl(\frac{D\!-\!2}8\Bigr) \kappa^2 \partial_{\mu}^{\p}
\Biggl\{ i\Bigl[S\Bigr](x;x^{\p}) \partial^{\mu} i\Delta_{\rm
cf}(x;x^{\p}) \Biggr\} }
\nonumber \\
& & \hspace{-.3cm} = \frac{i \kappa^2}{32 \pi^D}
\Gamma^2\Bigl(\frac{D}2\Bigr)
\partial_{\mu}^{\p} \Biggl\{ \frac{\gamma^{\nu} \Delta x_{\nu}}{\Delta x^D}
(a a^{\p})^{1-\frac{D}2} \Biggl[ -\frac{(D\!-\!2) \Delta
x^{\mu}}{\Delta x^D} + \frac{(D\!-\!2) H a \delta^{\mu}_0}{2 \Delta
x^{D-2}} \Biggr] \Biggr\} ,\;
\qquad \\
& & \hspace{-.3cm} = \frac{i \kappa^2}{32 \pi^D}
\Gamma^2\Bigl(\frac{D}2\Bigr) (a a^{\p})^{1-\frac{D}2} \Biggl\{
-\frac{(D\!-\!1)(D\!-\!2) \gamma^{\mu} \Delta x_{\mu}}{\Delta
x^{2D}} + \frac{(D\!-\!2) H a \gamma^0}{2 \Delta x^{2D-2}}
\qquad \nonumber \\
& & \hspace{2cm} + \frac{(D\!-\!2)^2 a^{\p} H \Delta \eta
\gamma^{\mu} \Delta x_{\mu}}{2 \Delta x^{2D}} - \frac{(D\!-\!1)
(D\!-\!2) a H \Delta
\eta \gamma^{\mu} \Delta x_{\mu}}{\Delta x^{2D}} \nonumber \\
& & \hspace{7.4cm} - \frac{(D\!-\!2)^2 a a^{\p} H^2 \gamma^{\mu}
\Delta x_{\mu}}{4 \Delta x^{2D-2}} \Biggr\} .\; \qquad \label{32b}
\end{eqnarray}

The first term of Equation \ref{32b} originates from both
derivatives acting on the conformal coordinate separation. It
belongs in Table~\ref{Dcfmost}. The next three terms come from a
single derivative acting on a scale factor, and the final term in
Equation \ref{32b} derives from both derivatives acting upon scale
factors. These last four terms belong in Table~\ref{Dcfless}. They
can be expressed as dimensionless functions of $D$, $a$ and $a^{\p}$
times three basic terms,
\begin{eqnarray}
\lefteqn{\frac{i \kappa^2}{16 \pi^D} \Gamma^2\Bigl(\frac{D}2\Bigr)
(a a^{\p})^{1 -\frac{D}2} \Biggl\{ -\frac18 (D\!-\!2)^2 \times
\frac{a a^{\p} H^2 \gamma^{\mu} \Delta x_{\mu}}{\Delta x^{2D-2}} +
\frac14 (D\!-\!2) a \times \frac{H
\gamma^0}{\Delta x^{2D-2}} } \nonumber \\
& & \hspace{2.7cm} + \Bigl[ \frac14 (D\!-\!2)^2 a^{\p} \!-\! \frac12
(D\!-\!1) (D\!-\!2) a \Bigr] \times \frac{H \Delta \eta \gamma^{\mu}
\Delta x_{\mu}}{ \Delta x^{2D}} \Biggr\} .\; \qquad
\end{eqnarray}

\begin{table}
\caption{Conformal $i\Delta_{\rm cf}$ terms in which some
derivatives act upon scale factors. All contributions are multiplied
by $\frac{i \kappa^2}{16 \pi^D} \, \Gamma^2(\frac{D}2) (a
a^{\p})^{1-\frac{D}2}$.}

\vbox{\tabskip=0pt \offinterlineskip
\def\tablerule{\noalign{\hrule}}
\halign to390pt {\strut#& \vrule#\tabskip=1em plus2em& \hfil#\hfil&
\vrule#& \hfil#\hfil& \vrule#& \hfil#\hfil& \vrule#& \hfil#\hfil&
\vrule#& \hfil#\hfil& \vrule#& \hfil#\hfil& \vrule#\tabskip=0pt\cr
\tablerule
\omit&height4pt&\omit&&\omit&&\omit&&\omit&&\omit&&\omit&\cr
&&$\!\!\!\!{\rm I}\!\!\!\!$ && $\!\!\!\!{\rm J}\!\!\!\!$ &&
$\!\!\!\!{\rm sub} \!\!\!\!$ && $\!\!\!\!\frac{a a^{\p}
H^2\gamma^{\mu} \Delta x_{\mu}}{\Delta x^{2D-2}}\!\!\!\!$ &&
$\!\!\!\!\frac{H \gamma^0}{\Delta x^{2D-2}}\!\!\!\!$ &&
$\!\!\!\!\frac{H \Delta \eta \, \gamma^{\mu} \Delta x_{\mu}}{\Delta
x^{2D}} \!\!\!\!$ & \cr
\omit&height4pt&\omit&&\omit&&\omit&&\omit&&\omit&&\omit&\cr
\tablerule
\omit&height2pt&\omit&&\omit&&\omit&&\omit&&\omit&&\omit&\cr && 2 &&
1 && \omit && $0$ && $0$ && $0$ & \cr
\omit&height2pt&\omit&&\omit&&\omit&&\omit&&\omit&&\omit&\cr
\tablerule
\omit&height2pt&\omit&&\omit&&\omit&&\omit&&\omit&&\omit&\cr && 2 &&
2 && a && $0$ && $0$ && $0$ & \cr
\omit&height2pt&\omit&&\omit&&\omit&&\omit&&\omit&&\omit&\cr
\tablerule
\omit&height2pt&\omit&&\omit&&\omit&&\omit&&\omit&&\omit&\cr && 2 &&
2 && b && $0$ && $-\frac12 (D\!-\!2) a^{\p}$ && $\frac12 (D\!-\!2) D
a^{\p}$ & \cr
\omit&height2pt&\omit&&\omit&&\omit&&\omit&&\omit&&\omit&\cr
\tablerule
\omit&height2pt&\omit&&\omit&&\omit&&\omit&&\omit&&\omit&\cr && 2 &&
3 && a && $0$ && $0$ && $0$ & \cr
\omit&height2pt&\omit&&\omit&&\omit&&\omit&&\omit&&\omit&\cr
\tablerule
\omit&height2pt&\omit&&\omit&&\omit&&\omit&&\omit&&\omit&\cr && 2 &&
3 && b && $0$ && $\frac14 (D\!-\!2) a^{\p}$ && $-\frac14 (D\!-\!2) D
a^{\p}$ & \cr
\omit&height2pt&\omit&&\omit&&\omit&&\omit&&\omit&&\omit&\cr
\tablerule
\omit&height2pt&\omit&&\omit&&\omit&&\omit&&\omit&&\omit&\cr && 3 &&
1 && \omit && $\frac12 (D\!-\!1)$ && $0$ && $0$ & \cr
\omit&height2pt&\omit&&\omit&&\omit&&\omit&&\omit&&\omit&\cr
\tablerule
\omit&height2pt&\omit&&\omit&&\omit&&\omit&&\omit&&\omit&\cr && 3 &&
2 && a && $-\frac14$ && $0$ && $0$ & \cr
\omit&height2pt&\omit&&\omit&&\omit&&\omit&&\omit&&\omit&\cr
\tablerule
\omit&height2pt&\omit&&\omit&&\omit&&\omit&&\omit&&\omit&\cr && 3 &&
2 && b && $-\frac18 (D\!-\!2)^2$ && $\frac14 (D\!-\!2) a$ &&
$\frac14 (D\!-\!2)^2 a^{\p}$ & \cr
\omit&height4pt&\omit&&\omit&&\omit&&\omit&&\omit&&\omit&\cr &&
\omit && \omit && \omit && \omit && \omit && $\!\!\!\!-\frac12
(D\!-\!2) (D\!-\!1) a\!\!\!\!$ & \cr
\omit&height2pt&\omit&&\omit&&\omit&&\omit&&\omit&&\omit&\cr
\tablerule
\omit&height2pt&\omit&&\omit&&\omit&&\omit&&\omit&&\omit&\cr && 3 &&
2 && c && $-\frac18 (D\!-\!2)$ && $0$ && $0$ & \cr
\omit&height2pt&\omit&&\omit&&\omit&&\omit&&\omit&&\omit&\cr
\tablerule
\omit&height2pt&\omit&&\omit&&\omit&&\omit&&\omit&&\omit&\cr && 3 &&
3 && a && $\frac1{16} (D\!-\!2)^2$ && $0$ && $\frac18 (D\!-\!2)^2 (a
\!-\! a^{\p})$ & \cr
\omit&height2pt&\omit&&\omit&&\omit&&\omit&&\omit&&\omit&\cr
\tablerule
\omit&height2pt&\omit&&\omit&&\omit&&\omit&&\omit&&\omit&\cr && 3 &&
3 && b && $-\frac18 (2D\!-\!3)$ && $0$ && $0$ & \cr
\omit&height2pt&\omit&&\omit&&\omit&&\omit&&\omit&&\omit&\cr
\tablerule
\omit&height2pt&\omit&&\omit&&\omit&&\omit&&\omit&&\omit&\cr && 3 &&
3 && c && $\frac1{16} (D\!-\!2)$ && $0$ && $0$ & \cr
\omit&height2pt&\omit&&\omit&&\omit&&\omit&&\omit&&\omit&\cr
\tablerule}}

\label{Dcfless}

\end{table}

These three terms turn out to be all we need, although intermediate
expressions sometimes show other kinds. An example is the
$[3\!\!-\!\!1]$ term,
\begin{eqnarray}
\lefteqn{\frac12 \Bigl(\frac{D\!-\!1}{D\!-\!2}\Bigr) \kappa^2
\partial_{\mu}^{\p} \Biggl\{\; \hspace{-.1cm}
\not{\hspace{-.1cm}\partial} \, i\Delta_{\rm cf}(x;x)
\, i\Bigl[S\Bigr](x;x^{\p}) \gamma^{\mu} \Biggr\} } \nonumber \\
& & \hspace{-.3cm} = \frac{i \kappa^2}{8 \pi^D}
\Gamma^2\Bigl(\frac{D}2\Bigr) \Bigl(\frac{D\!-\!1}{D\!-\!2}\Bigr)
\partial_{\mu}^{\p} \Biggl\{ (a a^{\p})^{1- \frac{D}2} \Biggl[
\frac{\gamma^{\alpha} \Delta x_{\alpha}}{\Delta x^D} \!+\! \frac{a H
\gamma^0}{2 \Delta x^{D-2}} \Biggr] \frac{\gamma^{\beta} \Delta
x_{\beta}}{\Delta x^D} \gamma^{\mu} \Biggr\} , \qquad \\
& & \hspace{-.3cm} = \frac{i \kappa^2}{8 \pi^D}
\Gamma^2\Bigl(\frac{D}2\Bigr) (a a^{\p})^{1-\frac{D}2} \Biggl\{ -2
\frac{(D\!-\!1)^2}{(D\!-\!2)} \frac{\gamma^{\mu} \Delta
x_{\mu}}{\Delta x^{2D}} \!-\! \frac12 (D\!-\!1)
\frac{a H \gamma^0}{\Delta x^{2D-2}} \nonumber \\
& & \hspace{2.9cm} + \frac12 (D\!-\!1) \frac{a^{\p} H
\gamma^0}{\Delta x^{2D-2}} -\frac14 (D\!-\!1) \frac{a a^{\p} H^2
\gamma^0 \gamma^{\mu} \Delta x_{\mu} \gamma^0}{\Delta x^{2D-2}}
\Biggr\} .\; \qquad \label{31}
\end{eqnarray}
As before, the first term in Equation \ref{31} belongs in
Table~{\ref{Dcfmost}. The second and third terms are of a type we
encountered in Equation \ref{32b} but the final term is not.
However, it is simple to bring this term to standard form by
anti-commuting the $\gamma^{\mu}$ through either $\gamma^0$,
\begin{eqnarray}
a a^{\p} H^2 \gamma^0 \gamma^{\mu} \Delta x_{\mu} \gamma^0 & = &
-a a^{\p} H^2 \gamma^{\mu} \Delta x_{\mu} - 2 a a^{\p} H^2 \Delta \eta \gamma^0 \; ,\\
& = & - a a^{\p} H^2 \gamma^{\mu} \Delta x_{\mu} - 2 (a \!-\!
a^{\p}) H \gamma^0 \; .
\end{eqnarray}
Note our use of the identity $(a \!-\! a^{\p}) = a a^{\p} H \Delta
\eta$.

When all terms in Table~\ref{Dcfless} are summed it emerges that a
factor of $ H^2 a a^{\p}$ can be extracted,
\begin{eqnarray}
\lefteqn{ -i \Bigl[\Sigma^{T\ref{Dcfless}} \Bigr](x;x^{\p}) =
\frac{i \kappa^2}{ 16 \pi^D} \Gamma^2\Bigl(\frac{D}2\Bigr) (a
a^{\p})^{1-\frac{D}2} \Biggl\{ -\frac1{16} (D^2 \!-\! 7D \!+\! 8)
\!\times\! \frac{a a^{\p} H^2 \gamma^{\mu}
\Delta x_{\mu}}{\Delta x^{2D-2}} } \nonumber \\
& & \hspace{-.4cm} + \frac14 (D\!-\!2) (a \!-\! a^{\p}) \!\times\!
\frac{H \gamma^0}{\Delta x^{2D-2}} \!-\! \frac18 (D\!-\!2)
(3D\!-\!2) (a \!-\! a^{\p}) \!\times\! \frac{H \!\Delta \eta
\gamma^{\mu} \Delta x_{\mu}}{\Delta x^{2D}}\!
\Biggr\} , \qquad \\
& & \hspace{-.7cm} = \frac{i \kappa^2 H^2}{16 \pi^D}
\Gamma^2\Bigl(\frac{D}2 \Bigr) (a a^{\p})^{2-\frac{D}2}
\Biggl\{-\frac1{16} (D^2 \!-\! 7D \!+\! 8) \times
\frac{\gamma^{\mu} \Delta x_{\mu}}{\Delta x^{2D-2}} \nonumber \\
& & \hspace{1.5cm} + \frac14 (D\!-\!2) \times \frac{\gamma^0 \Delta
\eta}{ \Delta x^{2D-2}} \!-\! \frac18 (D\!-\!2) (3D\!-\!2) \times
\frac{\Delta \eta^2 \gamma^{\mu} \Delta x_{\mu}}{\Delta x^{2D}}
\Biggr\} . \label{T8fin}
\end{eqnarray}
Note the fact that this expression is odd under interchange of
$x^{\mu}$ and $x^{\prime \mu}$. Although individual contributions to
the last two columns of Table~\ref{Dcfless} are not odd under
interchange, their sum always produces a factor of $a \!-\! a^{\p}
\!=\! a a^{\p} H \Delta \eta$ which makes Equation \ref{T8fin} odd.

Expression \ref{T8fin} can be simplified using the differential
identities,
\begin{eqnarray}
\frac{\Delta \eta^2 \gamma^{\mu} \Delta x_{\mu}}{\Delta x^{2D}} & =
& \frac{\partial_0^2}{4 (D\!-\!2) (D\!-\!1)} \Bigl(
\frac{\gamma^{\mu}
\Delta x_{\mu}}{\Delta x^{2D-4}}\Bigr) \nonumber \\
& & \hspace{2cm} - \frac1{2(D\!-\!1)} \, \frac{\gamma^{\mu} \Delta
x_{\mu}}{ \Delta x^{2D-2}} + \frac1{D\!-\!1} \, \frac{\gamma^0
\Delta \eta}{\Delta
x^{2D-2}} \; , \qquad \\
\frac{\gamma^0\Delta \eta}{\Delta x^{2D-2}} & = & \frac{\gamma^0
\partial_0}{2 (D\!-\!2)} \, \Bigl(\frac1{\Delta x^{2D-4}}\Bigr) \; .
\end{eqnarray}
The result is,
\begin{eqnarray}
\lefteqn{ -i \Bigl[\Sigma^{T\ref{Dcfless}} \Bigr](x;x^{\p}) =
\frac{i \kappa^2 H^2 }{16 \pi^D} \Gamma^2\Bigl(\frac{D}2\Bigr) (a
a^{\p})^{2-\frac{D}2} \Biggl\{- \frac{(D^3 \!-\! 11 D^2 \!+\! 23D
\!-\! 12)}{16 (D\!-\!1)} \,
\frac{\gamma^{\mu} \Delta x_{\mu}}{\Delta x^{2D-2}} } \nonumber \\
& & \hspace{1.7cm} - \frac{D}{16 (D\!-\!1)} \, \gamma^0 \partial_0
\Bigl(\frac1{\Delta x^{2D-4}}\Bigr) - \frac1{32} \Bigl(\frac{3D
\!-\! 2}{ D \!-\!1}\Bigr) \partial_0^2 \Bigl(\frac{\gamma^{\mu}
\Delta x_{\mu}}{\Delta x^{2D-4}}\Bigr) \Biggr\} . \;\qquad
\label{T8b}
\end{eqnarray}
We now exploit partial integration identities of the same type as
those previously used for Table~\ref{Dcfmost},
\begin{eqnarray}
\lefteqn{\frac{\gamma^{\mu} \Delta x_{\mu}}{\Delta x^{2D-4}} =
\frac{- \hspace{-.1cm} \not{\hspace{-.1cm}\partial}}{2 (D\!-\!3)} \,
\Bigl(\frac1{ \Delta x^{2D-6}}\Bigr) = -\frac{\hspace{-.1cm}
\not{\hspace{-.1cm}\partial}}2
\Bigl( \frac1{\Delta x^2} \Bigr) + O(D\!-\!4) \; , } \\
\lefteqn{\frac{\gamma^{\mu} \Delta x_{\mu}}{\Delta x^{2D-2}} =
\frac{- \hspace{-.1cm} \not{\hspace{-.1cm}\partial} \, \partial^2}{4
(D\!-\!2)
(D\!-\!3) (D\!-\!4)} \, \Bigl(\frac1{\Delta x^{2D-6}}\Bigr) \; , } \nonumber \\
& & = \frac{\hspace{-.1cm} \not{\hspace{-.1cm}\partial} \,
\partial^2}{16} \Biggl\{ \frac{\ln(\mu^2 \Delta x^2)}{\Delta x^2}
\Biggr\} + O(D\!-\!4) - \frac{i \pi^{\frac{D}2} \mu^{D-4}}{2
\Gamma(\frac{D}2)} \, \frac{\hspace{-.1cm}
\not{\hspace{-.1cm}\partial} \, \delta^D(x\!-\!x^{\p})}{
(D\!-\!3) (D\!-\!4)} \; , \\
\lefteqn{\frac1{\Delta x^{2D-4}} = \frac{\partial^2}{2 (D\!-\!3)
(D\!-\!4)}
\Bigl(\frac1{\Delta x^{2D-6}}\Bigr) \; ,} \nonumber \\
& & = -\frac{\partial^2}4 \Biggl\{\frac{\ln(\mu^2 \Delta
x^2)}{\Delta x^2} \Biggr\} \!+\! O(D\!-\!4) \!+\! \frac{i 2
\pi^{\frac{D}2} \mu^{D-4}}{ \Gamma(\frac{D}2 \!-\!1)} \,
\frac{\delta^D(x\!-\!x^{\p})}{(D\!-\!3)(D\!-\!4)} \; . \qquad
\end{eqnarray}
It is also useful to convert temporal derivatives to spatial ones
using,
\begin{equation}
\gamma^0 \partial_0 = \; \hspace{-.1cm} \not{\hspace{-.1cm}
\partial} - \hspace{-.1cm} \overline{\not{\hspace{-.1cm} \partial} }
\qquad {\rm and} \qquad \partial_0^2 = \nabla^2 - \partial^2 \; .
\end{equation}
Substituting these relations in Equation \ref{T8b} gives,
\begin{eqnarray}
\lefteqn{ -i \Bigl[\Sigma^{T\ref{Dcfless}}\Bigr](x;x^{\p}) =
\frac{\kappa^2 H^2 \mu^{D-4} \Gamma(\frac{D}2)\, (a
a^{\p})^{2-\frac{D}2} }{2^9 \pi^{\frac{D}2} (D\!-\!1) (D\!-\!3)
(D\!-\!4)} \Biggl\{-\Bigl(D^3 \!-\! 13 D^2 \!+\! 27 D \!-\! 12
\Bigr) \hspace{-.1cm} \not{\hspace{-.1cm} \partial}}
\nonumber \\
& & \hspace{1cm} -2 D (D\!-\!2) \, \hspace{-.1cm}
\overline{\not{\hspace{-.1cm}
\partial} } \Biggr\} \delta^D(x\!-\!x^{\p}) + \frac{i \kappa^2 H^2}{2^9 \!\cdot\!
3 \!\cdot\! \pi^4} \Biggl\{ \Bigl[6 \, \hspace{-.1cm}
\not{\hspace{-.1cm}
\partial} \, \partial^2 \!-\! 2 \; \hspace{-.1cm} \overline{\not{\hspace{-.1cm}
\partial} } \, \partial^2 \Bigr] \Bigl( \frac{\ln(\mu^2 \Delta x^2)}{\Delta
x^2} \Bigr) \nonumber \\
& & \hspace{5cm} + 5 \! \not{\hspace{-.1cm}\partial} \, (\nabla^2
\!-\!
\partial^2) \Bigl(\frac1{\Delta x^2} \Bigr) \Biggr\} \!+\! O(D\!-\!4) \; .
\qquad \label{3rdcon}
\end{eqnarray}

\subsection{Sub-Leading Contributions from $i \delta \Delta_{A} $ }

\begin{table}
\caption{Contractions from the $i\d\!\D_{A}$ part of the graviton
propagator}

\vbox{\tabskip=0pt \offinterlineskip
\def\tablerule{\noalign{\hrule}}
\halign to390pt {\strut#& \vrule#\tabskip=1em plus2em& \hfil#&
\vrule#& \hfil#& \vrule#& \hfil#& \vrule#& \hfil#\hfil&
\vrule#\tabskip=0pt\cr \tablerule
\omit&height4pt&\omit&&\omit&&\omit&&\omit&\cr &&\hidewidth {\rm I}
&&\hidewidth {\rm J} \hidewidth&& \hidewidth {\rm sub} \hidewidth&&
\hidewidth $iV_I^{\alpha\beta}(x) \, i[S](x;x^{\p}) \, i
V_J^{\rho\sigma}(x^{\p}) \, [\mbox{}_{\alpha\beta} T^A_{\rho\sigma}]
\, i\delta\!\Delta_A(x;x^{\p})$ \hidewidth&\cr
\omit&height4pt&\omit&&\omit&&\omit&&\omit&\cr \tablerule
\omit&height2pt&\omit&&\omit&&\omit&&\omit&\cr && 2 && 1 && \omit &&
$-\f{1}{(D-3)}\ka^2\del^{^{\p}}_{\mu}\{
\hspace{-.1cm}\not{\hspace{-.1cm}\bar{\del}}
i[S](x;x^{\p})\g^{\mu}i\d\!\D_{A}(x;x^{\p})\}$ & \cr
\omit&height2pt&\omit&&\omit&&\omit&&\omit&\cr \tablerule
\omit&height2pt&\omit&&\omit&&\omit&&\omit&\cr && 2 && 2 && a && $
\f{1}{4}\ka^2\hspace{-.1cm} \not{\hspace{-.1cm}\bar{\del}}\{\del_{k}
i[S](x;x^{\p})\g_{k}i\d\!\D_{A}(x;x^{\p})\}$ & \cr
\omit&height2pt&\omit&&\omit&&\omit&&\omit&\cr \tablerule
\omit&height2pt&\omit&&\omit&&\omit&&\omit&\cr && 2 && 2 && b && $ +
\f{1}{4}\ka^2\del_{\ell}\{\g_{k}\del_{\ell}
i[S](x;x^{\p})\g_{k}i\d\!\D_{A}(x;x^{\p}) \}$ & \cr
\omit&height2pt&\omit&&\omit&&\omit&&\omit&\cr \tablerule
\omit&height2pt&\omit&&\omit&&\omit&&\omit&\cr && 2 && 2 && c && $-
\f{1}{2 (D-3)}\ka^2\del_{k}\{
\hspace{-.1cm}\not{\hspace{-.1cm}\bar{\del}}
i[S](x;x^{\p})\g_{k}i\d\!\D_{A}(x;x^{\p})\}$ & \cr
\omit&height2pt&\omit&&\omit&&\omit&&\omit&\cr \tablerule
\omit&height2pt&\omit&&\omit&&\omit&&\omit&\cr && 2 && 3 && a &&
$\f1{2 (D-3)}\ka^2 \hspace{-.1cm}\not{\hspace{-.1cm}\bar{\del}}
i[S](x;x^{\p}) \, {\hspace{-.1cm}\not{\hspace{-.1cm}\del}}^{\p}
i\d\!\D_{A}(x;x^{\p}) $ & \cr
\omit&height2pt&\omit&&\omit&&\omit&&\omit&\cr \tablerule
\omit&height2pt&\omit&&\omit&&\omit&&\omit&\cr && 2 && 3 && b &&
$-\f{1}{4}\ka^2\g_{k}\del_{\ell}i[S](x;x^{\p})
\g_{(k}\del_{\ell)}i\d\!\D_{A}(x;x^{\p}) $ & \cr
\omit&height2pt&\omit&&\omit&&\omit&&\omit&\cr \tablerule
\omit&height2pt&\omit&&\omit&&\omit&&\omit&\cr && 2 && 3 && c && $ +
\f{1}{4 (D-3)}\ka^2
\hspace{-.1cm}\not{\hspace{-.1cm}\bar{\del}}i[S](x;x^{\p})
\hspace{-.1cm}\not{\hspace{-.1cm}\bar{\del}}i\d\!\D_{A}(x;x^{\p})$&\cr
\omit&height2pt&\omit&&\omit&&\omit&&\omit&\cr \tablerule
\omit&height2pt&\omit&&\omit&&\omit&&\omit&\cr && 3 && 1 && a &&
$\f{1}{2} (\f{D-1}{D-3}) \ka^2\del^{\p}_{\mu}
\{\hspace{-.1cm}\not{\hspace{-.1cm}\del}
i\d\!\D_{A}(x;x^{\p})i[S](x;x^{\p})\g^{\mu}\} $ &\cr
\omit&height2pt&\omit&&\omit&&\omit&&\omit&\cr \tablerule
\omit&height2pt&\omit&&\omit&&\omit&&\omit&\cr && 3 && 1 && b &&
$-\f{1}{2 (D-3)}\ka^2\del^{\p}_{\mu}
\{\hspace{-.1cm}\not{\hspace{-.1cm}\bar{\del}}
i\d\!\D_{A}(x;x^{\p})i[S](x;x^{\p})\g^{\mu}\} $ & \cr
\omit&height2pt&\omit&&\omit&&\omit&&\omit&\cr \tablerule
\omit&height2pt&\omit&&\omit&&\omit&&\omit&\cr && 3 && 2 && a &&
$\f{1}{2 (D-3)}\ka^2\del_{k}
\{\hspace{-.1cm}\not{\hspace{-.1cm}\del}
i\d\!\D_{A}(x;x^{\p})i[S](x;x^{\p})\g_{k}\}$ & \cr
\omit&height2pt&\omit&&\omit&&\omit&&\omit&\cr \tablerule
\omit&height2pt&\omit&&\omit&&\omit&&\omit&\cr && 3 && 2 && b &&
$-\f{1}{4 (D-3)}\ka^2\del_{k}
\{\hspace{-.1cm}\not{\hspace{-.1cm}\bar{\del}}
i\d\!\D_{A}(x;x^{\p})i[S](x;x^{\p})\g_{k}\} $ & \cr
\omit&height2pt&\omit&&\omit&&\omit&&\omit&\cr \tablerule
\omit&height2pt&\omit&&\omit&&\omit&&\omit&\cr && 3 && 2 && c && $ +
\f{1}{8}\ka^2\hspace{-.1cm}
\not{\hspace{-.1cm}\bar{\del}}\{i[S](x;x^{\p})
\hspace{-.1cm}\not{\hspace{-.1cm}\bar{\del}} i\d\!\D_{A}(x;x^{\p})\}
$ & \cr \omit&height2pt&\omit&&\omit&&\omit&&\omit&\cr \tablerule
\omit&height2pt&\omit&&\omit&&\omit&&\omit&\cr && 3 && 2 && d && $ +
\f{1}{8}\ka^2\del_{k}\{\g_{\ell}
i[S](x;x^{\p})\g_{\ell}\del_{k}i\d\!\D_{A}(x;x^{\p})\} $ & \cr
\omit&height2pt&\omit&&\omit&&\omit&&\omit&\cr \tablerule
\omit&height2pt&\omit&&\omit&&\omit&&\omit&\cr && 3 && 3 && a &&
$-\f{1}{4} (\f{D-1}{D-3}) \ka^2\g^{\mu} i[S](x;x^{\p})\del_{\mu} \,
{\hspace{-.1cm} \not{\hspace{-.1cm}\del}}^{\p} i\d\!\D_{A}(x;x^{\p})
$ &\cr \omit&height2pt&\omit&&\omit&&\omit&&\omit&\cr \tablerule
\omit&height2pt&\omit&&\omit&&\omit&&\omit&\cr && 3 && 3 && b &&
$-\f{1}{4 (D-3)}\ka^2\g^{\mu} i[S](x;x^{\p})\del_{\mu}\hspace{-.1cm}
\not{\hspace{-.1cm}\bar{\del}}i\d\!\D_{A}(x;x^{\p}) $ &\cr
\omit&height2pt&\omit&&\omit&&\omit&&\omit&\cr \tablerule
\omit&height2pt&\omit&&\omit&&\omit&&\omit&\cr && 3 && 3 && c && $
+\f{1}{4 (D-3)}\ka^2\g_{k} i[S](x;x^{\p})\del_{k}\hspace{-.1cm}
\not{\hspace{-.1cm}\del}^{\p} i\d\!\D_{A}(x;x^{\p}) $ &\cr
\omit&height2pt&\omit&&\omit&&\omit&&\omit&\cr \tablerule
\omit&height2pt&\omit&&\omit&&\omit&&\omit&\cr && 3 && 3 && d &&
$-\f{1}{16} (\f{D-5}{D-3}) \ka^2\g_{k}
i[S](x;x^{\p})\del_{k}\hspace{-.1cm}\not{\hspace{-.1cm}\bar{\del}}
i\d\!\D_{A}(x;x^{\p}) $ & \cr
\omit&height2pt&\omit&&\omit&&\omit&&\omit&\cr \tablerule
\omit&height2pt&\omit&&\omit&&\omit&&\omit&\cr && 3 && 3 && e &&
$-\f{1}{16}\ka^2\g_{k}i[S](x;x^{\p})\g_{k} \nabla^2
i\d\!\D_{A}(x;x^{\p}) $ & \cr
\omit&height2pt&\omit&&\omit&&\omit&&\omit&\cr \tablerule}}

\label{DAcon}

\end{table}

In this subsection we work out the contribution from substituting
the residual $A$-type part of the graviton propagator in
Table~\ref{gen3},
\begin{equation}
i\Bigl[{}_{\alpha\beta} \Delta_{\rho\sigma}\Bigr](x;x^{\p})
\longrightarrow \Bigl[ \overline{\eta}_{\alpha \rho}
\overline{\eta}_{\sigma \beta} \!+\! \overline{\eta}_{\alpha \sigma}
\overline{\eta}_{\rho \beta} \!-\! \frac2{D\!-\!3}
\overline{\eta}_{\alpha\beta} \overline{\eta}_{\rho\sigma} \Bigr]
i\delta\!\Delta_A(x;x^{\p}) \; . \label{DApart}
\end{equation}
As with the conformal contributions of the previous section we first
make the requisite contractions and then act the derivatives. The
result of this first step is summarized in Table~\ref{DAcon}. We
have sometimes broken the result for a single vertex pair into as
many as five terms because the three different tensors in Equation
\ref{DApart} can make distinct contributions, and because distinct
contributions also come from breaking up factors of $\gamma^{\alpha}
J^{\beta \mu}$. These distinct contributions are labeled by
subscripts $a$, $b$, $c$, etc. We have tried to arrange them so that
terms closer to the beginning of the alphabet have fewer purely
spatial derivatives.

The next step is to act the derivatives and it is of course
necessary to have an expression for $i\delta\!\Delta_A(x;x^{\p})$ at
this stage. From Equation \ref{DeltaA} one can infer,
\begin{eqnarray}
\lefteqn{i\delta\!\Delta_A(x;x^{\p}) = } \nonumber \\
& & \hspace{-.5cm} \frac{H^2}{16 \pi^{\frac{D}2}}
\frac{\Gamma(\frac{D}2 \!+\! 1)}{\frac{D}2 \!-\! 2} \frac{(a
a^{\p})^{2- \frac{D}2}}{\Delta x^{D-4}} +
\frac{H^{D-2}}{(4\pi)^\frac{D}2} \frac{\Gamma(D\!-\!1)}{\Gamma(
\frac{D}2
)} \Biggl\{- \pi\cot\Bigl(\frac{\pi}2 D\Bigr) + \ln(aa^{\p}) \Biggr\} \nonumber \\
& & \hspace{-.7cm} + \frac{H^{D-2}}{(4\pi)^{\frac{D}2}} \!
\sum_{n=1}^{\infty} \! \left\{\!\frac1{n} \frac{\Gamma(n \!+\!D\!-\!
1)}{\Gamma(n \!+\! \frac{D}2)} \Bigl(\frac{y}4 \Bigr)^n \!\!\!\! -
\frac1{n \!-\! \frac{D}2 \!+\! 2} \frac{\Gamma(n \!+\!  \frac{D}2
\!+\! 1)}{\Gamma(n \!+\! 2)} \Bigl(\frac{y}4 \Bigr)^{n - \frac{D}2
+2} \!\right\}  . \quad \label{dA}
\end{eqnarray}
In $D\!=\!4$ the most singular contributions to Equation
\ref{3ptloop} have the form, $i\delta\!\Delta_A/{\Delta x}^5$.
Because the infinite series terms in Equation \ref{dA} go like
positive powers of $\Delta x^2$ these terms make integrable
contributions to the quantum-corrected Dirac equation in Equation
\ref{Diraceq}. We can therefore take $D\!=\!4$ for those terms, at
which point all the infinite series terms drop. Hence it is only
necessary to keep the first line of Equation \ref{dA} and that is
all we shall ever use.

\begin{table}
\caption{Residual $i\d\!\D_{A}$ terms giving both powers of $\Delta
x^2$. The two coefficients are  $A_1 \equiv \frac{i \kappa^2
H^2}{2^6 \pi^D} \Gamma(\frac{D}2 {\scriptstyle +1})\times $ $
\Gamma(\frac{D}2) (a a^{\p})^{2- \frac{D}2}$ and $A_2 \equiv \frac{i
\kappa^2 H^{D-2}}{2^{D+2} \pi^D} \Gamma({\scriptstyle D-2})
[{\scriptstyle \ln(a a^{\p}) - \pi \cot}(\frac{D\pi}2)]$.}

\vbox{\tabskip=0pt \offinterlineskip
\def\tablerule{\noalign{\hrule}}
\halign to390pt {\strut#& \vrule#\tabskip=1em plus2em& \hfil#\hfil&
\vrule#& \hfil#\hfil& \vrule#& \hfil#\hfil& \vrule#\tabskip=0pt\cr
\tablerule \omit&height4pt&\omit&&\omit&&\omit&\cr
\omit&height2pt&\omit&&\omit&&\omit&\cr &&\omit\hidewidth {\rm
Function} \hidewidth &&\omit\hidewidth {\rm Vertex\ Pair\ 2-1}
\hidewidth && {\rm Vertex\ Pair\ 2-2} & \cr
\omit&height4pt&\omit&&\omit&&\omit&\cr \tablerule
\omit&height2pt&\omit&&\omit&&\omit&\cr && $A_1 \partial^2
\hspace{-.1cm} \not{\hspace{-.1cm}\del} (\frac1{\Delta x^{2D-6}})$
&& $\frac{(D-1)}{(D-2)(D-3)^2(D-4)}$ && $0$ & \cr
\omit&height2pt&\omit&&\omit&&\omit&\cr \tablerule
\omit&height2pt&\omit&&\omit&&\omit&\cr && $A_1 \partial^2 \;
\hspace{-.1cm} \overline{\not{\hspace{-.1cm}\del} } (\frac1{\Delta
x^{2D-6}})$ && $\frac{-D}{(D-2)(D-3)^2(D-4)}$ && $\frac{-1}{
(D-2)(D-3)^2(D-4)}$ & \cr \omit&height2pt&\omit&&\omit&&\omit&\cr
\tablerule \omit&height2pt&\omit&&\omit&&\omit&\cr && $A_2
\partial^2 \; \hspace{-.1cm} \overline{\not{\hspace{-.1cm}\del} }
(\frac1{\Delta x^{D-2}})$ && $\frac{-2}{D-3}$ && $0$ & \cr
\omit&height2pt&\omit&&\omit&&\omit&\cr \tablerule
\omit&height2pt&\omit&&\omit&&\omit&\cr && $A_1 \nabla^2
\hspace{-.1cm} \not{\hspace{-.1cm}\del} (\frac1{\Delta x^{2D-6}})$
&& $0$ && $\frac{D (D^2 -3D -2)}{4(D-2)(D-3)^2 (D-4)}$ & \cr
\omit&height2pt&\omit&&\omit&&\omit&\cr \tablerule
\omit&height2pt&\omit&&\omit&&\omit&\cr && $A_2 \nabla^2
\hspace{-.1cm} \not{\hspace{-.1cm}\del} (\frac1{\Delta x^{D-2}})$ &&
$0$ && $\frac{(D^2-3D-2)}{2(D-3)}$ & \cr
\omit&height2pt&\omit&&\omit&&\omit&\cr \tablerule
\omit&height2pt&\omit&&\omit&&\omit&\cr && $A_1 \nabla^2 \;
\hspace{-.1cm} \overline{\not{\hspace{-.1cm}\del}} (\frac1{\Delta
x^{2D-6}})$ && $0$ && $\frac{-D}{(D-2)(D-3)^2}$ & \cr
\omit&height2pt&\omit&&\omit&&\omit&\cr \tablerule
\omit&height2pt&\omit&&\omit&&\omit&\cr && $A_2\nabla^2 \;
\hspace{-.1cm} \overline{\not{\hspace{-.1cm}\del}} (\frac1{\Delta
x^{D-2}})$ && $0$ && $-2(\frac{D-4}{D-3})$ & \cr
\omit&height2pt&\omit&&\omit&&\omit&\cr \tablerule}}

\label{DAmostb}

\end{table}

The contributions from $i\delta\!\Delta_A$ are more complicated than
those from $i\Delta_{\rm cf}$ for several reasons. The fact that
there is a second series in Equation \ref{dA} occasions our
Table~\ref{DAmostb}. These contributions are distinguished by all
derivatives acting upon the conformal coordinate separation and by
both series making nonzero contributions. Because these terms are
special we shall explicitly carry out the reduction of the
$2\!\!-\!\!2$ contribution. All three $2\!\!-\!\!2$ contractions on
Table~\ref{DAcon} can be expressed as a certain tensor contracted
into a generic form,
\begin{equation}
\Bigl[\delta_{ij} \delta_{k\ell} \!+\! \delta_{ik} \delta_{j\ell}
\!-\! \frac2{D\!-\!3} \delta_{i\ell} \delta_{jk} \Bigr] \times
\frac{\kappa^2}4 \partial_i \gamma_j \Bigl\{
i\delta\!\Delta_A(x;x^{\p})
\partial_k i[S](x;x^{\p}) \gamma_{\ell} \Bigr\} \; . \label{generic}
\end{equation}
So we may as well work out the generic term and then do the
contractions at the end. Substituting the fermion propagator brings
this generic term to the form,
\begin{eqnarray}
{\rm Generic} & \equiv & \frac{\kappa^2}4 \partial_i \gamma_j
\Bigl\{
i\delta\!\Delta_A(x;x^{\p}) \partial_k i[S](x;x^{\p}) \gamma_{\ell} \Bigr\} \; , \\
& = & -\frac{i \kappa^2 \Gamma(\frac{D}2)}{8 \pi^{\frac{D}2}}
\partial_i \gamma_j \Bigl\{ i\delta\!\Delta_A(x;x^{\p}) \partial_k \Bigl(
\frac{\gamma^{\mu} \Delta x_{\mu}}{\Delta x^D} \Bigr) \gamma_{\ell}
\Bigr\}\; .
\end{eqnarray}

Now recall that there are two sorts of terms in the only part of
$i\delta\!\Delta_A(x;x^{\p})$ that can make a nonzero contribution
for $D\!=\!4$,
\begin{eqnarray}
i\delta\!\Delta_{A1}(x;x^{\p}) & \equiv & \frac{H^2}{16
\pi^{\frac{D}2}} \frac{\Gamma(\frac{D}2\!+\!1)}{\frac{D}2\!-\!2}
\frac{(a a^{\p})^{2-\frac{D}2}}{
\Delta x^{D-4}} \; , \\
i\delta\!\Delta_{A2}(x;x^{\p}) & \equiv & \frac{H^{D-2}}{(4
\pi)^{\frac{D}2}} \frac{\Gamma(D\!-\!1)}{\Gamma(\frac{D}2)}
\Bigl\{-\pi\cot\Bigl(\frac{\pi}2 D \Bigr) + \ln(a a^{\p})\Bigr\} \;
.
\end{eqnarray}
Because all the derivatives are spatial we can pass the scale
factors outside to obtain,
\begin{eqnarray}
\lefteqn{{\rm Generic}_1 } \nonumber \\
& & = -\frac{i \kappa^2 H^2}{2^6 \pi^D} \frac{\Gamma(\frac{D}2)
\Gamma(\frac{D}2\!+\!1)}{(D\!-\!4)} (a a^{\p})^{2-\frac{D}2}
\partial_i \gamma_j \Bigl\{ \frac1{\Delta x^{D-4}} \partial_k
\Bigl(\frac{\gamma^{\mu} \Delta
x_{\mu}}{\Delta x^D}\Bigr) \gamma_{\ell} \Bigr\} \; , \qquad \\
\lefteqn{{\rm Generic}_2 } \nonumber \\
& & = -\frac{i \kappa^2 H^{D-2}}{2^{D+3} \pi^D} \Gamma(D\!-\!1)
\Bigl\{-\pi\cot\Bigl(\frac{\pi}2 D \Bigr) \!+\! \ln(a a^{\p})\Bigr\}
\partial_i \gamma_j \partial_k \Bigl(\frac{\gamma^{\mu} \Delta
x_{\mu}}{\Delta x^D}\Bigr)
\gamma_{\ell} \; , \qquad \\
& & = \frac{i \kappa^2 H^{D-2}}{2^{D+3} \pi^D} \Gamma(D\!-\!2)
\Bigl\{-\pi\cot\Bigl(\frac{\pi}2 D \Bigr) \!+\! \ln(a a^{\p})\Bigr\}
\partial_i \gamma_j \partial_k \hspace{-.1cm}\not{\hspace{-.05cm}
\partial} \gamma_{\ell} \Bigl(\frac1{\Delta x^{D-2}}\Bigr) \; .
\qquad
\end{eqnarray}
To complete the reduction of the first generic term we note,
\begin{eqnarray}
\frac1{\Delta x^{D-4}} \partial_k \Bigl(\frac{\gamma^{\mu} \Delta
x_{\mu}}{ \Delta x^D} \Bigr) & = & \frac{\gamma_k}{\Delta x^{2D-4}}
\!-\! \frac{D
\gamma^{\mu} \Delta x_{\mu} \Delta x_k}{\Delta x^{2D-2}} \; , \\
& = & \frac12 \Bigl(\frac{D\!-\!4}{D\!-\!2}\Bigr)
\frac{\gamma_k}{\Delta x^{2D-4}} \!+\! \frac{D}{2(D\!-\!2)}
\partial_k \Bigl(\frac{\gamma^{\mu}
\Delta x_{\mu}}{\Delta x^{2D-4}}\Bigr) \; , \\
& = & \frac1{4 (D\!-\!3)(D\!-\!2)} \Bigl\{ \gamma_k \partial^2 - D
\partial_k \hspace{-.1cm}\not{\hspace{-.05cm} \partial} \Bigr\}
\frac1{\Delta x^{2D-6}} \; . \qquad
\end{eqnarray}
Hence the first generic term is,
\begin{eqnarray}
\lefteqn{{\rm Generic}_1 = \frac{i \kappa^2 H^2}{2^8 \pi^D}
\frac{\Gamma( \frac{D}2)
\Gamma(\frac{D}2\!+\!1)}{(D\!-\!4)(D\!-\!3)(D\!-\!2)} (a a^{\p})^{
2-\frac{D}2} } \nonumber \\
& & \hspace{4cm} \times \Bigl\{D \partial_i \gamma_j \partial_k
\hspace{-.1cm}\not{\hspace{-.05cm} \partial} \gamma_{\ell} -
\partial^2
\partial_i \gamma_j \gamma_k \gamma_{\ell} \Bigr\} \frac1{\Delta x^{2D-6}}
\; . \qquad
\end{eqnarray}

Now we contract the tensor prefactor of Equation \ref{generic} into
the appropriate spinor-differential operators. For the first generic
term this is,
\begin{eqnarray}
\lefteqn{\Bigl[\delta_{ij} \delta_{k\ell} \!+\! \delta_{ik}
\delta_{j\ell} \!-\! \frac2{D\!-\!3} \delta_{i\ell} \delta_{jk}
\Bigr] \times \Bigl\{D
\partial_i \gamma_j \partial_k \hspace{-.1cm}\not{\hspace{-.05cm} \partial}
\gamma_{\ell} - \partial^2 \partial_i \gamma_j \gamma_k
\gamma_{\ell} \Bigr\} }
\nonumber \\
& & \hspace{-.5cm} = D \Bigl(\frac{D\!-\!5}{D\!-\!3}\Bigr)
\hspace{-.1cm}\not{\hspace{-.05cm} \overline{\partial}}
\hspace{-.1cm}\not{\hspace{-.05cm} \partial}
\hspace{-.1cm}\not{\hspace{-.05cm} \overline{\partial}} \!+\! D
\nabla^2 \gamma_i \hspace{-.1cm}\not{\hspace{-.05cm} \partial}
\gamma_i \!-\!
\partial^2 \hspace{-.1cm}\not{\hspace{-.05cm} \overline{\partial}}
\gamma_i \gamma_i \!-\! \partial^2 \gamma_i
\hspace{-.1cm}\not{\hspace{-.05cm} \overline{\partial}} \gamma_i
\!+\! \frac2{D\!-\!3} \partial^2 \gamma_i \gamma_i
\hspace{-.1cm}\not{\hspace{-.05cm} \overline{\partial}} \; . \qquad
\end{eqnarray}
This term can be simplified using the identities,
\begin{eqnarray}
\hspace{-.1cm}\not{\hspace{-.05cm} \overline{\partial}}
\hspace{-.1cm}\not{\hspace{-.05cm} \partial}
\hspace{-.1cm}\not{\hspace{-.05cm} \overline{\partial}} & = & -
\hspace{-.1cm}\not{\hspace{-.05cm} \overline{\partial}}
\hspace{-.1cm}\not{\hspace{-.05cm} \overline{\partial}}
\hspace{-.1cm}\not{\hspace{-.05cm} \partial} - 2
\hspace{-.1cm}\not{\hspace{-.05cm} \overline{\partial}} \nabla^2 =
\nabla^2 \hspace{-.1cm}\not{\hspace{-.05cm} \partial} - 2
\hspace{-.1cm}\not{\hspace{-.05cm} \overline{\partial}} \nabla^2 =
-\nabla^2 \hspace{-.1cm}\not{\hspace{-.05cm} \partial} \!+\! 2
\nabla^2
\gamma^0 \partial_0 \; , \\
\gamma_i \hspace{-.1cm}\not{\hspace{-.05cm} \partial} \gamma_i & = &
-\gamma_i \gamma_i \hspace{-.1cm}\not{\hspace{-.05cm} \partial} -2
\hspace{-.1cm}\not{\hspace{-.05cm} \overline{\partial}} = (D\!-\!1)
\hspace{-.1cm}\not{\hspace{-.05cm} \partial} - 2
\hspace{-.1cm}\not{\hspace{-.05cm} \overline{\partial}} = (D\!-\!3)
\hspace{-.1cm}\not{\hspace{-.05cm} \partial} + 2 \gamma^0 \partial_0 \; , \\
\hspace{-.1cm}\not{\hspace{-.05cm} \overline{\partial}} \gamma_i
\gamma_i & = & -(D\!-\!1) \hspace{-.1cm}\not{\hspace{-.05cm}
\overline{\partial}} = \gamma_i \gamma_i
\hspace{-.1cm}\not{\hspace{-.05cm} \overline{\partial}}
\; , \\
\gamma_i \hspace{-.1cm}\not{\hspace{-.05cm} \overline{\partial}}
\gamma_i & = & -\gamma_i \gamma_i \hspace{-.1cm}\not{\hspace{-.05cm}
\overline{\partial}} -2 \hspace{-.1cm}\not{\hspace{-.05cm}
\overline{\partial}} = (D\!-\!3) \hspace{-.1cm}\not{\hspace{-.05cm}
\overline{\partial}} \; .
\end{eqnarray}
Applying these identities gives,
\begin{eqnarray}
\lefteqn{\Bigl[\delta_{ij} \delta_{k\ell} \!+\! \delta_{ik}
\delta_{j\ell} \!-\! \frac2{D\!-\!3} \delta_{i\ell} \delta_{jk}
\Bigr] \times \Bigl\{D
\partial_i \gamma_j \partial_k \hspace{-.1cm}\not{\hspace{-.05cm} \partial}
\gamma_{\ell} - \partial^2 \partial_i \gamma_j \gamma_k
\gamma_{\ell} \Bigr\} }
\nonumber \\
& & \hspace{2.6cm} = \Bigl(D^2 \!-\! \frac{2D}{D\!-\!3}\Bigr)
\nabla^2 \hspace{-.1cm}\not{\hspace{-.05cm} \partial} \!-\! 4D
\Bigl(\frac{D\!-\!4}{D\!- \!3}\Bigr) \nabla^2
\hspace{-.1cm}\not{\hspace{-.05cm} \overline{\partial}} \!-\!
\frac4{D\!-\!3} \partial^2 \hspace{-.1cm}\not{\hspace{-.05cm}
\overline{\partial}} \; . \;\qquad
\end{eqnarray}
For the second generic term the relevant contraction is,
\begin{eqnarray}
\lefteqn{\Bigl[\delta_{ij} \delta_{k\ell} \!+\! \delta_{ik}
\delta_{j\ell} \!-\! \frac2{D\!-\!3} \delta_{i\ell} \delta_{jk}
\Bigr] \times \partial_i \gamma_j \partial_k
\hspace{-.1cm}\not{\hspace{-.05cm} \partial} \gamma_{\ell}}
\nonumber \\
& & \hspace{4cm} = \Bigl(\frac{D\!-\!5}{D\!-\!3}\Bigr)
\hspace{-.1cm}\not{\hspace{-.05cm} \overline{\partial}}
\hspace{-.1cm}\not{\hspace{-.05cm} \partial}
\hspace{-.1cm}\not{\hspace{-.05cm} \overline{\partial}} \!+\!
\nabla^2
\gamma_i \hspace{-.1cm}\not{\hspace{-.05cm} \partial} \gamma_i \; , \\
& & \hspace{4cm} = \Bigl(D \!-\! \frac{2}{D\!-\!3}\Bigr) \nabla^2
\hspace{-.1cm}\not{\hspace{-.05cm} \partial} \!-\! 4
\Bigl(\frac{D\!-\!4}{D\!- \!3}\Bigr) \nabla^2
\hspace{-.1cm}\not{\hspace{-.05cm} \overline{\partial}} \; .
\end{eqnarray}

\begin{table}
\caption{Residual $i\d\!\D_{A}$ terms in which all derivatives act
upon $\Delta x^2(x;x^{\p})$. All contributions are multiplied by
$\frac{i \kappa^2 H^2}{2^6 \pi^D} \Gamma(\frac{D}2 {\scriptstyle
+1}) \Gamma(\frac{D}2) (a a^{\p})^{2- \frac{D}2}$. }

\vbox{\tabskip=0pt \offinterlineskip
\def\tablerule{\noalign{\hrule}}
\halign to390pt {\strut#& \vrule#\tabskip=1em plus2em& \hfil#\hfil&
\vrule#& \hfil#\hfil& \vrule#& \hfil#\hfil& \vrule#& \hfil#\hfil&
\vrule#& \hfil#\hfil& \vrule#& \hfil#\hfil& \vrule#& \hfil#\hfil&
\vrule#\tabskip=0pt\cr \tablerule
\omit&height4pt&\omit&&\omit&&\omit&&\omit&&\omit&&\omit&&\omit&\cr
&& $\!\!\!\!{\rm I}\!\!\!\!\!\!$ && $\!\!\!\!{\rm J}\!\!\!\!\!\!$ &&
$\!\!\!\!{\rm sub}\!\!\!\!\!\!$ && $\!\!\!\!\frac{\gamma^{\mu}
\Delta x_{\mu}}{\Delta x^{2D-2}}\!\!\!\!$ && $\!\!\!\!\frac{\gamma^i
\Delta x_i}{\Delta x^{2D-2}}\!\!\!\!$ && $\!\!\!\frac{\Vert \Delta
\vec{x}\Vert^2 \gamma^{\mu} \Delta x_{\mu}}{\Delta x^{2D}}\!\!\!\!$
&& $\!\!\!\!\frac{\Vert \Delta \vec{x}\Vert^2 \gamma^i \Delta
x_i}{\Delta x^{2D}}\!\!\!\!$ &\cr
\omit&height4pt&\omit&&\omit&&\omit&&\omit&&\omit&&\omit&&\omit&\cr
\tablerule
\omit&height2pt&\omit&&\omit&&\omit&&\omit&&\omit&&\omit&&\omit&\cr
&& 2 && 3 && a && $2(\frac{D-1}{D-3})$ &&
$\!\!\!\!-\frac{2D}{D-3}\!\!\!$ && $0$ && $0$ & \cr
\omit&height2pt&\omit&&\omit&&\omit&&\omit&&\omit&&\omit&&\omit&\cr
\tablerule
\omit&height2pt&\omit&&\omit&&\omit&&\omit&&\omit&&\omit&&\omit&\cr
&& 2 && 3 && b && $0$ && $1$ && $\frac{D^2}2$ && $-2 {\scriptstyle
D}$ & \cr
\omit&height2pt&\omit&&\omit&&\omit&&\omit&&\omit&&\omit&&\omit&\cr
\tablerule
\omit&height2pt&\omit&&\omit&&\omit&&\omit&&\omit&&\omit&&\omit&\cr
&& 2 && 3 && c && $0$ && $\!\!\!\!-(\frac{D-1}{D-3})\!\!\!\!$ &&
$-\frac{D}{D-3}$ && $\frac{2D}{D-3}$ & \cr
\omit&height2pt&\omit&&\omit&&\omit&&\omit&&\omit&&\omit&&\omit&\cr
\tablerule
\omit&height2pt&\omit&&\omit&&\omit&&\omit&&\omit&&\omit&&\omit&\cr
&& 3 && 1 && a && $\!\!\!\!-\frac{4(D-1)(D-2)}{D-3}\!\!\!\!$ && $0$
&& $0$ && $0$ & \cr
\omit&height2pt&\omit&&\omit&&\omit&&\omit&&\omit&&\omit&&\omit&\cr
\tablerule
\omit&height2pt&\omit&&\omit&&\omit&&\omit&&\omit&&\omit&&\omit&\cr
&& 3 && 1 && b && $2(\frac{D-1}{D-3})$ &&
$\!\!\!\!2(\frac{D-4}{D-3})\!\!\!\!$ && $0$ && $0$ & \cr
\omit&height2pt&\omit&&\omit&&\omit&&\omit&&\omit&&\omit&&\omit&\cr
\tablerule
\omit&height2pt&\omit&&\omit&&\omit&&\omit&&\omit&&\omit&&\omit&\cr
&& 3 && 2 && a && $0$ && $\!\!\!\! 4 (\frac{D-2}{D-3})\!\!\!\!$ &&
$0$ && $0$ & \cr
\omit&height2pt&\omit&&\omit&&\omit&&\omit&&\omit&&\omit&&\omit&\cr
\tablerule
\omit&height2pt&\omit&&\omit&&\omit&&\omit&&\omit&&\omit&&\omit&\cr
&& 3 && 2 && b && $-(\frac{D-1}{D-3})$ && $\!\!\!\!
(\frac{D+1}{D-3}) \!\!\!\!$ && $\!\!\!\!2 (\frac{D-1}{D-3})\!\!\!\!$
&& $-4 (\frac{D-1}{D-3})$ & \cr
\omit&height2pt&\omit&&\omit&&\omit&&\omit&&\omit&&\omit&&\omit&\cr
\tablerule
\omit&height2pt&\omit&&\omit&&\omit&&\omit&&\omit&&\omit&&\omit&\cr
&& 3 && 2 && c && $\frac12 ({\scriptstyle D-1})$ && $\!\!\!\!
-\frac12 ({\scriptstyle D+1})\!\!\!\!$ && $\!\!\!\!- ({\scriptstyle
D-1})\!\!\!\!$ && $2 ({\scriptstyle D-1})$ & \cr
\omit&height2pt&\omit&&\omit&&\omit&&\omit&&\omit&&\omit&&\omit&\cr
\tablerule
\omit&height2pt&\omit&&\omit&&\omit&&\omit&&\omit&&\omit&&\omit&\cr
&& 3 && 2 && d && $\frac12 ({\scriptstyle D-1})^2$ && $\!\!\!\!
-\frac12 ({\scriptstyle D+1})\!\!\!\!$ && $\!\!\!\!- ({\scriptstyle
D-1})^2\!\!\!\!$ && $2 ({\scriptstyle D-1})$ & \cr
\omit&height2pt&\omit&&\omit&&\omit&&\omit&&\omit&&\omit&&\omit&\cr
\tablerule
\omit&height2pt&\omit&&\omit&&\omit&&\omit&&\omit&&\omit&&\omit&\cr
&& 3 && 3 && a && $\!\!\!\!2\frac{(D-1)(D-2)}{(D-3)}\!\!\!\!$ && $0$
&& $0$ && $0$ & \cr
\omit&height2pt&\omit&&\omit&&\omit&&\omit&&\omit&&\omit&&\omit&\cr
\tablerule
\omit&height2pt&\omit&&\omit&&\omit&&\omit&&\omit&&\omit&&\omit&\cr
&& 3 && 3 && b && $-(\frac{D-1}{D-3})$ && $\!\!\!\!
-(\frac{D-4}{D-3})\!\!\!\!$ && $0$ && $0$ & \cr
\omit&height2pt&\omit&&\omit&&\omit&&\omit&&\omit&&\omit&&\omit&\cr
\tablerule
\omit&height2pt&\omit&&\omit&&\omit&&\omit&&\omit&&\omit&&\omit&\cr
&& 3 && 3 && c && $-(\frac{D-1}{D-3})$ && $\!\!\!\!
-(\frac{D-4}{D-3})\!\!\!\!$ && $0$ && $0$ & \cr
\omit&height2pt&\omit&&\omit&&\omit&&\omit&&\omit&&\omit&&\omit&\cr
\tablerule
\omit&height2pt&\omit&&\omit&&\omit&&\omit&&\omit&&\omit&&\omit&\cr
&& 3 && 3 && d && $\!\!\!\!-\frac{(D-1)(D-5)}{4 (D-3)}\!\!\!\!$ &&
$\!\!\!\! \frac12 (\frac{D-5}{D-3})\!\!\!\!$ && $\frac{(D-5)(D-2)}{4
(D-3)}$ && $\!\!\!\!-\frac{(D-5)(D-2)}{2(D-3)}\!\!\!\!$ & \cr
\omit&height2pt&\omit&&\omit&&\omit&&\omit&&\omit&&\omit&&\omit&\cr
\tablerule
\omit&height2pt&\omit&&\omit&&\omit&&\omit&&\omit&&\omit&&\omit&\cr
&& 3 && 3 && e && $-\frac14 ({\scriptstyle D-1})^2$ && $\!\!\!\!
\frac12 ({\scriptstyle D-1})\!\!\!\!$ && $\!\!\!\!\frac14
({\scriptstyle D-2}) ({\scriptstyle D-1})\!\!\!\!$ && $-\frac12
({\scriptstyle D-2})$ & \cr
\omit&height2pt&\omit&&\omit&&\omit&&\omit&&\omit&&\omit&&\omit&\cr
\tablerule}}

\label{DAmostc}

\end{table}

In summing the contributions from Table~\ref{DAmostb} it is best to
take advantage of cancellations between $A_1$ and $A_2$ terms. These
occur between the 2nd and 3rd terms in the second column, the 4th
and 5th terms of the 3rd column, and the 6th and 7th terms of the
3rd column. In each of these cases the result is finite; and it
actually vanishes in the final case! Only the first term of column 2
and the 2nd term of column 3 contribute divergences. The result for
the three contributions from $[2\!\!-\!\!1]$ in Table~\ref{DAmostb}
is,
\begin{eqnarray}
\lefteqn{-\frac{\kappa^2 H^2 \mu^{D-4}}{2^5 \pi^{\frac{D}2}}
\frac{(D\!-\!1) \Gamma(\frac{D}2\!+\!1)}{(D\!-\!3)^2 (D\!-\!4)} \,
(a a^{\p})^{2 -\frac{D}2} \hspace{-.1cm} \not{\hspace{-.1cm}
\partial} \, \delta^D(x\!-\!x^{\p}) }
\nonumber \\
& & + \frac{i \kappa^2 H^2}{2^6 \pi^4} \Biggl\{\! -\frac32
\partial^2 \hspace{-.1cm} \not{\hspace{-.1cm} \partial} \Bigl[
\frac{\ln(\mu^2 \Delta x^2)}{\Delta x^2}\Bigr] \!+\! \partial^2
\,\hspace{-.1cm} \overline{\not{\hspace{-.1cm} \partial}} \Bigl[
\frac{4 \!+\! 2 \ln(\frac14 H^2 \Delta x^2)}{\Delta x^2}\Bigr]
\!\Biggr\} \!+\! O(D\!-\!4) .\; \qquad
\end{eqnarray}
The result for the five contributions from $[2\!\!-\!\!2]$ in
Table~\ref{DAmostb} is,
\begin{eqnarray}
\lefteqn{\frac{\kappa^2 H^2 \mu^{D-4}}{2^5 \pi^{\frac{D}2}} \frac{
\Gamma(\frac{D}2\!+\!1)}{(D\!-\!3)^2 (D\!-\!4)} \, (a a^{\p})^{2
-\frac{D}2} \; \hspace{-.1cm} \overline{\not{\hspace{-.1cm}
\partial}} \, \delta^D(x\!-\!x^{\p}) }
\nonumber \\
& & + \frac{i \kappa^2 H^2}{2^6 \pi^4} \Biggl\{\! \frac12 \partial^2
\, \hspace{-.1cm} \overline{\not{\hspace{-.1cm} \partial}} \Bigl[
\frac{\ln(\mu^2 \Delta x^2)}{\Delta x^2}\Bigr] \!-\! \nabla^2
\hspace{-.1cm} \not{\hspace{-.1cm} \partial} \Bigl[ \frac{2 \!+\!
\ln(\frac14 H^2 \Delta x^2) }{\Delta x^2}\Bigr] \!\Biggr\} +
O(D\!-\!4) . \qquad
\end{eqnarray}
As might be expected from the similarities in their reductions,
these two terms combine together nicely in the total for
Table~\ref{DAmostb},
\begin{eqnarray}
\lefteqn{-i \Bigl[\Sigma^{T\ref{DAmostb}}\Bigr](x;x^{\p}) =
\frac{\kappa^2 H^2 \mu^{D-4}}{2^5 \pi^{\frac{D}2}}
\frac{\Gamma(\frac{D}2\!+\!1) (a a^{\p})^{2- \frac{D}2}}{(D\!-\!3)^2
(D\!-\!4)} \Bigl[-(D\!-\!1) \hspace{-.1cm} \not{\hspace{-.1cm}
\partial} + \hspace{-.1cm} \overline{\not{\hspace{-.1cm}
\partial}} \, \Bigr] \delta^D(x\!-\!x^{\p}) } \nonumber \\
& & \hspace{3cm} + \frac{i \kappa^2 H^2}{2^6 \pi^4} \Biggl\{\Bigl(
-\frac32 \hspace{-.1cm} \not{\hspace{-.1cm} \partial} \partial^2
\!+\! \frac12 \; \hspace{-.1cm} \overline{\not{\hspace{-.1cm}
\partial}} \, \partial^2 \Bigl)
\Bigl[ \frac{\ln(\mu^2 \Delta x^2)}{\Delta x^2}\Bigr] \nonumber \\
& & \hspace{3.5cm} + \Bigl(2 \; \hspace{-.1cm}
\overline{\not{\hspace{-.1cm}
\partial}} \, \partial^2 \!-\! \hspace{-.1cm} \not{\hspace{-.1cm} \partial} \,
\nabla^2 \Bigr) \Bigl[ \frac{2 \!+\! \ln(\frac14 H^2 \Delta
x^2)}{\Delta x^2} \Bigr] \Biggr\} + O(D\!-\!4) . \qquad
\label{4thcon}
\end{eqnarray}

The next class is comprised of terms in which only the first series
of $i\delta\!\Delta_A$ makes a nonzero contribution when all
derivatives act upon the conformal coordinate separation. The
results for this class of terms are summarized in
Table~\ref{DAmostc}. In reducing these terms the following
derivatives occur many times,
\begin{eqnarray}
\partial_i i \delta\!\Delta_A(x;x^{\p}) \!& = & \!-\frac{H^2}{8 \pi^{\frac{D}2}} \,
\Gamma\Bigl(\frac{D}2\!+\!1\Bigr) (a a^{\p})^{2-\frac{D}2} \,
\frac{\Delta x^i}{
\Delta x^{D-2}} = -\partial_i^{\p} i \delta\!\Delta_A(x;x^{\p})  ,\;  \\
\partial_0 i \delta\!\Delta_A(x;x^{\p}) & = & \frac{H^2}{8 \pi^{\frac{D}2}}
\, \Gamma\Bigl(\frac{D}2\!+\!1\Bigr) (a a^{\p})^{2-\frac{D}2}
\Biggl\{ \frac{\Delta \eta}{\Delta x^{D-2}} \!-\! \frac{a H}{2
\Delta x^{D-4}} \Biggr\}
\nonumber \\
& & \hspace{5cm} + \frac{H^{D-2}}{2^D \pi^{\frac{D}2}} \frac{\Gamma(
D\!-\!1)}{\Gamma(\frac{D}2)} \, a H \; , \qquad \\
\partial_0^{\p} i \delta\!\Delta_A(x;x^{\p}) & = & \frac{H^2}{8 \pi^{\frac{D}2}}
\, \Gamma\Bigl(\frac{D}2\!+\!1\Bigr) (a a^{\p})^{2-\frac{D}2}
\Biggl\{ -\frac{\Delta \eta}{\Delta x^{D-2}} \!-\! \frac{a^{\p} H}{2
\Delta x^{D-4}} \Biggr\}
\nonumber \\
& & \hspace{5cm} + \frac{H^{D-2}}{2^D \pi^{\frac{D}2}}
\frac{\Gamma(D\!-\!1)}{ \Gamma(\frac{D}2)} \, a^{\p} H \; . \qquad
\end{eqnarray}
We also make use of a number of gamma matrix identities,
\begin{eqnarray}
\gamma^{\mu} \gamma_{\mu} & = & - D \quad {\rm and} \quad \gamma^i
\gamma^i =
-(D\!-\!1) \; , \label{gammafirst} \\
\gamma^{\mu} \gamma^{\nu} \gamma_{\mu} & = & (D\!-\!2) \gamma^{\nu}
\quad {\rm and} \quad \gamma^i \gamma^{\nu} \gamma^i = (D\!-\!1)
\gamma^{\nu} -
2 \overline{\gamma}^{\nu} \; , \qquad \\
(\gamma^{\mu} \Delta x_{\mu})^2 & = & - \Delta x^2 \quad {\rm and}
\quad
(\gamma^i \Delta x^i)^2 = -\Vert \Delta \vec{x} \Vert^2 \; , \\
\gamma^i \gamma^{\mu} \Delta x_{\mu} \gamma^i & = & (D\!-\!1)
\gamma^{\mu}
\Delta x_{\mu} - 2 \gamma^i \Delta x^i \; , \\
\gamma^i \Delta x^i \gamma^{\mu} \Delta x_{\mu} \gamma^j \Delta x^j
& = & \Vert \Delta \vec{x} \Vert^2 \gamma^{\mu} \Delta x_{\mu} - 2
\Vert \Delta \vec{x} \Vert^2 \gamma^i \Delta x^i \; .
\label{gammalast}
\end{eqnarray}

In summing the many terms of Table~\ref{DAmostc} the constant $K
\equiv D - 2/(D\!-\!3)$ occurs suspiciously often,
\begin{eqnarray}
\lefteqn{-i \Bigl[\Sigma^{T\ref{DAmostc}}\Bigr](x;x^{\p}) = \frac{i
\kappa^2 H^2}{ 2^6 \pi^D} \Gamma\Bigl(\frac{D}2 \!+\! 1\Bigr)
\Gamma\Bigl(\frac{D}2\Bigr)
(a a^{\p})^{2-\frac{D}2} } \nonumber \\
& & \times \Biggl\{ \Bigl[-2 (D\!-\!1) \!+\! \Bigl(\frac{D\!-\!1}4
\Bigr) K \Bigr] \frac{\gamma^{\mu} \Delta x_{\mu}}{\Delta x^{2D-2}}
+ \Bigl[-(D\!-\!2)
\!+\! \frac{K}2 \Bigr] \frac{\gamma^i \Delta x_i}{\Delta x^{2D-2}} \nonumber \\
& & \hspace{1.4cm} - \Bigl(\frac{D\!-\!2}4 \Bigr) K \frac{\Vert
\Delta \vec{x} \Vert^2 \gamma^{\mu} \Delta x_{\mu}}{\Delta x^{2D}} +
\frac{(D\!-\!2)(D\!-\!4) }{ (D\!-\!3)} \frac{\Vert \Delta \vec{x}
\Vert^2 \gamma^i \Delta x_i}{ \Delta x^{2D}} \Biggr\} . \qquad
\label{S12-1}
\end{eqnarray}
The last two terms can be reduced using the identities,
\begin{eqnarray}
\frac{\Vert \Delta \vec{x} \Vert^2 \gamma^{\mu} \Delta
x_{\mu}}{\Delta x^{2D}} &\!\!\!\!=\!\!\!\!& \frac12
\frac{\gamma^{\mu} \Delta x_{\mu}}{\Delta x^{2D-2}} \!+\!
\frac1{D\!-\!1} \frac{\gamma^i \Delta x_i}{\Delta x^{2D-2}} \!+\!
\frac{\nabla^2}{4(D\!-\!2) (D\!-\!1)} \Bigl(\frac{\gamma^{\mu}
\Delta x_{\mu}}{\Delta x^{2D-4}} \Bigr) , \qquad \label{difID1} \\
\frac{\Vert \Delta \vec{x} \Vert^2 \gamma^i \Delta x_i}{\Delta
x^{2D}} &\!\!\!\!= \!\!\!\!&\frac12 \Bigl(\frac{D\!+\!1}{D\!-\!1}
\Bigr) \frac{\gamma^i \Delta x_i}{\Delta x^{2D-2}} +
\frac{\nabla^2}{4(D\!-\!2) (D\!-\!1)} \Bigl(\frac{\gamma^i \Delta
x_i}{\Delta x^{2D-4}} \Bigr) . \qquad \label{difID2}
\end{eqnarray}
Substituting these in Equation \ref{S12-1} gives,
\begin{eqnarray}
\lefteqn{-i \Bigl[\Sigma^{T\ref{DAmostc}}\Bigr](x;x^{\p}) = \frac{i
\kappa^2 H^2}{ 2^6 \pi^D} \Gamma\Bigl(\frac{D}2 \!+\! 1\Bigr)
\Gamma\Bigl(\frac{D}2\Bigr) (a a^{\p})^{2-\frac{D}2} \Biggl\{
\Bigl[-2 (D\!-\!1) \!+\! \frac{D K}8 \Bigr] }
\nonumber \\
& & \hspace{1cm} \times \frac{\gamma^{\mu} \Delta x_{\mu}}{\Delta
x^{2D-2}} + \Bigl[-\frac{(D\!-\!2)(D^2 \!-\! 5D \!+\! 10)}{2
(D\!-\!1)(D\!-\!3)} \!+\! \frac{D K}{4 (D\!-\!1)} \Bigr]
\frac{\gamma^i \Delta x_i}{\Delta x^{2D-2}}
\nonumber \\
& & \hspace{2.5cm} - \frac{K \nabla^2}{16 (D\!-\!1)}
\Bigl(\frac{\gamma^{\mu} \Delta x_{\mu}}{\Delta x^{2D-4}} \Bigr) +
\frac{(D\!-\!4) \nabla^2 }{4 (D\!-\!1) (D\!-\!3)} \frac{\gamma^i
\Delta x_i}{\Delta x^{2D-4}} \Biggr\} . \qquad \label{S12-2}
\end{eqnarray}
We then apply the same formalism as in the previous sub-section to
partially integrate, extract the local divergences and take
$D\!=\!4$ for the remaining, integrable and ultraviolet finite
nonlocal terms,
\begin{eqnarray}
\lefteqn{-i \Bigl[\Sigma^{T\ref{DAmostc}}\Bigr](x;x^{\p}) =
\frac{\kappa^2 H^2 \mu^{D-4}}{2^7 \pi^{\frac{D}2}}
\frac{\Gamma(\frac{D}2 \!+\! 1) (a a^{\p})^{2 -
\frac{D}2}}{(D\!-\!3) (D\!-\!4)} } \nonumber \\
& & \hspace{-.5cm} \times \Biggl\{ \Bigl[\frac{D K}8 \!-\!
2(D\!-\!1)\Bigr] \hspace{-.1cm} \not{\hspace{-.1cm} \partial} \!+\!
\Bigl[\frac{D K}{4(D\!-\!1)} \!-\! \frac{(D\!-\!2)(D^2 \!-\! 5D
\!+\! 10)}{2 (D\!-\! 1) (D\!-\!3)} \Bigr] \; \hspace{-.1cm}
\overline{\not{\hspace{-.1cm} \partial} } \Biggr\}
\delta^D(x\!-\!x^{\p}) \nonumber \\
& & \hspace{-.5cm} +\frac{i \kappa^2 H^2}{2^9 \!\cdot\! 3 \!\cdot\!
\pi^4} \Biggl\{\!\Bigl[-15 \hspace{-.1cm} \not{\hspace{-.1cm}
\partial} \, \partial^2 \!-\! 4 \; \hspace{-.1cm}
\overline{\not{\hspace{-.1cm} \partial}} \,\partial^2 \Bigr]
\Bigl(\frac{\ln(\mu^2 \Delta x^2)}{\Delta x^2} \Bigr) \!+\! \hspace{
-.1cm} \not{\hspace{-.1cm} \partial} \, \nabla^2 \Bigl(\frac1{\Delta
x^2} \Bigr)\!\Biggr\} \!+\! O(D\!-\!4) . \qquad \label{5thcon}
\end{eqnarray}

The final class is comprised of terms in which one or more
derivatives act upon a scale factor. Within this class we report
contributions from the first series in Table~\ref{DAlessa} and
contributions from the second series in Table~\ref{DAlessb}. Each
nonzero entry in the 4th and 5th columns of Table~\ref{DAlessa}
diverges logarithmically like $1/\Delta x^{2D-4}$. However, the sum
in each case results in an additional factor of $a\!-\!a^{\p} \!=\!
a a^{\p} H \Delta \eta$ which makes the contribution from
Table~\ref{DAlessa} integrable,

\begin{table}
\caption{Residual $i\d\!\D_{A}$ terms in which some derivatives act
upon the scale factors of the first series. The factor $\frac{i
\kappa^2 H^2}{2^6 \pi^D} \Gamma(\frac{D}2 {\scriptstyle +1})
\Gamma(\frac{D}2) (a a^{\p})^{2- \frac{D}2}$ multiplies all
contributions.}

\vbox{\tabskip=0pt \offinterlineskip
\def\tablerule{\noalign{\hrule}}
\halign to390pt {\strut#& \vrule#\tabskip=1em plus2em& \hfil#\hfil&
\vrule#& \hfil#\hfil& \vrule#& \hfil#\hfil& \vrule#& \hfil#\hfil&
\vrule#& \hfil#\hfil& \vrule#& \hfil#\hfil& \vrule#\tabskip=0pt\cr
\tablerule
\omit&height4pt&\omit&&\omit&&\omit&&\omit&&\omit&&\omit&\cr &&
$\!\!\!\!{\rm I}\!\!\!\!\!\!$ && $\!\!\!\!{\rm J}\!\!\!\!\!\!$ &&
$\!\!\!\!{\rm sub}\!\!\!\!\!\!$ && $\!\!\!\!\frac{H \gamma^0}{\Delta
x^{2D-4}}\!\!\!\!$ && $\!\!\!\!\frac{H \gamma^i \Delta x_i
\gamma^{\mu} \Delta x_{\mu} \gamma^0}{ \Delta x^{2D-2}}\!\!\!\!$ &&
$\!\!\!\frac{H^2 a a^{\p} \gamma^{\mu} \Delta x_{\mu} }{\Delta
x^{2D-4}}\!\!\!\!$ &\cr
\omit&height4pt&\omit&&\omit&&\omit&&\omit&&\omit&&\omit&\cr
\tablerule
\omit&height2pt&\omit&&\omit&&\omit&&\omit&&\omit&&\omit&\cr && 2 &&
1 && \omit && $\!\!\!\!2(\frac{D-1}{D-3}) a^{\p}\!\!\!\!$ &&
$\!\!\!\!(\frac{2 D}{D-3}) a^{\p} \!\!\!\!$ && $0$ & \cr
\omit&height2pt&\omit&&\omit&&\omit&&\omit&&\omit&&\omit&\cr
\tablerule
\omit&height2pt&\omit&&\omit&&\omit&&\omit&&\omit&&\omit&\cr && 2 &&
3 && a && $\!\!\!\!-(\frac{D-1}{D-3}) a^{\p}\!\!\!\!$ &&
$\!\!\!\!(\frac{-D}{D-3}) a^{\p} \!\!\!\!$ && $0$ & \cr
\omit&height2pt&\omit&&\omit&&\omit&&\omit&&\omit&&\omit&\cr
\tablerule
\omit&height2pt&\omit&&\omit&&\omit&&\omit&&\omit&&\omit&\cr && 3 &&
1 && a && $0$ && $0$ && $\!\!\!\!\frac{(D-1)(D-4)}{2 (D-3)}\!\!\!\!$
& \cr \omit&height2pt&\omit&&\omit&&\omit&&\omit&&\omit&&\omit&\cr
\tablerule
\omit&height2pt&\omit&&\omit&&\omit&&\omit&&\omit&&\omit&\cr && 3 &&
1 && b && $0$ && $\!\!\!\!(\frac{D-4}{D-3}) a^{\p}\!\!\!\!$ && $0$ &
\cr \omit&height2pt&\omit&&\omit&&\omit&&\omit&&\omit&&\omit&\cr
\tablerule
\omit&height2pt&\omit&&\omit&&\omit&&\omit&&\omit&&\omit&\cr && 3 &&
2 && a && $\!\!\!\!-(\frac{D-1}{D-3}) a\!\!\!\!$ && $\!\!\!\! -2
(\frac{D-2}{D-3}) a\!\!\!\!$ && $0$ & \cr
\omit&height2pt&\omit&&\omit&&\omit&&\omit&&\omit&&\omit&\cr
\tablerule
\omit&height2pt&\omit&&\omit&&\omit&&\omit&&\omit&&\omit&\cr && 3 &&
3 && a && $0$ && $0$ && $\!\!\!\!-\frac{(D-1)(D-4)}{4
(D-3)}\!\!\!\!$ & \cr
\omit&height2pt&\omit&&\omit&&\omit&&\omit&&\omit&&\omit&\cr
\tablerule
\omit&height2pt&\omit&&\omit&&\omit&&\omit&&\omit&&\omit&\cr && 3 &&
3 && b && $0$ && $\!\!\!\! \frac12 (\frac{D-4}{D-3}) a\!\!\!\!$ &&
$0$ & \cr
\omit&height2pt&\omit&&\omit&&\omit&&\omit&&\omit&&\omit&\cr
\tablerule
\omit&height2pt&\omit&&\omit&&\omit&&\omit&&\omit&&\omit&\cr && 3 &&
3 && c && $0$ && $\!\!\!\! -\frac12 (\frac{D-4}{D-3})
a^{\p}\!\!\!\!$ && $0$ & \cr
\omit&height2pt&\omit&&\omit&&\omit&&\omit&&\omit&&\omit&\cr
\tablerule}}

\label{DAlessa}

\end{table}

\begin{table}
\caption{Residual $i\d\!\D_{A}$ terms in which some derivatives act
upon the scale factors of the second series. All contributions are
multiplied by $\frac{i \kappa^2 H^{D-2}}{2^{D+2} \pi^D}
\Gamma({\scriptstyle D-1})$.}

\vbox{\tabskip=0pt \offinterlineskip
\def\tablerule{\noalign{\hrule}}
\halign to390pt {\strut#& \vrule#\tabskip=1em plus2em& \hfil#\hfil&
\vrule#& \hfil#\hfil& \vrule#& \hfil#\hfil& \vrule#& \hfil#\hfil&
\vrule#& \hfil#\hfil& \vrule#& \hfil#\hfil& \vrule#\tabskip=0pt\cr
\tablerule
\omit&height4pt&\omit&&\omit&&\omit&&\omit&&\omit&&\omit&\cr &&
$\!\!\!\!{\rm I}\!\!\!\!\!\!$ && $\!\!\!\!{\rm J}\!\!\!\!\!\!$ &&
$\!\!\!\!{\rm sub}\!\!\!\!\!\!$ && $\!\!\!\frac{H \gamma^0}{\Delta
x^{D}} \!\!\!\!$ && $\!\!\!\!\frac{H \gamma^i \Delta x_i
\gamma^{\mu} \Delta x_{\mu} \gamma^0}{\Delta x^{D+2}}\!\!\!\!$ &&
$\!\!\!\! \partial^2 (\frac{H \gamma^0}{\Delta x^{D-2}}) \!\!\!\!$
&\cr \omit&height4pt&\omit&&\omit&&\omit&&\omit&&\omit&&\omit&\cr
\tablerule
\omit&height2pt&\omit&&\omit&&\omit&&\omit&&\omit&&\omit&\cr && 2 &&
1 && \omit && $\!\!\!\!-2(\frac{D-1}{D-3}) a^{\p}\!\!\!\!$ &&
$\!\!\!\!-(\frac{2D}{D-3}) a^{\p}\!\!\!\!$ && $0$ & \cr
\omit&height2pt&\omit&&\omit&&\omit&&\omit&&\omit&&\omit&\cr
\tablerule
\omit&height2pt&\omit&&\omit&&\omit&&\omit&&\omit&&\omit&\cr && 2 &&
3 && a && $(\frac{D-1}{D-3}) a^{\p}$ && $(\frac{D}{D-3}) a^{\p}$ &&
$0$ & \cr
\omit&height2pt&\omit&&\omit&&\omit&&\omit&&\omit&&\omit&\cr
\tablerule
\omit&height2pt&\omit&&\omit&&\omit&&\omit&&\omit&&\omit&\cr && 3 &&
1 && a && $0$ && $0$ && $\!\!\!\!\frac{(D-1) \,
a}{(D-2)(D-3)}\!\!\!\!$ &\cr
\omit&height2pt&\omit&&\omit&&\omit&&\omit&&\omit&&\omit&\cr
\tablerule
\omit&height2pt&\omit&&\omit&&\omit&&\omit&&\omit&&\omit&\cr && 3 &&
2 && a && $(\frac{D-1}{D-3}) a$ && $\!\!\!\!(\frac{D}{D-3})
a\!\!\!\!$ && $0$ & \cr
\omit&height2pt&\omit&&\omit&&\omit&&\omit&&\omit&&\omit&\cr
\tablerule}}

\label{DAlessb}

\end{table}

\begin{eqnarray}
\lefteqn{-i\Bigl[\Sigma^{T\ref{DAlessa}}\Bigr](x;x^{\p}) = \frac{i
\kappa^2 H^4}{2^6 \pi^D} \Gamma\Bigl(\frac{D}2 \!+\! 1\Bigr)
\Gamma\Bigl(\frac{D}2\Bigr) (a a^{\p})^{3-\frac{D}2} \Biggl\{
-\Bigl(\frac{D\!-\!1
}{D\!-\!3}\Bigr) \frac{\gamma^0 \Delta \eta}{\Delta x^{2D-4}} } \nonumber \\
& & \hspace{1.3cm} -\frac12 \Bigl(\frac{3D\!-\!4}{D\!-\!3}\Bigr)
\frac{\gamma^i \Delta x_i \gamma^{\mu} \Delta x_{\mu} \gamma^0
\Delta \eta}{ \Delta x^{2D-2}} +
\frac{(D\!-\!1)(D\!-\!4)}{4(D\!-\!3)} \frac{\gamma^{\mu} \Delta
x_{\mu}}{\Delta x^{2D-4}} \Biggr\} . \;\qquad
\end{eqnarray}
This is another example of the fact that the self-energy is odd
under interchange of $x^{\mu}$ and $x^{\prime \mu}$.

The same thing happens with the contribution from
Table~\ref{DAlessb},
\begin{eqnarray}
\lefteqn{-i\Bigl[\Sigma^{T\ref{DAlessb}}\Bigr](x;x^{\p}) = \frac{i
\kappa^2 H^D}{2^{D+2} \pi^D} \Gamma(D-1) a a^{\p} \Biggl\{
\Bigl(\frac{D\!-\!1}{D\!-\!3}\Bigr) \frac{\gamma^0 \Delta
\eta}{\Delta x^{D}} }
\nonumber \\
& & \hspace{1.3cm} + \Bigl(\frac{D}{D\!-\!3}\Bigr) \frac{\gamma^i
\Delta x_i \gamma^{\mu} \Delta x_{\mu} \gamma^0 \Delta \eta}{\Delta
x^{D+2}} + \gamma^0 \Bigl(\frac{D\!-\!1}{D\!-\!3}\Bigr) \frac{i 2
\pi^{\frac{D}2}}{ \Gamma(\frac{D}2)} \frac{\delta^D(x\!-\!x^{\p})}{H
a} \Biggr\} . \qquad
\end{eqnarray}
We can therefore set $D\!=\!4$, at which point the two Tables cancel
except for the delta function term,
\begin{equation}
-i\Bigl[\Sigma^{T\ref{DAlessa}+\ref{DAlessb}}\Bigr](x;x^{\p}) =
\frac{\kappa^2 H^{D-2}}{(4\pi)^{\frac{D}2}} \frac{\Gamma(D\!-\!1)}{
\Gamma(\frac{D}2)} \times -\frac12
\Bigl(\frac{D\!-\!1}{D\!-\!3}\Bigr) a H \gamma^0
\delta^D(x\!-\!x^{\p}) \!+\! O(D\!-\!4) . \label{6thcon}
\end{equation}
It is worth commenting that this term violates the reflection
symmetry of Equation \ref{refl}. In $D\!=\!4$ it cancels the similar
term in Equation \ref{1stcon}.

\subsection{Sub-Leading Contributions from $i \delta \Delta_{B} $ }

In this subsection we work out the contribution from substituting
the residual $B$-type part of the graviton propagator in
Table~\ref{gen3},
\begin{equation}
i\Bigl[{}_{\alpha\beta} \Delta_{\rho\sigma}\Bigr] \longrightarrow
-\Bigl[ \delta^0_{\alpha} \delta^0_{\sigma} \overline{\eta}_{\beta
\rho} + \delta^0_{\alpha} \delta^0_{\rho} \overline{\eta}_{\beta
\sigma} + \delta^0_{\beta} \delta^0_{\sigma} \overline{\eta}_{\alpha
\rho} + \delta^0_{\beta} \delta^0_{\rho} \overline{\eta}_{\alpha
\sigma} \Bigr] i\delta\!\Delta_B \; . \label{DBpart}
\end{equation}
As in the two previous sub-sections we first make the requisite
contractions and then act the derivatives. The result of this first
step is summarized in Table~\ref{DBcon}. We have sometimes broken
the result for a single vertex pair into parts because the four
different tensors in (\ref{DBpart}) can make distinct contributions,
and because distinct contributions also come from breaking up
factors of $\gamma^{\alpha} J^{\beta \mu}$. These distinct
contributions are labeled by subscripts $a$, $b$, $c$, etc.

$i\delta\!\Delta_B(x;x^{\p})$ is the residual of the $B$-type
propagator of Equation \ref{DeltaB} after the conformal contribution
has been subtracted,
\begin{eqnarray}
\lefteqn{i\delta\!\Delta_B(x;x^{\p}) = \frac{H^2
\Gamma(\frac{D}2)}{16 \pi^{\frac{D}2}} \frac{(a
a^{\p})^{2-\frac{D}2}}{\Delta x^{D-4}}
-\frac{H^{D-2}}{(4\pi)^\frac{D}2}
\frac{\Gamma(D\!-\!2)}{\Gamma\Bigl(\frac{D}2
\Bigr)} } \nonumber \\
& & \hspace{1.8cm} + \frac{H^{D-2}}{(4 \pi)^{\frac{D}2}}
\sum_{n=1}^{\infty} \left\{ \frac{\Gamma(n \!+\!
\frac{D}2)}{\Gamma(n \!+\! 2)} \Bigl( \frac{y}4 \Bigr)^{n -
\frac{D}2 +2} - \frac{\Gamma(n \!+\! D \!-\! 2)}{\Gamma(n \!+\!
\frac{D}2)} \Bigl(\frac{y}4 \Bigr)^n \right\}\! . \qquad \label{dB}
\end{eqnarray}

\begin{table}
\caption{Contractions from the $i\d\!\D_B$ part of the graviton
propagator.}

\vbox{\tabskip=0pt \offinterlineskip
\def\tablerule{\noalign{\hrule}}
\halign to390pt {\strut#& \vrule#\tabskip=1em plus2em& \hfil#\hfil&
\vrule#& \hfil#\hfil& \vrule#& \hfil#\hfil& \vrule#& \hfil#\hfil&
\vrule#\tabskip=0pt\cr \tablerule
\omit&height4pt&\omit&&\omit&&\omit&&\omit&\cr &&$\!\!\!\!{\rm
I}\!\!\!\!$ && $\!\!\!\!{\rm J} \!\!\!\!$ && $\!\!\!\! {\rm sub}
\!\!\!\!$ && $iV_I^{\alpha\beta}(x) \, i[S](x;x^{\p}) \, i
V_J^{\rho\sigma}(x^{\p}) \, [\mbox{}_{\alpha\beta} T^B_{\rho\sigma}]
\, i\delta\!\Delta_B(x;x^{\p})$ &\cr
\omit&height4pt&\omit&&\omit&&\omit&&\omit&\cr \tablerule
\omit&height2pt&\omit&&\omit&&\omit&&\omit&\cr && 2 && 1 && \omit &&
$0$ & \cr \omit&height2pt&\omit&&\omit&&\omit&&\omit&\cr \tablerule
\omit&height2pt&\omit&&\omit&&\omit&&\omit&\cr && 2 && 2 && a &&
$-\f{1}{2}\ka^2\del^{^{\p}}_0 \{
 \g^{(0}\del^{k)} i[S](x;x^{\p})\g_{k} i\d\!\D_{B}(x;x^{\p}) \} $ & \cr
\omit&height2pt&\omit&&\omit&&\omit&&\omit&\cr \tablerule
\omit&height2pt&\omit&&\omit&&\omit&&\omit&\cr && 2 && 2 && b &&
$-\f{1}{2}\ka^2\del_k \{ \g^{(0}\del^{k)} i[S](x;x^{\p})\g^{0}
i\d\!\D_{B}(x;x^{\p})\}$ & \cr
\omit&height2pt&\omit&&\omit&&\omit&&\omit&\cr \tablerule
\omit&height2pt&\omit&&\omit&&\omit&&\omit&\cr && 2 && 3 && a &&
$-\f{1}{8}\ka^2\g_k\del_{0}i[S](x;x^{\p}) \g^k \,
\del^{^{\p}}_{0}i\d\!\D_{B}(x;x^{\p})$ & \cr
\omit&height2pt&\omit&&\omit&&\omit&&\omit&\cr \tablerule
\omit&height2pt&\omit&&\omit&&\omit&&\omit&\cr && 2 && 3 && b &&
$\f{1}{8}\ka^2\g ^0\del^{\p}_{0} i\d\!\D_{B}(x;x^{\p}) \,
\del_{k}i[S](x;x^{\p})\g^{k} $ & \cr
\omit&height2pt&\omit&&\omit&&\omit&&\omit&\cr \tablerule
\omit&height2pt&\omit&&\omit&&\omit&&\omit&\cr && 2 && 3 && c &&
$-\f{1}{8}\ka^2\g^k\del_{k} i\d\!\D_{B}(x;x^{\p}) \,
\del_{0}i[S](x;x^{\p})\g^{0} $ & \cr
\omit&height2pt&\omit&&\omit&&\omit&&\omit&\cr \tablerule
\omit&height2pt&\omit&&\omit&&\omit&&\omit&\cr && 2 && 3 && d &&
$\f{1}{8}\ka^2\g^0\del^{k}i[S](x;x^{\p}) \g^0 \,
\del_{k}i\d\!\D_{B}(x;x^{\p}) $ & \cr
\omit&height2pt&\omit&&\omit&&\omit&&\omit&\cr \tablerule
\omit&height2pt&\omit&&\omit&&\omit&&\omit&\cr && 3 && 1 && \omit &&
$0$ & \cr \omit&height2pt&\omit&&\omit&&\omit&&\omit&\cr \tablerule
\omit&height2pt&\omit&&\omit&&\omit&&\omit&\cr && 3 && 2 && a &&
$\f{1}{8}\ka^2\del^{\p}_{0}\{\g^{k} i[S](x;x^{\p}) \g_k \,
\del_{0}i\d\!\D_{B}(x;x^{\p})\}$ & \cr
\omit&height2pt&\omit&&\omit&&\omit&&\omit&\cr \tablerule
\omit&height2pt&\omit&&\omit&&\omit&&\omit&\cr && 3 && 2 && b &&
$\f{1}{8}\ka^2\g^k\del_{k}\{ i[S](x;x^{\p}) \g^0 \,
\del_{0}i\d\!\D_{B}(x;x^{\p})\} $ & \cr
\omit&height2pt&\omit&&\omit&&\omit&&\omit&\cr \tablerule
\omit&height2pt&\omit&&\omit&&\omit&&\omit&\cr && 3 && 2 && c &&
$-\f{1}{8}\ka^2\g^0\del^{\p}_{0}\{ i[S](x;x^{\p}) \g^k \,
\del_{k}i\d\!\D_{B}(x;x^{\p})\} $ & \cr
\omit&height2pt&\omit&&\omit&&\omit&&\omit&\cr \tablerule
\omit&height2pt&\omit&&\omit&&\omit&&\omit&\cr && 3 && 2 && d &&
$-\f{1}{8}\ka^2\del_{k}\{\g^{0} i[S](x;x^{\p}) \g^0 \,
\del^{k}i\d\!\D_{B}(x;x^{\p})\}$ & \cr
\omit&height2pt&\omit&&\omit&&\omit&&\omit&\cr \tablerule
\omit&height2pt&\omit&&\omit&&\omit&&\omit&\cr && 3 && 3 && a &&
$-\f{1}{16}\ka^2\g_{k}i[S](x;x^{\p})\g^{k}
\del_0\del^{\p}_{0}i\d\!\D_{B}(x;x^{\p}) $ & \cr
\omit&height2pt&\omit&&\omit&&\omit&&\omit&\cr \tablerule
\omit&height2pt&\omit&&\omit&&\omit&&\omit&\cr && 3 && 3 && b &&
$\f{1}{16}\ka^2\g^{0}i[S](x;x^{\p})\g^k \,
\del_k\del^{\p}_{0}i\d\!\D_{B}(x;x^{\p}) $ & \cr
\omit&height2pt&\omit&&\omit&&\omit&&\omit&\cr \tablerule
\omit&height2pt&\omit&&\omit&&\omit&&\omit&\cr && 3 && 3 && c &&
$-\f{1}{16}\ka^2\g^{k}i[S](x;x^{\p})\g^0 \,
\del_0\del_{k}i\d\!\D_{B}(x;x^{\p}) $ & \cr
\omit&height2pt&\omit&&\omit&&\omit&&\omit&\cr \tablerule
\omit&height2pt&\omit&&\omit&&\omit&&\omit&\cr && 3 && 3 && d &&
$\f{1}{16}\ka^2\g^{0}i[S](x;x^{\p})\g^0 \nabla^2
i\d\!\D_{B}(x;x^{\p}) $ & \cr
\omit&height2pt&\omit&&\omit&&\omit&&\omit&\cr \tablerule}}

\label{DBcon}

\end{table}

As was the case for the $i\delta\!\Delta_A(x;x^{\p})$ contributions
considered in the previous sub-section, this diagram is not
sufficiently singular for the infinite series terms from
$i\delta\!\Delta_B(x;x^{\p})$ to make a nonzero contribution in the
$D\!=\!4$ limit. Unlike $i\delta\!\Delta_A(x;x^{\p})$, even the
$n\!=\!0$ terms of $i\delta\!\Delta_B(x;x^{\p})$ vanish for
$D\!=\!4$. This means they can only contribute when multiplied by a
divergence.

\begin{table}
\caption{Residual $i\delta\!\Delta_B$ terms in which all derivatives
act upon $\Delta x^2(x;x^{\p})$. All contributions are multiplied by
$\frac{i \kappa^2 H^2}{2^8 \pi^D} \Gamma^2(\frac{D}2) (D\!-\!4) (a
a^{\p})^{2- \frac{D}2}$. }

\vbox{\tabskip=0pt \offinterlineskip
\def\tablerule{\noalign{\hrule}}
\halign to390pt {\strut#& \vrule#\tabskip=1em plus2em& \hfil#\hfil&
\vrule#& \hfil#\hfil& \vrule#& \hfil#\hfil& \vrule#& \hfil#\hfil&
\vrule#& \hfil#\hfil& \vrule#& \hfil#\hfil& \vrule#& \hfil#\hfil&
\vrule#\tabskip=0pt\cr \tablerule
\omit&height4pt&\omit&&\omit&&\omit&&\omit&&\omit&&\omit&&\omit&\cr
&& $\!\!\!\!{\rm I}\!\!\!\!\!\!$ && $\!\!\!\!{\rm J}\!\!\!\!\!\!$ &&
$\!\!\!\!{\rm sub}\!\!\!\!\!\!$ && $\!\!\!\!\frac{\gamma^{\mu}
\Delta x_{\mu}}{\Delta x^{2D-2}}\!\!\!\!$ && $\!\!\!\!\frac{\gamma^i
\Delta x_i}{\Delta x^{2D-2}}\!\!\!\!$ && $\!\!\!\frac{\Vert \Delta
\vec{x}\Vert^2 \gamma^{\mu} \Delta x_{\mu}}{\Delta x^{2D}}\!\!\!\!$
&& $\!\!\!\!\frac{\Vert \Delta \vec{x}\Vert^2 \gamma^i \Delta
x_i}{\Delta x^{2D}}\!\!\!\!$ &\cr
\omit&height4pt&\omit&&\omit&&\omit&&\omit&&\omit&&\omit&&\omit&\cr
\tablerule
\omit&height2pt&\omit&&\omit&&\omit&&\omit&&\omit&&\omit&&\omit&\cr
&& 2 && 3 && a && ${\scriptstyle (D-1)^2}$ && ${\scriptstyle
-(D+1)}$ && ${\scriptstyle -D (D-1)}$ && ${\scriptstyle 2D}$ & \cr
\omit&height2pt&\omit&&\omit&&\omit&&\omit&&\omit&&\omit&&\omit&\cr
\tablerule
\omit&height2pt&\omit&&\omit&&\omit&&\omit&&\omit&&\omit&&\omit&\cr
&& 2 && 3 && b && ${\scriptstyle (D-1)}$ && ${\scriptstyle -2D + 1}$
&& ${\scriptstyle -D}$ && ${\scriptstyle 2D}$ & \cr
\omit&height2pt&\omit&&\omit&&\omit&&\omit&&\omit&&\omit&&\omit&\cr
\tablerule
\omit&height2pt&\omit&&\omit&&\omit&&\omit&&\omit&&\omit&&\omit&\cr
&& 2 && 3 && c && ${\scriptstyle 0}$ && ${\scriptstyle -(D-1)}$ &&
${\scriptstyle -D}$ && ${\scriptstyle 2D}$ & \cr
\omit&height2pt&\omit&&\omit&&\omit&&\omit&&\omit&&\omit&&\omit&\cr
\tablerule
\omit&height2pt&\omit&&\omit&&\omit&&\omit&&\omit&&\omit&&\omit&\cr
&& 2 && 3 && d && ${\scriptstyle 0}$ && ${\scriptstyle -1}$ &&
${\scriptstyle -D}$ && ${\scriptstyle 2D}$ & \cr
\omit&height2pt&\omit&&\omit&&\omit&&\omit&&\omit&&\omit&&\omit&\cr
\tablerule
\omit&height2pt&\omit&&\omit&&\omit&&\omit&&\omit&&\omit&&\omit&\cr
&& 3 && 2 && a && $\!\!\!\!{\scriptstyle -2(D-1)(D-2)}\!\!\!\!$ &&
${\scriptstyle 3D-5}$ && ${\scriptstyle 2 (D-1)^2}$ &&
$\!\!\!\!{\scriptstyle -4(D-1)}$ & \cr
\omit&height2pt&\omit&&\omit&&\omit&&\omit&&\omit&&\omit&&\omit&\cr
\tablerule
\omit&height2pt&\omit&&\omit&&\omit&&\omit&&\omit&&\omit&&\omit&\cr
&& 3 && 2 && b && $\!\!\!\!{\scriptstyle -(D-1)}\!\!\!\!$ &&
${\scriptstyle 3 (D-1)}$ && ${\scriptstyle 2 (D-1)}$ &&
$\!\!\!\!{\scriptstyle -4(D-1)}$ & \cr
\omit&height2pt&\omit&&\omit&&\omit&&\omit&&\omit&&\omit&&\omit&\cr
\tablerule
\omit&height2pt&\omit&&\omit&&\omit&&\omit&&\omit&&\omit&&\omit&\cr
&& 3 && 2 && c && $\!\!\!\!{\scriptstyle 0}\!\!\!\!$ &&
${\scriptstyle 2D -3}$ && ${\scriptstyle 2 (D-1)}$ &&
$\!\!\!\!{\scriptstyle -4(D-1)}$ & \cr
\omit&height2pt&\omit&&\omit&&\omit&&\omit&&\omit&&\omit&&\omit&\cr
\tablerule
\omit&height2pt&\omit&&\omit&&\omit&&\omit&&\omit&&\omit&&\omit&\cr
&& 3 && 2 && d && $\!\!\!\!{\scriptstyle -(D-1)}\!\!\!\!$ &&
${\scriptstyle 2D -1}$ && $\!\!\!\!{\scriptstyle 2 (D-1)}\!\!\!\!$
&& $\!\!\!\!{\scriptstyle -4(D-1)}$ & \cr
\omit&height2pt&\omit&&\omit&&\omit&&\omit&&\omit&&\omit&&\omit&\cr
\tablerule
\omit&height2pt&\omit&&\omit&&\omit&&\omit&&\omit&&\omit&&\omit&\cr
&& 3 && 3 && a && $\!\!\!\!{\scriptstyle \frac12
(D-1)(D-3)}\!\!\!\!$ && $\!\!\!\!{\scriptstyle -(D-3)}\!\!\!\!$ &&
$\!\!\!\!{\scriptstyle -\frac12 (D-1)(D-2)}\!\!\!\!$ &&
$\!\!\!\!{\scriptstyle (D-2)}\!\!\!\!$ & \cr
\omit&height2pt&\omit&&\omit&&\omit&&\omit&&\omit&&\omit&&\omit&\cr
\tablerule
\omit&height2pt&\omit&&\omit&&\omit&&\omit&&\omit&&\omit&&\omit&\cr
&& 3 && 3 && b && $\!\!\!\!{\scriptstyle 0}\!\!\!\!$ && $\!\!\!\!
{\scriptstyle -\frac12 (D-2)}\!\!\!\!$ && $\!\!\!\!{\scriptstyle
-\frac12 (D-2)}\!\!\!\!$ && $\!\!\!\!{\scriptstyle (D-2)}\!\!\!\!$ &
\cr
\omit&height2pt&\omit&&\omit&&\omit&&\omit&&\omit&&\omit&&\omit&\cr
\tablerule
\omit&height2pt&\omit&&\omit&&\omit&&\omit&&\omit&&\omit&&\omit&\cr
&& 3 && 3 && c && $\!\!\!\!{\scriptstyle 0}\!\!\!\!$ && $\!\!\!\!
{\scriptstyle -\frac12 (D-2)}\!\!\!\!$ && $\!\!\!\!{\scriptstyle
-\frac12 (D-2)}\!\!\!\!$ && $\!\!\!\!{\scriptstyle (D-2)}\!\!\!\!$ &
\cr
\omit&height2pt&\omit&&\omit&&\omit&&\omit&&\omit&&\omit&&\omit&\cr
\tablerule
\omit&height2pt&\omit&&\omit&&\omit&&\omit&&\omit&&\omit&&\omit&\cr
&& 3 && 3 && d && ${\scriptstyle \frac12 (D-1)}$ && ${\scriptstyle
-(D-1)}$ && ${\scriptstyle -\frac12 (D-2)}$ && ${\scriptstyle
(D-2)}$ & \cr
\omit&height2pt&\omit&&\omit&&\omit&&\omit&&\omit&&\omit&&\omit&\cr
\tablerule
\omit&height2pt&\omit&&\omit&&\omit&&\omit&&\omit&&\omit&&\omit&\cr
\tablerule
\omit&height2pt&\omit&&\omit&&\omit&&\omit&&\omit&&\omit&&\omit&\cr
&& $\!\!\!\!{\rm Total}\!\!\!\!$ && \omit && \omit &&
$\!\!\!\!{\scriptstyle -\frac12 (D-1)(D-2)} \!\!\!\!$ &&
$\!\!\!\!{\scriptstyle 3 (D-2)}\!\!\!\!$ && $\!\!\!\! {\scriptstyle
\frac12 (D+2)(D-2)}\!\!\!\!$ && $\!\!\!\! {\scriptstyle
-4(D-2)}\!\!\!\!$ &\cr
\omit&height2pt&\omit&&\omit&&\omit&&\omit&&\omit&&\omit&&\omit&\cr
\tablerule}}

\label{DBmost}

\end{table}

Contributions from the $[2\!\!-\!\!2]$ vertex pair require special
treatment to take advantage of the cancelation between the two
series. We will work out the ``a'' term from Table~\ref{DBcon},
\begin{eqnarray}
\lefteqn{\Bigl[2\!\!-\!\!2\Bigr]_a = -\frac{i \kappa^2
\Gamma(\frac{D}2\!-\!1) }{16 \pi^{\frac{D}2}} \partial_0^{\p}
\Biggl\{ i\delta\!\Delta_B(x;x^{\p}) \Bigl( \gamma^0 \hspace{-.1cm}
\not{\hspace{-.1cm} \partial} \hspace{-.1cm} \;
\overline{\not{\hspace{-.1cm} \partial}} \!-\! \gamma^i
\hspace{-.1cm} \not{\hspace{-.1cm} \partial} \gamma^i
\partial_0\Bigr) \Bigl[\frac1{\Delta
x^{D-2}} \Bigr] \Biggr\} , } \\
& & \hspace{-.5cm} = \frac{i \kappa^2 \Gamma(\frac{D}2\!-\!1)}{16
\pi^{\frac{D}2}} \partial_0^{\p} \Biggl\{
i\delta\!\Delta_B(x;x^{\p}) \Bigl(-3
\partial_0 \; \hspace{-.1cm} \overline{\not{\hspace{-.1cm} \partial}} \!+\!
\gamma^0 \nabla^2 \!+\! (D\!-\!1) \hspace{-.1cm} \not{\hspace{-.1cm}
\partial}
\partial_0\Bigr) \Bigl[\frac1{\Delta x^{D-2}} \Bigr] \Biggr\} . \qquad
\end{eqnarray}
A key identity for reducing the $[2\!-\!2]$ terms involves commuting
two derivatives through $1/\Delta x^{D-4}$,
\begin{equation}
\frac1{\Delta x^{D-4}} \partial_{\mu} \partial_{\nu}
\Bigl[\frac1{\Delta x^{D-2}}\Bigr] = \frac1{4(D\!-\!3)} \Bigl(
-\eta_{\mu\nu} \partial^2 \!+\! D \partial_{\mu} \partial_{\nu}
\Bigr) \Bigl[\frac1{\Delta x^{2D-6}}\Bigr] \; . \label{keyID}
\end{equation}
This can be used to extract the derivatives from the first term of
$i\delta\!\Delta_B(x;x^{\p})$, at which point the result is
integrable and we can take $D\!=\!4$,
\begin{eqnarray}
\lefteqn{\Bigl[2\!\!-\!\!2\Bigr]^1_a = \frac{i \kappa^2 H^2}{2^8
\pi^D} \Gamma\Bigl(\frac{D}2\Bigr) \Gamma\Bigl(\frac{D}2\!-\!1\Bigr)
}
\nonumber \\
& & \hspace{2cm} \times \partial_0^{\p} \Biggl\{ \frac{(a
a^{\p})^{2-\frac{D}2}}{\Delta x^{D-4}} \Bigl(-3 \partial_0 \;
\hspace{-.1cm} \overline{\not{\hspace{-.1cm}
\partial}} \!+\! \gamma^0 \nabla^2 \!+\! (D\!-\!1) \hspace{-.1cm} \not{
\hspace{-.1cm} \partial} \partial_0\Bigr) \Bigl[\frac1{\Delta
x^{D-2}} \Bigr] \Biggr\} , \qquad \\
& & = \frac{i \kappa^2 H^2}{2^9 \pi^D} \frac{\Gamma(\frac{D}2
\!+\!1) \Gamma(
\frac{D}2\!-\!1)}{D\!-\!3} (a a^{\p})^{2-\frac{D}2} \nonumber \\
& & \hspace{.5cm} \times \Bigl(-\partial_0 \!-\! \frac12 (D\!-\!4) H
a^{\p}\Bigr) \Bigl(-3 \partial_0 \; \hspace{-.1cm}
\overline{\not{\hspace{-.1cm}
\partial}} \!+\! \gamma^0 \nabla^2 \!+\! (D\!-\!1) \hspace{-.1cm} \not{
\hspace{-.1cm} \partial} \partial_0\Bigr) \Bigl[\frac1{\Delta
x^{2D-6}}
\Bigr] , \qquad \\
& & = -\frac{i \kappa^2 H^2}{2^8 \pi^4} \gamma^0 \partial_0 \Bigl(3
\partial_0^2 \!+\! \nabla^2\Bigr) \Bigl[\frac1{\Delta x^2}\Bigr] +
O(D\!-\!4) \; .
\end{eqnarray}
Of course the second term of $i\delta\!\Delta_B$ is constant so the
derivatives are already extracted,
\begin{eqnarray}
\Bigl[2\!\!-\!\!2\Bigr]^2_a \!& = &\! \frac{i \kappa^2
H^{D-2}}{2^{D+3} \pi^D} \frac{\Gamma(D\!-\!2)}{D\!-\!2} \partial_0
\Bigl(-3
\partial_0 \; \hspace{-.1cm} \overline{\not{\hspace{-.1cm}
\partial}} \!+\! \gamma^0 \nabla^2 \!+\! (D\!-\!1) \hspace{-.1cm}
\not{\hspace{-.1cm} \partial} \partial_0\Bigr)
\Bigl[\frac1{\Delta x^{D-2}} \Bigr] , \qquad \\
\!& = & \!\frac{i \kappa^2 H^2}{2^8 \pi^4} \gamma^0 \partial_0
\Bigl(3
\partial_0^2 \!+\! \nabla^2\Bigr) \Bigl[\frac1{\Delta x^2}\Bigr] +
O(D\!-\!4) \; .
\end{eqnarray}
Hence the total for $[2\!\!-\!\!2]_a$ is zero in $D\!=\!4$
dimensions!

The analogous result for the initial reduction of the other
$[2\!\!-\!\!2]$ term is,
\begin{eqnarray}
\lefteqn{\Bigl[2\!\!-\!\!2\Bigr]_b = \frac{i \kappa^2
\Gamma(\frac{D}2\!-\!1)}{
16 \pi^{\frac{D}2}} } \nonumber \\
& & \times \partial_k \Biggl\{ i\delta\!\Delta_B(x;x^{\p})
\Bigl(-\gamma^0
\partial_0 \partial_k \!+\! \hspace{-.1cm} \overline{\not{\hspace{-.1cm}
\partial}} \, \partial_k \!+\! \gamma_k \partial_0^2 \!+\! \gamma_k \;
\hspace{-.1cm} \overline{\not{\hspace{-.1cm} \partial}} \gamma^0
\partial_0 \Bigr) \Bigl[\frac1{\Delta x^{D-2}}\Bigr] \Biggr\} .
\qquad
\end{eqnarray}
The results for each of the two terms of $i\delta\!\Delta_B$ are,
\begin{eqnarray}
\Bigl[2\!\!-\!\!2\Bigr]^1_b & = & \frac{i \kappa^2 H^2}{2^9 \pi^D}
\frac{\Gamma(\frac{D}2 \!+\!1) \Gamma(\frac{D}2\!-\!1)}{D\!-\!3}
(a a^{\p})^{2-\frac{D}2} \nonumber \\
& & \hspace{3.5cm} \times \Bigl(-2 \gamma^0 \partial_0 \nabla^2 +
\hspace{-.1cm} \overline{\not{\hspace{-.1cm} \partial}} \, \nabla^2
+ \hspace{-.1cm} \overline{\not{\hspace{-.1cm} \partial}} \,
\partial_0^2 \Bigr)
\Bigl[\frac1{\Delta x^{2D-6}} \Bigr] , \qquad \\
& = & \frac{i \kappa^2 H^2}{2^8 \pi^4} \Bigl(-2 \gamma^0 \partial_0
\nabla^2 + \hspace{-.1cm} \overline{\not{\hspace{-.1cm} \partial}}
\, \nabla^2 + \hspace{-.1cm} \overline{\not{\hspace{-.1cm}
\partial}} \, \partial_0^2 \Bigr)
\Bigl[\frac1{\Delta x^2}\Bigr] + O(D\!-\!4) \; , \\
\Bigl[2\!\!-\!\!2\Bigr]^2_b & = & \frac{i \kappa^2 H^{D-2}}{2^{D+3}
\pi^D} \frac{\Gamma(D\!-\!2)}{D\!-\!2} \Bigl(2 \gamma^0 \partial_0
\nabla^2 - \hspace{-.1cm} \overline{\not{\hspace{-.1cm} \partial}}
\, \nabla^2 - \hspace{-.1cm} \overline{\not{\hspace{-.1cm}
\partial}} \, \partial_0^2 \Bigr)
\Bigl[\frac1{\Delta x^{D-2}} \Bigr] , \qquad \\
& = & \frac{i \kappa^2 H^2}{2^8 \pi^4} \Bigl(2 \gamma^0 \partial_0
\nabla^2 - \hspace{-.1cm} \overline{\not{\hspace{-.1cm} \partial}}
\, \nabla^2 - \hspace{-.1cm} \overline{\not{\hspace{-.1cm}
\partial}} \, \partial_0^2 \Bigr) \Bigl[\frac1{\Delta x^2}\Bigr] +
O(D\!-\!4) \; .
\end{eqnarray}
Hence the entire contribution from $[2\!\!-\!\!2]$ vanishes in
$D\!=\!4$.

The lower vertex pairs all involve at least one derivative of
$i\delta\!\Delta_B$,
\begin{eqnarray}
\partial_i i \delta\!\Delta_B(x;x^{\p}) & = & -\frac{H^2 \Gamma(\frac{D}2)}{16
\pi^{\frac{D}2}} (D\!-\!4) (a a^{\p})^{2-\frac{D}2} \, \frac{\Delta
x^i}{
\Delta x^{D-2}} = -\partial_i^{\p} \delta\!\Delta_B(x;x^{\p}) \; , \qquad \\
\partial_0 i \delta\!\Delta_B(x;x^{\p}) & = & \frac{H^2 \Gamma(\frac{D}2)}{16
\pi^{\frac{D}2}} (D\!-\!4) (a a^{\p})^{2-\frac{D}2} \Biggl\{
\frac{\Delta \eta}{
\Delta x^{D-2}} \!-\! \frac{a H}{2 \Delta x^{D-4}} \Biggr\} , \\
\partial_0^{\p} i \delta\!\Delta_B(x;x^{\p}) & = & \frac{H^2 \Gamma(\frac{D}2)}{16
\pi^{\frac{D}2}} (D\!-\!4) (a a^{\p})^{2-\frac{D}2}
\Biggl\{-\frac{\Delta \eta}{ \Delta x^{D-2}} \!-\! \frac{a^{\p} H}{2
\Delta x^{D-4}} \Biggr\} .
\end{eqnarray}
These reductions are very similar to those of the analogous
$i\delta\!\Delta_A$ terms. We make use of the same gamma matrix
identities of Equations \ref{gammafirst}-\ref{gammalast} that were
used in the previous sub-section. The only really new feature is
that one sometimes encounters factors of $\Delta \eta^2$ which we
always resolve as,
\begin{equation}
\Delta \eta^2 = -\Delta x^2 + \Vert \Delta \vec{x} \Vert^2 \; .
\end{equation}
Table~\ref{DBmost} gives our results for the most singular
contributions, those in which all derivatives act upon the conformal
coordinate separation $\Delta x^2$.

The only really unexpected thing about Table~\ref{DBmost} is the
overall factor of $(D\!-\!2)$ common to each of the four sums,
\begin{eqnarray}
\lefteqn{-i \Bigl[\Sigma^{T\ref{DBmost}}\Bigr](x;x^{\p}) = \frac{i
\kappa^2 H^2}{ 2^8 \pi^D} \Gamma^2\Bigl(\frac{D}2\Bigr) (D\!-\!2)
(D\!-\!4) (a a^{\p})^{2-\frac{D}2} \Biggl\{ -\frac12 (D\!-\!1)
\frac{\gamma^{\mu}
\Delta x_{\mu}}{\Delta x^{2D-2}} } \nonumber \\
& & \hspace{2cm} + 3 \frac{\gamma^i \Delta x_i}{\Delta x^{2D-2}} +
\frac12 (D\!+\!2) \frac{\Vert \Delta \vec{x} \Vert^2 \gamma^{\mu}
\Delta x_{\mu}}{ \Delta x^{2D}} - 4 \frac{\Vert \Delta \vec{x}
\Vert^2 \gamma^i \Delta x_i}{ \Delta x^{2D}} \Biggr\} . \qquad
\end{eqnarray}
As with the result of Table~\ref{DAmostc}, we use the differential
identities \ref{difID1}-\ref{difID2} to prepare the last two terms
for partial integration,
\begin{eqnarray}
\lefteqn{-i \Bigl[\Sigma^{T\ref{DBmost}}\Bigr](x;x^{\p}) = \frac{i
\kappa^2 H^2}{ 2^8 \pi^D} \Gamma^2\Bigl(\frac{D}2\Bigr) (D\!-\!2)
(D\!-\!4)
(a a^{\p})^{2-\frac{D}2} } \nonumber \\
& & \times \Biggl\{ -\frac14 (D\!-\!4) \frac{\gamma^{\mu} \Delta
x_{\mu}}{ \Delta x^{2D-2}} + \frac12 \Bigl( \frac{3 D \!-\!
8}{D\!-\!1}\Bigr)
\frac{\gamma^i \Delta x_i}{\Delta x^{2D-2}} \nonumber \\
& & \hspace{1cm} + \frac{(D\!+\!2) \, \nabla^2}{8 (D\!-\!1)
(D\!-\!2)} \Bigl(\frac{\gamma^{\mu} \Delta x_{\mu}}{\Delta x^{2D
-4}} \Bigr) - \frac{\nabla^2} {(D\!-\!1)(D\!-\!2)}
\Bigl(\frac{\gamma^i \Delta x_i}{ \Delta x^{2D-4}} \Bigr) \Biggr\} ,
\qquad
\nonumber \\
& & = \frac{i \kappa^2 H^2}{2^8 \pi^D}
\Gamma^2\Bigl(\frac{D}2\Bigr) (a a^{\p})^{2-\frac{D}2} \Biggl\{
\frac1{16} \Bigl(\frac{D\!-\!4}{D\!-\!3}\Bigr) \hspace{-.1cm}
\not{\hspace{-.1cm} \partial} \partial^2 - \frac18 \frac{(3D
\!-\!8)}{(D\!-\!1) (D\!-\!3)} \; \hspace{-.1cm}
\overline{\not{\hspace{-.1cm}
\partial}} \, \partial^2 \nonumber \\
& & \hspace{.9cm} - \frac1{16} \frac{(D\!+\!2) (D\!-\!4)}{(D\!-\!1)
(D\!-\!3)} \hspace{-.1cm} \not{\hspace{-.1cm} \partial} \nabla^2 +
\frac12 \frac{(D \!-\!4)}{(D\!-\!1) (D\!-\!3)} \; \hspace{-.1cm}
\overline{\not{\hspace{-.1cm}
\partial}} \, \nabla^2 \Biggr\} \frac1{\Delta x^{2D-6}} .\;
\end{eqnarray}
The expression is now integrable so we can take $D\!=\!4$,
\begin{equation}
-i \Bigl[\Sigma^{T\ref{DBmost}}\Bigr](x;x^{\p}) = \frac{i \kappa^2
H^2}{2^8 \pi^4} \Bigl\{-\frac16 \; \hspace{-.1cm}
\overline{\not{\hspace{-.1cm}
\partial}} \, \partial^2 \Bigr\} \frac1{\Delta x^2} + O(D\!-\!4) \; .
\label{7thcon}
\end{equation}

Unlike the $i\delta\!\Delta_A$ terms there is no net contribution
when one or more of the derivatives acts upon a scale factor. If
both derivatives act on scale factors the result is integrable in
$D\!=\!4$ dimensions, and vanishes owing to the factor of
$(D\!-\!4)^2$ from differentiating both $a^{2-\frac{D}2}$ and
$a^{\prime 2 - \frac{D}2}$. If a single derivative acts upon a scale
factor, the result is a factor of either $(D\!-\!4) a$ or $(D\!-\!4)
a^{\p}$ times a term which is logarithmically divergent and {\it
even} under interchange of $x^{\mu}$ and $x^{\prime \mu}$. As we
have by now seen many times, the sum of all such terms contrives to
obey reflection symmetry of Equation \ref{refl} by the separate
extra factors of $(D\!-\!4) a$ and $(D\!-\!4) a^{\p}$ combining to
give,
\begin{equation}
(D\!-\!4) (a - a^{\p}) = (D\!-\!4) a a^{\p} H \Delta \eta \; .
\end{equation}
Of course this makes the sum integrable in $D\!=\!4$ dimensions, at
which point we can take $D\!=\!4$ and the result vanishes on account
of the overall factor of $(D\!-\!4)$.

\subsection{Sub-Leading Contributions from $i \delta \Delta_{C} $ }

The point of this subsection is to compute the contribution from
replacing the graviton propagator in Table~\ref{gen3} by its
residual $C$-type part,
\begin{equation}
i\Bigl[{}_{\alpha\beta} \Delta_{\rho\sigma}\Bigr] \!\rightarrow \!2
\Biggl[ \frac{\eta_{\alpha\beta}
\eta_{\rho\sigma}}{(D\!-\!2)(D\!-\!3)} \!+\! \frac{\delta^0_{\alpha}
\delta^0_{\beta} \eta_{\rho\sigma} \!+\! \eta_{\alpha\beta}
\delta^0_{\rho} \delta^0_{\sigma}}{D\!-\!3} \!+\!
\Bigl(\frac{D\!-\!2}{D\!-\!3}\Bigr) \delta^0_{\alpha}
\delta^0_{\beta} \delta^0_{\rho} \delta^0_{\sigma} \Biggr]
i\delta\!\Delta_C . \label{DCpart}
\end{equation}
As in the previous sub-sections we first make the requisite
contractions and then act the derivatives. The result of this first
step is summarized in Table~\ref{DCcon}. We have sometimes broken
the result for a single vertex pair into parts because the four
different tensors in Equation \ref{DCpart} can make distinct
contributions, and because distinct contributions also come from
breaking up factors of $\gamma^{\alpha} J^{\beta \mu}$. These
distinct contributions are labeled by subscripts $a$, $b$, $c$, etc.

\begin{table}
\caption{Contractions from the $i\delta\!\Delta_C$ part of the
graviton propagator.}

\vbox{\tabskip=0pt \offinterlineskip
\def\tablerule{\noalign{\hrule}}
\halign to390pt {\strut#& \vrule#\tabskip=1em plus2em& \hfil#\hfil&
\vrule#& \hfil#\hfil& \vrule#& \hfil#\hfil& \vrule#& \hfil#\hfil&
\vrule#\tabskip=0pt\cr \tablerule
\omit&height4pt&\omit&&\omit&&\omit&&\omit&\cr &&$\!\!\!\! {\rm
I}\!\!\!\!$ && $\!\!\!\! {\rm J} \!\!\!\!$ && $\!\!\!\! {\rm sub}
\!\!\!\!$ && $iV_I^{\alpha\beta}(x) \, i[S](x;x^{\p}) \, i
V_J^{\rho\sigma}(x^{\p}) \, [\mbox{}_{\alpha\beta} T^C_{\rho\sigma}]
\, i\delta\!\Delta_C(x;x^{\p})$ &\cr
\omit&height4pt&\omit&&\omit&&\omit&&\omit&\cr \tablerule
\omit&height2pt&\omit&&\omit&&\omit&&\omit&\cr && 2 && 1 && a &&
$-\frac1{(D-3)(D-2)} \kappa^2 \hspace{-.1cm} \not{\hspace{ -.1cm}
\partial} \, \delta^D(x\!-\!x^{\p}) \, i\delta\!\Delta_C(x;x)$ & \cr
\omit&height2pt&\omit&&\omit&&\omit&&\omit&\cr \tablerule
\omit&height2pt&\omit&&\omit&&\omit&&\omit&\cr && 2 && 1 && b &&
$-\frac1{D-3} \kappa^2 \partial^{\p}_{\mu} \{ \gamma^0
\partial_0 \, i[S](x;x^{\p}) \gamma^{\mu} \, i\delta\!\Delta_C(x;x^{\p}) \}$ & \cr
\omit&height2pt&\omit&&\omit&&\omit&&\omit&\cr \tablerule
\omit&height2pt&\omit&&\omit&&\omit&&\omit&\cr && 2 && 2 && a &&
$\frac1{2(D-3)(D-2)} \kappa^2 \hspace{-.1cm} \not{\hspace{-.1cm}
\partial} \, \delta^D(x\!-\!x^{\p}) \, i\delta\!\Delta_C(x;x)$ & \cr
\omit&height2pt&\omit&&\omit&&\omit&&\omit&\cr \tablerule
\omit&height2pt&\omit&&\omit&&\omit&&\omit&\cr && 2 && 2 && b &&
$-\frac1{2(D-3)} \kappa^2 \gamma^0 \partial_0 \,
\delta^D(x\!-\!x^{\p}) \, i\delta\!\Delta_C(x;x)$ & \cr
\omit&height2pt&\omit&&\omit&&\omit&&\omit&\cr \tablerule
\omit&height2pt&\omit&&\omit&&\omit&&\omit&\cr && 2 && 2 && c &&
$+\frac1{2(D-3)} \kappa^2 \partial^{\p}_{\mu} \{ \gamma^0
\partial_0 \, i[S](x;x^{\p}) \gamma^{\mu} \, i\delta\!\Delta_C(x;x^{\p}) \}$ & \cr
\omit&height2pt&\omit&&\omit&&\omit&&\omit&\cr \tablerule
\omit&height2pt&\omit&&\omit&&\omit&&\omit&\cr && 2 && 2 && d &&
$-\frac12 (\frac{D-2}{D-3}) \kappa^2 \partial^{\p}_0 \{ \gamma^0
\partial_0 \, i[S](x;x^{\p}) \gamma^0 \, i\delta\!\Delta_C(x;x^{\p}) \}$ &
\cr \omit&height2pt&\omit&&\omit&&\omit&&\omit&\cr \tablerule
\omit&height2pt&\omit&&\omit&&\omit&&\omit&\cr && 2 && 3 && a &&
$-\frac{(D-1)}{4(D-3)(D-2)} \kappa^2 \, \delta^D(x\!-\!x^{\p}) \,
\hspace{-.1cm} \not{\hspace{-.1cm} \partial}^{\p} \,i
\delta\!\Delta_C(x;x^{\p})$ & \cr
\omit&height2pt&\omit&&\omit&&\omit&&\omit&\cr \tablerule
\omit&height2pt&\omit&&\omit&&\omit&&\omit&\cr && 2 && 3 && b &&
$+\frac1{4(D-3)} \kappa^2 \, \delta^D(x\!-\!x^{\p}) \gamma^i
\partial_i^{\p} \, i\delta\!\Delta_C(x;x^{\p})$ & \cr
\omit&height2pt&\omit&&\omit&&\omit&&\omit&\cr \tablerule
\omit&height2pt&\omit&&\omit&&\omit&&\omit&\cr && 2 && 3 && c &&
$+\frac14 (\frac{D-1}{D-3}) \kappa^2 \gamma^0 \partial_0 \,
i[S](x;x^{\p}) \, \hspace{-.1cm} \not{\hspace{-.1cm} \partial}^{\p}
\,i \delta\!\Delta_C(x;x^{\p})$ & \cr
\omit&height2pt&\omit&&\omit&&\omit&&\omit&\cr \tablerule
\omit&height2pt&\omit&&\omit&&\omit&&\omit&\cr && 2 && 3 && d &&
$-\frac14 (\frac{D-2}{D-3}) \kappa^2 \gamma^0 \partial_0 \,
i[S](x;x^{\p}) \gamma^i \partial_i^{\p} \,
i\delta\!\Delta_C(x;x^{\p})$ & \cr
\omit&height2pt&\omit&&\omit&&\omit&&\omit&\cr \tablerule
\omit&height2pt&\omit&&\omit&&\omit&&\omit&\cr && 3 && 1 && a &&
$-\frac{(D-1)}{2(D-3)(D-2)} \kappa^2 \partial^{\p}_{\mu} \{ \,
\hspace{-.1cm} \not{\hspace{-.1cm} \partial} \, i
\delta\!\Delta_C(x;x^{\p}) \, i[S](x;x^{\p}) \gamma^{\mu} \}$ & \cr
\omit&height2pt&\omit&&\omit&&\omit&&\omit&\cr \tablerule
\omit&height2pt&\omit&&\omit&&\omit&&\omit&\cr && 3 && 1 && b &&
$+\frac1{2(D-3)} \kappa^2 \partial^{\p}_{\mu} \{ \gamma^i
\partial_i \,i \delta\!\Delta_C(x;x^{\p}) \, i[S](x;x^{\p}) \gamma^{\mu} \}$ & \cr
\omit&height2pt&\omit&&\omit&&\omit&&\omit&\cr \tablerule
\omit&height2pt&\omit&&\omit&&\omit&&\omit&\cr && 3 && 2 && a &&
$\frac{(D-1)}{4(D-3)(D-2)} \kappa^2 \partial^{\p}_{\mu} \{ \,
\hspace{-.1cm} \not{\hspace{-.1cm} \partial} \, i
\delta\!\Delta_C(x;x^{\p}) \, i[S](x;x^{\p}) \gamma^{\mu} \}$ & \cr
\omit&height2pt&\omit&&\omit&&\omit&&\omit&\cr \tablerule
\omit&height2pt&\omit&&\omit&&\omit&&\omit&\cr && 3 && 2 && b &&
$-\frac14 (\frac{D-1}{D-3}) \kappa^2 \partial^{\p}_0 \{ \,
\hspace{-.1cm} \not{\hspace{-.1cm} \partial} \, i
\delta\!\Delta_C(x;x^{\p}) \, i[S](x;x^{\p}) \gamma^0 \}$ & \cr
\omit&height2pt&\omit&&\omit&&\omit&&\omit&\cr \tablerule
\omit&height2pt&\omit&&\omit&&\omit&&\omit&\cr && 3 && 2 && c && $-
\frac1{4(D-3)} \kappa^2 \partial^{\p}_{\mu} \{ \gamma^i
\partial_i \,i \delta\!\Delta_C(x;x^{\p}) \, i[S](x;x^{\p}) \gamma^{\mu} \}$ & \cr
\omit&height2pt&\omit&&\omit&&\omit&&\omit&\cr \tablerule
\omit&height2pt&\omit&&\omit&&\omit&&\omit&\cr && 3 && 2 && d &&
$+\frac14 (\frac{D-2}{D-3}) \kappa^2 \partial^{\p}_0 \{ \gamma^i
\partial_i \,i \delta\!\Delta_C(x;x^{\p}) \, i[S](x;x^{\p}) \gamma^0 \}$ &
\cr \omit&height2pt&\omit&&\omit&&\omit&&\omit&\cr \tablerule
\omit&height2pt&\omit&&\omit&&\omit&&\omit&\cr && 3 && 3 && a &&
$\frac{(D-1)^2}{8(D-3)(D-2)} \kappa^2 \gamma^{\mu} i[S](x;x^{\p})
\partial_{\mu} \hspace{-.1cm} \not{\hspace{-.1cm} \partial}^{\p} \, i
\delta\!\Delta_C(x;x^{\p})$ & \cr
\omit&height2pt&\omit&&\omit&&\omit&&\omit&\cr \tablerule
\omit&height2pt&\omit&&\omit&&\omit&&\omit&\cr && 3 && 3 && b &&
$-\frac18 (\frac{D-1}{D-3}) \kappa^2 \gamma^{\mu} i[S](x;x^{\p})
\partial_{\mu} \gamma^j \partial_j^{\p} \, i \delta\!\Delta_C(x;x^{\p})$ & \cr
\omit&height2pt&\omit&&\omit&&\omit&&\omit&\cr \tablerule
\omit&height2pt&\omit&&\omit&&\omit&&\omit&\cr && 3 && 3 && c &&
$-\frac18 (\frac{D-1}{D-3}) \kappa^2 \gamma^i i[S](x;x^{\p})
\partial_i \, \hspace{-.1cm} \not{\hspace{-.1cm} \partial}^{\p} \,
i \delta\!\Delta_C(x;x^{\p})$ & \cr
\omit&height2pt&\omit&&\omit&&\omit&&\omit&\cr \tablerule
\omit&height2pt&\omit&&\omit&&\omit&&\omit&\cr && 3 && 3 && d &&
$+\frac18 (\frac{D-2}{D-3}) \kappa^2 \gamma^i i[S](x;x^{\p})
\partial_i \gamma^j \partial_j^{\p} \,i \delta\!\Delta_C(x;x^{\p})$ & \cr
\omit&height2pt&\omit&&\omit&&\omit&&\omit&\cr \tablerule}}

\label{DCcon}

\end{table}

Here $i\delta\!\Delta_C(x;x^{\p})$ is the residual of the $C$-type
propagator of Equation \ref{DeltaC} after the conformal contribution
has been subtracted,
\begin{eqnarray}
\lefteqn{i \delta\!\Delta_C(x;x^{\p}) = \frac{H^2}{16
\pi^{\frac{D}2}} \Bigl( \frac{D}2 \!-\! 3\Bigr)
\Gamma\Bigl(\frac{D}2 \!-\! 1\Bigr) \frac{(a
a^{\p})^{2-\frac{D}2}}{\Delta x^{D-4}}+
\frac{H^{D-2}}{(4\pi)^{\frac{D}2}}
\frac{\Gamma(D \!-\! 3)}{\Gamma(\frac{D}2)} } \nonumber \\
& & \hspace{-.7cm} - \frac{H^{D-2}}{(4\pi)^{\frac{D}2}}
\!\!\sum_{n=1}^{\infty} \!\!\left\{ \!\!\Bigl(n \!-\! \frac{D}2
\!+\! 3\Bigr) \frac{\Gamma(n \!+\! \frac{D}2 \!-\! 1)}{\Gamma(n
\!+\! 2)} \Bigl(\frac{y}4 \Bigr)^{n -\frac{D}2 +2} \!\!\!\!\!\!\! -
(n\!+\!1) \frac{\Gamma(n \!+\! D \!-\! 3)}{\Gamma(n \!+\!
\frac{D}2)} \Bigl(\frac{y}4 \Bigr)^n \!\right\} \!\!. \qquad
\label{dC}
\end{eqnarray}
As with the contributions from $i\delta\!\Delta_B(x;x^{\p})$
considered in the previous sub-section, the only way
$i\delta\!\Delta_C(x;x^{\p})$ can give a nonzero contribution in
$D\!=\!4$ dimensions is for it to multiply a singular term. That
means only the $n\!=\!0$ term can possibly contribute. Even for the
$n\!=\!0$ term, both derivatives must act upon a $\Delta x^2$ to
make a nonzero contribution in $D\!=\!4$ dimensions.

Those of the $[2\!\!-\!\!1]$ and $[2\!\!-\!\!2]$ vertex pairs which
are not proportional to delta functions after the initial
contraction of Table~\ref{DCcon} all contrive to give delta
functions in the end. This happens through the same key identity
\ref{keyID} which was used to reduce the analogous terms in the
previous subsection. In each case we have finite constants times
different contractions of the following tensor function,
\begin{eqnarray}
\lefteqn{ \partial_{\mu}^{\p} \Biggl\{ i\delta\!\Delta_C(x;x^{\p})
\partial_{\alpha}
\partial_{\beta} \Bigl[\frac1{\Delta x^{D-2}}\Bigr] \Biggr\} = \frac{H^{D-2}}{
(4\pi)^{\frac{D}2}} \frac{\Gamma(D\!-\!3)}{\Gamma(\frac{D}2)}\,
\partial_{\mu}^{\p}
\partial_{\alpha} \partial_{\beta} \Bigl[\frac1{\Delta x^{D-2}}\Bigr] }
\nonumber \\
& & \hspace{2cm} + \frac{H^2}{16 \pi^{\frac{D}2}} \Bigl(\frac{D}2
\!-\!3\Bigr) \Gamma\Bigl(\frac{D}2 \!-\!1\Bigr) \partial_{\mu}^{\p}
\Biggl\{ \frac{(a a^{\p})^{2- \frac{D}2}}{\Delta x^{D-4}} \,
\partial_{\alpha} \partial_{\beta} \Bigl[\frac1{
\Delta x^{D-2}} \Bigr]\Biggr\} , \\
& & \hspace{-.5cm} = \frac{H^{D-2}}{(4\pi)^{\frac{D}2}}
\frac{\Gamma(D\!-\!3)}{ \Gamma(\frac{D}2)} \, \partial_{\mu}^{\p}
\partial_{\alpha} \partial_{\beta} \Bigl[\frac1{\Delta
x^{D-2}}\Bigr] + \frac{H^{D-2}}{16 \pi^{\frac{D}2}} \Bigl(\frac{D}2
\!-\!3\Bigr) \Gamma\Bigl(\frac{D}2 \!-\!1\Bigr)
(a a^{\p})^{2-\frac{D}2} \nonumber \\
& & \hspace{2cm} \times \Bigl(\partial_{\mu}^{\p} \!-\! \frac12
(D\!-\!4) H a^{\p} \Bigr) \Biggl\{ \frac{D \, \partial_{\alpha}
\partial_{\beta}}{4 (D\!-\!3)} - \frac{\eta_{\alpha\beta}
\partial^2}{4(D\!-\!3)} \Biggr\} \Bigl[\frac1{\Delta
x^{2D-6}} \Bigr] , \qquad \\
& & \hspace{-.5cm} = \frac{H^2}{16 \pi^2} \partial_{\mu}^{\p}
\partial_{\alpha}
\partial_{\beta} \Bigl[\frac1{\Delta x^2}\Bigr] - \frac{H^2}{16 \pi^2}
\partial_{\mu}^{\p} \Bigl( \partial_{\alpha} \partial_{\beta} \!-\! \frac14
\eta_{\alpha\beta} \partial^2\Bigr) \Bigl[\frac1{\Delta x^2}\Bigr] +
O(D\!-\!4) , \qquad \\
& & \hspace{-.5cm} = -\frac{i H^2}{16} \, \eta_{\alpha\beta}
\partial_{\mu} \delta^4(x\!-\!x^{\p}) + O(D\!-\!4) \; . \label{deltalim}
\end{eqnarray}

It remains to multiply Equation \ref{deltalim} by the appropriate
prefactors and take the appropriate contraction. For example, the
$[2\!\!-\!\!1]_b$ contribution is,
\begin{eqnarray}
\lefteqn{-\frac{\kappa^2}{D\!-\!3} \times \frac{i \Gamma(\frac{D}2
\!-\!1)}{4 \pi^{\frac{D}2}} \times \gamma^0 \delta^{\alpha}_0
\gamma^{\beta} \gamma^{\mu} \times -\frac{i H^2}{16} \, \eta_{\alpha
\beta} \partial_{\mu}
\delta^4(x\!-\!x^{\p}) } \nonumber \\
& & \hspace{4cm} = \frac{\kappa^2 H^2}{16 \pi^2} \times \frac14
\hspace{-.1cm} \not{\hspace{-.1cm} \partial} \delta^4(x\!-\!x^{\p})
+ O(D\!-\!4) \; . \qquad
\end{eqnarray}
We have summarized the results in Table~\ref{DCdelta}, along with
all terms for which the initial contractions of Table~\ref{DCcon}
produced delta functions. The sum of all such terms is,
\begin{equation}
-i \Bigl[\Sigma^{T\ref{DCdelta}}\Bigr](x;x^{\p}) = \frac{\kappa^2
H^2}{16 \pi^2} \Bigl\{ -\frac38 \hspace{-.1cm} \not{\hspace{-.1cm}
\partial} - \frac14 \; \hspace{-.1cm} \overline{\not{\hspace{-.1cm}
\partial} \, } \Bigr\} \delta^4(x\!-\!x^{\p}) + O(D\!-\!4) \; .
\label{8thcon}
\end{equation}

\begin{table}
\caption{Delta functions from the $i\delta\!\Delta_C$ part of the
graviton propagator.}

\vbox{\tabskip=0pt \offinterlineskip
\def\tablerule{\noalign{\hrule}}
\halign to390pt {\strut#& \vrule#\tabskip=1em plus2em& \hfil#\hfil&
\vrule#& \hfil#\hfil& \vrule#& \hfil#\hfil& \vrule#& \hfil#\hfil&
\vrule#& \hfil#\hfil& \vrule#\tabskip=0pt\cr \tablerule
\omit&height4pt&\omit&&\omit&&\omit&&\omit&&\omit&\cr
&&$\!\!\!\!{\rm I}\!\!\!\!$ && $\!\!\!\!{\rm J} \!\!\!\!$ &&
$\!\!\!\! {\rm sub} \!\!\!\!$ && $\!\!\!\! \frac{\kappa^2 H^2}{16
\pi^2} \hspace{-.1cm} \not{\hspace{-.1cm} \partial} \,
\delta^4(x\!-\!x^{\p}) \!\!\!\!$ && $\!\!\!\! \frac{\kappa^2 H^2}{16
\pi^2} \; \hspace{-.1cm} \overline{\not{ \hspace{-.1cm} \partial}}
\, \delta^4(x\!-\!x^{\p}) \!\!\!\!$ &\cr
\omit&height4pt&\omit&&\omit&&\omit&&\omit&&\omit&\cr \tablerule
\omit&height2pt&\omit&&\omit&&\omit&&\omit&&\omit&\cr && 2 && 1 && a
&& $-\frac12$ && $0$ & \cr
\omit&height2pt&\omit&&\omit&&\omit&&\omit&&\omit&\cr \tablerule
\omit&height2pt&\omit&&\omit&&\omit&&\omit&&\omit&\cr && 2 && 1 && b
&& $\frac14$ && $0$ & \cr
\omit&height2pt&\omit&&\omit&&\omit&&\omit&&\omit&\cr \tablerule
\omit&height2pt&\omit&&\omit&&\omit&&\omit&&\omit&\cr && 2 && 2 && a
&& $\frac14$ && $0$ & \cr
\omit&height2pt&\omit&&\omit&&\omit&&\omit&&\omit&\cr \tablerule
\omit&height2pt&\omit&&\omit&&\omit&&\omit&&\omit&\cr && 2 && 2 && b
&& $-\frac12$ && $\frac12$ & \cr
\omit&height2pt&\omit&&\omit&&\omit&&\omit&&\omit&\cr \tablerule
\omit&height2pt&\omit&&\omit&&\omit&&\omit&&\omit&\cr && 2 && 2 && c
&& $-\frac18$ && $0$ & \cr
\omit&height2pt&\omit&&\omit&&\omit&&\omit&&\omit&\cr \tablerule
\omit&height2pt&\omit&&\omit&&\omit&&\omit&&\omit&\cr && 2 && 2 && d
&& $\frac14$ && $-\frac14$ & \cr
\omit&height2pt&\omit&&\omit&&\omit&&\omit&&\omit&\cr \tablerule
\omit&height2pt&\omit&&\omit&&\omit&&\omit&&\omit&\cr && 2 && 3 && a
&& $0$ && $0$ & \cr
\omit&height2pt&\omit&&\omit&&\omit&&\omit&&\omit&\cr \tablerule
\omit&height2pt&\omit&&\omit&&\omit&&\omit&&\omit&\cr && 2 && 3 && b
&& $0$ && $0$ & \cr
\omit&height2pt&\omit&&\omit&&\omit&&\omit&&\omit&\cr \tablerule
\omit&height2pt&\omit&&\omit&&\omit&&\omit&&\omit&\cr \tablerule
\omit&height2pt&\omit&&\omit&&\omit&&\omit&&\omit&\cr && $\!\!\!\!
{\rm Total} \!\!\!\!$ && \omit && \omit && $-\frac38$ && $-\frac14$
& \cr \omit&height2pt&\omit&&\omit&&\omit&&\omit&&\omit&\cr
\tablerule}}

\label{DCdelta}

\end{table}

\begin{table}
\caption{Residual $i\delta\!\Delta_C$ terms in which all derivatives
act upon $\Delta x^2(x;x^{\p})$. All contributions are multiplied by
$\frac{i \kappa^2 H^2}{2^8 \pi^D} \Gamma(\frac{D}2) \Gamma(\frac{D}2
\!-\!1) \frac{(D-4)(D-6)}{ D-3} (a a^{\p})^{2- \frac{D}2}$. }

\vbox{\tabskip=0pt \offinterlineskip
\def\tablerule{\noalign{\hrule}}
\halign to390pt {\strut#& \vrule#\tabskip=1em plus2em& \hfil#\hfil&
\vrule#& \hfil#\hfil& \vrule#& \hfil#\hfil& \vrule#& \hfil#\hfil&
\vrule#& \hfil#\hfil& \vrule#& \hfil#\hfil& \vrule#& \hfil#\hfil&
\vrule#\tabskip=0pt\cr \tablerule
\omit&height4pt&\omit&&\omit&&\omit&&\omit&&\omit&&\omit&&\omit&\cr
&& $\!\!\!\!\!\!{\rm I}\!\!\!\!\!\!$ && $\!\!\!\!\!\!{\rm
J}\!\!\!\!\!\!$ && $\!\!\!\!\!\!{\rm sub}\!\!\!\!\!\!$ &&
$\!\!\!\!\frac{\gamma^{\mu} \Delta x_{\mu}}{\Delta
x^{2D-2}}\!\!\!\!$ && $\!\!\!\!\frac{\gamma^i \Delta x_i}{\Delta
x^{2D-2}}\!\!\!\!$ && $\!\!\!\frac{\Vert \vec{x}\Vert^2 \gamma^{\mu}
\Delta x_{\mu}}{\Delta x^{2D}}\!\!\!\!$ && $\!\!\!\!\frac{\Vert
\vec{x}\Vert^2 \gamma^i \Delta x_i}{\Delta x^{2D}}\!\!\!\!$ &\cr
\omit&height4pt&\omit&&\omit&&\omit&&\omit&&\omit&&\omit&&\omit&\cr
\tablerule
\omit&height2pt&\omit&&\omit&&\omit&&\omit&&\omit&&\omit&&\omit&\cr
&& 2 && $\!\!\!\!3\!\!\!\!$ && $\!\!\!\!{\rm c}\!\!\!\!$ &&
${\scriptstyle -(D-1)^2}$ && ${\scriptstyle D (D-1)}$ &&
${\scriptstyle 0}$ && ${\scriptstyle 0}$ & \cr
\omit&height2pt&\omit&&\omit&&\omit&&\omit&&\omit&&\omit&&\omit&\cr
\tablerule
\omit&height2pt&\omit&&\omit&&\omit&&\omit&&\omit&&\omit&&\omit&\cr
&& 2 && $\!\!\!\!3\!\!\!\!$ && $\!\!\!\!{\rm d}\!\!\!\!$ &&
${\scriptstyle 0}$ && $\!\!\!\!{\scriptstyle (D-1)(D-2)} \!\!\!\!$
&& ${\scriptstyle D (D-2)}$ && $\!\!\!\!{\scriptstyle -2 D (D-2)}
\!\!\!\!$ & \cr
\omit&height2pt&\omit&&\omit&&\omit&&\omit&&\omit&&\omit&&\omit&\cr
\tablerule
\omit&height2pt&\omit&&\omit&&\omit&&\omit&&\omit&&\omit&&\omit&\cr
&& 3 && $\!\!\!\!1\!\!\!\!$ && $\!\!\!\!{\rm a}\!\!\!\!$ &&
${\scriptstyle 4 (D-1)}$ && ${\scriptstyle 0}$ && ${\scriptstyle 0}$
&& ${\scriptstyle 0}$ & \cr
\omit&height2pt&\omit&&\omit&&\omit&&\omit&&\omit&&\omit&&\omit&\cr
\tablerule
\omit&height2pt&\omit&&\omit&&\omit&&\omit&&\omit&&\omit&&\omit&\cr
&& 3 && $\!\!\!\!1\!\!\!\!$ && $\!\!\!\!{\rm b}\!\!\!\!$ &&
${\scriptstyle -2 (D-1)}$ && ${\scriptstyle -2 (D-4)}$ &&
${\scriptstyle 0}$ && ${\scriptstyle 0}$ & \cr
\omit&height2pt&\omit&&\omit&&\omit&&\omit&&\omit&&\omit&&\omit&\cr
\tablerule
\omit&height2pt&\omit&&\omit&&\omit&&\omit&&\omit&&\omit&&\omit&\cr
&& 3 && $\!\!\!\!2\!\!\!\!$ && $\!\!\!\!{\rm a}\!\!\!\!$ &&
$\!\!\!\!{\scriptstyle -2(D-1)}\!\!\!\!$ && ${\scriptstyle 0}$ &&
${\scriptstyle 0}$ && $\!\!\!\!{\scriptstyle 0}\!\!\!\!$ & \cr
\omit&height2pt&\omit&&\omit&&\omit&&\omit&&\omit&&\omit&&\omit&\cr
\tablerule
\omit&height2pt&\omit&&\omit&&\omit&&\omit&&\omit&&\omit&&\omit&\cr
&& 3 && $\!\!\!\!2\!\!\!\!$ && $\!\!\!\!{\rm b}\!\!\!\!$ &&
$\!\!\!\!{\scriptstyle 2 (D-1) (D-2) }\!\!\!\!$ && $\!\!\!\!
{\scriptstyle -2 (D-1) (D-2)}\!\!\!\!$ && ${\scriptstyle 0}$ &&
$\!\!\!\!{\scriptstyle 0}\!\!\!\!$ & \cr
\omit&height2pt&\omit&&\omit&&\omit&&\omit&&\omit&&\omit&&\omit&\cr
\tablerule
\omit&height2pt&\omit&&\omit&&\omit&&\omit&&\omit&&\omit&&\omit&\cr
&& 3 && $\!\!\!\!2\!\!\!\!$ && $\!\!\!\!{\rm c}\!\!\!\!$ &&
$\!\!\!\!{\scriptstyle (D-1)}\!\!\!\!$ && ${\scriptstyle (D -4)}$ &&
${\scriptstyle 0}$ && $\!\!\!\!{\scriptstyle 0}\!\!\!\!$ & \cr
\omit&height2pt&\omit&&\omit&&\omit&&\omit&&\omit&&\omit&&\omit&\cr
\tablerule
\omit&height2pt&\omit&&\omit&&\omit&&\omit&&\omit&&\omit&&\omit&\cr
&& 3 && $\!\!\!\!2\!\!\!\!$ && $\!\!\!\!{\rm d}\!\!\!\!$ &&
$\!\!\!\!{\scriptstyle 0}\!\!\!\!$ && $\!\!\!\!{\scriptstyle -(2D-3)
(D-2)}\!\!\!\!$ && $\!\!\!\!{\scriptstyle -2 (D-1)
(D-2)}\!\!\!\!\!\!$ && $\!\!\!\!{\scriptstyle 4 (D-1)
(D-2)}\!\!\!\!$ & \cr
\omit&height2pt&\omit&&\omit&&\omit&&\omit&&\omit&&\omit&&\omit&\cr
\tablerule
\omit&height2pt&\omit&&\omit&&\omit&&\omit&&\omit&&\omit&&\omit&\cr
&& 3 && $\!\!\!\!3\!\!\!\!$ && $\!\!\!\!{\rm a}\!\!\!\!$ &&
$\!\!\!\!{\scriptstyle -(D-1)^2}\!\!\!\!$ && $\!\!\!\!{\scriptstyle
0} \!\!\!\!$ && $\!\!\!\!{\scriptstyle 0}\!\!\!\!$ &&
$\!\!\!\!{\scriptstyle 0} \!\!\!\!$ & \cr
\omit&height2pt&\omit&&\omit&&\omit&&\omit&&\omit&&\omit&&\omit&\cr
\tablerule
\omit&height2pt&\omit&&\omit&&\omit&&\omit&&\omit&&\omit&&\omit&\cr
&& 3 && $\!\!\!\!3\!\!\!\!$ && $\!\!\!\!{\rm b}\!\!\!\!$ &&
$\!\!\!\!{\scriptstyle \frac12 (D-1)^2}\!\!\!\!$ && $\!\!\!\!
{\scriptstyle \frac12 (D-1) (D-4)}\!\!\!\!$ &&
$\!\!\!\!{\scriptstyle 0} \!\!\!\!$ && $\!\!\!\!{\scriptstyle
0}\!\!\!\!$ & \cr
\omit&height2pt&\omit&&\omit&&\omit&&\omit&&\omit&&\omit&&\omit&\cr
\tablerule
\omit&height2pt&\omit&&\omit&&\omit&&\omit&&\omit&&\omit&&\omit&\cr
&& 3 && $\!\!\!\!3\!\!\!\!$ && $\!\!\!\!{\rm c}\!\!\!\!$ &&
$\!\!\!\!{\scriptstyle \frac12 (D-1)^2}\!\!\!\!$ && $\!\!\!\!
{\scriptstyle \frac12 (D-1) (D-4)}\!\!\!\!$ &&
$\!\!\!\!{\scriptstyle 0} \!\!\!\!$ && $\!\!\!\!{\scriptstyle
0}\!\!\!\!$ & \cr
\omit&height2pt&\omit&&\omit&&\omit&&\omit&&\omit&&\omit&&\omit&\cr
\tablerule
\omit&height2pt&\omit&&\omit&&\omit&&\omit&&\omit&&\omit&&\omit&\cr
&& 3 && $\!\!\!\!3\!\!\!\!$ && $\!\!\!\!{\rm d}\!\!\!\!$ &&
${\scriptstyle -\frac12 (D-1) (D-2)}$ && ${\scriptstyle (D-2)}$ &&
${\scriptstyle \frac12 (D-2)^2}$ && ${\scriptstyle -(D-2)^2}$ & \cr
\omit&height2pt&\omit&&\omit&&\omit&&\omit&&\omit&&\omit&&\omit&\cr
\tablerule
\omit&height2pt&\omit&&\omit&&\omit&&\omit&&\omit&&\omit&&\omit&\cr
\tablerule
\omit&height2pt&\omit&&\omit&&\omit&&\omit&&\omit&&\omit&&\omit&\cr
&& $\!\!\!\!{\rm Total}\!\!\!\!$ && \omit && \omit &&
$\!\!\!\!{\scriptstyle \frac12 (D-1)(D-2)} \!\!\!\!$ &&
$\!\!\!\!{\scriptstyle 2(D-1) -D (D-2)} \!\!\!\!$ && $\!\!\!\!
{\scriptstyle -\frac12 (D-2)^2}\!\!\!\!$ && $\!\!\!\! {\scriptstyle
(D-2)^2}\!\!\!\!$ &\cr
\omit&height2pt&\omit&&\omit&&\omit&&\omit&&\omit&&\omit&&\omit&\cr
\tablerule}}

\label{DCmost}

\end{table}

All the lower vertex pairs involve one or more derivatives of
$i\delta\!\Delta_C$,
\begin{eqnarray}
\partial_i i \delta\!\Delta_C & = & -\frac{H^2 \Gamma(\frac{D}2\!-\!1)}{32
\pi^{\frac{D}2}} (D\!-\!6) (D\!-\!4) (a a^{\p})^{2-\frac{D}2} \,
\frac{\Delta x^i}{
\Delta x^{D-2}} = -\partial_i^{\p} i\delta\!\Delta_C \; , \qquad \\
\partial_0 i \delta\!\Delta_C \!& = &\! \frac{H^2 \Gamma(\frac{D}2\!-\!1)}{32
\pi^{\frac{D}2}} (D\!-\!6)(D\!-\!4) (a a^{\p})^{2-\frac{D}2}
\Biggl\{ \frac{\Delta
\eta}{\Delta x^{D-2}} \!-\! \frac{a H}{2 \Delta x^{D-4}} \Biggr\} , \qquad \\
\partial_0^{\p} i \delta\!\Delta_C \!& = &\! \frac{H^2 \Gamma(\frac{D}2\!-\!1)}{32
\pi^{\frac{D}2}} (D\!-\!6)(D\!-\!4) (a a^{\p})^{2-\frac{D}2}
\Biggl\{-\frac{\Delta \eta}{\Delta x^{D-2}} \!-\! \frac{a^{\p} H}{2
\Delta x^{D-4}} \Biggr\} .\; \qquad
\end{eqnarray}
Their reduction follows the same pattern as in the previous two
sub-sections. Table~\ref{DCmost} summarizes the results for the case
in which all derivatives act upon the conformal coordinate
separation $\Delta x^2$.

When summed, three of the columns of Table~\ref{DCmost} reveal a
factor of $(D\!-\!2)$ which we extract,
\begin{eqnarray}
\lefteqn{-i \Bigl[\Sigma^{T\ref{DCmost}}\Bigr](x;x^{\p}) = \frac{i
\kappa^2 H^2}{ 2^8 \pi^D} \Gamma\Bigl(\frac{D}2\Bigr)
\Gamma\Bigl(\frac{D}2 \!-\!1\Bigr) \frac{(D\!-\!2) (D\!-\!4)
(D\!-\!6)}{(D\!-\!3)} (a a^{\p})^{2-\frac{D}2} }
\nonumber \\
& & \times \Biggl\{ \frac12 (D\!-\!1) \frac{\gamma^{\mu} \Delta
x_{\mu}}{ \Delta x^{2D-2}} + \Bigl[
2\Bigl(\frac{D\!-\!1}{D\!-\!2}\Bigr) \!-\! D \Bigr]
\frac{\gamma^i \Delta x_i}{\Delta x^{2D-2}} \nonumber \\
& & \hspace{3cm} - \frac12 (D\!-\!2) \frac{\Vert \Delta \vec{x}
\Vert^2 \gamma^{\mu} \Delta x_{\mu}}{\Delta x^{2D}} + (D\!-\!2)
\frac{\Vert \Delta \vec{x} \Vert^2 \gamma^i \Delta x_i}{\Delta
x^{2D}} \Biggr\} . \qquad \label{T18}
\end{eqnarray}
We partially integrate Equation \ref{T18} with the aid of Equations
\ref{difID1}-\ref{difID2} and then take $D\!=\!4$, just as we did
for the sum of Table~\ref{DBmost},
\begin{eqnarray}
\lefteqn{-i \Bigl[\Sigma^{T\ref{DCmost}}\Bigr](x;x^{\p}) = \frac{i
\kappa^2 H^2}{ 2^8 \pi^D} \Gamma\Bigl(\frac{D}2\Bigr)
\Gamma\Bigl(\frac{D}2 \!-\!1\Bigr) \frac{(D\!-\!2) (D\!-\!4)
(D\!-\!6)}{(D\!-\!3)} (a a^{\p})^{2-\frac{D}2} }
\nonumber \\
& & \times \Biggl\{ \frac{D}4 \frac{\gamma^{\mu} \Delta
x_{\mu}}{\Delta x^{2D-2}} + \Bigl[2
\Bigl(\frac{D\!-\!1}{D\!-\!2}\Bigr) \!-\! \frac{D^2}{2
(D\!-\!1)} \Bigr] \frac{\gamma^i \Delta x_i}{\Delta x^{2D-2}} \nonumber \\
& & \hspace{3.5cm} - \frac{\nabla^2}{8 (D\!-\!1)}
\Bigl(\frac{\gamma^{\mu} \Delta x_{\mu}}{\Delta x^{2D-4}}\Bigr) +
\frac{\nabla^2}{4(D\!-\!1)} \Bigl(\frac{\gamma^i \Delta x_i}{\Delta
x^{2D-4}}\Bigr) \Biggr\} , \qquad \\
& & \hspace{-.5cm} = \frac{i \kappa^2 H^2}{2^8 \pi^D}
\Gamma\Bigl(\frac{D}2 \Bigr) \Gamma\Bigl(\frac{D}2 \!-\!1\Bigr)
\frac{(D\!-\!2) (D\!-\!6)}{(D\!-\!1) (D\!-\!3)^2} (a
a^{\p})^{2-\frac{D}2} \Biggl\{- \frac{D (D\!-\!1)}{16 (D\!-\!2)}
\hspace{-.1cm} \not{\hspace{-.1cm} \partial} \partial^2 \nonumber \\
& & + \frac{(D^3 \!-\! 6D^2 \!+\! 8 D \!-\! 4)}{8 (D\!-\!2)^2} \;
\hspace{-.1cm} \overline{\not{\hspace{ -.1cm} \partial}} \,
\partial^2 \!+\! \Bigl(\frac{D\!-\!4}{16}\Bigr) \hspace{-.1cm}
\not{\hspace{-.1cm}
\partial} \nabla^2\!-\!\Bigl(\frac{D\!-\!4}8 \Bigr) \, \hspace{-.1cm}
\overline{\not{\hspace{-.1cm} \partial}} \, \nabla^2 \Biggr\}
\frac1{\Delta x^{2D-6}} , \qquad \\
& & \hspace{-.5cm} = \frac{i
\kappa^2 H^2}{2^8 \pi^4} \Bigl\{ \frac12 \hspace{-.1cm}
\not{\hspace{-.1cm} \partial} \,
\partial^2 +\frac16 \; \hspace{-.1cm} \overline{\not{\hspace{-.1cm}
\partial}} \,
\partial^2 \Bigr\} \frac1{\Delta x^2} + O(D\!-\!4) \; .
\label{9thcon}
\end{eqnarray}
As already explained, terms for which one or more derivative acts
upon a scale factor make no contribution in $D\!=\!4$ dimensions, so
this is the final nonzero contribution.

\subsection{Renormalized Result}
The regulated result we have worked so hard to compute derives from
summing expressions \ref{1stcon}, \ref{2ndcon}, \ref{3rdcon},
\ref{4thcon}, \ref{5thcon}, \ref{6thcon}, \ref{7thcon}, \ref{8thcon}
and \ref{9thcon},
\begin{eqnarray}
\lefteqn{-i\Bigl[\Sigma\Bigr](x;x^{\p}) = \kappa^2 \Biggl\{\beta_1
(a a^{\p})^{1- \frac{D}2} \hspace{-.1cm} \not{\hspace{-.1cm}
\partial}
\partial^2 + \beta_2 (a a^{\p})^{2-\frac{D}2} H^2 \hspace{-.1cm}
\not{\hspace{-.1cm} \partial} + \beta_3 (a a^{\p})^{2-\frac{D}2} H^2
\; \hspace{-.1cm} \overline{\not{
\hspace{-.1cm} \partial}} } \nonumber \\
& & \hspace{2cm} + b_2 H^2 \hspace{-.1cm} \not{\hspace{-.1cm}
\partial} + b_3 H^2 \; \hspace{-.1cm} \overline{\not{\hspace{-.1cm}
\partial}} \Biggr\} \delta^D(x\!-\!x^{\p}) + \frac{\kappa^2 H^2}{16
\pi^2} \times -3 \ln(a) \; \hspace{-.1cm}
\overline{\not{\hspace{-.1cm} \partial}} \delta^4(x\!-\!x^{\p})
\nonumber \\
& & \hspace{.2cm} -\frac{i \kappa^2}{2^8 \pi^4} (a a^{\p})^{-1}
\hspace{-.1cm} \not{\hspace{-.1cm} \partial} \partial^4 \Bigl[
\frac{\ln(\mu^2 \Delta x^2)}{ \Delta x^2} \Bigr] + \frac{i \kappa^2
H^2}{2^8 \pi^4} \Biggl\{ \Bigl( -\frac{15}2 \hspace{-.1cm}
\not{\hspace{-.1cm} \partial} \, \partial^2 + \hspace{-.1cm}
\overline{\not{\hspace{-.1cm} \partial}} \, \partial^2 \Bigr)
\Bigl[ \frac{\ln(\mu^2 \Delta x^2)}{\Delta x^2} \Bigr] \nonumber \\
& & \hspace{1cm} + \Bigl(8 \; \hspace{-.1cm}
\overline{\not{\hspace{-.1cm}
\partial}} \partial^2 \!-\! 4 \hspace{-.1cm} \not{\hspace{-.1cm} \partial}
\nabla^2 \Bigr) \Bigl[ \frac{\ln(\frac14 H^2\Delta x^2)}{\Delta x^2}
\Bigr] \!-\! 7 \hspace{-.1cm} \not{\hspace{-.1cm} \partial} \,
\nabla^2 \Bigl[ \frac1{\Delta x^2} \Bigr]\!\Biggr\} \!+\! O(D\!-\!4)
. \qquad \label{regres}
\end{eqnarray}
The various $D$-dependent constants in Equation \ref{regres} are,
\begin{eqnarray}
\beta_1 & \!\!=\!\! & \frac{\mu^{D-4}}{2^8 \pi^{\frac{D}2}}
\frac{\Gamma( \frac{D}2 \!-\!1)}{(D\!-\!3) (D\!-\!4)} \Biggl\{ -2 D
\!+\! 1 \!-\!
\frac2{D\!-\!2} \Biggr\} , \\
\beta_2 &\!\! =\!\! & \frac{\mu^{D-4}}{2^9 \pi^{\frac{D}2}}
\frac{\Gamma(\frac{D}2 \!+\! 1)}{(D\!-\!3) (D\!-\!4)}
\Biggl\{\frac12 D^2 \!-\! 10 D \!+\! 15 \!-\! \frac{24}{D} \!-\!
\frac6{D\!-\!1} \!-\! \frac{35}{D\!-\!3}
\Biggr\} , \qquad \\
\beta_3 & \!\!= \!\!& \frac{\mu^{D-4}}{2^9 \pi^{\frac{D}2}}
\frac{\Gamma(\frac{D}2 \!+\!1)}{(D\!-\!3) (D\!-\!4)} \Biggl\{-D
\!+\! 3 \!+\!
\frac9{D\!-\!3} \Biggr\} , \\
b_2 &\!\!=\!\!& \frac{H^{D-4}}{(4 \pi)^{\frac{D}2}}
\frac{\Gamma(D\!-\!1)}{ \Gamma(\frac{D}2)} \Biggl\{ -\frac{(D\!+\!1)
(D\!-\!1) (D\!-\!4)}{2 (D\!-\!3)}
\times \frac{\pi}2 \cot\Bigl(\frac{\pi D}2\Bigr) \nonumber \\
& & \hspace{4.5cm} - \frac{(D\!-\!1)(D^3 \!-\! 8 D^2 \!+\! 23 D
\!-\! 32)}{8
(D\!-\!2)^2 (D\!-\!3)^2} - \frac7{48}\Biggr\} , \qquad \\
b_3 &\!\!=\!\!& \frac{H^{D-4}}{(4 \pi)^{\frac{D}2}}
\frac{\Gamma(D\!-\!1)}{ \Gamma(\frac{D}2)} \Biggl\{ \frac34 \Bigl(D
\!-\! \frac2{D\!-\!3}\Bigr)
\times \frac{\pi}2 \cot\Bigl(\frac{\pi D}2\Bigr) \nonumber \\
& & \hspace{6cm} + \frac34 \frac{(D^2 \!-\! 6 D \!+\!
8)}{(D\!-\!2)^2 (D\!-\!3)^2} - \frac52 \Biggr\} .
\end{eqnarray}
In obtaining these expressions we have always chosen to convert
finite, $D\!=\!4$ terms with $\partial^2$ acting on $1/\Delta x^2$,
into delta functions,
\begin{equation}
\partial^2 \Bigl[\frac1{\Delta x^2}\Bigr] = i 4 \pi^2 \delta^4(x\!-\!x^{\p}) \; .
\end{equation}
All such terms have then been included in $b_2$ and $b_3$.

\begin{figure}

\includegraphics[width=4 in]{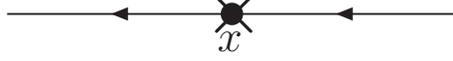}
%%\centerline{\epsfig{file=ffig3.eps}}
\caption{Contribution from
counterterms.}
\label{ffig3}
\end{figure}

The local divergences in this expression are canceled by the BPHZ
counterterms enumerated at the end of section 3. The generic diagram
topology is depicted in Figure \ref{ffig3}, and the analytic form
is,
\begin{eqnarray}
\lefteqn{-i\Bigl[\Sigma^{\rm ctm}\Bigr](x;x^{\p}) = \sum_{I=1}^3 i
C_{Iij} \,
\delta^D(x-x^{\p}) \; , } \\
& & = -\kappa^2 \Bigl\{\alpha_1 (a a^{\p})^{-1} \hspace{-.1cm}
\not{\hspace{-.1cm}
\partial} \partial^2 + \alpha_2 D (D\!-\!1) H^2 \hspace{-.1cm} \not{\hspace{
-.1cm} \partial} + \alpha_3 H^2 \; \hspace{-.1cm}
\overline{\not{\hspace{-.1cm}
\partial}} \Bigr\} \delta^D(x\!-\!x^{\p})  . \label{genctm}
\end{eqnarray}
In comparing Equation \ref{regres} and Equation \ref{genctm} it
would seem that the simplest choice for the coefficients $\alpha_i$
is,
\begin{equation}
\alpha_1 = \beta_1 \quad , \quad \alpha_2 = \frac{\beta_2 \!+\!
b_2}{D (D\!-\!1)} \quad {\rm and} \quad \alpha_3 = \beta_3 \!+\! b_3
\; . \label{sim}
\end{equation}
This choice absorbs all local constants but one is of course left
with time dependent terms proportional to $\ln(a a^{\p})$,
\begin{eqnarray}
\beta_1 (a a^{\p})^{1-\frac{D}2} - \alpha_1 (a a^{\p})^{-1} & = &
+\frac1{2^6 \pi^2}
\frac{\ln(a a^{\p})}{a a^{\p}} + O(D\!-\!4) \; , \\
\beta_2 (a a^{\p})^{2-\frac{D}2} + b_2 - D (D\!-\! 1) \alpha_2 & = &
+\frac{7.5}{2^6 \pi^2} \ln(a a^{\p}) + O(D\!-\!4) \; , \\
\beta_3 (a a^{\p})^{2-\frac{D}2} + b_3 - \alpha_3 & = & -\frac1{2^6
\pi^2} \ln(a a^{\p}) + O(D\!-\!4) \; .
\end{eqnarray}
Our final result for the renormalized self-energy is,
\begin{eqnarray}
\lefteqn{-i\Bigl[\Sigma^{\rm ren}\Bigr](x;x^{\p})
\!=\!\!\frac{\kappa^2}{2^6 \pi^2} \Biggl\{\!\frac{\ln(a a^{\p})}{a
a^{\p}} \hspace{-.1cm} \not{\hspace{-.1cm} \partial}
\partial^2 \!+\! \frac{15}2 \ln(a a^{\p}) H^2\!\hspace{-.1cm} \not{\hspace{-.1cm}
\partial} \!-\! 7 \ln(a a^{\p}) H^2 \; \hspace{-.1cm} \overline{\not{\hspace{-.1cm}
\partial}} \!\Biggr\} \delta^4(x \!-\! x^{\p}) } \nonumber \\
& & \hspace{.2cm} - \frac{i \kappa^2}{2^8 \pi^4} (a a^{\p})^{-1}
\hspace{-.1cm} \not{\hspace{-.1cm} \partial} \partial^4 \Bigl[
\frac{\ln(\mu^2 \Delta x^2)}{ \Delta x^2} \Bigr] + \frac{i \kappa^2
H^2}{2^8 \pi^4} \Biggl\{\Bigl(-\frac{15}2 \hspace{-.1cm}
\not{\hspace{-.1cm} \partial} \, \partial^2 + \hspace{-.1cm}
\overline{\not{\hspace{-.1cm} \partial}} \, \partial^2 \Bigr)
\Bigl[ \frac{\ln(\mu^2 \Delta x^2)}{\Delta x^2} \Bigr] \nonumber \\
& & \hspace{3cm} + \Bigl(8 \; \hspace{-.1cm}
\overline{\not{\hspace{-.1cm}
\partial}} \partial^2 \!-\! 4 \hspace{-.1cm} \not{\hspace{-.1cm} \partial}
\nabla^2 \Bigr) \Bigl[ \frac{\ln(\frac14 H^2\Delta x^2)}{\Delta x^2}
\Bigr] \!-\! 7 \hspace{-.1cm} \not{\hspace{-.1cm} \partial} \,
\nabla^2 \Bigl[ \frac1{\Delta x^2} \Bigr]\!\Biggr\} . \qquad
\label{ren}
\end{eqnarray}

\section{CORRECTING THE MODES}

It is worth summarizing the conventions used in computing the
fermion self-energy. We worked on de Sitter background in conformal
coordinates,
\begin{equation}
ds^2 = a^2(\eta) \Bigl(-d\eta^2 + d\vec{x} \!\cdot\! d\vec{x}\Bigr)
\qquad {\rm where} \qquad a(\eta) = -\frac1{H \eta} = e^{H t} \; .
\end{equation}
We used dimensional regularization and obtained the self-energy for
the conformally re-scaled fermion field,
\begin{equation}
\Psi(x) \equiv a^{(\frac{D-1}2)} \psi(x) \; .
\end{equation}
The local Lorentz gauge was fixed to allow an algebraic expression
for the vierbein in terms of the metric \cite{RPW1}. The general
coordinate gauge was fixed to make the tensor structure of the
graviton propagator decouple from its spacetime dependence
\cite{TW1,RPW2}. The result we obtained is,
\begin{eqnarray}
\lefteqn{\Bigl[\Sigma^{\rm ren}\Bigr](x;x{\p}) \!=\!\frac{i \kappa^2
H^2}{2^6 \pi^2} \Biggl\{\frac{\ln(a a^{\p})}{H^2 a a^{\p}}
\hspace{-.1cm} \not{\hspace{-.1cm}
\partial} \partial^2 \!+\! \frac{15}2 \ln(a a^{\p}) \hspace{-.1cm} \not{\hspace{
-.1cm} \partial} \!-\! 7 \ln(a a^{\p}) \; \hspace{-.1cm}
\overline{\not{\hspace{
-.1cm} \partial}} \Biggr\} \delta^4(x \!-\! x^{\p}) } \nonumber \\
& & \hspace{.2cm} + \frac{\kappa^2}{2^8 \pi^4} (a a^{\p})^{-1}
\hspace{-.1cm} \not{\hspace{-.1cm} \partial} \partial^4 \Bigl[
\frac{\ln(\mu^2 \Delta x^2)}{ \Delta x^2} \Bigr] + \frac{\kappa^2
H^2}{2^8 \pi^4} \Biggl\{\Bigl(\frac{15}2 \hspace{-.1cm}
\not{\hspace{-.1cm} \partial} \, \partial^2 - \hspace{-.1cm}
\overline{\not{\hspace{-.1cm} \partial}} \, \partial^2 \Bigr)
\Bigl[ \frac{\ln(\mu^2 \Delta x^2)}{\Delta x^2} \Bigr] \nonumber \\
& & \hspace{1.5cm} + \Bigl(-8 \; \hspace{-.1cm}
\overline{\not{\hspace{-.1cm}
\partial}} \partial^2 \!+\! 4 \hspace{-.1cm} \not{\hspace{-.1cm} \partial}
\nabla^2 \Bigr) \Bigl[ \frac{\ln(\frac14 H^2\Delta x^2)}{\Delta x^2}
\Bigr] \!+\! 7 \hspace{-.1cm} \not{\hspace{-.1cm} \partial} \,
\nabla^2 \Bigl[ \frac1{\Delta x^2} \Bigr]\!\Biggr\} + O(\kappa^4).
\qquad
\end{eqnarray}
where $\kappa^2 \equiv 16 \pi G$ is the loop counting parameter of
quantum gravity. The various differential and spinor-differential
operators are,
\begin{equation}
\partial^2 \equiv \eta^{\mu\nu} \partial_{\mu} \partial_{\nu} \;\; , \;\;
\nabla^2 \equiv \partial_i \partial_i \;\; , \;\; \hspace{-.1cm}
\not{\hspace{-.1cm} \partial} \equiv \gamma^{\mu} \partial_{\mu}
\;\; {\rm and} \;\; \hspace{-.1cm} \overline{\not{\hspace{-.1cm}
\partial}} \, \equiv \gamma^i
\partial_i \; ,
\end{equation}
where $\eta^{\mu\nu}$ is the Lorentz metric and $\gamma^{\mu}$ are
the gamma matrices. The conformal coordinate interval is basically
$\Delta x^2 \equiv (x\!-\!x^{\p})^{\mu} (x\!-\!x^{\p})^{\nu}
\eta_{\mu\nu}$, up to a subtlety about the imaginary part which will
be explained shortly.

The linearized, effective Dirac equation we will solve is,
\begin{equation}
i\hspace{-.1cm}\not{\hspace{-.08cm} \partial}_{ij} \Psi_{j}(x) -
\int d^4x^{\p} \, \Bigl[\mbox{}_i \Sigma_j \Bigr](x;x^{\p}) \,
\Psi_{j}(x^{\p}) = 0 \; . \label{Diraceqn}
\end{equation}
In judging the validity of this exercise it is important to answer
five questions:
\begin{enumerate}
\item{How do solutions to Equation \ref{Diraceqn} depend upon the finite parts
of counterterms?}
\item{What is the imaginary part of $\Delta x^2$?}
\item{What can we do without the higher loop contributions to the
fermion self-energy?}
\item{What is the relation between the $\comp$-number, effective field
Equation \ref{Diraceqn} and the Heisenberg operator equations of
Dirac + Einstein? and}
\item{How do solutions to Equation \ref{Diraceqn} change when different gauges
are used?}
\end{enumerate}
In next section we will comment on issues 1-3. Issues
4 and 5 are closely related, and require a lengthy digression that
we have consigned to section 2 of this chapter.

\subsection{The Linearized Effective Dirac Equation}
Dirac + Einstein is not perturbatively renormalizable \cite{DVN}, so
we could only obtain a finite result by absorbing divergences in the
BPHZ sense \cite{BP,H,Z1,Z2} using three counterterms involving
either higher derivatives or the curvature $R=12H^{2}$,
\begin{equation}
-\kappa^2 H^2 \Bigl\{\frac{\alpha_1}{H^2 a a^{\p}} \hspace{-.1cm}
\not{\hspace{-.1cm} \partial} \partial^2 + \alpha_2 D (D\!-\!1)
\hspace{-.1cm} \not{\hspace{-.1cm} \partial} + \alpha_3 \;
\hspace{-.1cm} \overline{\not{ \hspace{-.1cm} \partial}} \Bigr\}
\delta^D(x\!-\!x^{\p}) \; .
\end{equation}
No physical principle seems to fix the finite parts of these
counterterms so any result which derives from their values is
arbitrary. We chose to null local terms at the beginning of
inflation ($a = 1$), but any other choice could have been made and
would have affected the solution to Equation \ref{Diraceqn}. Hence
there is no point in solving the equation exactly. However, each of
the three counterterms is related to a term in Equation \ref{ren}
which carries a factor of $\ln(a a^{\p})$,
\begin{equation}
\frac{\alpha_1}{H^2 a a^{\p}} \hspace{-.1cm} \not{\hspace{-.1cm}
\partial}
\partial^2  \Longleftrightarrow  \frac{\ln(a a^{\p})}{H^2 a a^{\p}} \hspace{-.1cm}
\not{\hspace{-.1cm} \partial} \partial^2 \; , \label{1stlog}
\end{equation}
\begin{equation}
 \alpha_2 D (D\!-\!1) \hspace{-.1cm}
\not{\hspace{-.1cm} \partial}  \Longleftrightarrow  \frac{15}2 \ln(a
a^{\p}) \hspace{-.1cm} \not{\hspace{-.1cm}
\partial} \; , \label{2ndlog}
\end{equation}
\begin{equation}
\alpha_3 \; \hspace{-.1cm} \overline{\not{\hspace{-.1cm} \partial}}
 \Longleftrightarrow  -7 \ln(a a^{\p}) \; \hspace{-.1cm}
\overline{\not{\hspace{- .1cm} \partial}} \; . \label{3rdlog}
\end{equation}
Unlike the $\alpha_i$'s, the numerical coefficients of the right
hand terms are uniquely fixed and completely independent of
renormalization. The factors of $\ln(a a^{\p})$ on these right hand
terms mean that they dominate over any finite change in the
$\alpha_i$'s at late times. It is in this late time regime that we
can make reliable predictions about the effect of quantum
gravitational corrections.

The analysis we have just made is a standard feature of low energy
effective field theory, and has many distinguished antecedents
\cite{BN,SW,FS,HS,CDH,CD,DMC1,DL,JFD1,JFD2,MV,HL,ABS,KK1,KK2}. Loops
of massless particles make finite, nonanalytic contributions which
cannot be changed by counterterms and which dominate the far
infrared. Further, these effects must occur as well, with precisely
the same numerical values, in whatever fundamental theory ultimately
resolves the ultraviolet problems of quantum gravity.

We must also clarify what is meant by the conformal coordinate
interval $\Delta x^2(x;x^{\p})$ which appears in Equation \ref{ren}.
The in-out effective field equations correspond to the replacement,
\begin{equation}
\Delta x^2(x;x^{\p}) \longrightarrow \Delta x^2_{\scriptscriptstyle
++}(x;x^{\p}) \equiv \Vert \vec{x} - \vec{x}^{\p} \Vert^2 -
(\mid\e-\e^{\p}\mid-i\d)^2 \; . \label{D++}
\end{equation}
These equations govern the evolution of quantum fields under the
assumption that the universe begins in free vacuum at asymptotically
early times and ends up the same way at asymptotically late times.
This is valid for scattering in flat space but not for cosmological
settings in which particle production prevents the in vacuum from
evolving to the out vacuum. Persisting with the in-out effective
field equations would result in quantum correction terms which are
dominated by events from the infinite future! This is the correct
answer to the question being asked, which is, ``what must the field
be in order to make the universe to evolve from in vacuum to out
vacuum?'' However, that question is not very relevant to any
observation we can make.

A more realistic question is, ``what happens when the universe is
released from a prepared state at some finite time and allowed to
evolve as it will?'' This sort of question can be answered using the
Schwinger-Keldysh formalism \cite{JS,KTM,BM,LVK,CSHY,RDJ,CH,FW}.
Here we digress to briefly derive it. To sketch the derivation,
consider a real scalar field, $\varphi(x)$ whose Lagrangian (not
Lagrangian density) at time $t$ is $L[\varphi(t)]$. The well-known
functional integral expression for the matrix element of an operator
$\mathcal{O}_1[ \varphi]$ between states whose wave functionals are
given at a starting time $s$ and a last time $\ell$ is
\begin{equation}
\Bigl\langle \Phi \Bigl\vert T^*\Bigl(\mathcal{O}_1[\varphi]\Bigr)
\Bigr\vert \Psi \Bigr\rangle = \Fint [d\varphi] \,
\mathcal{O}_1[\varphi] \, \Phi^*[\varphi(\ell)] \, e^{i
\int_{s}^{\ell} dt L[\varphi(t)]} \, \Psi[\varphi(s)] \; .
\label{1stint}
\end{equation}
The $T^*$-ordering symbol in the matrix element indicates that the
operator $\mathcal{O}_1[\varphi]$ is time-ordered, except that any
derivatives are taken {\it outside} the time-ordering. We can use
Equation \ref{1stint} to obtain a similar expression for the matrix
element of the {\it anti}-time-ordered product of some operator
$\mathcal{O}_2[\varphi]$ in the presence of the reversed states,
\begin{eqnarray}
\Bigl\langle \Psi \Bigl\vert
\overline{T}^*\Bigl(\mathcal{O}_2[\varphi]\Bigr) \Bigr\vert \Phi
\Bigr\rangle & = & \Bigl\langle \Phi \Bigl\vert T^*\Bigl(
\mathcal{O}_2^{\dagger}[\varphi]\Bigr) \Bigr\vert \Psi \Bigr\rangle^* \; , \\
& = & \Fint [d\varphi] \, \mathcal{O}_2[\varphi] \,
\Phi[\varphi(\ell)] \, e^{-i \int_{s}^{\ell} dt L[\varphi(t)]} \,
\Psi^*[\varphi(s)] \; . \label{2ndint}
\end{eqnarray}

Now note that summing over a complete set of states $\Phi$ gives a
delta functional,
\begin{equation}
\sum_{\Phi} \Phi\Bigl[\varphi_-(\ell)\Bigr] \,
\Phi^*\Bigl[\varphi_+(\ell) \Bigr] = \delta\Bigl[\varphi_-(\ell)
\!-\! \varphi_+(\ell) \Bigr] \; . \label{sum}
\end{equation}
Taking the product of Equation \ref{1stint} and Equation
\ref{2ndint}, and using Equation \ref{sum}, we obtain a functional
integral expression for the expectation value of any
anti-time-ordered operator $\mathcal{O}_2$ multiplied by any
time-ordered operator $\mathcal{O}_1$,
\begin{eqnarray}
\lefteqn{\Bigl\langle \Psi \Bigl\vert
\overline{T}^*\Bigl(\mathcal{O}_2[ \varphi]\Bigr)
T^*\Bigl(\mathcal{O}_1[\varphi]\Bigr) \Bigr\vert \Psi \Bigr\rangle =
\Fint [d\varphi_+] [d\varphi_-] \, \delta\Bigl[\varphi_-(\ell)
\!-\! \varphi_+(\ell)\Bigr] } \nonumber \\
& & \hspace{1.5cm} \times \mathcal{O}_2[\varphi_-]
\mathcal{O}_1[\varphi_+] \Psi^*[\varphi_-(s)] e^{i \int_s^{\ell} dt
\Bigl\{L[\varphi_+(t)] - L[\varphi_-(t)]\Bigr\}} \Psi[\varphi_+(s)]
\; . \qquad \label{fund}
\end{eqnarray}
This is the fundamental relation between the canonical operator
formalism and the functional integral formalism in the
Schwinger-Keldysh formalism.

The Feynman rules follow from Equation \ref{fund} in close analogy
to those for in-out matrix elements. Because the same field is
represented by two different dummy functional variables,
$\varphi_{\pm}(x)$, the endpoints of lines carry a $\pm$ polarity.
External lines associated with the operator $\mathcal{O}_2[
\varphi]$ have $-$ polarity whereas those associated with the
operator $\mathcal{O}_1[\varphi]$ have $+$ polarity. Interaction
vertices are either all $+$ or all $-$. Vertices with $+$ polarity
are the same as in the usual Feynman rules whereas vertices with the
$-$ polarity have an additional minus sign. Propagators can be $++$,
$-+$, $+-$ and $--$.

The four propagators can be read off from the fundamental relation
\ref{fund} when the free Lagrangian is substituted for the full one.
It is useful to denote canonical expectation values in the free
theory with a subscript $0$. With this convention we see that the
$++$ propagator is just the ordinary Feynman propagator,
\begin{equation}
i\Delta_{\scriptscriptstyle ++}(x;x^{\p}) = \Bigl\langle \Omega
\Bigl\vert T\Bigl(\varphi(x) \varphi(x^{\p}) \Bigr) \Bigr\vert
\Omega \Bigr\rangle_0 = i\Delta(x;x^{\p}) \; . \label{++}
\end{equation}
The other cases are simple to read off and to relate to the Feynman
propagator,
\begin{eqnarray}
i\Delta_{\scriptscriptstyle -+}(x;x^{\p}) \!\!\! & = & \!\!\!
\Bigl\langle \Omega \Bigl\vert \varphi(x) \varphi(x^{\p}) \Bigr\vert
\Omega \Bigr\rangle_0 \nonumber \\
& = & \theta(t\!-\!t^{\p})
i\Delta(x;x^{\p}) \!+\! \theta(t^{\p}\!-\!t) \Bigl[i\Delta(x;x^{\p})
\Bigr]^* \! , \quad \label{-+} \\
i\Delta_{\scriptscriptstyle +-}(x;x^{\p}) \!\!\! & = & \!\!\!
\Bigl\langle \Omega \Bigl\vert \varphi(x^{\p}) \varphi(x) \Bigr\vert
\Omega \Bigr\rangle_0 \nonumber \\
& = & \theta(t\!-\!t^{\p})
\Bigl[i\Delta(x;x^{\p})\Bigr]^* \!\!+\! \theta(t^{\p}\!-\!t)
i\Delta(x;x^{\p}) , \quad \label{+-} \\
i\Delta_{\scriptscriptstyle --}(x;x^{\p}) \!\!\! & = & \!\!\!
\Bigl\langle \Omega \Bigl\vert \overline{T}\Bigl(\varphi(x)
\varphi(x^{\p}) \Bigr) \Bigr\vert \Omega \Bigr\rangle_0 =
\Bigl[i\Delta(x;x^{\p})\Bigr]^* . \label{--}
\end{eqnarray}
Therefore we can get the four propagators of the Schwinger-Keldysh
formalism from the Feynman propagator once that is known.

Because external lines can be either $+$ or $-$ every N-point 1PI
function of the in-out formalism gives rise to $2^N$ 1PI functions
in the Schwinger-Keldysh formalism. For example, the 1PI 2-point
function of the in-out formalism --- which is known as the
self-mass-squared $M^2(x;x^{\p})$ for our scalar example ---
generalizes to four self-mass-squared functions,
\begin{equation}
M^2(x;x^{\p}) \longrightarrow M^2_{\scriptscriptstyle \pm
\pm}(x;x^{\p}) \; .
\end{equation}
The first subscript denotes the polarity of the first position
$x^{\mu}$ and the second subscript gives the polarity of the second
position $x^{\prime \mu}$.

Recall that the in-out effective action is the generating functional
of 1PI functions. Hence its expansion in powers of the background
field $\phi(x)$ takes the form,
\begin{equation}
\Gamma[\phi] = S[\phi] - \frac12 \int d^4x \int d^4x^{\p} \phi(x)
M^2(x;x^{\p}) \phi(x^{\p}) + O(\phi^3) \; ,
\end{equation}
where $S[\phi]$ is the classical action. In contrast, the
Schwinger-Keldysh effective action must depend upon two fields ---
call them $\phi_+(x)$ and $\phi_-(x)$ --- in order to access the
different polarities. At lowest order in the weak field expansion we
have,
\begin{eqnarray}
\lefteqn{\Gamma[\phi_{\scriptscriptstyle +},\phi_{\scriptscriptstyle
-}] = S[\phi_{\scriptscriptstyle +}] - S[\phi_{\scriptscriptstyle
-}]} \nonumber \\
& &  -\frac12 \int \!\! d^4x \! \int \!\! d^4x^{\p} \Biggl\{
\phi_{\scriptscriptstyle +}(x) M^2_{\scriptscriptstyle
++}\!(x;x^{\p}) \phi_{\scriptscriptstyle +}(x^{\p}) \!+\!
\phi_{\scriptscriptstyle +}(x) M^2_{
\scriptscriptstyle+-}\!(x;x^{\p}) \phi_{\scriptscriptstyle
-}(x^{\p})
\nonumber \\ && \!+\! \phi_{\scriptscriptstyle -}(x)
M^2_{\scriptscriptstyle -+}\!(x;x^{\p}) \phi_{\scriptscriptstyle
+}(x^{\p}) \!+\! \phi_{\scriptscriptstyle -}(x) M^2_{
\scriptscriptstyle --}\!(x;x^{\p}) \phi_{\scriptscriptstyle -}(x^{
\p}) \!\Biggr\} \!+\! O(\phi^3_{\pm}). \, \qquad
\end{eqnarray}

The effective field equations of the in-out formalism are obtained
by varying the in-out effective action,
\begin{equation}
\frac{\delta \Gamma[\phi]}{\delta \phi(x)} = \frac{\delta
S[\phi]}{\delta \phi(x)} - \! \int \! d^4x^{\p} M^2(x;x^{\p})
\phi(x^{\p}) + O(\phi^2) .
\end{equation}
Note that these equations are not causal in the sense that the
integral over $x^{\prime \mu}$ receives contributions from points to
the future of $x^{\mu}$. No initial value formalism is possible for
these equations. Note also that even a Hermitian field operator such
as $\varphi(x)$ will not generally admit purely real effective field
solutions $\phi(x)$ because 1PI functions have imaginary parts. This
makes the in-out effective field equations quite unsuitable for
applications in cosmology.

The Schwinger-Keldysh effective field equations are obtained by
varying with respect to $\phi_+$ and then setting both fields equal,
\begin{equation}
\frac{\delta \Gamma[\phi_{\scriptscriptstyle
+},\phi_{\scriptscriptstyle -}] }{\delta \phi_{\scriptscriptstyle
+}(x)} \Biggl\vert_{\phi_{\scriptscriptstyle \pm} = \phi} \!\!\! =
\frac{\delta S[\phi]}{\delta \phi(x)} - \! \int \! d^4x^{\p}
\Bigl[M^2_{\scriptscriptstyle ++}\!(x;x^{\p}) +
M^2_{\scriptscriptstyle +-}\!(x;x^{\p})\Bigr] \phi(x^{\p}) +
O(\phi^2) .
\end{equation}
The sum of $M^2_{\scriptscriptstyle ++}\!(x;x^{\p})$ and
$M^2_{\scriptscriptstyle +-}\!(x;x^{\p})$ is zero unless $x^{\prime
\mu}$ lies on or within the past light-cone of $x^{\mu}$. So the
Schwinger-Keldysh field equations admit a well-defined initial value
formalism in spite of the fact that they are nonlocal. Note also
that the sum of $M^2_{\scriptscriptstyle ++}\!(x;x^{\p})$ and
$M^2_{\scriptscriptstyle +-}\!(x;x^{\p})$ is real, which neither 1PI
function is separately.

From the preceding discussion we can infer these simple rules:
\begin{itemize}
\item{The linearized effective Dirac equation of the Schwinger-Keldysh
formalism takes the form Equation \ref{Diraceqn} with the
replacement,
\begin{equation}
\Bigl[\mbox{}_i \Sigma_j\Bigr](x;x^{\p}) \longrightarrow
\Bigl[\mbox{}_i \Sigma_j\Bigr]_{\scriptscriptstyle
++}\!\!\!\!(x;x^{\p}) + \Bigl[\mbox{}_i
\Sigma_j\Bigr]_{\scriptscriptstyle +-}\!\!\!\!(x;x^{\p}) \; ;
\end{equation}}
\item{The ${\scriptscriptstyle ++}$ fermion self-energy is Equation \ref{ren} with
the replacement Equation \ref{D++}; and}
\item{The ${\scriptscriptstyle +-}$ fermion self-energy is,
\begin{eqnarray}
\lefteqn{ - \frac{\kappa^2}{2^8 \pi^4 a a^{\p}} \hspace{-.1cm}
\not{\hspace{-.1cm} \partial}
\partial^4 \Bigl[\frac{\ln(\mu^2 \Delta x^2)}{ \Delta x^2} \Bigr] -
\frac{\kappa^2 H^2}{2^8 \pi^4} \Biggl\{\Bigl(\frac{15}2
\hspace{-.1cm} \not{\hspace{-.1cm} \partial} \, \partial^2 -
\hspace{-.1cm} \overline{\not{\hspace{-.1cm} \partial}} \,
\partial^2 \Bigr)
\Bigl[ \frac{\ln(\mu^2 \Delta x^2)}{\Delta x^2} \Bigr]} \nonumber \\
& & \hspace{.3cm} + \Bigl(-8 \; \hspace{-.1cm}
\overline{\not{\hspace{-.1cm}
\partial}} \partial^2 \!+\! 4 \hspace{-.1cm} \not{\hspace{-.1cm} \partial}
\nabla^2 \Bigr) \Bigl[ \frac{\ln(\frac14 H^2\Delta x^2)}{\Delta x^2}
\Bigr] \!+\! 7 \hspace{-.1cm} \not{\hspace{-.1cm} \partial} \,
\nabla^2 \Bigl[ \frac1{\Delta x^2} \Bigr]\!\Biggr\} + O(\kappa^4) ,
\qquad
\end{eqnarray}
with the replacement,
\begin{equation}
\Delta x^2(x;x^{\p}) \longrightarrow \Delta x^2_{\scriptscriptstyle
+-}(x;x^{\p}) \equiv \Vert \vec{x} - \vec{x}^{\p} \Vert^2 -
(\e-\e^{\p} + i\d)^2 \; . \label{D+-}
\end{equation}}
\end{itemize}
The difference of the ${\scriptscriptstyle ++}$ and
${\scriptscriptstyle +-}$ terms leads to zero contribution in
Equation \ref{Diraceqn} unless the point $x^{\prime \mu}$ lies on or
within the past light-cone of $x^{\mu}$.

We can only solve for the one loop corrections to the field because
we lack the higher loop contributions to the self-energy. The
general perturbative expansion takes the form,
\begin{equation}
\Psi(x) = \sum_{\ell = 0}^{\infty} \kappa^{2\ell} \Psi^{\ell}(x)
\qquad {\rm and}\,\,\, \Bigl[\Sigma\Bigr](x;x^{\p}) =
\sum_{\ell=1}^{\infty} \kappa^{2\ell}
\Bigl[\Sigma^{\ell}\Bigr](x;x^{\p}) \; .
\end{equation}
One substitutes these expansions into the effective Dirac equation
in Equation \ref{Diraceqn} and then segregates powers of $\kappa^2$,
\begin{equation}
i\hspace{-.1cm}\not{\hspace{-.08cm} \partial} \Psi^0(x) = 0 \qquad ,
\qquad i\hspace{-.1cm}\not{\hspace{-.08cm} \partial} \Psi^1(x) =
\int d^4x^{\p} \Bigl[\Sigma^1\Bigr](x;x^{\p}) \Psi^0(x^{\p}) \qquad
{\rm et\ cetera.}
\end{equation}
We shall work out the late time limit of the one loop correction
$\Psi^1_i(\eta,\vec{x};\vec{k},s)$ for a spatial plane wave of
helicity $s$,
\begin{equation}
\Psi^0_i(\eta,\vec{x};\vec{k},s) = \frac{e^{-i k \eta}}{\sqrt{2k}}
u_i(\vec{k},s) e^{i \vec{k} \cdot \vec{x}} \qquad {\rm where} \qquad
k^{\ell} \gamma^{\ell}_{ij} u_j(\vec{k},s) = k \gamma^0_{ij}
u_j(\vec{k},s) \; .\label{freefun}
\end{equation}

\subsection{Effective Field Equations}

The purpose of this section is to elucidate the relation between the
Heisenberg operators of Dirac + Einstein --- $\overline{\psi}_i(x)$,
$\psi_i(x)$ and $h_{\mu\nu}(x)$ --- and the $\comp$-number plane
wave mode solutions $\Psi_i(x;\vec{k},s)$ of the linearized,
effective Dirac equation in Equation \ref{Diraceqn}. After
explaining the relation we work out an example, at one loop order,
in a simple scalar analogue model. Finally, we return to Dirac +
Einstein to explain how $\Psi_i(x;\vec{k},s)$ changes with
variations of the gauge.

One solves the gauge-fixed Heisenberg operator equations
perturbatively,
\begin{eqnarray}
h_{\mu\nu}(x) & = & h^0_{\mu\nu}(x) + \kappa h^1_{\mu\nu}(x) +
\kappa^2
h^2_{\mu\nu}(x) + \ldots \; , \qquad \\
\psi_i(x) & = & \psi^0_i(x) + \kappa \psi^1_i(x) + \kappa^2
\psi^2_i(x) + \ldots \; .
\end{eqnarray}
Because our state is released in free vacuum at $t=0$ ($\eta =
-1/H$), it makes sense to express the operator as a functional of
the creation and annihilation operators of this free state. So our
initial conditions are that $h_{\mu\nu}$ and its first time
derivative coincide with those of $h^0_{\mu\nu}(x)$ at $t=0$, and
also that $\psi_i(x)$ coincides with $\psi^0_i(x)$. The zeroth order
solutions to the Heisenberg field equations take the form,
\begin{eqnarray}
h^0_{\mu\nu}(x) & = & \int \frac{d^{D-1}k}{(2\pi)^{D-1}}
\sum_{\lambda} \Bigl\{ \epsilon_{\mu\nu}(\eta;\vec{k},\lambda) e^{i
\vec{k} \cdot \vec{x}}
\alpha(\vec{k},\lambda) \nonumber \\
& & \hspace{5cm} + \epsilon^*_{\mu\nu}(\eta;\vec{k},\lambda) e^{-i
\vec{k} \cdot \vec{x}} \alpha^{\dagger}(\vec{k},\lambda) \Bigr\} \;
,
\qquad \\
\psi^0_i(x) & = & a^{-(\frac{D-1}2)} \int
\frac{d^{D-1}k}{(2\pi)^{D-1}} \sum_s \Bigl\{ \frac{e^{-ik
\eta}}{\sqrt{2 k}} u_i(\vec{k},s) e^{i \vec{k}
\cdot \vec{x}} b(\vec{k},s) \nonumber \\
& & \hspace{5cm} + \frac{e^{ik \eta}}{\sqrt{2 k}}
v_i(\vec{k},\lambda) e^{-i \vec{k} \cdot \vec{x}}
c^{\dagger}(\vec{k},s) \Bigr\} \; . \qquad
\end{eqnarray}
The graviton mode functions are proportional to Hankel functions
whose precise specification we do not require. The Dirac mode
functions $u_i(\vec{k}, s)$ and $v_i(\vec{k},s)$ are precisely those
of flat space by virtue of the conformal invariance of massless
fermions. The canonically normalized creation and annihilation
operators obey,
\begin{eqnarray}
\Bigl[\alpha(\vec{k},\lambda),
\alpha^{\dagger}(\vec{k}^{\p},\lambda^{\p})\Bigr] & = &
\delta_{\lambda \lambda^{\p}} (2\pi)^{D-1} \delta^{D-1}(\vec{k}
\!-\!
\vec{k}^{\p}) \label{alpop} \; , \\
\Bigl\{b(\vec{k},s), b^{\dagger}(\vec{k}^{\p},s^{\p})\Bigr\} & = &
\delta_{s s^{\p}} (2\pi)^{D-1} \delta^{D-1}(\vec{k} \!-\!
\vec{k}^{\p}) = \Bigl\{c(\vec{k},s),
c^{\dagger}(\vec{k}^{\p},s^{\p})\Bigr\} \; . \qquad \label{bcop}
\end{eqnarray}

The zeroth order Fermi field $\psi^0_i(x)$ is an anti-commuting
operator whereas the mode function $\Psi^0(x;\vec{k},s)$ is a
$\comp$-number. The latter can be obtained from the former by
anti-commuting with the fermion creation operator,
\begin{equation}
\Psi^0_i(x;\vec{k},s) = a^{\frac{D-1}2} \Bigl\{\psi^0_i(x),
b^{\dagger}(\vec{k},s)\Bigr\} = \frac{e^{-i k \eta}}{\sqrt{2k}}
u_i(\vec{k},s) e^{i \vec{k} \cdot \vec{x}} \; .
\end{equation}
The higher order contributions to $\psi_i(x)$ are no longer linear
in the creation and annihilation operators, so anti-commuting the
full solution $\psi_i(x)$ with $b^{\dagger}(\vec{k},s)$ produces an
operator. The quantum-corrected fermion mode function we obtain by
solving Equation \ref{Diraceqn} is the expectation value of this
operator in the presence of the state which is free vacuum at $t=0$,
\begin{equation}
\Psi_i(x;\vec{k},s) = a^{\frac{D-1}2} \Bigl\langle \Omega \Bigl\vert
\Bigl\{ \psi_i(x), b^{\dagger}(\vec{k},s) \Bigr\} \Bigr\vert \Omega
\Bigr\rangle \; . \label{SKop}
\end{equation}
This is what the Schwinger-Keldysh field equations give. The more
familiar, in-out effective field equations obey a similar relation
except that one defines the free fields to agree with the full ones
in the asymptotic past, and one takes the in-out matrix element
after anti-commuting.

\subsection{A Worked-Out Example}
It is perhaps worth seeing a worked-out example, at one loop order,
of the relation \ref{SKop} between the Heisenberg operators and the
Schwinger-Keldysh field equations. To simplify the analysis we will
work with a model of two scalars in flat space,
\begin{equation}
\mathcal{L} = - \partial_{\mu} \varphi^* \partial^{\mu} \varphi -
m^2 \varphi^* \varphi - \lambda \chi \!:\! \varphi^* \varphi \!:\! -
\frac12 \partial_{\mu} \chi \partial^{\mu} \chi \; . \label{2sL}
\end{equation}
In this model $\varphi$ plays the role of our fermion $\psi_i$, and
$\chi$ plays the role of the graviton $h_{\mu\nu}$. Note that we
have normal-ordered the interaction term to avoid the harmless but
time-consuming digression that would be required to deal with $\chi$
developing a nonzero expectation value. We shall also omit
discussion of counterterms.

The Heisenberg field equations for Equation \ref{2sL} are,
\begin{eqnarray}
\partial^2 \chi - \lambda \!:\! \varphi^* \varphi \!:\! & = & 0 \; , \\
(\partial^2 - m^2) \varphi - \lambda \chi \varphi & = & 0 \; .
\end{eqnarray}
As with Dirac + Einstein, we solve these equations perturbatively,
\begin{eqnarray}
\chi(x) & = & \chi^0(x) + \lambda \chi^1(x) + \lambda^2 \chi^2(x) +
\ldots \; , \\
\varphi(x) & = & \varphi^0(x) + \lambda \varphi^1(x) + \lambda^2
\varphi^2(x) + \ldots \; .
\end{eqnarray}
The zeroth order solutions are,
\begin{eqnarray}
\chi^0(x) & = & \int \frac{d^{D-1}k}{(2\pi)^{D-1}} \Bigl\{
\frac{e^{-ikt}}{ \sqrt{2k}} e^{i \vec{k} \cdot \vec{x}}
\alpha(\vec{k}) + \frac{e^{ikt}}{ \sqrt{2k}} e^{-i \vec{k} \cdot
\vec{x}} \alpha^{\dagger}(\vec{k}) \Bigr\}
\; , \\
\varphi^0(x) & = & \int \frac{d^{D-1}k}{(2\pi)^{D-1}} \Bigl\{
\frac{e^{-i \omega t}}{\sqrt{2 \omega}} e^{i \vec{k} \cdot \vec{x}}
b(\vec{k}) + \frac{e^{i \omega t}}{\sqrt{2 \omega}} e^{-i \vec{k}
\cdot \vec{x}} c^{\dagger}(\vec{k}) \Bigr\} \; .
\end{eqnarray}
Here $k \equiv \Vert \vec{k} \Vert$ and $\omega \equiv \sqrt{k^2 +
m^2}$. The creation and annihilation operators are canonically
normalized,
\begin{equation}
\Bigl[\alpha(\vec{k}),\alpha^{\dagger}(\vec{k}^{\p})\Bigr] =
\Bigl[b(\vec{k}),b^{\dagger}(\vec{k}^{\p})\Bigr] =
\Bigl[c(\vec{k}),c^{\dagger}(\vec{k}^{\p})\Bigr] = (2\pi)^{D-1}
\delta^{D-1}( \vec{k} - \vec{k}^{\p}) \; .
\end{equation}
We choose to develop perturbation theory so that all the operators
and their first time derivatives agree with the zeroth order
solutions at $t=0$. The first few higher order terms are,
\begin{eqnarray}
\chi^1(x) & \!\!\!\!\! = \!\!\!\!\! & \int_0^t \!\! dt^{\p} \!\!
\int d^{D-1}x^{\p} \, \Bigl\langle x \Bigl\vert \frac1{\partial^2}
\Bigr\vert x^{\p}
\Bigr\rangle_{\rm ret} \!:\! \varphi^{0*}(x^{\p}) \varphi^0(x^{\p}) \!:\! \; , \\
\varphi^1(x) & \!\!\!\!\! = \!\!\!\!\! & \int_0^t \!\! dt^{\p} \!\!
\int d^{D-1}x^{\p} \, \Bigl\langle x \Bigl\vert \frac1{\partial^2
\!-\! m^2} \Bigr\vert x^{\p}
\Bigr\rangle_{\rm ret} \chi^0(x^{\p}) \varphi^0(x^{\p}) \; , \\
\varphi^2(x) & \!\!\!\!\! = \!\!\!\!\! & \int_0^t \!\! dt^{\p} \!\!
\int d^{D-1}x^{\p} \Bigl\langle x \Bigl\vert \frac1{\partial^2 \!-\!
m^2} \Bigr\vert x^{\p} \Bigr\rangle_{\rm ret} \Bigl\{\chi^1(x^{\p})
\varphi^0(x^{\p}) \!+\! \chi^0(x^{\p}) \varphi^1(x^{\p}) \Bigr\} .
\qquad
\end{eqnarray}

The commutator of $\varphi^0(x)$ with $b^{\dagger}(\vec{k})$ is a
$\comp$-number,
\begin{equation}
\Bigl[ \varphi^0(x) , b^{\dagger}(\vec{k}) \Bigr] = \frac{e^{-i
\omega t}}{ \sqrt{2 \omega}} \, e^{i \vec{k} \cdot \vec{x}} \equiv
\Phi^0(x;\vec{k}) \; . \label{Phi^0}
\end{equation}
However, commuting the full solution with $b^{\dagger}(\vec{k})$
leaves operators,
\begin{eqnarray}
\lefteqn{\Bigl[ \varphi(x) , b^{\dagger}(\vec{k}) \Bigr] =
\Phi^0(x;\vec{k}) + \lambda \int_0^t \!\! dt^{\p} \!\! \int
d^{D-1}x^{\p} \, \Bigl\langle x \Bigl\vert \frac1{\partial^2 \!-\!
m^2} \Bigr\vert x^{\p}
\Bigr\rangle_{\rm ret} \chi^0(x^{\p}) \Phi^0(x^{\p};\vec{k}) } \nonumber \\
& & \hspace{-.5cm} + \lambda^2 \int_0^t \!\! dt^{\p} \!\! \int
d^{D-1}x^{\p} \, \Bigl\langle x \Bigl\vert \frac1{\partial^2 \!-\!
m^2} \Bigr\vert x^{\p} \Bigr\rangle_{\rm ret} \Biggl\{
\Bigl[\chi^1(x^{\p}) , b^{\dagger}(\vec{k})\Bigr]
\varphi^0(x^{\p}) + \chi^1(x^{\p}) \Phi^0(x^{\p};\vec{k}) \nonumber \\
& & \hspace{5.8cm} + \chi^0(x^{\p}) \Bigl[ \varphi^1(x^{\p}) ,
b^{\dagger}(\vec{k}) \Bigr] \Biggr\} + O(\lambda^3) \; . \qquad
\label{com}
\end{eqnarray}
The commutators in Equation \ref{com} are easily evaluated,
\begin{eqnarray}
\lefteqn{\Bigl[\chi^1(x^{\p}) , b^{\dagger}(\vec{k})\Bigr]
\varphi^0(x^{\p}) }
\nonumber \\
& & \hspace{1.5cm} = \int_0^{t^{\p}} \!\! dt^{\prime\prime} \!\!
\int d^{D-1}x^{\prime\prime} \, \Bigl\langle x^{\p} \Bigl\vert
\frac1{\partial^2} \Bigr\vert x^{\prime\prime} \Bigr\rangle_{\rm
ret}
\varphi^{0*}(x^{\prime\prime}) \varphi^0(x^{\p})
\Phi^0(x^{\prime\prime};\vec{k}) \; , \qquad \\
\lefteqn{\chi^0(x^{\p}) \Bigl[\varphi^1(x^{\p}) ,
b^{\dagger}(\vec{k}) \Bigr] }
\nonumber \\
& & \hspace{1.4cm} = \int_0^{t^{\p}} \!\! dt^{\p\p}\!\!\int
d^{D-1}x^{\p\p} \, \Bigl\langle x^{\p} \Bigl\vert \frac1{\partial^2
\!-\! m^2} \Bigr\vert x^{\p\p} \Bigr\rangle_{\rm ret} \chi^0(x^{\p})
\chi^0(x^{\p\p}) \Phi^0(x^{\p\p};\vec{k})  \;.\; \qquad
\end{eqnarray}
Hence the expectation value of Equation \ref{com} gives,
\begin{eqnarray}
\lefteqn{\Bigl\langle \Omega \Bigl\vert \Bigl[ \varphi(x),
b^{\dagger}(\vec{k}) \Bigr] \Bigr\vert \Omega \Bigr\rangle =
\Phi^0(x;\vec{k}) + \lambda^2 \int_0^t \!\! dt^{\p} \!\! \int
d^{D-1}x^{\p} \, \Bigl\langle x \Bigl\vert \frac1{\partial^2
\!-\! m^2} \Bigr\vert x^{\p} \Bigr\rangle_{\rm ret} } \nonumber \\
& & \times \int_0^{t^{\p}} \!\! dt^{\p\p} \!\! \int d^{D-1}x^{\p\p}
\, \Biggl\{ \Bigl\langle x^{\p} \Bigl\vert \frac1{\partial^2}
\Bigr\vert x^{\p\p} \Bigr\rangle_{\rm ret} \Bigl\langle \Omega
\Bigl\vert \varphi^{0*}(x^{\p\p}) \varphi^0(x^{\p}) \Bigr\vert
\Omega \Bigr\rangle \nonumber \\
& & \hspace{1.2 cm} + \Bigl\langle x^{\p} \Bigl\vert
\frac1{\partial^2 \!-\! m^2} \Bigr\vert x^{\p\p} \Bigr\rangle_{\rm
ret} \Bigl\langle \Omega \Bigl\vert \chi^0(x^{\p}) \chi^0(x^{\p\p})
\Bigr\vert \Omega \Bigr\rangle \Biggr\} \Phi^0(x^{\p\p}; \vec{k}) +
O(\lambda^4)  .\; \qquad \label{expcom}
\end{eqnarray}

To make contact with the effective field equations we must first
recognize that the retarded Green's functions can be written in
terms of expectation values of the free fields,
\begin{eqnarray}
\lefteqn{\Bigl\langle x^{\p} \Bigl\vert \frac1{\partial^2}
\Bigr\vert x^{\p\p} \Bigr\rangle_{\rm ret} = -i \theta(t^{\p} \!-\!
t^{\p\p}) \Bigl[ \chi^0(x^{\p}) ,
\chi^0(x^{\p\p})\Bigr] } \\
& & \hspace{1cm} = -i \theta(t^{\p} \!-\! t{\p\p}) \Biggl\{
\Bigl\langle \Omega \Bigl\vert \chi^0(x^{\p}) \chi^0(x^{\p\p})
\Bigr\vert \Omega \Bigr\rangle - \Bigl\langle \Omega \Bigl\vert
\chi^0(x^{\p\p}) \chi^0(x^{\p}) \Bigr\vert \Omega \Bigr\rangle
\Biggr\} \; , \qquad \\
\lefteqn{\Bigl\langle x^{\p} \Bigl\vert \frac1{\partial^2 \!-\! m^2}
\Bigr\vert x^{\p\p} \Bigr\rangle_{\rm ret} = -i \theta(t^{\p} \!-\!
t^{\p\p}) \Bigl[ \varphi^0(x^{\p}) ,
\varphi^{0*}(x^{\p\p})\Bigr] } \\
& & \hspace{.9cm} = -i \theta(t^{\p} \!-\! t^{\p\p}) \Biggl\{
\Bigl\langle \Omega \Bigl\vert \varphi^0(x^{\p})
\varphi^{0*}(x^{\p\p}) \Bigr\vert \Omega \Bigr\rangle - \Bigl\langle
\Omega \Bigl\vert \varphi^{*0}(x^{\p\p}) \varphi^0(x^{\p})
\Bigr\vert \Omega \Bigr\rangle \Biggr\}  . \;\qquad
\end{eqnarray}
Substituting these relations into Equation \ref{expcom} and
canceling some terms gives the expression we have been seeking,
\begin{eqnarray}
\lefteqn{\Bigl\langle \Omega \Bigl\vert \Bigl[ \varphi(x),
b^{\dagger}(\vec{k}) \Bigr] \Bigr\vert \Omega \Bigr\rangle =
\Phi^0(x;\vec{k}) -i \lambda^2 \int_0^t \!\! dt^{\p} \!\! \int
d^{D-1}x^{\p} \, \Bigl\langle x \Bigl\vert \frac1{\partial^2
\!-\! m^2} \Bigr\vert x^{\p} \Bigr\rangle_{\rm ret} } \nonumber \\
& & \times \int_0^{t^{\p}} \!\! dt^{\p\p} \!\! \int d^{D-1}x^{\p\p}
\, \Biggl\{ \Bigl\langle \Omega\Bigl\vert \chi^0(x^{\p})
\chi^0(x^{\p\p}) \Bigr\vert \Omega \Bigr\rangle \Bigl\langle
\Omega\Bigl\vert \varphi^0(x^{\p}) \varphi^{0*}(x^{\p\p}) \Bigr\vert
\Omega \Bigr\rangle
\nonumber \\
& & \hspace{1.3cm} - \Bigl\langle\Omega \Bigl\vert \chi^0(x^{\p\p})
\chi^0(x^{\p}) \Bigr\vert \Omega \Bigr\rangle \Bigl\langle\Omega
\Bigl\vert \varphi^{0*}(x^{\p\p}) \varphi^0(x^{\p}) \Bigr\vert
\Omega \Bigr\rangle \Biggr\} \Phi^0(x^{\p\p};\vec{k}) + O(\lambda^4)
 .\; \qquad \label{fexp}
\end{eqnarray}

We turn now to the effective field equations of the
Schwinger-Keldysh formalism. The $\comp$-number field corresponding
to $\varphi(x)$ at linearized order is $\Phi(x)$. If the state is
released at $t=0$ then the equation $\Phi(x)$ obeys is,
\begin{equation}
(\partial^2 - m^2) \Phi(x) - \int_0^t \!\! dt^{\p} \!\! \int
d^{D-1}x^{\p} \Bigl\{ M^2_{\scriptscriptstyle ++}(x;x^{\p}) +
M^2_{\scriptscriptstyle +-}(x;x^{\p}) \Bigr\} \Phi(x^{\p}) = 0 \; .
\label{Phieqn}
\end{equation}
The one loop diagram for the self-mass-squared of $\varphi$ is
depicted in Figure \ref{ffig4}.

\begin{figure}
%\centerline{\epsfig{file=ffig4.eps}}
\includegraphics[width=4 in]{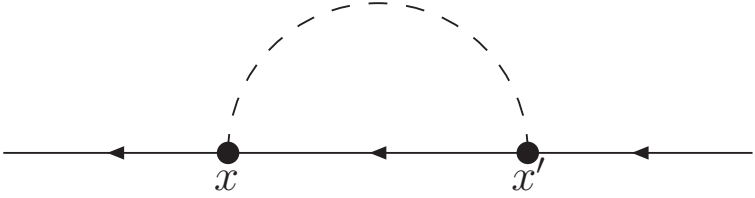}
\caption{Self-mass-squared for $\varphi$ at one loop order. Solid
lines stands for $\varphi$ propagators while dashed lines represent
$\chi$ propagators.} \label{ffig4}
\end{figure}

Because the self-mass-squared has two external lines, there are $2^2
= 4$ polarities in the Schwinger-Keldysh formalism. The two we
require are \cite{DW,FW},
\begin{eqnarray}
-i M^2_{\scriptscriptstyle ++}(x;x^{\p}) & = & (-i \lambda)^2
\Bigl\langle x \Bigl\vert \frac{i}{\partial^2} \Bigr\vert x^{\p}
\Bigr\rangle_{\scriptscriptstyle ++} \Bigl\langle x \Bigl\vert
\frac{i}{\partial^2 \!-\! m^2} \Bigr\vert x^{\p}
\Bigr\rangle_{\scriptscriptstyle ++} + O(\lambda^4) \; , \label{M++} \\
-i M^2_{\scriptscriptstyle +-}(x;x^{\p}) & = & (-i \lambda) (+i
\lambda) \Bigl\langle x \Bigl\vert \frac{i}{\partial^2} \Bigr\vert
x^{\p} \Bigr\rangle_{ \scriptscriptstyle +-} \Bigl\langle x
\Bigl\vert \frac{i}{\partial^2 \!-\! m^2} \Bigr\vert x^{\p}
\Bigr\rangle_{\scriptscriptstyle +-} + O(\lambda^4) \; . \qquad
\label{M+-}
\end{eqnarray}

To recover Equation \ref{fexp} we must express the various
Schwinger-Keldysh propagators in terms of expectation values of the
free fields. The ${\scriptscriptstyle ++}$ polarity gives the usual
Feynman propagator \cite{FW},
\begin{eqnarray}
\lefteqn{\Bigl\langle x \Bigl\vert \frac{i}{\partial^2} \Bigr\vert
x^{\p} \Bigr\rangle_{\scriptscriptstyle ++} = \theta(t \!-\! t^{\p})
\Bigl\langle \Omega \Bigl\vert \chi^0(x) \chi^0(x^{\p}) \Bigr\vert
\Omega \Bigr\rangle \!+\! \theta(t^{\p} \!-\! t) \Bigl\langle \Omega
\Bigl\vert \chi^0(x^{\p}) \chi^0(x) \Bigr\vert \Omega
\Bigr\rangle \; , \qquad } \\
\lefteqn{\Bigl\langle x \Bigl\vert \frac{i}{\partial^2 \!-\! m^2}
\Bigr\vert x^{\p}
\Bigr\rangle_{\scriptscriptstyle ++} } \nonumber \\
& & \hspace{1.3cm} = \theta(t \!-\! t^{\p}) \Bigl\langle \Omega
\Bigl\vert \varphi^0(x) \varphi^{0*}(x^{\p}) \Bigr\vert \Omega
\Bigr\rangle \!+\! \theta(t^{\p} \!-\! t) \Bigl\langle \Omega
\Bigl\vert \varphi^{0*}(x^{\p}) \varphi^0(x) \Bigr\vert \Omega
\Bigr\rangle  . \qquad
\end{eqnarray}
The ${\scriptscriptstyle +-}$ polarity propagators are \cite{FW},
\begin{eqnarray}
\Bigl\langle x \Bigl\vert \frac{i}{\partial^2} \Bigr\vert x^{\p}
\Bigr\rangle_{\scriptscriptstyle +-} & = & \Bigl\langle \Omega
\Bigl\vert \chi^0(x^{\p}) \chi^0(x) \Bigr\vert \Omega
\Bigr\rangle \; , \\
\Bigl\langle x \Bigl\vert \frac{i}{\partial^2 \!-\! m^2} \Bigr\vert
x^{\p} \Bigr\rangle_{\scriptscriptstyle +-} & = & \Bigl\langle
\Omega \Bigl\vert \varphi^{0*}(x^{\p}) \varphi^0(x) \Bigr\vert
\Omega \Bigr\rangle \; . \qquad
\end{eqnarray}
Substituting these relations into Equation \ref{M++} and Equation
\ref{M+-} and making use of the identity $1 = \theta(t \!-\! t^{\p})
\!+\! \theta(t^{\p} \!-\! t)$ gives,
\begin{eqnarray}
\lefteqn{ M^2_{\scriptscriptstyle ++}(x;x^{\p}) +
M^2_{\scriptscriptstyle +-}(x;x^{\p}) = -i \lambda^2 \theta(t \!-\!
t^{\p}) \Biggl\{ \Bigl\langle \Omega \Bigl\vert \chi^0(x)
\chi^0(x^{\p}) \Bigr\vert \Omega
\Bigr\rangle } \nonumber \\
& & \hspace{-.7cm} \times \Bigl\langle \Omega \Bigl\vert
\varphi^0(x) \varphi^{0*}(x^{\p}) \Bigr\vert \Omega \Bigr\rangle
\!-\! \Bigl\langle \Omega \Bigl\vert \chi^0(x^{\p}) \chi^0(x)
\Bigr\vert \Omega \Bigr\rangle \Bigl\langle \Omega \Bigl\vert
\varphi^{0*}(x^{\p}) \varphi^0(x) \Bigr\vert \Omega \Bigr\rangle
\!\Biggr\} \!+\! O(\lambda^4)  . \; \qquad \label{oneloop}
\end{eqnarray}

We now solve Equation \ref{Phieqn} perturbatively. The free plane
wave mode function \ref{Phi^0} is of course a solution at order
$\lambda^0$. With Equation \ref{oneloop} we easily recognize its
perturbative development as,
\begin{eqnarray}
\lefteqn{\Phi(x;\vec{k}) = \Phi^0(x;\vec{k}) -i \lambda^2 \int_0^t
\!\! dt^{\p} \!\! \int d^{D-1}x^{\p} \, \Bigl\langle x \Bigl\vert
\frac1{\partial^2
\!-\! m^2} \Bigr\vert x^{\p} \Bigr\rangle_{\rm ret} } \nonumber \\
& & \times \int_0^{t^{\p}} \!\! dt^{\p\p} \!\! \int d^{D-1}x^{\p\p}
\, \Biggl\{ \Bigl\langle \Omega\Bigl\vert \chi^0(x^{\p})
\chi^0(x^{\p\p}) \Bigr\vert \Omega \Bigr\rangle \Bigl\langle\Omega
\Bigl\vert \varphi^0(x^{\p}) \varphi^{0*}(x^{\p\p}) \Bigr\vert
\Omega \Bigr\rangle
\nonumber \\
& & \hspace{1.5cm} - \Bigl\langle\Omega \Bigl\vert \chi^0(x^{\p\p})
\chi^0(x^{\p}) \Bigr\vert \Omega \Bigr\rangle \Bigl\langle\Omega
\Bigl\vert \varphi^{0*}(x^{\p\p}) \varphi^0(x^{\p}) \Bigr\vert
\Omega \Bigr\rangle \Biggr\} \Phi^0(x^{\p\p};\vec{k}) + O(\lambda^4)
. \; \qquad
\end{eqnarray}
That agrees with Equation \ref{fexp}, so we have established the
desired connection,
\begin{equation}
\Phi(x;\vec{k}) = \Bigl\langle \Omega \Bigl\vert \Bigl[ \varphi(x),
b^{\dagger}(\vec{k}) \Bigr] \Bigr\vert \Omega \Bigr\rangle \; ,
\end{equation}
at one loop order.

\subsection{Gauge Issues}

The preceding discussion has made clear that we are working in a
particular local Lorentz and general coordinate gauge. We are also
doing perturbation theory. The function $\Psi^0_i(x;\vec{k},s)$
describes how a free fermion of wave number $\vec{k}$ and helicity
$s$ propagates through classical de Sitter background in our gauge.
What $\Psi^1_i(x;\vec{k},s)$ gives is the first quantum correction
to this mode function. It is natural to wonder how the effective
field $\Psi_i(x;\vec{k},s)$ changes if a different gauge is used.

The operators of the original, invariant Lagrangian transform as
follows under diffeomorphisms ($x^{\mu} \rightarrow x^{\prime \mu}$)
and local Lorentz rotations ($\Lambda_{ij}$),\footnote{Of course the
spinor and vector representations of the local Lorentz
transformation are related as usual, with same parameters
$\omega_{cd}(x)$ contracted into the appropriate representation
matrices,
\begin{eqnarray}
\Lambda_{ij} \equiv \delta_{ij} - \frac{i}2 \omega_{cd}
J^{cd}_{~~ij} + \ldots \qquad {\rm and} \qquad \Lambda_b^{~c} =
\delta_b^{~c} - \omega_b^{~c} + \ldots \; . \nonumber
\end{eqnarray}}
\begin{eqnarray}
\psi^{\p}_i(x) & = & \Lambda_{ij}\Bigl(x^{\prime -1}(x)\Bigr)
\psi_j\Bigl(
x^{\prime -1}(x)\Bigr) \; , \\
e^{\p}_{\mu b}(x) & = & \frac{\partial x^{\nu}}{\partial x^{\prime
\mu}} \Lambda_b^{~c}\Bigl(x^{\prime -1}(x)\Bigr) e_{\nu c}\Bigl(
x^{\prime -1}(x) \Bigr) \; .
\end{eqnarray}
The invariance of the theory guarantees that the transformation of
any solution is also a solution. Hence the possibility of performing
local transformations precludes the existence of a unique initial
value solution. This is why no Hamiltonian formalism is possible
until the gauge has been fixed sufficiently to eliminate
transformations which leave the initial value surface unaffected.

Different gauges can be reached using field-dependent gauge
transformations \cite{TW0}. This has a relatively simple effect upon
the Heisenberg operator $\psi_i(x)$, but a complicated one on the
linearized effective field $\Psi_i(x;\vec{k},s)$. Because local
Lorentz and diffeomorphism gauge conditions are typically specified
in terms of the gravitational fields, we assume $x^{\prime \mu}$ and
$\Lambda_{ij}$ depend upon the graviton field $h_{\mu\nu}$. Hence so
too does the transformed field,
\begin{equation}
\psi^{\p}_i[h](x) = \Lambda_{ij}[h]\Bigl(x^{\prime -1}[h](x)\Bigr)
\psi_j\Bigl(x^{\prime -1}[h](x)\Bigr) \; .
\end{equation}
In the general case that the gauge changes even on the initial value
surface, the creation and annihilation operators also transform,
\begin{equation}
b^{\p}[h](\vec{k},s) = \frac1{\sqrt{2k}} u^*_i(\vec{k},s) \int
d^{D-1}x \, e^{-i \vec{k} \cdot \vec{x}}
\psi^{\p}_i[h](\eta_i,\vec{x}) \; ,
\end{equation}
where $\eta_i \equiv -1/H$ is the initial conformal time. Hence the
linearized effective field transforms to,
\begin{equation}
\Psi^{\p}_i(x;\vec{k},s) = a^{\frac{D-1}2} \Bigl\langle \Omega
\Bigl\vert \Bigl\{ \psi^{\p}_i[h](x),
b^{\prime\dagger}[h](\vec{k},s) \Bigr\} \Bigr\vert \Omega
\Bigr\rangle \; . \label{SKprime}
\end{equation}
This is quite a complicated relation. Note in particular that the
$h_{\mu\nu}$ dependence of $x^{\prime \mu}[h]$ and $\Lambda_{ij}[h]$
means that $\Psi^{\p}_i(x;\vec{k},s)$ is not simply a Lorentz
transformation of the original function $\Psi_i(x;\vec{k},s)$
evaluated at some transformed point.

\section{ENHANCED FERMION MODE FUNCTION}

We first modify our regularized result for the fermion self energy
by the employing Schwinger-Keldysh formalism to make it causal and
real. We then solve the quantum corrected Dirac equation and find
the fermion mode function at late times. Our result is that it grows
without bound as if there were a time-dependent field strength
renormalization of the free field mode function. If inflation lasts
long enough, perturbation theory must break down. The same result
occurs in the Hartree approximation although the numerical
coefficients differ.
\subsection{Some Key Reductions}

The purpose of this section is to derive three results that are used
repeatedly in reducing the nonlocal contributions to the effective
field equations.  We observe that the nonlocal terms of Equation
\ref{ren} contain $1/\Delta x^2$. We can avoid denominators by
extracting another derivative,
\begin{equation}
\f{1}{\D x^2}=\frac{\del^2}4 \ln(\D x^2) \qquad {\rm and} \qquad
\f{\ln(\D x^2)}{\D x^2} = \frac{\del^2}8 \Bigl[\ln^2(\D x^{2}) - 2
\ln(\D x^2) \Bigr] \; . \label{id2}
\end{equation}
The Schwinger-Keldysh field equations involve the difference of
${\scriptscriptstyle ++}$ and ${\scriptscriptstyle +-}$ terms, for
example, \bea \hspace{-1cm} && \f{\ln(\mu^2\D
x^2_{\scriptscriptstyle ++})}{\D x^2_{ \scriptscriptstyle ++}} -
\f{\ln(\mu^2 \D x^2_{\scriptscriptstyle +-})}{\D
x^2_{\scriptscriptstyle +-}} \nn \\
\hspace{-1.3cm} & & = \f{\del^2}{8} \Biggl\{ \ln^2(\mu^2\D
x^2_{\scriptscriptstyle ++}) - 2 \ln(\mu^2\D x^2_{\scriptscriptstyle
++}) - \ln^2(\mu^2\D x^2_{ \scriptscriptstyle +-})+2\ln(\mu^2\D
x^2_{\scriptscriptstyle +-}) \Biggr\}. \;\; \label{r1} \eea We now
define the coordinate intervals $\D\eta \equiv \e \!-\! \e^{\p}$ and
$\D x\equiv \Vert \vec{x} \!-\! \vec{x}^{\p} \Vert$ in terms of
which the ${\scriptscriptstyle ++}$ and ${\scriptscriptstyle +-}$
intervals are, \be \D x^2_{\scriptscriptstyle ++} = \D x^2 - (\vert
\D\e \vert \!-\! i\d)^2 \,\,\, {\rm and} \,\,\, \D
x^2_{\scriptscriptstyle +-} = \D x^2 - (\D\e \!+\! i\d)^2 \,\, . \ee
When $\e^{\p}>\e$ we have $\D x^2_{\scriptscriptstyle ++} = \D x^2_{
\scriptscriptstyle +-}$ , so the ${\scriptscriptstyle ++}$ and
${\scriptscriptstyle +-}$ terms in Equation \ref{r1} cancel. This
means there is no contribution from the future. When $\e^{\p}<\e$
and $\D x>\D\e$ (past spacelike separation) we can take $\delta =
0$,
\begin{equation}
\ln(\mu^2\D x^2_{\scriptscriptstyle ++}) = \ln[\mu^2(\D x^2 \!-\! \D
\e^2 )] = \ln(\mu^2\D x^2_{\scriptscriptstyle +-}) \qquad (\Delta x
> \Delta \eta > 0) \,\, . \label{ln1}
\end{equation}
So the ${\scriptscriptstyle ++}$ and ${\scriptscriptstyle +-}$ terms
again cancel. Only for $\eta^{\p}<\eta$ and $\D x<\D\eta$ (past
timelike separation) are the two logarithms different,
\begin{equation}
\ln(\mu^2\D x^2_{\scriptscriptstyle +\pm}) = \ln[\mu^2(\D\eta^2
\!-\! \D x^2)] \pm i \pi \qquad (\Delta \eta > \Delta x > 0) \,\, .
\end{equation}
Hence Equation \ref{r1} can be written as,
\begin{eqnarray}
\f{\ln(\mu^2\D x^2_{\scriptscriptstyle ++})}{\D
x^2_{\scriptscriptstyle ++}} - \f{\ln(\mu^2 \D
x^2_{\scriptscriptstyle +-})}{\D x^2_{\scriptscriptstyle +-}} = \f{i
\pi}2 \del^2 \Biggl\{\theta(\D\eta \!-\! \D x)
\Bigl[\ln(\mu^2(\D\e^2 \!-\! \D x^2) \!-\! 1\Bigr] \Biggr\} .\;\;
\label{r2}
\end{eqnarray}
This step shows how the Schwinger-Kledysh formalism achieves
causality.

To integrate expression \ref{r2} up against the plane wave mode
function \ref{freefun} we first pull the $x^{\mu}$ derivatives
outside the integration, then make the change of variables
$\vec{x}^{\p} \!=\! \vec{x} \!+\! \vec{r}$ and perform the angular
integrals,
\begin{eqnarray}
\lefteqn{\int d^4x^{\p} \Biggl\{ \f{\ln(\mu^2\D x^2_{++})}{\D
x^2_{++}} - \f{\ln(\mu^2\D x^2_{+-})}{\D x^2_{+-}} \Biggr\}
\Psi^0_i(\e^{\p},\vec{x}^{\p},\vec{k},s)}
\nn \\
&& = \f{i 2\pi^2}{k} u_i(\vec{k},s) \del^2 e^{i\vec{k}\cdot\vec{x}}
\int^{\e}_{\e_{i}} d\e^{\p} \frac{e^{-ik \eta^{\p}}}{\sqrt{2k}} \!\!
\int^{\D\e}_{0}
\!\!\! dr r \sin(kr) \Bigl\{ \ln[\mu^2(\D\e^2 \!-\! r^2)] \!-\! 1\Bigr\} \nn \\
&& = \f{i2\pi^2}{k \sqrt{2k}} e^{i\vec{k} \cdot \vec{x}}
u_i(\vec{k},s) [-\del^2_{0} \!-\! k^2] \int^{\e}_{\e_{i}} \!\!
d\e^{\p}
e^{-ik \eta^{\p}} \Delta\eta^{2} \nn \\
&& \hspace{4.5cm} \times \int_{0}^{1} \!\!\! dz z \sin(\a z) \Bigl\{
\ln(1 \!-\! z^2) \!+\! 2 \ln(\f{\mu\a}{k}) \!-\! 1 \Bigr\} .
\;\qquad \label{r3}
\end{eqnarray}
Here $\alpha \equiv k \Delta \eta$ and $\e_{i} \equiv -1/H$ is the
initial conformal time, corresponding to physical time $t=0$. The
integral over $z$ is facilitated by the special function,
\begin{eqnarray}
\lefteqn{\xi(\a) \equiv \int_{0}^{1} \!\! dz z \sin(\a z) \ln(1
\!-\! z^2) = \f{2}{\a^2} \sin(\a) - \f{1}{\a^2} \Bigl[\cos(\a) \!+\!
\a \sin(\a) \Bigr] }
\nn \\
& & \hspace{2cm} \times \Bigl[{\rm si}(2\a) \!+\! \f{\pi}{2} \Bigr]
+ \f{1}{\a^2}\Bigl[\sin(\a) \!-\! \a \cos(\a)\Bigr] \Bigl[{\rm
ci}(2\a) \!-\! \gamma \!-\! \ln(\f{\a}{2})\Bigr] .\;\; \qquad
\end{eqnarray}
Here $\gamma$ is the Euler-Mascheroni constant and the sine and
cosine integrals are,
\begin{eqnarray}
{\rm si}(x) & \equiv & -\int_{x}^{\infty} \!\! dt \, \f{\sin(t)}{t}
=
- \f{\pi}{2} + \int_{0}^{x} \!\! dt \, \f{\sin t}{t} \,\, , \\
{\rm ci}(x) & \equiv & -\int_{x}^{\infty} \!\! dt \, \f{\cos t}{t} =
\g + \ln(x) + \int_{0}^{x} \!\! dt \, \Bigl[\f{\cos(t) - 1}{t}\Bigr]
\,\, .
\end{eqnarray}
After substituting the $\xi$ function and performing the elementary
integrals, Equation \ref{r3} becomes,
\begin{eqnarray}
\lefteqn{\int d^4x^{\p} \Biggl\{ \f{\ln(\mu^2\D x^2_{++})}{\D
x^2_{++}} - \f{\ln(\mu^2\D x^2_{+-})}{\D x^2_{+-}} \Biggr\}
\Psi^0_i(\e^{\p},\vec{x}^{\p},\vec{k},s)
= -\f{i 2 \pi^2}{k \sqrt{2k}} e^{i \vec{k} \cdot \vec{x}} u_i(\vec{k},s) } \nn \\
& & \hspace{.5cm} \times (\del^2_{k \eta} \!+\! 1) \hspace{-.1cm}
\int_{\e_i}^{ \e} \hspace{-.3cm} d\e^{\p} e^{-ik\e^{\p}}
\Biggl\{\a^2\xi(\a) \!+\! \Bigl[2\ln(\f{ \mu\a}{k}) \!-\!
1\Bigr]\Bigl[\sin(\a) \!-\! \a \cos(\a) \Bigr] \Biggr\} . \;\;\qquad
\label{r4}
\end{eqnarray}

One can see that the integrand is of order $\a^3 \ln(\alpha)$ for
small $\alpha$, which means we can pass the derivatives through the
integral. After some rearrangements, the first key identity emerges,
\begin{eqnarray}
\lefteqn{ \int d^4x^{\p} \Biggl\{ \frac{\ln(\mu^2 \Delta
x^2_{\scriptscriptstyle ++})}{\Delta x^2_{\scriptscriptstyle ++}} -
\frac{\ln(\mu^2 \Delta x^2_{ \scriptscriptstyle +-})}{\Delta
x^2_{\scriptscriptstyle +-}} \Biggr\}
\Psi^0(\eta^{\p},\vec{x}^{\p};\vec{k},s) } \nonumber \\
& & = -i 4 \pi^2 k^{-1} \Psi^0(\eta,\vec{x};\vec{k},s)
\int_{\eta_i}^{\eta}\!\!\! d\eta^{\p} e^{i k \Delta \eta}
\Biggl\{-\cos(k \Delta \eta) \int_{\scriptscriptstyle 0}^{2 k \Delta
\eta} \!\!\!\!\!\! dt \, \frac{\sin(t)}{t} \nonumber \\
& & \hspace{2.8cm} + \sin(k \Delta \eta) \Biggl[
\int_{\scriptscriptstyle 0}^{2 k \Delta \eta} \!\!\!\!\!\! dt \,
\Bigl(\frac{\cos(t) \!-\! 1}{t}\Bigr) \!+\! 2 \ln(2 \mu \Delta
\eta)\Biggr] \Biggr\} .\;\; \label{key1}
\end{eqnarray}
Note that we have written $e^{-i k \eta^{\p}} = e^{-i k \eta} \times
e^{+i k \Delta \eta}$ and extracted the first phase to reconstruct
the full tree order solution $\Psi^0(\eta,\vec{x};\vec{k},s) =
\frac{e^{-i k \eta}}{ \sqrt{2k}} u_i(\vec{k},s) e^{i \vec{k} \cdot
\vec{x}}$.

The second identity derives from acting a d'Alembertian on Equation
\ref{key1}. The d'Alembertian passes through the tree order solution
to give,
\begin{equation}
\partial^2 \Psi^0(\eta,\vec{x};\vec{k},s) = \Psi^0(\eta,\vec{x};\vec{k},s)
\partial_{\eta} (\partial_{\eta} \!-\! 2 i k) \; .
\end{equation}
Because the integrand goes like $\alpha \ln(\alpha)$ for small
$\alpha$, we can pass the first derivative through the integral to
give,
\begin{eqnarray}
\lefteqn{ \partial^2 \int d^4x^{\p} \Biggl\{ \frac{\ln(\mu^2 \Delta
x^2_{\scriptscriptstyle ++})}{\Delta x^2_{\scriptscriptstyle ++}} -
\frac{\ln(\mu^2 \Delta x^2_{ \scriptscriptstyle +-})}{\Delta
x^2_{\scriptscriptstyle +-}} \Biggr\}
\Psi^0(\eta^{\p},\vec{x}^{\p};\vec{k},s) } \nonumber \\
& & \hspace{1.5cm} = i 4 \pi^2 \Psi^0(\eta,\vec{x};\vec{k},s)
\partial_{\eta} \int_{\e_i}^{\e} \!\! d\e^{\p} \Biggl\{
\int_{\scriptscriptstyle 0}^{2\a} \!\! dt \Bigl(\f{e^{it} \!-\!
1}{t} \Bigr) + 2 \ln(\f{2\mu\a}{k}) \Biggr\}  .\;\; \qquad
\end{eqnarray}
We can pass the final derivative through the first integral but, for
the second, we must carry out the integration. The result is our
second key identity,
\begin{eqnarray}
\lefteqn{ \partial^2 \int d^4x^{\p} \Biggl\{ \frac{\ln(\mu^2 \Delta
x^2_{ \scriptscriptstyle ++})}{\Delta x^2_{\scriptscriptstyle ++}} -
\frac{\ln(\mu^2 \Delta x^2_{ \scriptscriptstyle +-})}{\Delta
x^2_{\scriptscriptstyle +-}}
\Biggr\} \Psi^0(\eta^{\p},\vec{x}^{\p};\vec{k},s) } \nonumber \\
& & \hspace{1.5cm} = i 4 \pi^2 \Psi^0(\eta,\vec{x};\vec{k},s)
\Biggl\{ 2 \ln\Bigl[\frac{2\mu}{H} (1 \!+\! H \eta)\Bigr] \!+\!
\int_{\eta_i}^{\eta} \!\!\! d\eta^{\p} \, \Bigl(\frac{e^{i 2 k
\Delta \eta} \!-\! 1}{\Delta \eta}\Bigr) \Biggr\} . \;\;\qquad
\label{key2}
\end{eqnarray}

The final key identity is derived through the same procedures.
Because they should be familiar by now we simply give the result,
\begin{eqnarray}
\lefteqn{ \int d^4x^{\p} \Biggl\{ \frac1{\Delta
x^2_{\scriptscriptstyle ++}} - \frac1{\Delta x^2_{\scriptscriptstyle
+-}} \Biggr\}
\Psi^0(\eta^{\p},\vec{x}^{\p}; \vec{k},s) } \nonumber \\
& & \hspace{3cm} = - i 4 \pi^2 k^{-1} \Psi^0(\eta,\vec{x};\vec{k},s)
\int_{\e_i}^{\e} \!\! d\e^{\p} \, e^{ik\D\e}\sin(k\D\e) \,\, .
\qquad \label{key3}
\end{eqnarray}

\subsection{Solving the Effective Dirac Equation\label{NL}}
In this section we first evaluate the various nonlocal contributions
using the three identities of the previous section. Then we evaluate
the vastly simpler and, as it turns out, more important, local
contributions. Finally, we solve for
$\Psi^1(\eta,\vec{x};\vec{k},s)$ at late times.

The various nonlocal contributions to Equation \ref{Diraceqn} take
the form,
\begin{eqnarray}
\lefteqn{\int \!\! d^4x^{\p} \sum_{I=1}^5 U^I_{ij}
\Biggr\{\f{\ln(\a_{I}^2\D x^2_{ \scriptscriptstyle ++})}{\D
x^2_{\scriptscriptstyle ++}} - \f{\ln(\a_{I}^2\D
x^2_{\scriptscriptstyle +-})}{\D x^2_{\scriptscriptstyle +-}}
\Biggl\}
\Psi^0_j(\eta^{\p},\vec{x}^{\p};\vec{k},s) } \nonumber \\
& & \hspace{3.8cm} + \int \!\! d^4x^{\p} U^6_{ij} \Biggr\{\f1{\D
x^2_{ \scriptscriptstyle ++}} - \f1{\D x^2_{\scriptscriptstyle +-}}
\Biggl\} \Psi^0_j(\eta^{\p},\vec{x}^{\p};\vec{k},s)\;  . \qquad
\label{analyticform}
\end{eqnarray}
The spinor differential operators $U^I_{ij}$ are listed in Table
\ref{nond}. The constants $\a_{I}$ are $\mu$ for $I = 1,2,3$, and
$\f12 H$ for $I=4,5$.

\begin{table}
\caption{Derivative operators $U^I_{ij}$: Their common prefactor is
$\f{\kappa^2 H^2}{2^8\pi^4}$.}

\vbox{\tabskip=0pt \offinterlineskip
\def\tablerule{\noalign{\hrule}}
\halign to390pt {\strut#& \vrule#\tabskip=1em plus2em& \hfil#&
\vrule#& \hfil#\hfil& \vrule#& \hfil#& \vrule#& \hfil#\hfil&
\vrule#\tabskip=0pt\cr \tablerule
\omit&height4pt&\omit&&\omit&&\omit&&\omit&\cr &&\omit\hidewidth $I$
&&\omit\hidewidth $U^I_{ij}$ \hidewidth&& \omit\hidewidth $I$
\hidewidth&& \omit\hidewidth $U^I_{ij}$ \hidewidth&\cr
\omit&height4pt&\omit&&\omit&&\omit&&\omit&\cr \tablerule
\omit&height2pt&\omit&&\omit&&\omit&&\omit&\cr && 1 && $(H^2
aa^{\p})^{-1}\not{\hspace{-.08cm}\del}\del^4$ && 4 &&
$\hspace{0.5cm}-8\not{\hspace{-.08cm}\bar{\del}}\del^2\hspace{0.5cm}
$ &\cr \omit&height2pt&\omit&&\omit&&\omit&&\omit&\cr \tablerule
\omit&height2pt&\omit&&\omit&&\omit&&\omit&\cr && 2 && $\f{15}{2}
\not{\hspace{-.08cm}\del}\del^2$ && 5 &&
$4\not{\hspace{-.08cm}\del}\nabla^2$ &\cr
\omit&height2pt&\omit&&\omit&&\omit&&\omit&\cr \tablerule
\omit&height2pt&\omit&&\omit&&\omit&&\omit&\cr && 3 &&
$-\not{\hspace{-.08cm}\bar{\del}}\del^2$ && 6 &&
$7\not{\hspace{-.08cm}\del}\nabla^2$ &\cr
\omit&height2pt&\omit&&\omit&&\omit&&\omit&\cr \tablerule}}

\label{nond}

\end{table}

As an example, consider the contribution from $U^2_{ij}$:
\begin{eqnarray}
\lefteqn{\f{15}{2} \f{\kappa^2H^2}{2^8\pi^4}
\not{\hspace{-.08cm}\del} \del^2 \!\! \int \!\! d^4x^{\p}
\Biggr\{\f{\ln(\mu^2\D x^2_{\scriptscriptstyle ++})}{\D
x^2_{\scriptscriptstyle ++}} - \f{\ln(\mu^2\D
x^2_{\scriptscriptstyle +-})}{
\D x^2_{\scriptscriptstyle +-}}\Biggl\} \Psi^0(\eta^{\p},\vec{x}^{\p};\vec{k},s) }\nn\\
&& \hspace{-.5cm} = \f{15}{2} \f{\kappa^2H^2}{2^8\pi^4}
\!\not{\hspace{-.1cm} \del} \!\times\! i 4 \pi^2
\Psi^{0}(\e,\vec{x};\vec{k},s)\Biggl\{\!2\ln\Bigl[ \f{2\mu}{H}(1
\!+\! H\e) \Bigr] \!+\!\! \int_{\e_i}^{\e} \!\! d\e^{\p} \Bigl(\f
{e^{2ik\D\e} \!-\! 1}{\D\e} \Bigr)\! \Biggr\} ,\; \qquad \\
&& \hspace{-.5cm} = \f{\kappa^2H^2}{2^6\pi^2} i H \g^0
\Psi^0(\eta,\vec{x}; \vec{k},s) \times \f{15}{2} \f{1}{ 1 \!+\! H\e}
\Biggl\{e^{2i\f{k}{H}(1+H\e)} \!+\! 1\Biggr\} . \label{non2}
\end{eqnarray}
In these reductions we have used $i \hspace{-.1cm}
\not{\hspace{-.1cm}\del} \Psi^0(\eta,\vec{x};\vec{k},s) \!=\! i
\gamma^0 \Psi^0(\eta,\vec{x};\vec{k},s) \, \partial_{\eta}$ and the
second key identity \ref{key2}. Recall from the Introduction that
reliable predictions are only possible for late times, which
corresponds to $\eta \rightarrow 0^-$. We therefore take this limit,
\begin{eqnarray}
\lefteqn{\f{15}{2} \f{\kappa^2H^2}{2^8\pi^4}
\not{\hspace{-.08cm}\del} \del^2 \!\! \int \!\! d^4x^{\p}
\Biggr\{\f{\ln(\mu^2\D x^2_{\scriptscriptstyle ++})}{\D
x^2_{\scriptscriptstyle ++}} - \f{\ln(\mu^2\D
x^2_{\scriptscriptstyle +-})}{
\D x^2_{\scriptscriptstyle +-}}\Biggl\} \Psi^0(\eta^{\p},\vec{x}^{\p};\vec{k},s)}\nn \\
& & \hspace{3cm} \longrightarrow \f{\kappa^2H^2}{2^6\pi^2} i H \g^0
\Psi^0(\eta,\vec{x};\vec{k},s) \times \f{15}{2}
\Bigl\{\exp(2i\f{k}{H}) + 1 \Bigr\} . \qquad
\end{eqnarray}

The other five nonlocal terms have very similar reductions. Each of
them also goes to $\frac{\kappa^2 H^2}{2^6 \pi^2} \times i H
\gamma^0 \Psi^0(\eta, \vec{x};\vec{k},s)$ times a finite constant at
late times. We summarize the results in Table \ref{noncon} and
relegate the details to an appendix.
\begin{table}
\caption{Nonlocal contributions to $\int d^4x^{\p}
[\Sigma](x;x^{\p}) \Psi^0(\eta^{\p}, \vec{x^{\p}};\vec{k},s)$ at
late times. Multiply each term by $\f{\kappa^2H^2}{2^6 \pi^2} \times
iH \g^0 \Psi^0(\eta,\vec{x};\vec{k},s)$.}

\vbox{\tabskip=0pt \offinterlineskip
\def\tablerule{\noalign{\hrule}}
\halign to390pt {\strut#& \vrule#\tabskip=1em plus2em& \hfil#\hfil&
\vrule#& \hfil#\hfil& \vrule#\tabskip=0pt\cr \tablerule
\omit&height6pt&\omit&&\omit&\cr && ${\rm I}$ && Coefficient\ of\
the\ late\ time\ contribution\ from\ each\ $U^I_{ij}$ &\cr
\omit&height6pt&\omit&&\omit&\cr \tablerule
\omit&height4pt&\omit&&\omit&\cr && 1 && $0$ & \cr
\omit&height4pt&\omit&&\omit&\cr \tablerule
\omit&height4pt&\omit&&\omit&\cr && 2 &&
$\f{15}{2}\Bigl\{\exp(2i\f{k}{H})+1\Bigr\}$ & \cr
\omit&height4pt&\omit&&\omit&\cr \tablerule
\omit&height4pt&\omit&&\omit&\cr && 3 &&
$-i\f{k}{H}\Bigl\{2\ln(\f{2\mu}{H})-\int_{\e_i}^{0}d\e^{\p}
\Bigl(\f{\exp(-2ik\e^{\p})-1}{\e^{\p}}\Bigr)\Bigr\}$ & \cr
\omit&height4pt&\omit&&\omit&\cr \tablerule
\omit&height4pt&\omit&&\omit&\cr && 4 &&
$8i\f{k}{H}\int_{\e_i}^{0}d\e^{\p}
\Bigl(\f{\exp(-2ik\e^{\p})-1}{\e^{\p}}\Bigr)$ & \cr
\omit&height4pt&\omit&&\omit&\cr \tablerule
\omit&height4pt&\omit&&\omit&\cr && 5 &&
$4\f{k^2}{H}\int_{\e_i}^{0}d\e^{\p}e^{-2ik\e^{\p}}
\Bigl\{\int_{0}^{-2k\e^{\p}}dt\Bigl(\f{\exp(-it)-1}{t}\Bigr)
+2\ln(H\e^{\p})\Bigr\}$ & \cr \omit&height4pt&\omit&&\omit&\cr
\tablerule \omit&height4pt&\omit&&\omit&\cr && 6 &&
$-\f{7}{2}i\f{k}{H}\Bigl\{\exp(2i\f{k}{H})-1\Bigr\}$ & \cr
\omit&height4pt&\omit&&\omit&\cr \tablerule}}

\label{noncon}

\end{table}

The next step is to evaluate the local contributions. This is a
straightforward exercise in calculus, using only the properties of
the tree order solution \ref{freefun} and the fact that
$\partial_{\mu} a = H a^2 \delta^0_{\mu}$. The result is,
\begin{eqnarray}
\lefteqn{\f{i\kappa^2 H^2}{2^6\pi^2} \!\!\int \!\! d^4x^{\p}
\Biggl\{ \f{\ln(aa^{\p})}{ H^2 aa^{\p}} \! \not{\hspace{-.1cm}\del}
\del^2 \!+\! \f{15}{2} \ln(aa^{\p}) \! \not{ \hspace{-.1cm}\del}
\!-\! 7 \ln(aa^{\p}) \!\! \not{\hspace{-.08cm} \bar{\del}}
\Biggr\} \d^4(x \!-\! x^{\p}) \Psi^0(\eta^{\p},\vec{x}^{\p};\vec{k},s) } \nonumber \\
& & = \frac{i \kappa^2 H^2}{2^6 \pi^2} \Biggl\{ \frac{\ln(a)}{H^2 a}
\!\not{\hspace{-.1cm}\del} \del^2 \Bigl(\frac1{a}
\Psi^0(\eta,\vec{x};\vec{k},s) \Bigr) + \frac1{H^2 a} \!
\not{\hspace{-.1cm}\del} \del^2 \Bigl(\frac{\ln(a)}{
a} \Psi^0(\eta,\vec{x};\vec{k},s)\Bigr) \nonumber \\
& & \hspace{0.9cm} + \frac{15}2 \Bigl(\ln(a) \!
\not{\hspace{-.1cm}\del} \!+\! \not{\hspace{-.1cm}\del} \ln(a)\Bigr)
\Psi^0(\eta,\vec{x};\vec{k},s) - 14 \ln(a) \!\not{\hspace{-.1cm}
\bar{\del}}
\Psi^0(\eta,\vec{x};\vec{k},s) \Biggr\} , \qquad \label{localgen} \\
& & = \frac{\kappa^2 H^2}{2^6 \pi^2} iH \gamma^0
\Psi^0(\eta,\vec{x};\vec{k},s) \times \Biggl\{ \frac{17}2 a - 14 i
\frac{k}{H} \ln(a) - 2 i \frac{k}{H} \Biggr\} . \label{localtotal}
\end{eqnarray}

The local quantum corrections \ref{localtotal} are evidently much
stronger than their nonlocal counterparts in Table \ref{noncon}!
Whereas the nonlocal terms approach a constant, the leading local
contribution grows like the inflationary scale factor, $a = e^{H
t}$. Even factors of $\ln(a)$ are negligible by comparison. We can
therefore write the late time limit of the one loop field equation
as,
\begin{eqnarray}
i\hspace{-.1cm}\not{\hspace{-.08cm}\del} \kappa^2
\Psi^{1}(\eta,\vec{x}; \vec{k},s) \longrightarrow \f{\kappa^2
H^2}{2^6\pi^2} \f{17}{2} i H a \g^0 \Psi^0(\eta,\vec{x};\vec{k},s)
\,\, .
\end{eqnarray}
The only way for the left hand side to reproduce such rapid growth
is for the time derivative in
$i\hspace{-.1cm}\not{\hspace{-.08cm}\del}$ to act on a factor of
$\ln(a)$,
\begin{equation}
i\gamma^{\mu}\partial_{\mu}\ln(a)
=i\gamma^{\mu}\frac{Ha^2}{a}\delta^0_{\mu}=iHa\gamma^0\;.
\end{equation}
We can therefore write the late time limit of the tree plus one loop
mode functions as,
\begin{equation}
\Psi^{0}(\eta,\vec{x};\vec{k},s) + \kappa^2
\Psi^{1}(\eta,\vec{x};\vec{k},s) \longrightarrow \Biggl\{1 \!+\!
\f{\kappa^2 H^2}{2^6 \pi^2} \f{17}{2} \ln(a) \Biggr\}
\Psi^0(\eta,\vec{x};\vec{k},s) \,\, . \label{modefun}
\end{equation}
All other corrections actually fall off at late times. For example,
those from the $\ln(a)$ terms in Equation \ref{localtotal} go like
$\ln(a)/a$.

There is a clear physical interpretation for the sort of solution we
see in Equation \ref{modefun}. When the corrected field goes to the
free field times a constant, that constant represents a field
strength renormalization. When the quantum corrected field goes to
the free field times a function of time that is independent of the
form of the free field solution, it is natural to think in terms of
a {\it time dependent field strength renormalization},
\begin{equation}
\Psi(\eta,\vec{x};\vec{k},s) \longrightarrow
\frac{\Psi^0(\eta,\vec{x}; \vec{k},s)}{\sqrt{Z_2(t)}} \quad {\rm
where} \quad Z_2(t) = 1 \!-\! \frac{17 \kappa^2 H^2}{2^6 \pi^2}
\ln(a) \!+\! O(\kappa^4) \; .\label{fieldrenor}
\end{equation}
Of course we only have the order $\kappa^2$ correction, so one does
not know if this behavior persists at higher orders. If no higher
loop correction supervenes, the field would switch from positive
norm to negative norm at $\ln(a) = 2^6 \pi^2/17 \kappa^2 H^2$. In
any case, it is safe to conclude that perturbation theory must break
down near this time.

\subsection{Hartree Approximation}

The appearance of a time-dependent field strength renormalization is
such a surprising result that it is worth noting we can understand
it on a simple, qualitative level using the Hartree, or mean-field,
approximation. This technique has proved useful in a wide variety of
problems from atomic physics \cite{DRH} and statistical mechanics
\cite{RKP}, to nuclear physics \cite{HoS} and quantum field theory
\cite{HJS}. Of particular relevance to our work is the insight the
Hartree approximation provides into the generation of photon mass by
inflationary particle production in SQED \cite{DDPT,DPTD,PW2}.

The idea is that we can approximate the dynamics of Fermi fields
interacting with the graviton field operator, $h_{\mu\nu}$, by
taking the expectation value of the Dirac Lagrangian in the graviton
vacuum. To the order we shall need it, the Dirac Lagrangian is
Equation \ref{Dexp},
\begin{eqnarray}
\lefteqn{\mathcal{L}_{\rm Dirac} = \overline{\Psi}
i\hspace{-.1cm}\not{ \hspace{-.08cm} \partial} \Psi + \frac{\kappa}2
\Bigl\{h \overline{\Psi} i\hspace{-.1cm} \not{\hspace{-.08cm}
\partial} \Psi \!-\! h^{\mu\nu} \overline{\Psi} \gamma_{\mu} i
\partial_{\nu} \Psi \!-\! h_{\mu\rho , \sigma}
\overline{\Psi} \gamma^{\mu} J^{\rho\sigma} \Psi\Bigr\} } \nonumber \\
& & + \kappa^2 \Bigl[\frac18 h^2 \!-\! \frac14 h^{\rho\sigma}
h_{\rho\sigma} \Bigr] \overline{\Psi} i\hspace{-.1cm}
\not{\hspace{-.08cm} \partial} \Psi + \kappa^2 \Bigl[-\frac14 h h
^{\mu\nu} \!+\! \frac38 h^{\mu\rho} h_{\rho}^{~\nu}
\Bigr] \overline{\Psi} \gamma_{\mu} i \partial_{\nu} \Psi \nonumber \\
& & \hspace{-.6cm} + \kappa^2 \Bigl[-\frac14 h h_{\mu\rho , \sigma}
\!+\! \frac18 h^{\nu}_{~\rho} h_{\nu \sigma , \mu} \!+\! \frac14
(h^{\nu}_{~\mu} h_{\nu \rho})_{,\sigma} \!+\! \frac14
h^{\nu}_{~\sigma} h_{\mu\rho , \nu}\Bigr] \overline{\Psi}
\gamma^{\mu} J^{\rho\sigma} \Psi + O(\kappa^3) .\;
\qquad\label{DiracL}
\end{eqnarray}
Of course the expectation value of a single graviton field is zero,
but the expectation value of the product of two fields is the
graviton propagator in Equation \ref{gprop0},
\begin{eqnarray}
\lefteqn{\langle\Om\mid T\Bigl[ h_{\mu\nu}(x) h_{\rho\sigma}(x^{\p})
\Bigr] \mid\Om\rangle }
\nonumber \\
& & \hspace{.9cm} =
i\D_{A}(x;x^{\p})\Bigl[\mbox{}_{\mu\nu}T^{A}_{\rho\sigma}\Bigr]+
i\D_{B}(x;x^{\p})\Bigl[\mbox{}_{\mu\nu}T^{B}_{\rho\sigma}\Bigr]+
i\D_{C}(x;x^{\p})\Bigl[\mbox{}_{\mu\nu}T^{C}_{\rho\sigma}\Bigr]  .\;
\qquad \label{gprop}
\end{eqnarray}
Recall the index factors from Equations \ref{A}-\ref{C},
\begin{eqnarray}
\lefteqn{\Bigl[\mbox{}_{\mu\nu}T^{A}_{\rho\sigma}\Bigr] =
2\bar{\e}_{\mu(\rho}\bar{\e}_{\sigma)\nu} - \frac{2}{D \!-\! 3}
\bar{\e}_{\mu\nu}\bar{\e}_{\rho\sigma} \quad , \quad
\Bigl[\mbox{}_{\mu\nu}T^{B}_{\rho\sigma}\Bigr] =
-4\d^0\mbox{}_{(\mu}\bar{\e}_{\nu)}\mbox{}_{(\rho}\d^0_{\sigma)}
\,\, , } \\
& & \Bigl[\mbox{}_{\mu\nu}T^{C}_{\rho\sigma}\Bigr] = \frac2{(D \!-\!
2) (D \!-\! 3)} \Bigl[(D \!-\! 3) \d^{0}_{\mu}\d^{0}_{\nu} +
\bar{\e}_{\mu\nu}\Bigr] \Bigl[(D \!-\! 3)
\d^{0}_{\rho}\d^{0}_{\sigma} + \bar{\e}_{\rho\sigma}\Bigr] \,\, .
\qquad \label{gtensor}
\end{eqnarray}
Recall also that parenthesized indices are symmetrized and that a
bar over a common tensor such as the Kronecker delta function
denotes that its temporal components have been nulled,
\begin{equation}
\overline{\delta}^{\mu}_{\nu} \equiv \delta^{\mu}_{\nu} -
\delta^{\mu}_{\scriptscriptstyle 0} \delta^0_{\nu} \qquad , \qquad
\overline{\eta}_{\mu\nu} \equiv \eta_{\mu\nu} + \delta^0_{\mu}
\delta^0_{\nu} \; .
\end{equation}

The three scalar propagators that appear in Equation \ref{gprop}
have complicated expressions \ref{DeltaA}-\ref{DeltaC} which imply
the following results for their coincidence limits and for the
coincidence limits of their first derivatives,
\begin{eqnarray}
\lim_{x^{\p} \rightarrow x} \, {i\Delta}_A(x;x^{\p}) & = &
\frac{H^{D-2}}{(4\pi)^{ \frac{D}2}}
\frac{\Gamma(D-1)}{\Gamma(\frac{D}2)} \left\{-\pi \cot\Bigl(
\frac{\pi}2 D \Bigr) + 2 \ln(a) \right\} , \\
\lim_{x^{\p} \rightarrow x} \, \partial_{\mu} {i\Delta}_A(x;x^{\p})
& = & \frac{H^{D-2}}{(4\pi)^{\frac{D}2}}
\frac{\Gamma(D-1)}{\Gamma(\frac{D}2)} \times H a \delta^0_{\mu} =
\lim_{x^{\p} \rightarrow x} \, \partial_{\mu}^{\p}
{i\Delta}_A(x;x^{\p}) \; , \qquad \\
\lim_{x^{\p} \rightarrow x} \, {i\Delta}_B(x;x^{\p}) & = &
\frac{H^{D-2}}{(4\pi)^{
\frac{D}2}} \frac{\Gamma(D-1)}{\Gamma(\frac{D}2)}\times -\frac1{D\!-\!2} \; ,\\
\lim_{x^{\p} \rightarrow x} \, \partial_{\mu} {i\Delta}_B(x;x^{\p})
& = & 0 =
\lim_{x^{\p} \rightarrow x} \, \partial_{\mu}^{\p} {i\Delta}_B(x;x^{\p}) \; , \\
\lim_{x^{\p} \rightarrow x} \, {i\Delta}_C(x;x^{\p}) & = &
\frac{H^{D-2}}{(4\pi)^{ \frac{D}2}}
\frac{\Gamma(D-1)}{\Gamma(\frac{D}2)}\times \frac1{(D\!-\!2)
(D\!-\!3)} \; ,\\
\lim_{x^{\p} \rightarrow x} \, \partial_{\mu} {i\Delta}_C(x;x^{\p})
& = & 0 = \lim_{x^{\p} \rightarrow x} \, \partial_{\mu}^{\p}
{i\Delta}_C(x;x^{\p}) \; .
\end{eqnarray}
We are interested in terms which grow at late times. Because the
$B$-type and $C$-type propagators go to constants, and their
derivatives vanish, they can be neglected. The same is true for the
divergent constant in the coincidence limit of the $A$-type
propagator. In the full theory it would be absorbed into a constant
counterterm. Because the remaining, time dependent terms are finite,
we may as well take $D \!=\! 4$. Our Hartree approximation therefore
amounts to making the following replacements in Equation
\ref{DiracL},
\begin{eqnarray}
h_{\mu\nu} h_{\rho\sigma} & \longrightarrow & \frac{H^2}{4 \pi^2}
\ln(a) \Bigl[ \overline{\eta}_{\mu\rho} \overline{\eta}_{ \nu\sigma}
\!+\! \overline{\eta}_{\mu\sigma} \overline{\eta}_{\nu\rho} \!-\! 2
\overline{\eta}_{\mu\nu} \overline{\eta}_{\rho\sigma} \Bigr] \; ,
\label{Atensor1} \\
h_{\mu\nu} h_{\rho\sigma, \alpha} & \longrightarrow & \frac{H^2}{8
\pi^2} H a \delta^0_{\alpha} \, \Bigl[\overline{\eta}_{\mu\rho}
\overline{\eta}_{\nu\sigma} \!+\! \overline{\eta}_{\mu\sigma}
\overline{\eta}_{\nu\rho} \!-\! 2 \overline{\eta}_{\mu\nu}
\overline{\eta}_{\rho\sigma} \Bigr] \; . \label{Atensor2}
\end{eqnarray}

It is now just a matter of contracting Equations
\ref{Atensor1}-\ref{Atensor2} appropriately to produce each of the
quadratic terms in Equation \ref{DiracL}. For example, the first
term gives,
\begin{eqnarray}
\f{\kappa^2}{8} h^2 \overline{\Psi} i \hspace{-.1cm} \not{
\hspace{-.08cm}\del} \Psi &\!\!\!\! \longrightarrow \!\!\!\!&
\f{\kappa^2 H^2}{2^5\pi^2} \ln(a)
\Bigl[\e^{\mu\nu}\e^{\rho\sigma}\Bigr]\Bigl[\bar{\e}_{\mu\rho}
\bar{\e}_{\nu\sigma} +\bar{\e}_{\mu\sigma}\bar{\e}_{\nu\rho}
-2\bar{\e}_{\mu\nu}\bar{\e}_{\rho\sigma}\Bigr]
\overline{\Psi}i\hspace{-.1cm}\not{\hspace{-.08cm}\del}\Psi ,
\qquad \\
& \!\!\!\! = \!\!\!\! & \f{\kappa^2
H^2}{2^5\pi^2}\ln(a)\Bigl[3+3-18\Bigr]
\overline{\Psi}i\hspace{-.1cm}\not{\hspace{-.08cm}\del}\Psi \,\,.
\end{eqnarray}
The second quadratic term gives a proportional result,
\begin{eqnarray}
\f{-\kappa^2}{4}h^{\rho\sigma}h_{\rho\sigma}
\overline{\Psi}i\hspace{-.1cm}\not{\hspace{-.08cm}\del}\Psi
\longrightarrow\f{-\kappa^2 H^2}{2^4\pi^2}\ln(a)\bigl[9+3-6\Bigr]
\overline{\Psi}i\hspace{-.1cm}\not{\hspace{-.08cm}\del}\Psi \,\, .
\end{eqnarray}
The total for these first two terms is $\f{-3\kappa^2 H^2}{4\pi^2}
\ln(a) \overline{\Psi}i\hspace{-.1cm}
\not{\hspace{-.08cm}\del}\Psi$.

The third and fourth of the quadratic terms in Equation \ref{DiracL}
result in only spatial derivatives,
\begin{eqnarray}
\f{-\kappa^2}{4}hh^{\mu\nu}\overline{\Psi}\g_{\mu}i\del_{\nu}\Psi
& \longrightarrow & \f{-\kappa^2
H^2}{2^4\pi^2}\ln(a)\Bigl[1+1-6\Bigr]
\overline{\Psi}i\hspace{-.1cm}\not{\hspace{-.08cm}\bar{\del}}\Psi
\,\, , \\
\f{3}{8}\kappa^2h^{\mu\rho}h_{\,\rho}^{\nu}
\overline{\Psi}\g_{\mu}i\del_{\nu}\Psi & \longrightarrow &
\f{3\kappa^2H^2}{2^5\pi^2}\ln(a)\Bigl[3+1-2\Bigr]
\overline{\Psi}i\hspace{-.1cm}\not{\hspace{-.08cm}\bar{\del}}\Psi
\,\, .
\end{eqnarray}
The total for this type of contribution is $\f{7\kappa^2
H^2}{2^4\pi^2} \ln(a) \overline{\Psi} i \hspace{-.1cm}
\not{\hspace{-.08cm} \bar{\del}} \Psi$.

The final four quadratic terms in Equation \ref{DiracL} involve
derivatives acting on at least one of the two graviton fields,
\begin{eqnarray}
-\f{\kappa^2}{4}hh_{\mu\rho,\si}\overline{\Psi}\g^{\mu}J^{\rho\si}\Psi
& \longrightarrow & \f{-\kappa^2 H^2}{2^5\pi^2}Ha\Bigl[1+1-6\Bigr]
\bar{\e}_{\mu\rho}\overline{\Psi}\g^{\mu}J^{\rho0}\Psi \; , \\
\f{\kappa^2}{8}h^{\nu}_{\,\rho}h_{\nu\si,\mu}
\overline{\Psi}\g^{\mu}J^{\rho\si}\Psi & \longrightarrow &
\f{\kappa^2H^2}{2^6\pi^2}Ha\Bigl[3+1-2\Bigr] \bar{\e}_{\rho\si}
\overline{\Psi}\g^0J^{\rho\si}\Psi \; , \\
\f{\kappa^2}{4}\Bigl(h^{\nu}_{\,\mu}h_{\nu\rho}\Bigr)_{,\sigma}
\overline{\Psi}\g^{\mu}J^{\rho\si}\Psi & \longrightarrow &
\f{\kappa^2H^2}{2^4\pi^2}Ha\Bigl[3+1-2\Bigr]\bar\e_{\mu\rho}
\overline{\Psi}\g^{\mu}J^{\rho0}\Psi \; , \\
\f{\kappa^2}{4}h^{\nu}_{\,\si}h_{\mu\rho,\nu}
\overline{\Psi}\g^{\mu}J^{\rho\si}\Psi & \longrightarrow & 0 \; .
\end{eqnarray}
The second of these contributions vanishes owing to the
anti\-sym\-met\-ry of the Lorentz representation matrices,
$J^{\mu\nu} \equiv \frac{i}4 [\gamma^{\mu},\gamma^{\nu}]$, whereas
$\overline{\eta}_{\mu \rho} \gamma^{\mu} J^{\rho 0} = -\frac{3i}2
\gamma^0$. Hence the sum of all four terms is
$\f{-3\kappa^2H^2}{8\pi^2} H a \overline{\Psi} i \g^0 \Psi$.

Combining these results gives,
\begin{eqnarray}
\lefteqn{\Bigl\langle \mathcal{L}_{\rm Dirac} \Bigr\rangle =
\overline{\Psi} i\!\hspace{-.1cm}\not{\hspace{-.08cm} \partial} \Psi
-\frac{3 \kappa^2 H^2}{4 \pi^2} \ln(a) \overline{\Psi}
i\hspace{-.1cm} \not{\hspace{-.08cm} \partial}
\Psi } \nonumber \\
& & \hspace{3cm} - \frac{3 \kappa^2 H^2}{8 \pi^2} H a
\overline{\Psi} i\gamma^0 \Psi \!+\! \frac{7 \kappa^2 H^2}{16 \pi^2}
\ln(a) \overline{\Psi} i \, \hspace{-.1cm}
\overline{\not{\hspace{-.08cm} \partial}} \Psi \!+\! O(\kappa^4)
, \\
& & \hspace{-.7cm} = \!\overline{\Psi} \Bigl[1 \!-\! \frac{3
\kappa^2 H^2}{8 \pi^2} \ln(a)\Bigr]
i\!\hspace{-.1cm}\not{\hspace{-.08cm} \partial} \Bigl[1 \!-\!
\frac{3 \kappa^2 H^2}{8 \pi^2} \ln(a)\Bigr] \Psi \!+\! \frac{7
\kappa^2 H^2}{16 \pi^2} \ln(a) \overline{\Psi} i \, \hspace{-.15cm}
\overline{\not{ \hspace{-.08cm} \partial}} \Psi \!+\! O(\kappa^4)
.\; \qquad\label{Hartreesum}
\end{eqnarray}
If we express the equations associated with Equation
\ref{Hartreesum} according to the perturbative scheme of Section 2,
the first order equation is,
\begin{eqnarray}
i\hspace{-.1cm}\not{\hspace{-.08cm}\del} \kappa^2
\Psi^{1}(\eta,\vec{x}; \vec{k},s) = \f{\kappa^2 H^2}{2^6\pi^2} i H
\g^0 \Psi^0(\eta,\vec{x};\vec{k},s) \Bigl\{ 24 a - 28 i
\frac{k}{H}\ln(a) \Bigr\} \,\, . \label{Hareqn}
\end{eqnarray}
This is similar, but not identical to, what we got in expression
\ref{localtotal} from the delta function terms of the actual one
loop self-energy in Equation \ref{ren}. In particular, the exact
calculation gives $\frac{17}{2} a \!-\! 14 i \frac{k}H \ln(a)$,
rather than the Hartree approximation of $24 a \!-\! 28 i \frac{k}H
\ln(a)$. Of course the $\ln(a)$ terms make corrections to $\Psi^1$
which fall like $\ln(a)/a$, so the real disagreement between the two
methods is limited to the differing factors of $\frac{17}{2}$ versus
$24$.

We are pleased that such a simple technique comes so close to
recovering the result of a long and tedious calculation. The slight
discrepancy is no doubt due to terms in the Dirac Lagrangian by
Equation \ref{DiracL} which are linear in the graviton field
operator. As described in relation \ref{SKop} of section 2, the
linearized effective field $\Psi_i(x;\vec{k},s)$ represents
$a^{\frac{D-1}2}$ times the expectation value of the anti-commutator
of the Heisenberg field operator $\psi_i(x)$ with the free fermion
creation operator $b(\vec{k},s)$. At the order we are working,
quantum corrections to $\Psi_i(x;\vec{k},s)$ derive from
perturbative corrections to $\psi_i(x)$ which are quadratic in the
free graviton creation and annihilation operators. Some of these
corrections come from a single $h h \overline{\psi} \psi$ vertex,
while others derive from two $h \overline{\psi} \psi$ vertices. The
Hartree approximation recovers corrections of the first kind, but
not the second, which is why we believe it fails to agree with the
exact result. Yukawa theory presents a fully worked-out example
\cite{PW,GP,MW2} in which the {\it entire} lowest-order correction
to the fermion mode functions derives from the product of two such
linear terms, so the Hartree approximation fails completely in that
case.
\section{CONCLUSIONS}%
We have used dimensional regularization to compute quantum
gravitational corrections to the fermion self-energy at one loop
order in a locally de Sitter background. Our regulated result is
Equation \ref{regres}. Although Dirac $+$ Einstein is not
perturbatively renormalizable \cite{DVN} we obtained a finite result
shown by Equation \ref{ren} by absorbing the divergences with BPHZ
counterterms.

For this 1PI function, and at one loop order, only three
counterterms are necessary. None of them represents redefinitions of
terms in the Lagrangian of Dirac $+$ Einstein. Two of the required
counterterms of Equation \ref{invctms} are generally coordinate
invariant fermion bilinears of dimension six. The third counterterm
of Equation \ref{nictm} is the only other fermion bilinear of
dimension six which respects the symmetries shown by Equations
\ref{homot}-\ref{dilx} of our de Sitter noninvariant gauge shown in
Equation \ref{GR} and also obeys the reflection property shown in
Equation \ref{refl} of the self-energy for massless fermions.

Although parts of this computation are quite intricate we have good
confidence that Equation \ref{ren} is correct for three reasons.
First, there is the flat space limit of taking $H$ to zero while
taking the conformal time to be $\eta = -e^{-H t}/H$ with $t$ held
fixed. This checks the leading conformal contributions. Our second
reason for confidence is the fact that all divergences can be
absorbed using just the three counterterms we have inferred in
chapter 2 on the basis of symmetry. This was by no means the case
for individual terms; many separate pieces must be added to
eliminate other divergences. The final check comes from the fact
that the self-energy of a massless fermion must be odd under
interchange of its two coordinates. This was again not true for
separate contributions, yet it emerged when terms were summed.

Although our renormalized result could be changed by altering the
finite parts of the three BPHZ counterterms, this does not affect
its leading behavior in the far infrared. It is simple to be
quantitative about this. Were we to make finite shifts $\Delta
\alpha_i$ in our counterterms Equation \ref{sim} the induced change
in the renormalized self-energy would be,
\begin{equation}
-i \Bigl[\Delta \Sigma^{\rm ren}\Bigr](x;x^{\p}) = -\kappa^2
\Biggl\{ \frac{\Delta \alpha_1}{a a^{\p}} \hspace{-.1cm}
\not{\hspace{-.1cm}
\partial}
\partial^2 + 12 \Delta \alpha_2 H^2 \hspace{-.1cm} \not{\hspace{-.1cm}
\partial} + \Delta \alpha_3 H^2 \; \hspace{-.1cm} \overline{\not{\hspace{-.1cm}
\partial}} \Biggr\} \delta^4(x\!-\!x^{\p}) \; . \label{arb}
\end{equation}
No physical principle seems to fix the $\Delta \alpha_i$ so any
result that derives from their values is arbitrary. This is why BPHZ
renormalization does not yield a complete theory. However, at late
times (which accesses the far infrared because all momenta are
redshifted by $a(t) = e^{Ht}$) the local part of the renormalized
self-energy of Equation \ref{ren} is dominated by the large
logarithms,
\begin{equation}
\frac{\kappa^2}{2^6 \pi^2} \Biggl\{\frac{\ln(a a^{\p})}{a a^{\p}}
\hspace{-.1cm} \not{\hspace{-.1cm} \partial} \partial^2 + \frac{15}2
\ln(a a^{\p}) H^2 \hspace{-.1cm} \not{\hspace{-.1cm} \partial} - 7
\ln(a a^{\p}) H^2 \; \hspace{-.1cm} \overline{\not{\hspace{-.1cm}
\partial}} \Biggr\} \delta^4(x \!-\! x^{\p}) \; . \label{fixed}
\end{equation}
The coefficients of these logarithms are finite and completely fixed
by our calculation. As long as the shifts $\Delta \alpha_i$ are
finite, their impact Equation \ref{arb} must eventually be dwarfed
by the large logarithms in Equation \ref{fixed}.

None of this should seem surprising, although it does with
disturbing regularity. The comparison we have just made is a
standard feature of low energy effective field theory and has a very
old and distinguished pedigree
\cite{BN,SW,FS,HS,CDH,CD,DMC1,DL,JFD1,JFD2,MV,HL,ABS,KK1,KK2}. Loops
of massless particles make finite, nonanalytic contributions which
cannot be changed by local counterterms and which dominate the far
infrared. Further, these effects must occur as well, with precisely
the same numerical values, in whatever fundamental theory ultimately
resolves the ultraviolet problem of quantum gravity. That is why
Feinberg and Sucher got exactly the same long range force from the
exchange of massless neutrinos using Fermi theory \cite{FS,HS} as
one would get from the Standard Model \cite{HS}.

So we can use Equation \ref{ren} reliably in the far infrared. Our
motivation for undertaking this exercise was to search for a
gravitational analogue of what Yukawa-coupling a massless, minimally
coupled scalar does to massless ferm\-i\-ons during inflation
\cite{PW}. Obtaining Equation \ref{ren} completes the first part in
that program. In the second stage we used the Schwinger-Keldysh
formalism to include one loop, quantum gravitational corrections to
the Dirac equation. Because Dirac + Einstein is not perturbatively
renormalizable, it makes no sense to solve this equation generally.
However, the equation should give reliable predictions at late times
when the arbitrary finite parts of the BPHZ counterterms Equation
\ref{genctm} are insignificant compared to the completely determined
factors of $\ln(a a^{\p})$ on terms of Equations
\ref{1stlog}-\ref{3rdlog} which otherwise have the same structure.
In this late time limit we find that the one loop corrected, spatial
plane wave mode functions behave as if the tree order mode functions
were simply subject to a time-dependent field strength
renormalization,
\begin{eqnarray}
Z_2(t) = 1 - \f{17}{4\pi} G H^2 \ln(a) + O(G^2) \,\,\,\,{\rm
where}\,\,\,\, G=16\pi\kappa^2\,\,.
\end{eqnarray}
If unchecked by higher loop effects, this would vanish at $\ln(a)
\simeq 1/G H^2$. What actually happens depends upon higher order
corrections, but there is no way to avoid perturbation theory
breaking down at this time, at least in this gauge.

Might this result be a gauge artifact? One reaches different gauges
by making field dependent transformations of the Heisenberg
operators. We have worked out the change in Equation \ref{SKprime}
this induces in the linearized effective field, but the result is
not simple. Although the linearized effective field obviously
changes when different gauge conditions are employed to compute it,
we believe (but have not proven) that the late time factors of
$\ln(a)$ do not change.

It is important to realize that the 1PI functions of a gauge theory
in a fixed gauge are not devoid of physical content by virtue of
depending upon the gauge. In fact, they encapsulate the physics of a
quantum gauge field every bit as completely as they do when no gauge
symmetry is present. One extracts this physics by forming the 1PI
functions into gauge independent and physically meaningful
combinations. The S-matrix accomplishes this in flat space quantum
field theory. Unfortunately, the S-matrix fails to exist for Dirac +
Einstein in de Sitter background, nor would it correspond to an
experiment that could be performed if it did exist
\cite{EW,AS,TW9}.

If it is conceded that we know what it means to release the universe
in a free state then it would be simple enough --- albeit tedious
--- to construct an analogue of $\psi_i(x)$ which is invariant under
gauge transformations that do not affect the initial value surface.
For example, one might extend to fermions the treatment given for
pure gravity by \cite{TW10}:
\begin{itemize}
\item{Propagate an operator-valued geodesic a fixed invariant time from the
initial value surface;}
\item{Use the spin connection $A_{\mu cd} J^{cd}$ to parallel transport along
the ge\-o\-des\-ic; and}
\item{Evaluate $\psi$ at the operator-valued geodesic, in the Lorentz frame
which is transported from the initial value surface.}
\end{itemize}
This would make an invariant, as would any number of
other constructions \cite{GMH}. For that matter, the gauge-fixed 1PI
functions also correspond to the expectation values of invariant
operators \cite{TW0}. Mere invariance does not guarantee physical
significance, nor does gauge dependence preclude it.

What is needed is for the community to agree upon a relatively
simple set of operators which stand for experiments that could be
performed in de Sitter space. There is every reason to expect a
successful outcome because the last few years have witnessed a
resolution of the similar issue of how to measure quantum
gravitational back-reaction during inflation, driven either by a
scalar inflaton \cite{WU,AW1,AW2,GB} or by a bare cosmological
constant \cite{TW11}. That process has begun for quantum field
theory in de Sitter space \cite{EW,AS,GMH,TW11} and one must wait for it
to run its course. In the meantime, it is safest to stick with what
we have actually shown: perturbation theory must break down for
Dirac + Einstein in the simplest gauge.

This is a surprising result but we were able to understand it
qualitatively using the Hartree approximation in which one takes the
expectation value of the Dirac Lagrangian in the graviton vacuum.
The physical interpretation seems to be that fermions propagate
through an effective geometry whose ever-increasing deviation from
de Sitter is controlled by inflationary graviton production. At one
loop order the fermions are passive spectators to this effective
geometry.

It is significant that inflationary graviton production enhances
fermion mode functions by a factor of $\ln(a)$ at one loop. Similar
factors of $\ln(a)$ have been found in the graviton vacuum energy
\cite{TW4,TW5}. These infrared logarithms also occur in the vacuum
energy and mode functions of a massless, minimally coupled scalar with a quartic
self-interaction \cite{OW1,OW2,KO}, and in the VEV's of almost all
operators in Yukawa theory \cite{MW2} and SQED \cite{PTsW,PTW}. A recent
all orders analysis was not even able to exclude the possibility
that they might contaminate the power spectrum of primordial density
fluctuations \cite{SW2,SW3,KC}!

The fact that infrared logarithms grow without bound raises the
exciting possibility that quantum gravitational corrections may be
significant during inflation, in spite of the minuscule coupling
constant of $G H^2 \ltwid 10^{-12}$. However, the only thing one can
legitimately conclude from the perturbative analysis is that
infrared logarithms cause perturbation theory to break down, in our
gauge, if inflation lasts long enough. Inferring what happens after
this breakdown requires a nonperturbative technique.

Starobinski\u{\i} has long advocated that a simple stochastic
formulation of scalar potential models serves to reproduce the
leading infrared logarithms of these models at each order in
perturbation theory \cite{AAS}. This fact has recently been proved
to all orders \cite{RPW4,TW8}. When the scalar potential is bounded
below it is even possible to sum the series of leading infrared
logarithms and infer their net effect at asymptotically late times
\cite{SY}! Applying Starobinski\u{\i}'s technique to more
complicated theories which also show infrared logarithms is a
formidable problem, but solutions have recently been obtained for
Yukawa theory \cite{MW2} and for SQED \cite{PTW}. It would be very
interesting to see what this technique gives for the infrared
logarithms we have exhibited, to lowest order, in Dirac + Einstein.
And it should be noted that even the potentially complicated,
invariant operators which might be required to settle the gauge
issue would be straightforward to compute in such a stochastic
formulation.
\section{NONLOCAL TERMS FROM TABLE \ref{NL} }

It is important to establish that the nonlocal terms make no
significant contribution at late times, so we will derive the
results summarized in Table \ref{noncon}. For simplicity we denote
as $[U^I]$ the contribution from each operator $U^I_{ij}$ in Table
\ref{nond}. We also abbreviate $\Psi^0(\e,\vec{x};\vec{k},s)$ as
$\Psi^0(x)$.

Owing to the factor of $1/a^{\p}$ in $U^1_{ij}$, and to the larger
number of derivatives, the reduction of $[U^1]$ is atypical,
\begin{eqnarray}
\lefteqn{[{\rm U^1}] \equiv \f{\kappa^2}{2^8\pi^4} \f{1}{a}
\hspace{-.1cm} \not{\hspace{-.08cm}\del}\del^4\int
d^4x^{\p}\f{1}{a^{\p}} \Biggr\{\f{\ln(\mu^2\D
x^2_{\scriptscriptstyle ++})}{\D x^2_{ \scriptscriptstyle
++}}-\f{\ln(\mu^2\D x^2_{\scriptscriptstyle +-})}{
\D x^2_{\scriptscriptstyle +-}}\Biggl\}\Psi^0(x^{\p}) \; , } \\
&&=\f{-i\kappa^2}{2^6\pi^2a}\g^0\Psi^0(x)\Bigl[-2ik\del_{\e}
+\del_{\e}^{2}\Bigr]\Biggl\{\del_{\e}\int_{\e_i}^{\e} \!\!
d\eta^{\p}
(-H\e^{\p}) \Bigl(\f{e^{2ik\D\e}-1}{\Delta \eta}\Bigr) \nonumber \\
&& \hspace{6cm} + \del_{\e}^2 \int_{\e_i}^{\e} \!\! d\eta^{\p}
(-2H\e^{\p})\ln(2\mu\D\e)\Biggr\}  , \\
&& = \f{-i\kappa^2}{2^6\pi^2a}\g^0\Psi^0\Bigl(-2ik+\del_{\e}\Bigr)
\Biggl\{-\f{e^{2ik(\e+\f{1}{H})}-1}{(\e+\f{1}{H})^2}\nn\\
&& \hspace{2cm} +\f{(2ik-H)e^{2ik(\e+\f{1}{H})}}{\e+\f{1}{H}}
-\f{3H^2}{(1+H\e)}+\f{2H^3\e}{(1+H\e)^2}\Biggr\}\; , \\
&& = \f{\kappa^2H^2}{2^6\pi^2}(H\e)iH\g^0\Psi\Biggl\{\f{2\Bigl[
e^{\f{2ik}{H}(1+H\e)}-1-2H\e\Bigr]}{(1+H\e)^3}+\f{(1-\f{2ik}{H})
e^{\f{2ik}{H}(1+H\e)}}{(1+H\e)^2}\nn\\
&& \hspace{6cm} + \f{5-4ik\e-\f{2ik}{H}}{(1+H\e)^2}+
\f{\f{6ik}{H}}{1+H\e}\Biggr\} \,\, .
\end{eqnarray}
This expression actually vanishes in the late time limit of $\eta \!
\rightarrow \! 0^-$.

$[U^2]$ was reduced in Section 4 so we continue with $[U^3]$,
\begin{eqnarray}
\lefteqn{[U^3] \equiv -\f{\kappa^2H^2}{2^8\pi^4} \hspace{-.1cm}
\not{\hspace{-.08cm}\bar{\del}} \del^2 \!\! \int \!
d^4x^{\p}\Biggr\{ \f{\ln(\mu^2\D x^2_{\scriptscriptstyle ++})}{\D
x^2_{\scriptscriptstyle ++}} - \f{\ln(\mu^2\D
x^2_{\scriptscriptstyle +-})}{\D x^2_{
\scriptscriptstyle +-}} \Biggl\} \Psi^0(x^{\p}) \; , } \\
&& = -\f{\kappa^2H^2}{2^8\pi^4}\hspace{-.13cm}
\not{\hspace{-.08cm}\bar{\del}}i4\pi^2\Psi^0(x)\Biggl\{
2\ln\Bigl[\f{2\mu}{H}(1+H\e)\Bigr] + \int_{\e_i}^{\e} d\e^{\p}
\Bigl(\f{e^{2ik\D\e}-1}{\D\e}\Bigr)\Biggr\} , \\
&& = \f{\kappa^2H^2}{2^6\pi^2}k\g^0\Psi^0(x)\Biggl\{2\ln\Bigl[
\f{2\mu}{H}(1+H\e)\Bigr]+\int_{\e_i}^{\e}d\e^{\p}
\Bigl(\f{e^{2ik\D\e}-1}{\D\e}\Bigr)\Biggr\} \; , \\
&& \longrightarrow \f{\kappa^2H^2}{2^6\pi^2}iH\g^0\Psi^0(x) \times
-\f{i k}{H} \Biggl\{2\ln(\f{2\mu}{H})-\int_{\e_i}^{0}d\e^{\p}
\Bigl(\f{e^{-2ik\e^{\p}}-1}{\e^{\p}}\Bigr)\Biggr\} . \qquad
\label{U3}
\end{eqnarray}
$U^4_{ij}$ has the same derivative structure as $U^3_{ij}$, so
$[U^4]$ follows from Equation \ref{U3},
\begin{eqnarray}
\lefteqn{[U^4] \equiv -\f{\kappa^2H^2}{2^8\pi^4} \times 8
\hspace{-.1cm} \not{\hspace{-.08cm}\bar{\del}} \del^2 \!\! \int \!
d^4x^{\p}\Biggr\{ \f{\ln(\frac14 H^2 \D x^2_{\scriptscriptstyle
++})}{\D x^2_{ \scriptscriptstyle ++}} - \f{\ln(\frac14 H^2 \D
x^2_{\scriptscriptstyle
+-})}{\D x^2_{\scriptscriptstyle +-}} \Biggl\} \Psi^0(x^{\p}) \; , } \\
&& \hspace{1.3cm} =
\f{\kappa^2H^2}{2^6\pi^2}8k\g^0\Psi^0(x)\Biggl\{2 \ln\Bigl[
(1+H\e)\Bigr]+\int_{\e_i}^{\e}d\e^{\p}
\Bigl(\f{e^{2ik\D\e}-1}{\D\e}\Bigr)\Biggr\} \; , \qquad \\
&& \hspace{1.3cm} \longrightarrow \f{\kappa^2H^2}{2^6\pi^2} i H \g^0
\Psi^0(x) \times 8 i \f{k}{H} \int_{\e_i}^{0} \!\! d\e^{\p}
\Bigl(\f{e^{-2ik\e^{\p}}-1}{\e^{\p}}\Bigr) \; .
\end{eqnarray}

$U^5_{ij}$ has a Laplacian rather than a d'Alembertian so we use
identity \ref{key1} for $[U^5]$. We also employ the abbreviation
$k\D\e \!=\! \a$,
\begin{eqnarray}
\lefteqn{[U^5] \equiv 4 \f{\kappa^2H^2}{2^8\pi^4} \hspace{-.1cm}
\not{\hspace{-.08cm} \del} \nabla^2 \!\! \int \!
d^4x^{\p}\Biggr\{\f{ \ln(\mu^2\D x^2_{\scriptscriptstyle ++})}{\D
x^2_{\scriptscriptstyle ++}} - \f{\ln(\mu^2\D
x^2_{\scriptscriptstyle +-})}{\D x^2_{
\scriptscriptstyle +-}} \Biggl\} \Psi^0(x^{\p}) \; , } \\
&& \hspace{-.5cm} = 4\f{\kappa^2H^2}{2^8\pi^4}\hspace{-.1cm}
\not{\hspace{-.08cm}\del} \nabla^2 \Bigl( \f{-4i\pi^2}{k} \Bigr)
\Psi^0(x) \int_{\e_i}^{\e}d\e^{\p} e^{i\a} \nn \\
&& \hspace{0cm} \times \Biggl\{\!-\cos(\a) \! \int_{0}^{2\a} \!\!\!
dt \, \f{\sin(t)}{t} + \sin(\a) \Bigl[\int_{0}^{2\a} \!\!\! dt
\Bigl( \f{\cos(t) \!-\! 1}{t}\Bigr) +
2\ln\Bigl(\f{H\a}{k}\Bigr)\Bigr] \!
\Biggr\} , \qquad \\
&& \hspace{-.5cm} = \f{\kappa^2H^2}{2^6\pi^2} i H \g^0 \Psi^0(x)
\times 4\f{k^2}{H} \int_{\e_i}^{\e} \!\!
d\e^{\p}e^{2i\a}\Bigl[\int_{0}^{2\a}
\!\!\! dt\Bigl( \f{e^{-it}-1}{t}\Bigr)+\ln(H\D\e)^2\Bigr] \; , \qquad \\
&& \hspace{-.5cm} \longrightarrow \f{\kappa^2H^2}{2^6\pi^2}iH\g^0
\Psi^0(x) \times 4 \f{k^2}{H} \int_{\e_i}^{0} \!\! d\e^{\p}e^{2i\a}
\Bigl[ \int_{0}^{2\a} \!\!\! dt
\Bigl(\f{e^{-it}-1}{t}\Bigr)+\ln(H\e^{\p})^2\Bigr] .
\end{eqnarray}
$U^6_{ij}$ has the same derivative structure as $U^5_{ij}$ but it
acts on a different integrand. We therefore apply identity
\ref{key3} for $[U^6]$,
\begin{eqnarray}
\lefteqn{[U^6] \equiv 7 \f{\kappa^2H^2}{2^8\pi^4} \hspace{-.1cm}
\not{\hspace{-.08cm}\del} \nabla^2 \!\! \int \! d^4x^{\p} \Biggr\{
\f{1}{\D x^2_{++}}-\f{1}{\D x^2_{+-}}\Biggl\}\Psi^0(x^{\p}) \; , } \\
&& = 7\f{\kappa^2H^2}{2^8\pi^4}\hspace{-.1cm}
\not{\hspace{-.08cm}\del}\nabla^2\times(-i4\pi^2)k^{-1}
\Psi^0(x)\int_{\e_i}^{\e}d\e^{\p}e^{ik\D\e}\sin(k\D\e) \; , \\
&&= \f{\kappa^2H^2}{2^6\pi^2}iH\g^0\Psi^0(x)\times-\f{7}{2}\f{ik}{H}
\Bigl[e^{\f{2ik}{H}(1+H\e)}-1\Bigr] \; , \\
&&\longrightarrow\f{\kappa^2H^2}{2^6\pi^2}iH\g^0\Psi^0(x)
\times-\f{7}{2}\f{ik}{H} \Bigl[e^{\f{2ik}{H}}-1\Bigr] \; .
\end{eqnarray}

\newpage

\centerline{BIOGRAPHICAL SKETCH}

Shun-Pei Miao came from Taiwan. She took her undergraduate degree in
physics at National Taiwan Normal University (NTNU) in 1997. After
that, she got a teaching job in a senior high school. Two years
later she went back to school and in 2001 took a master's degree
under the direction of Professor Pei-Ming Ho at National Taiwan
University (NTU). Her research led to a published paper entitled,
``Noncommutative Differential Calculus for D-Brane in Nonconstant
B-Field Background,'' Phys. Rev, {\bf D64}: 126002, 2001,
hep-th/0105191. After completing her master's degree, she was
fortunate to get a job at National Taiwan Normal University (NTNU)
and she planned to study abroad.

Miao came to the UF in the fall of 2002 and passed the Preliminary
Exam on her first attempt. She passed the graduate core courses
during her first year. In 2003-4 she took particle physics, quantum
field theory and Professor Fry's cosmology special topics course.
She took general relativity in 2004-5. In 2005-6 she took the
standard model. In the fall of 2006 she won a Marie-Curie Fellowship
to attend a trimester at the Institute of Henri Poincar\'{e}
entitled, ``Gravitational Waves, Relativistic Astrophysics and
Cosmology.'' In the spring of 2007 she took Professor Sikivie's dark
matter course.

Miao received her Ph.D. in the summer of 2007. After graduating she
took a postdoc position at the University of Utrecht, but hopes to
find a faculty job in her home country.


\begin{thebibliography}{100}
\pagestyle{plain}

\bibitem{DN} D.N. Spergel et al, Astrophys. J. Suppl 148 (2003)
175, astro-ph/0302209.

\bibitem{cos} J.L. Tonry, et al., Astrophys. J. 594 (2003) 1,
 astro-ph/0305008.

\bibitem{LP} L. Parker, Phys. Rev. 183 (1969) 1057.

\bibitem{LPG} L. P. Grishchuk, Sov. Phys. JETP 40 (1975) 409.

\bibitem{TW12} N. C. Tsamis, R. P. Woodard, Class. Quant. Grav. 20
(2003) 5205, astro-ph/0206010.

\bibitem{MC} V. F. Mukhanov, G. V. Chibisov, JETP Letters 33
(1981) 532.

\bibitem{AAS2} A. A. Starobinski\u{\i}, JETP Letters 30 (1979) 682.

\bibitem{PTW1} T. Prokopec, O. T\"ornkvist, R. P. Woodard, Phys. Rev.
Lett. 89 (2002) 101301, astro-ph/0205331.

\bibitem{PTW2} T. Prokopec, O. T\"ornkvist, R. P. Woodard, Ann. Phys.
303 (2003) 251, gr-qc/0205130.

\bibitem{PW3} T. Prokopec, R. P. Woodard, Annals Phys. {\bf 312} (2004) 1,
gr-qc/\-0310056.

\bibitem{PW} T. Prokopec, R. P. Woodard, JHEP 0310 (2003) 059,
astro-ph/0309593.

\bibitem{GP} B. Garbrecht, T. Prokopec, Phys. Rev. D73 (2006) 064036,
gr-qc/0602011.

\bibitem{TW6} N. C. Tsamis, R. P. Woodard, Phys. Rev. D54 (1996)
2621, hep-ph/9602317.

\bibitem{KW} E. O. Kahya, R. P. Woodard, Phys. Rev. D72 (2005) 104001,
gr-qc/0508015.

\bibitem{DW} L. D. Duffy, R. P. Woodard, Phys. Rev. D72 (2005)
024023, hep-ph/0505156.

\bibitem{MW1} S. P. Miao, R. P. Woodard, Class. Quant. Grav. 23
(2006) 1721, gr-qc/0511140.

\bibitem{BOW} T. Brunier, V. K. Onemli and R. P. Woodard, Class. Quant. Grav.
22 (2005) 59, gr-qc/0408080.

\bibitem{DVN} S. Deser, P. van Nieuwenhuizen, Phys. Rev. D10
(1974) 411.

\bibitem{BP} N. N. Bogoliubov, O. Parasiuk, Acta Math. 97 (1957) 227.

\bibitem{H} K. Hepp, Commun. Math. Phys. 2 (1966) 301.

\bibitem{Z1} W. Zimmermann, Commun. Math. Phys. 11 (1968) 1; 15
(1969) 208.

\bibitem{Z2} W. Zimmermann, in Lectures on Elementary Particles and
Quantum Field Theory, S. Deser, M. Grisaru and H. Pendleton (Eds.),
MIT Press, Cambridge, 1971, Vol. I.

\bibitem{BN} F. Bloch, H. Nordsieck, Phys. Rev. 52 (1937) 54.

\bibitem{SW} S. Weinberg, Phys. Rev. 140 (1965) B516.

\bibitem{FS} G. Feinberg, J. Sucher, Phys. Rev. 166 (1968) 1638.

\bibitem{HS} S. D. H. Hsu, P. Sikivie, Phys. Rev. D49 (1994) 4951,
hep-ph/9211301.

\bibitem{CDH} D. M. Capper, M. J. Duff, L. Halperin, Phys. Rev. D10
(1974) 461.

\bibitem{CD} D. M. Capper, M. J. Duff, Nucl. Phys. B84 (1974) 147.

\bibitem{DMC1} D. M. Capper, Nuovo Cimento A25 (1975) 29.

\bibitem{DL} M. J. Duff, J. T. Liu, Phys. Rev. Lett. 85 (2000) 2052,
hep-th/0003237.

\bibitem{JFD1} J. F. Donoghue, Phys. Rev. Lett. 72 (1994) 2996,
gr-qc/9310024.

\bibitem{JFD2} J. F. Donoghue, Phys. Rev. D50 (1994) 3874,
gr-qc/9405057.

\bibitem{MV} I. J. Muzinich, S. Kokos, Phys. Rev. D52 (1995) 3472,
hep-th/9501083.

\bibitem{HL} H. Hamber, S. Liu, Phys. Lett. B357 (1995) 51,
hep-th/9505182.

\bibitem{ABS} A. Akhundov, S. Belucci, A. Shiekh, Phys. Lett. B395
(1998) 16, gr-qc/9611018.

\bibitem{KK1} I. B. Kriplovich, G. G. Kirilin, J. Exp. Theor. Phys. 98
(2004) 1063, gr-qc/0402018.

\bibitem{KK2} I. B. Kriplovich, G. G. Kirilin, J. Exp. Theor. Phys. 95
(2002) 981, gr-qc/0207118.

\bibitem{BG1} F. A. Berends, R. Gastmans, Phys. Lett. B55 (1975) 311.

\bibitem{BG2} F. A. Berends, R. Gastmans, Ann. Phys. 98 (1976) 225.

\bibitem{RPW1} R. P. Woodard, Phys. Lett. B148 (1984) 440.

\bibitem{TW1} N. C. Tsamis, R. P. Woodard, Commun. Math. Phys.
162 (1994) 217.

\bibitem{AJ} B. Allen, T. Jacobson, Commun. Math. Phys. 103 (1986)
103.

\bibitem{AT} B. Allen, M. Turyn, Nucl. Phys. B292 (1987) 813.

\bibitem{HHT} S. W. Hawking, T. Hertog, N. Turok, Phys. Rev. D62
(2000) 063502, hep-th/0003016.

\bibitem{HK} A. Higuchi, S. S. Kouris, Class. Quant. Grav. 18 (2001)
4317, gr-qc/0107036.

\bibitem{AM} I. Antoniadis, E. Mottola, J. Math. Phys. 32 (1991)
1037.

\bibitem{HW} A. Higuchi, R. H. Weeks, Class. Quant. Grav. 20 (2003)
3005, gr-qc/0212031.

\bibitem{RP} R. Penrose, in Relativity, Groups and Topology, Les Houches
1963, C. DeWitt, B. DeWitt (Eds.), Gordon and Breach, New York,
1964.

\bibitem{BK} J. Bi\v{c}\'ak, P. Krtou\v{s}, Phys. Rev. D64 (2001),
124020, gr-qc/0107078.

\bibitem{RPW2} R. P. Woodard, De Sitter Breaking in Field
Theory, in: J.T. Liu, M. J. Duff, K.S. Stelle, R.P. Woodard (Eds.),
Deser fest: A Celebration of Life and Works of Stanley Deser, World
Scientific, Hackusack, 2006, pp. 339-351.

\bibitem{RJ} R. Jackiw, Phys. Rev. D9 (1974) 1686.

\bibitem{Lam} Y.-M. P. Lam, Phys. Rev. D7 (1973) 2943.

\bibitem{BD} N. D. Birrell, P. C. W. Davies, Quantum Fields in Curved
Space Cambridge University Press, Cambridge, 1982.

\bibitem{AF} B. Allen, A. Folacci, Phys.Rev. D35 (1987) 3771.

\bibitem{BA} B. Allen, Phys. Rev. D32 (1985) 3136.

\bibitem{OW1} V. K. Onemli, R. P. Woodard, Class. Quant. Grav. 19
(2002) 4607, gr-qc/0204065.

\bibitem{OW2} V. K. Onemli, R. P. Woodard, Phys. Rev. D70 (2004)
107301, gr-qc/0406098.

\bibitem{GK} G. Kleppe, Phys. Lett. B317 (1993) 305.

\bibitem{CR} P. Candelas, D. J. Raine, Phys. Rev. D12 (1975) 965.

\bibitem{DC} J. S. Dowker, R. Critchley, Phys. Rev. D13 (1976) 3224.

\bibitem{TW2} N. C. Tsamis, R. P. Woodard, Phys. Lett. B292 (1992)
269.

\bibitem{TW3} N. C. Tsamis, R. P. Woodard,  Annals Phys. 321
(2006) 875, gr-qc/0506056.

\bibitem{LHF} L. H. Ford, Phys. Rev. D31 (1985) 710.

\bibitem{FMVV} F. Finelli, G. Marozzi, G. P. Vacca, G. Venturi, Phys.
Rev. D71 (2005) 023522, gr-qc/0407101.

\bibitem{TW4} N. C. Tsamis, R. P. Woodard, Nucl. Phys. B474 (1996)
235, hep-ph/9602315.

\bibitem{TW5} N. C. Tsamis, R. P. Woodard, Annals Phys.  253 (1997)
1, hep-ph/9602316.

\bibitem{DMC2} D. M. Capper, J. Phys. A A13 (1980) 199.

\bibitem{TW7} N. C. Tsamis, R. P. Woodard, Phys. Lett. B426 (1998)
21, hep-ph/9710466.

\bibitem{BSD1} B. S. DeWitt, Dynamical Theory of Groups and Fields,
Gordon and Breach, New York, 1965.

\bibitem{BSD2} B. S. DeWitt, Phys. Rev. D162 (1967) 1195.

\bibitem{BSD3} B. S. DeWitt, A Gauge Invariant Effective Action,
in: C. J. Isham, R. Penrose , D. W. Sciama (Eds.), Quantum Gravity
2: Proceedings, Clarendon Press, Oxford, 1981, pp. 449.

\bibitem{LFA} L. F. Abbott, Acta Phys. Polon. B13 (1982) 33.

\bibitem{SRC} S. Coleman, Aspects of Symmetry, Cambridge University
Press, Cambridge, 1985.

\bibitem{JS} J. Schwinger, J. Math. Phys. 2 (1961) 407.

\bibitem{KTM} K. T. Mahanthappa, Phys. Rev. 126 (1962) 329.

\bibitem{BM} P. M. Bakshi, K. T. Mahanthappa, J. Math. Phys. 4 (1963)
1; J. Math. Phys. 4 (1963) 12.

\bibitem{LVK} L. V. Keldysh, Sov. Phys. JETP 20 (1965) 1018.

\bibitem{CSHY} K. C. Chou, Z. B. Su, B. L. Hao, L. Yu, Phys. Rept. 118
(1985) 1.

\bibitem{RDJ} R. D. Jordan, Phys. Rev. D33 (1986) 444.

\bibitem{CH} E. Calzetta, B. L. Hu, Phys. Rev. D35 (1987) 495.

\bibitem{FW} L. H. Ford, R. P. Woodard, Class. Quant. Grav. 22 (2005)
1637, gr-qc/0411003.

\bibitem{TW0} N. C. Tsamis, R. P. Woodard, Class. Quant. Grav. 2
(1985) 841.

\bibitem{DRH} D. R. Hartree, The Calculation of Atomic Structures,
Wiley, New York, 1957.

\bibitem{RKP} R. K. Pathria, Statistical Physics, 2nd ed.
Butterworth-Heinemann, Oxford, 1996, pp. 321-358.

\bibitem{HoS} C. J. Horowitz, B. D. Serot, Nucl. Phys. A368 (1981)
503.

\bibitem{HJS} H. J. Schnitzer, Phys. Rev. D10 (1974) 2042.

\bibitem{DDPT} A. C. Davis, K. Dimopoulos, T. Prokopec, O. T\"ornkvist,
Phys. Lett. B501 (2001) 165, astro-ph/0007214.

\bibitem{DPTD} K. Dimopoulos, T. Prokopec, O. T\"ornkvist, A. C. Davis,
Phys. Rev. D65 (2002) 063505, astro-ph/0108093.

\bibitem{PW2} T. Prokopec, R. P. Woodard, Am. J. Phys. 72 (2004) 60,
astro-ph/0303358.

\bibitem{MW2} S. P. Miao, R. P. Woodard, Phys. Rev. D74 (2006) 044019,
gr-qc/0602110.

\bibitem{EW} E. Witten, Quantum gravity in de Sitter space, in: A
Zichichi (Eds.), New Fields and Strings in Subnuclear Physics:
Proceedings of the International School of Subnuclear Physics
,Subnuclear Series 39, World Scientific, Singapore, 2002,
hep-th/01060109.

\bibitem{AS} A. Strominger, JHEP 0110 (2001) 034, hep-th/0106113.

\bibitem{TW9} N. C. Tsamis, R. P. Woodard, Class. Quant. Grav. 11
(1994) 2969.

\bibitem{TW10} N. C. Tsamis, R. P. Woodard, Ann. Phys. 215 (1992) 96.

\bibitem{GMH} S. B. Giddings, D. Marolf, J. B. Hartle, Phys. Rev.
D74 (2006) 064018, hep-th/0512200.

\bibitem{WU} W. Unruh, Cosmological long wavelength perturbations,
astro-ph/\-9802323.

\bibitem{AW1} L. R. Abramo, R. P. Woodard, Phys. Rev. D65 (2002)
043507, astro-ph/0109271.

\bibitem{AW2} L. R. Abramo, R. P. Woodard, Phys. Rev. D65 (2002)
063515, astro-ph/0109272.

\bibitem{GB} G. Geshnizjani, R. H. Br,enberger, Phys. Rev. D66
(2002) 123507, gr-qc/0204074.

\bibitem{TW11} N. C. Tsamis, R. P. Woodard, Class. Quant. Grav. 22
(2005) 4171, gr-qc/0506089.

\bibitem{KO} E. O. Kahya and V. K. Onemli, ``Quantum Stability of
$w <-1$ Phase of Cosmic Acceleration,'' gr-qc/0612026.

\bibitem{PTsW} T. Prokopec, N. C. Tsamis, R. P. Woodard, Class. Quant. Grav.
24 (2007) 201, gr-qc/0607094.

\bibitem{PTW} R. P. Woodard,  Generalizing Starobinski\u{\i}'s
Formalism to Yukawa Theory and to Scalar QED, gr-qc/0608037.

\bibitem{SW2} S. Weinberg, Phys. Rev. D72 (2005) 043514, hep-th/0506236.

\bibitem{SW3} S. Weinberg, Phys. Rev. D74 (2006) 023508, hep-th/0605244.

\bibitem{KC} K. Chaicherdsakul, Quantum Cosmological Correlations
in an inflating Universe: Can fermion and gauge fields loops give a
scale free spectrum? hep-th/0611352.

\bibitem{AAS} A. A. Starobinski\u{\i}, Stochastic de Sitter (inflationary)
stage in the early universe, in: H. J. de Vega, N. Sanchez (Eds.),
Field Theory, Quantum Gravity and Strings, Springer-Verlag, Berlin,
1986 pp. 107-126.

\bibitem{RPW4} R. P. Woodard, Nucl. Phys. Proc. Suppl. 148 (2005) 108,
astro-ph/0502556.

\bibitem{TW8} N. C. Tsamis, R. P. Woodard, Nucl. Phys. B724 (2005)
295, gr-qc/0505115.

\bibitem{SY} A. A. Starobinski\u{\i}, J. Yokoyama, Phys. Rev. D50
(1994) 6357, astro-ph/9407016.

\end{thebibliography}
\end{document}